\PassOptionsToPackage{unicode}{hyperref}
\PassOptionsToPackage{hyphens}{url}
\PassOptionsToPackage{dvipsnames,svgnames,x11names}{xcolor}
\documentclass[
  12pt]{article}

\usepackage{amsmath,amssymb}
\usepackage{iftex}
\usepackage[T1]{fontenc}
\ifPDFTeX
  \usepackage[utf8]{inputenc}
\fi
\usepackage{textcomp} 
\usepackage{lmodern}
\IfFileExists{upquote.sty}{\usepackage{upquote}}{}
\IfFileExists{microtype.sty}{
  \usepackage[]{microtype}
  \UseMicrotypeSet[protrusion]{basicmath} 
}{}
\makeatletter
\@ifundefined{KOMAClassName}{
  \IfFileExists{parskip.sty}{%
    \usepackage{parskip}
  }{
    \setlength{\parindent}{0pt}
    \setlength{\parskip}{6pt plus 2pt minus 1pt}}
}{
  \KOMAoptions{parskip=half}}
\makeatother
\usepackage{xcolor}
\makeatletter
\ifx\paragraph\undefined\else
  \let\oldparagraph\paragraph
  \renewcommand{\paragraph}{
    \@ifstar
      \xxxParagraphStar
      \xxxParagraphNoStar
  }
  \newcommand{\xxxParagraphStar}[1]{\oldparagraph*{#1}\mbox{}}
  \newcommand{\xxxParagraphNoStar}[1]{\oldparagraph{#1}\mbox{}}
\fi
\ifx\subparagraph\undefined\else
  \let\oldsubparagraph\subparagraph
  \renewcommand{\subparagraph}{
    \@ifstar
      \xxxSubParagraphStar
      \xxxSubParagraphNoStar
  }
  \newcommand{\xxxSubParagraphStar}[1]{\oldsubparagraph*{#1}\mbox{}}
  \newcommand{\xxxSubParagraphNoStar}[1]{\oldsubparagraph{#1}\mbox{}}
\fi
\makeatother

\usepackage{longtable,booktabs,array}
\usepackage{calc} 
\usepackage{etoolbox}
\makeatletter
\patchcmd\longtable{\par}{\if@noskipsec\mbox{}\fi\par}{}{}
\makeatother
\IfFileExists{footnotehyper.sty}{\usepackage{footnotehyper}}{\usepackage{footnote}}
\makesavenoteenv{longtable}
\usepackage{graphicx}
\makeatletter
\def\maxwidth{\ifdim\Gin@nat@width>\linewidth\linewidth\else\Gin@nat@width\fi}
\def\maxheight{\ifdim\Gin@nat@height>\textheight\textheight\else\Gin@nat@height\fi}
\makeatother
\setkeys{Gin}{width=\maxwidth,height=\maxheight,keepaspectratio}
\makeatletter
\def\fps@figure{htbp}
\makeatother

\makeatletter
\@ifpackageloaded{caption}{}{\usepackage{caption}}
\AtBeginDocument{%
\ifdefined\contentsname
  \renewcommand*\contentsname{Table of contents}
\else
  \newcommand\contentsname{Table of contents}
\fi
\ifdefined\listfigurename
  \renewcommand*\listfigurename{List of Figures}
\else
  \newcommand\listfigurename{List of Figures}
\fi
\ifdefined\listtablename
  \renewcommand*\listtablename{List of Tables}
\else
  \newcommand\listtablename{List of Tables}
\fi
\ifdefined\figurename
  \renewcommand*\figurename{Figure}
\else
  \newcommand\figurename{Figure}
\fi
\ifdefined\tablename
  \renewcommand*\tablename{Table}
\else
  \newcommand\tablename{Table}
\fi
}
\@ifpackageloaded{float}{}{\usepackage{float}}
\floatstyle{ruled}
\@ifundefined{c@chapter}{\newfloat{codelisting}{h}{lop}}{\newfloat{codelisting}{h}{lop}[chapter]}
\floatname{codelisting}{Listing}

\makeatother
\makeatletter
\@ifpackageloaded{caption}{}{\usepackage{caption}}
\@ifpackageloaded{subcaption}{}{\usepackage{subcaption}}
\makeatother

\usepackage[]{natbib}
\usepackage{bookmark}

\IfFileExists{xurl.sty}{\usepackage{xurl}}{} 
\hypersetup{
  pdftitle={Finite-Smoothing Efficiency and Inference in Single-Index Models},
  pdfauthor={},
  pdfkeywords={dimension reduction; generated residual; semiparametric efficiency; score inversion; variable screening},
  colorlinks=true,
  linkcolor={blue},
  filecolor={Maroon},
  citecolor={Blue},
  urlcolor={Blue},
  pdfcreator={LaTeX via pandoc}}

\usepackage{amsthm,mathtools,bm,tabularx,placeins}
\providecommand{\harvardurl}[1]{}
\renewcommand{\harvardurl}[1]{\url{#1}}
\newtheorem{theorem}{Theorem}
\newtheorem{proposition}[theorem]{Proposition}

\newtheorem{corollary}[theorem]{Corollary}
\theoremstyle{definition}

\theoremstyle{remark}

\newcommand{\E}{\mathbb E}
\newcommand{\Pp}{\mathbb P}

\newcommand{\Var}{\operatorname{Var}}

\newcommand{\supp}{\operatorname{supp}}

\newcommand{\tr}{\operatorname{tr}}

\newcommand{\dP}{d_{\mathrm P}}

\newcolumntype{L}[1]{>{\raggedright\arraybackslash}p{#1}}
\allowdisplaybreaks

\usepackage{xr}
\usepackage{placeins}

\newcommand{\anon}{1}
\if1\anon\hypersetup{pdfauthor={Jiayang Xu; Yusen Jia; Yangzhou Chen; Xin Chen}}\fi

\begin{document}

\def\spacingset#1{\renewcommand{\baselinestretch}%
{#1}\small\normalsize} \spacingset{1}


\if1\anon
{
  \title{\bf Finite-Smoothing Efficiency and Inference in Single-Index Models}
\author{%
  \large
  Jiayang Xu\textsuperscript{1,}\thanks{Jiayang Xu, Yusen Jia and Yangzhou Chen contributed equally to this work.}\quad
  Yusen Jia\textsuperscript{2,}\footnotemark[1]\quad
  Yangzhou Chen\textsuperscript{1,}\footnotemark[1]\quad
  Xin Chen\textsuperscript{1,}\thanks{Corresponding author. Email: \texttt{chenx8@sustech.edu.cn}.}\\[0.8em]
  {\normalsize \textsuperscript{1}Department of Statistics and Data Science,}\\
  {\normalsize Southern University of Science and Technology,}\\
  {\normalsize Shenzhen, Guangdong 518055, China}\\[0.5em]
  {\normalsize \textsuperscript{2}Thrust of Financial Technology,}\\
  {\normalsize The Hong Kong University of Science and Technology (Guangzhou),}\\
  {\normalsize Guangzhou, Guangdong 511453, China}
}
\date{}
\maketitle
} \fi

\if0\anon
{
  \bigskip
  \bigskip
  \bigskip
  \begin{center}
    {\LARGE\bf Finite-Smoothing Efficiency and Inference in Single-Index Models}
\end{center}
  \medskip
} \fi

\bigskip
\begin{abstract}
\spacingset{1.8}

When the error is independent of the predictors, its distribution can carry information about a single-index direction beyond the conditional mean. We develop Single-Index Kernel Density Regression (SI-KDR): at each candidate direction, it refits the link and evaluates a leave-one-out kernel log-likelihood of the residuals. Under Gaussian design, the population score vanishes at the true line at every positive smoothing tuple. At a fixed positive-curvature tuple, the leading covariance in the sample expansion is the conditional-mean efficiency bound times $\Lambda/G$, where $G$ is the information gain from independence and $\Lambda$ is the variance inflation from smoothing. We derive $\Lambda$ by following each observation through link refitting, the dual role of its residual as evaluation point and kernel center, and training standardization. Case-weight covariance estimates guide tuning for point estimation. Tuning separately at each candidate gives an exact conditional $F$ test. When predictors outnumber observations, independent screening that retains the support preserves the covariance comparison and the exact test, and $G/\Lambda$ becomes the squared-noncentrality ratio of a retained-coordinate test relative to the conditional-mean oracle. In four non-Gaussian error distribution designs at $n=600$, covariance-trace tuning reduced direction risk by $56$--$74\%$ relative to reference bandwidths, and score-inversion coverage was $93.8$--$96.6\%$.

\end{abstract}

\noindent%
\vfill

\newpage
\spacingset{1.8} 

\section{Introduction}
Consider the independent-error single-index model
\begin{equation}\label{eq:model}
Y=m_0(\beta_0^TX)+\varepsilon,\qquad\|\beta_0\|=1,\quad\varepsilon\perp X,\quad \E \varepsilon=0.
\end{equation}

Semiparametric least squares \citep{ichimura1993semiparametric} and minimum average variance estimation (MAVE) \citep{xia2002adaptive} estimate the line spanned by $\beta_0$ from the conditional mean. Independence supplies more information when errors are non-Gaussian, for example, skewed or multimodal errors from an unobserved discrete factor. For a given homoscedastic law, the directional information ratio is $G=\Var(\varepsilon)I_\varepsilon$, where $I_\varepsilon$ is the Fisher information for error location (Supplement Section~\ref*{supp-sec:core-efficiency}). Single-Index Kernel Density Regression (SI-KDR) exploits that information by refitting the link at each direction and evaluating the likelihood of the fitted residuals. Oversmoothing the link leaves signal in those residuals, so even a fine density estimate can miss modes (the population illustration in Figure~\ref{fig:pop-loss}).

We ask how much survives at the bandwidths actually used. One observation changes other observations' fitted links and contributes a kernel center to their density estimates. Tracking these effects and training standardization gives the influence of the refitted score. Under Gaussian design, rotational symmetry reduces the covariance comparison with the efficient influence function to a scalar variance inflation factor $\Lambda_k$, determined by the smoothing tuple. Under the sample expansion, SI-KDR's leading covariance is below the conditional-mean bound at the same sample size precisely when $\Lambda_k<G$.

The existing literature on single-index models has not adequately addressed the efficiency cost of estimating both the link function and the error density under finite smoothing. Regular estimators based only on the conditional mean, such as semiparametric least squares and MAVE, cannot have asymptotic covariance below the conditional-mean bound. The first-order theory of estimators that use the response density or the error law, such as the efficient score of \citet{ma2013efficient}, lets all bandwidths vanish and so assigns no cost to the bandwidths used in practice. In the single-index model with Gaussian design, the directions transverse to the index are rotationally symmetric, so curvature and score variance are multiples of the identity and the cost of smoothing is a single number, $\Lambda_k$. We take the bandwidths as given and make four contributions. First, under Gaussian design, the true line is a zero of the population SI-KDR score at every positive tuple and a local maximum at positive curvature, so smoothing changes the variance of SI-KDR but not its target, and $\Lambda_k$ measures this change (Section~\ref{sec:population}). Second, the influence function behind $\Lambda_k$ gives a case-weight covariance estimate whose trace ranks bandwidth pairs (Sections~\ref{sec:sample} and~\ref{sec:variance}). Third, tuning at each candidate index preserves an exact conditional $F$ test whose inversion approaches the efficient ellipsoid (Section~\ref{sec:variance}). Fourth, when predictors outnumber observations, independent screening preserves $G$, $\Lambda_k$ and the exact test within the retained model: SI-KDR coincides with the known-support fit on exact recovery, the retained dimension can grow polynomially in $n$, and $G/\Lambda_k$ is the squared-noncentrality ratio of a coordinate test relative to the conditional-mean oracle (Section~\ref{sec:highdim}). Section~\ref{sec:experiments} begins with a population illustration, simulations with up to 10,000 predictors, and a building-energy benchmark, where SI-KDR's test error was $15.4\%$ below mean MAVE's.

\paragraph*{Related work.}
\citet{delecroix2003efficient} construct a pseudo-maximum likelihood estimator for single-index models by estimating the conditional density of the response given a candidate index, and establish its asymptotic efficiency. \citet{ma2013efficient} derive the efficient score for a central subspace from the conditional response density. \citet{delecroix2006semiparametric} select a single-index direction and link bandwidth by a common objective. Other single-index methods use a kernel quadratic contrast \citep{li2017new} or impose convexity on the link \citep{kuchibhotla2023semiparametric}. Residual-density likelihoods have been used with parametric means \citep{yuan2007semiparametric} and in linear kernel density regression \citep{chen2020kernel}. \citet{linton2008estimation} refit both regression and residual density in a response transformation model. For an index direction, differentiating the full score with respect to the data distribution at fixed bandwidths \citep{ichimura2022influence} gives the variance cost of the fitted criterion.

With many predictors, sure independence screening \citep{fan2008sure} and its distance-correlation version \citep{li2012distance} reduce the dimension before fitting. Sparse sliced inverse regression \citep{lin2018consistency,x1Lin2019,zhang2026sparse}, $\ell_1$-regularized least squares \citep{sbNeykov2016} and penalized single-index estimation \citep{radchenko2015high} recover a sparse index direction from inverse-regression or conditional-mean information, without using the error density. For the retained coordinates, Gaussian residual-space rotation \citep{perry2010rotation} and conditional randomization with nuisance-only fitting \citep{sbLiu2022} provide precedents for calibration. Here the restricted SI-KDR score specifies which function of the reduced data to test against the omitted predictor, and its influence gives the local power.

\section{The estimator}\label{sec:method}
Let $\bar X$ and $\bar Y$ be the training means, and let $\widehat\Sigma$ and $\widehat\sigma_Y$ be the predictor covariance and the response standard deviation, both computed with divisor $n$. Define $X_i^w=\widehat\Sigma^{-1/2}(X_i-\bar X)$ and $Y_i^w=(Y_i-\bar Y)/\widehat\sigma_Y$. For a unit direction $\gamma$ in standardized coordinates, set $T_i=\gamma^TX_i^w$ and
\begin{align}
\widehat m_i&=\frac{\sum_{j\ne i}\phi_g(T_i-T_j)Y_j^w}
 {\sum_{j\ne i}\phi_g(T_i-T_j)+(n-1)\lambda},\quad r_i=Y_i^w-\widehat m_i,\label{eq:ridge}\\
d_i&=(n-1)^{-1}\sum_{j\ne i}\phi_h(r_i-r_j),\qquad
 L_n(\gamma)=n^{-1}\sum_i\log(d_i+\rho).\label{eq:objective}
\end{align}

Here $\phi_h$ is the normalized Gaussian kernel of bandwidth $h$. The ridge $\lambda$ protects the leave-one-out Nadaraya--Watson (NW) denominator and the floor $\rho$ protects the logarithm. Optimization holds $k=(g,h,\lambda,\rho)$ fixed and refits both smoothers at every direction. The raw estimate is $\widehat\beta=\widehat A^T\widehat\gamma/\|\widehat A^T\widehat\gamma\|$ with $\widehat A=\widehat\Sigma^{-1/2}$.

\section{Finite-smoothing target, influence and information}\label{sec:population}

At the population level, this section identifies the SI-KDR target, the influence of one observation, and the smoothing cost $\Lambda_k$ of the Introduction. Fix $p$ and suppose $X\sim N_p(\mu,\Sigma)$ with $\Sigma$ positive definite. The nonconstant link is bounded and $C^6$ with bounded derivatives through order six. The independent error has a fixed finite Gaussian-mixture density with positive component weights and variances. Fitting requires no mixture parameters. In population-standardized coordinates $X=(T,W)$ with $T\sim N(0,1)$, $W\sim N(0,I_{p-1})$ independent of $(T,\varepsilon)$, and $Y=m(T)+\varepsilon$. Write $f$ for the standardized error density, $\psi=f'/f$, and $\beta(\theta)=(1,\theta^T)^T/\sqrt{1+\|\theta\|^2}$.

\subsection{The population target}
We distinguish the conditional-mean profile $\E (Y\mid\beta^TX=t)$, the population smoothed profile
\[
M_{\beta,g,\lambda}(t)=\frac{\E \{\phi_g(t-\beta^TX)Y\}}{\E \phi_g(t-\beta^TX)+\lambda},
\]
and its sample counterpart in \eqref{eq:ridge}. Let $R_\theta=Y-M_{\beta(\theta),g,\lambda}(\beta(\theta)^TX)$ and let $p_\theta$ be its density. The finite-smoothing target is
\begin{equation}\label{eq:target}
Q_k(\theta)=\int p_\theta(r)\log\{(\phi_h*p_\theta)(r)+\rho\}\,dr,
\quad J_k=-\nabla^2Q_k(0),
\end{equation}
where the subscript is the tuning tuple $k=(g,h,\lambda,\rho)$, which may depend on $n$. At truth write $M=M_{\beta(0),g,\lambda}$, $q_g(t)=\phi_{\sqrt{1+g^2}}(t)$, $R=Y-M(T)$, $p_R$ for its density and $d_h=\phi_h*p_R$. The residual coefficient
\[
a(r)=\frac{d'_h(r)}{d_h(r)+\rho}
       +\left[\phi_h'*\frac{p_R}{d_h+\rho}\right](r)
\]
includes the evaluation point and center contributions and satisfies $\E  a(R)=0$.
\begin{proposition}[Population score and curvature]\label{prop:target}
For all positive $g,h,\lambda,\rho$, Gaussian rotation gives
\begin{equation}\label{eq:J}
 \nabla Q_k(0)=0,\qquad J_k=j_k I_{p-1} .
\end{equation}
\end{proposition}
To evaluate $j_k$, put $A_l(t)=\E  m^{(l)}\{(t+g\sqrt{1+g^2}\,\xi)/(1+g^2)\}$ for standard normal $\xi$ and $s_g=q_g/(q_g+\lambda)$. Then $M=s_gA_0$. Supplement~(\ref*{supp-inf:curvature}) gives the scalar formula
\[
 j_k=-\E \{a'(R)M'(T)^2+a(R)N(T)\},\quad
 N=-M''+TM'+s_g(TA_1-A_2)/(1+g^2).
\]
Without whitening, regression smoothing can shift the population score under correlated design (Supplement Section~\ref*{supp-supp:rawbias}).

\subsection{Influence function and information projection}
A prime denotes an independent copy, and we set
\begin{align}
\kappa(t,y)={}&-a\{y-M(t)\}M'(t)\nonumber\\
 &+\E \left[\frac{a(R')\phi_g'(T'-t)\{y-M(T')\}}{q_g(T')+\lambda}\right],\label{eq:kappa}\\
C_k&=\E \{T\kappa(T,Y)\},\qquad
\eta_k=W\{\kappa(T,Y)-C_kT\},\label{eq:eta}\\
B_k&=\Var(\eta_k)=b_kI_{p-1},\qquad
 b_k=\E \{\kappa(T,Y)-C_kT\}^2.\nonumber
\end{align}

\begin{proposition}[Influence function]\label{prop:influence}
With fixed preprocessing, the derivative of the population score with respect to a contaminating observation is $W\kappa(T,Y)$. When the training means, covariance, response scale, and back-transformation of the direction are included, it is \eqref{eq:eta}. In particular, the whitening correction $-C_kTW$ need not vanish at finite smoothing.
\end{proposition}

The expectation in \eqref{eq:kappa} records how an observation changes other fitted links. Conditional centering removes density-law perturbations multiplied by $W$. The off-diagonal training covariance contributes $-C_kTW$. Response centering and scaling contribute zero because the population direction score vanishes for every deterministic response shift and positive scale (Supplement Proposition~\ref*{supp-inf:influence}). If $D_0$ maps standardized tangent coordinates to an orthonormal raw tangent frame, then
\[
 \eta_{k,\rm raw}=D_0^{-T}\eta_k,\quad
 B_{k,\rm raw}=D_0^{-T}B_kD_0^{-1},\quad
 J_{k,\rm raw}=D_0^{-T}J_kD_0^{-1},\quad V_{k,\rm raw}=D_0V_kD_0^T .
\]
The last quantity measures uncertainty in angles on the original predictor scale.

Let $S_{\rm eff}=-Wm'(T)\psi(\varepsilon)$ and $\mathcal I=i_{\rm eff}I_{p-1}$, where $i_{\rm eff}=\E \{m'(T)^2\}I_f$ and $I_f=\E \psi(\varepsilon)^2$. This is the efficient information in the model with independent errors and unrestricted design. For $j_k>0$ put $V_k=J_k^{-1}B_kJ_k^{-T}$.
\begin{proposition}[Information loss]\label{prop:loss}
In the same tangent space,
\begin{align}
\E (\eta_k S_{\rm eff}^T)&=J_k,\nonumber\\
V_k-\mathcal I^{-1}
 &=\E \bigl[(J_k^{-1}\eta_k-\mathcal I^{-1}S_{\rm eff})
 (J_k^{-1}\eta_k-\mathcal I^{-1}S_{\rm eff})^T\bigr]\succeq0.\label{eq:loss}
\end{align}
\end{proposition}
Differentiating the zero-score identity along a rotation gives the cross moment in \eqref{eq:loss}. The excess variance is the squared discrepancy between the SI-KDR and efficient parameter influences. Even density smoothing alone can produce a strict loss. Refitting the link changes both $b_k$ and $j_k$ (Supplement Section~\ref*{supp-inf:theory}).

For a deterministic tuple $k$ with $j_k>0$, define $\Lambda_k=i_{\rm eff}b_k/j_k^2$. On the standardized response scale, $G=\Var(\varepsilon)I_f$ has the same value as on the original scale. At the same law the conditional-mean efficiency bound is $V_{\rm mean}=\Var(\varepsilon)I_{p-1}/\E \{m'(T)^2\}=G\mathcal I^{-1}$. Thus \eqref{eq:loss} gives
\begin{equation}\label{eq:gain-loss}
 V_k=\Lambda_k\mathcal I^{-1}
     =\frac{\Lambda_k}{G}\,V_{\rm mean},\qquad \Lambda_k\ge1 .
\end{equation}
Under the sample expansion of Theorem~\ref{thm:expansion}, the corresponding covariance coefficients are divided by the common estimation sample size. The strict gain in every nonzero tangent contrast is therefore equivalent to $\Lambda_k<G$.

\section{Sample theory at finite smoothing}\label{sec:sample}
This section shows that the estimator follows the population influence of Section~\ref{sec:population} at the bandwidths used, so that $\Lambda_k$ governs its covariance. For $F(r)=n^{-1}\sum_i\log(d_i(r)+\rho)$, the residual coefficient is
\begin{equation}\label{eq:coef}
a_i=n\partial_{r_i}F=\frac1{n-1}\sum_{j\ne i}\phi_h'(r_i-r_j)
\left\{(d_i+\rho)^{-1}+(d_j+\rho)^{-1}\right\}.
\end{equation}

The second reciprocal is the contribution from residual $i$ as a kernel center. Oddness of $\phi'_h$ gives $\sum_i a_i=0$. Supplement Section~\ref*{supp-supp:operator} bounds the change in $a$ and in the third derivative of $F$ when the fitted residuals replace those at the population-smoothed profile, which transfers the score and curvature to the fitted links.

\subsection{Transfer of the score and the Hessian}\label{sec:transfer}
At a fixed input coordinate system write $T_i=\beta_0^TX_i$, let $U$ be an orthonormal tangent frame, and write $Q_{ij}$ for the NW weight with its kernel numerator replaced by $\phi'_g$. Set $A_i=\sum_j Q_{ij}(Y_j-\widehat m_i)$. Exchanging the order of differentiation and summation gives the complete coefficients
\begin{equation}\label{eq:score-coef}
 \nabla L_n(0)=n^{-1}\sum_i c_iU^TX_i,\qquad
 c_i=-a_i A_i+\sum_j a_jQ_{ji}(Y_i-\widehat m_j).
\end{equation}
The incoming sum records how case $i$ changes other fitted links. For correlated inputs, $\mu_i=\E (U^TX_i\mid T_i)$ may be nonzero. Its contribution has the exact form
\[
 \sum_i\mu_i c_i=-\sum_i a_i\sum_j Q_{ij}(\mu_i-\mu_j)(Y_j-\widehat m_i).
\]
The remaining tangent residuals $U^TX_i-\mu_i$ are conditionally centered. Supplement Sections~\ref*{supp-sec:score} and~\ref*{supp-inf:theory} control their contribution by conditioning on the indices and bounding a zero-diagonal quadratic form. Equation~\eqref{eq:cs-gradient} gives the score after training standardization.

\subsection{Rates at finite smoothing scales}
\label{sec:rates}

Use $k$ to index tuning tuples and write $O_i=(X_i,Y_i)$. Centering the fitted profile at $M_{\theta,k}$ puts its smoothing bias into $J_k$ and $\eta_k$. The remaining stochastic jet errors have scales $\sigma_j=(ng^{2j+1})^{-1/2}$, $j=0,1,2$, up to logarithms. Let $\ell_n=1+\log n+\sum_{b\in k}\log_+(1/b)$. Assume $n^{-A}\le g,h,\lambda,\rho\le1$ eventually for finite $A$, and $ng\lambda/\ell_n^{12}\to\infty$. For $t_H=n^{-\nu_H}$ and $r_n=c_rn^{-\nu_r}$ with $\nu_H>0$ and $0<\nu_r<1/2$, put
\begin{align}
 D_S&=\ell_n^{32}n^{-1/2}g^{-3/2}h^{-2},\label{eq:rate-s}\\
 D_H&=\ell_n^{32}\{\sigma_0/h^3+\sigma_1/h^2+\sigma_2/h
 +(nh^5t_H)^{-1/2}+(nh^3t_H)^{-1}+t_H/h^2+n^{-1/2}+r_n\}.
 \label{eq:rate-h}
\end{align}

Write $H_{n,k}=\nabla^2 L_{n,k}$. The floor enters the rates only through $\ell_n$.

\begin{theorem}[Sample expansion at finite smoothing]\label{thm:expansion}
Under the Gaussian-design, bounded $C^6$ link and fixed Gaussian-mixture conditions of Section~\ref{sec:population}, if $D_S+D_H\to0$, then simultaneously over a fixed finite grid, including all same-sample preprocessing,
\begin{align}
 \sqrt n\nabla L_{n,k}(0)&=n^{-1/2}\sum_i\eta_k(O_i)+O_P(D_S),\label{eq:score}\\
 \sup_{\|\theta\|\le r_n}\|-H_{n,k}(\theta)-J_k\|&=O_P(D_H).
 \label{eq:Hessian}
\end{align}
If $\inf_n\lambda_{\min}(J_k)>0$, the finite Armijo algorithm of Section~\ref{sec:tuning}, started from the Fourier initial estimator and run with score tolerance $\tau_n=o(n^{-1/2})$, gives
\begin{equation}
 \sqrt n\widehat\theta_k=J_k^{-1}n^{-1/2}\sum_i\eta_k(O_i)
       +O_P(D_S+D_H+\sqrt n\tau_n).\label{eq:output}
\end{equation}
For fixed positive $k$ this yields a normal limit with covariance $V_k$. For every path on which all four parameters tend to zero, whatever their relative rates, the population curvature tends to $\mathcal I$ and the influences tend in $L^2$ to $S_{\rm eff}$.
\end{theorem}

Supplement Section~\ref*{supp-dir:ranges} gives sufficient polynomial rates for the four parameters. On the numerical grid, $h\asymp n^{-1/15}$, $g\asymp n^{-1/6}$, $\lambda\asymp n^{-3/4}$, $\rho=1/n$ and $t_H=n^{-1/5}$ give $D_S=O(\ell_n^{32}n^{-7/60})$ and $D_H=O\{\ell_n^{32}(n^{-1/60}+r_n)\}$. All tuples on a shrinking grid share the efficient limit. Supplement Section~\ref*{supp-dir:ranges} also gives an example with a sine link and Gaussian error in which $V_k-\mathcal I^{-1}$ is of exact order $h^8$ while the expansion and covariance-estimation errors are $o_P(h^8)$, so the finite-smoothing loss is the leading discrepancy from the efficient variance.

\section{Variance estimation and confidence sets}\label{sec:variance}

Point estimation selects bandwidths by covariance trace. Testing each candidate requires a separate selection rule that preserves its conditional null law.

\subsection{A case-weight covariance estimator}\label{sec:caseweight}

Give each observation weight $\bar w_i=w_i/\sum_jw_j$, recomputing training means, scales, both smoothers and the outer average. For $y_i(w)$ the weighted-standardized response and $r_i(w)=y_i(w)-\widehat m_i(w)$, the weighted smoothers are
\[
 \widehat m_i(w)=
 \frac{\sum_{j\ne i}\bar w_j\phi_g(t_i-t_j)y_j(w)}
 {\sum_{j\ne i}\bar w_j\phi_g(t_i-t_j)+(1-\bar w_i)\lambda},
 \qquad d_i(w)=\frac{\sum_{j\ne i}\bar w_j\phi_h(r_i-r_j)}{1-\bar w_i}.
\] In raw coordinates the weighted index is exactly
\begin{equation}\label{eq:weighted-index}
t_i(\beta,w)=\frac{\beta^T(X_i-\bar X_w)}{\sqrt{\beta^T\widehat\Sigma_w\beta}},
\end{equation}
so the dependence on the estimated covariance can be differentiated without eigenvector derivatives.

At the estimated direction choose an orthonormal frame $F$ and the chart $\beta(v)=(\widehat\beta+Fv)/\sqrt{1+\|v\|^2}$, and define
\begin{align}
\widehat\eta_i&=n\left.\frac{\partial}{\partial w_i}\nabla_v L_w\{\beta(v)\}\right|_{w=\boldsymbol1,v=0},\nonumber\\
\widehat J&=-\nabla_v^2 L_{\boldsymbol1}\{\beta(v)\}|_0,\quad
\widehat B=\frac1n\sum_i(\widehat\eta_i-\bar\eta)(\widehat\eta_i-\bar\eta)^T,\label{eq:IJ}\\
\widehat V&=\widehat J^{-1}\widehat B\widehat J^{-T}.
\nonumber
\end{align}

Here $\widehat V/n$ estimates the leading estimator covariance. The derivative holds $k$ fixed and includes the shared fits in every row of the score.

\begin{theorem}[Case-weight covariance estimation]\label{thm:variance}
Under the conditions of Theorem~\ref{thm:expansion}, suppose that $nht_D/\ell_n^{32}\to\infty$ for an auxiliary level $t_D$ and that $q_D\to0$, where $q_D$ is the explicit bound for the shared fits about the population smoothed profile given in Supplement Section~\ref*{supp-inf:theory}. Then, simultaneously over the grid and at the computed estimates, in matching raw tangent frames and with the raw matrices of Section~\ref{sec:population},
\[
\widehat B-B_{k,\rm raw}=O_P(q_D+q_D^2+n^{-1/2}),\quad
\widehat J-J_{k,\rm raw}=O_P(D_H),\quad \widehat V-V_{k,\rm raw}=o_P(1),
\]
where the last statement is for tuples with $\inf_n\lambda_{\min}(J_k)>0$. The rates of the tuning grid satisfy these conditions with $t_D=n^{-2/3}$, and the covariance error is then of order $\ell_n^{32}(n^{-1/60}+r_n)$. At fixed positive scales, all three errors are $O_P(n^{-1/2})$.
\end{theorem}
Under positive smoothing, a unique population minimizer of $\operatorname{tr}V_k$ gives consistent tuning and a normal limit; with tied minimizers, the joint score and trace fluctuations govern selection when they resolve the tie (Supplement Section~\ref*{supp-up:selection}).
\subsection{Tuning and computation}\label{sec:tuning}
The initial direction uses the equally weighted Fourier matrices of \citet{zhu2006fourier}. A preliminary fit selects $g$ by protected leave-one-out squared error and $h$ by residual log score at the start, then optimizes with that pair fixed. At this preliminary fit, choose the positive-curvature tuple minimizing $\tr\widehat V$ among the twenty candidates, breaking ties in grid order. If none has positive curvature, use the two leave-one-out criteria. The final projected Armijo ascent restarts at the initial direction with the chosen tuple fixed. Its chart radius is $\min\{1.5,2n^{-1/20}\}$ and its score tolerance is $n^{-1}$. Supplement Section~\ref*{supp-sec:algorithm} gives the grids, iteration limits and interior, score and positive-definiteness checks. A failed fit or diagnostic makes the confidence set empty.

\subsection{Confidence sets by score inversion}\label{sec:cs-inference}

At finite bandwidths a Wald region may undercover when curvature at the fitted maximum exceeds curvature at truth. Instead we invert a candidate-wise test using the Gaussian residual-space argument of \citet{perry2010rotation}.

At a candidate unit direction $\beta$ in the original coordinates, choose a deterministic orthonormal tangent frame $F_\beta$. Write $x_\beta=X\beta$, $W=XF_\beta$, $s^2=P_n(x_\beta-P_nx_\beta)^2$, $t=(x_\beta-P_nx_\beta)/s$ and
\[
\Pi=I_n-\boldsymbol1\boldsymbol1^T/n-tt^T/n,\qquad Z=\Pi W.
\]
Let $c$ be the coefficient vector in \eqref{eq:score-coef}, evaluated at $t$ with the training-standardized response $y$. Differentiating the standardized index gives the score in the original chart,
\begin{equation}\label{eq:cs-gradient}
S(\beta)=\frac{Z^T\Pi c}{ns}.
\end{equation}

The projection off $t$ accounts for whitening. At each candidate select $g$ by protected leave-one-out squared error, then $h$ by protected leave-one-out log score, breaking ties in grid order. These choices depend only on $(t,Y)$ and are held fixed in \eqref{eq:cs-gradient}. The point estimator uses the tangent design through the covariance trace.

Put $\bar u=\Pi c/\|\Pi c\|$, $z=Z^T\bar u$ and $A=Z^TZ-zz^T$.
\begin{theorem}[Exact conditional null distribution]\label{thm:cs-pivot}
Let $p\ge2$ and $n>p+1$, and suppose the observations are iid with Gaussian predictors of positive definite covariance. Under $Y\perp XF_\beta\mid X\beta$, conditionally on $(x_\beta,Y)$ and when both sample scales are positive and $\Pi c\ne0$,
\begin{equation}\label{eq:cs-F}
F_{\rm CS}(\beta)=\frac{n-p-1}{p-1}z^TA^{-1}z
       \sim F_{p-1,n-p-1}.
\end{equation}
The result holds for the candidate-wise tuning rule just defined and does not require accurate fits at every observation.
\end{theorem}

Conditionally, $z\sim N_{p-1}(0,\Omega)$ and $A\sim\operatorname{Wishart}_{p-1}(\Omega,n-3)$ independently, where $\Omega=\Var(XF_\beta\mid X\beta)$. Hence the calibration holds for any single-index conditional response law under Gaussian design, including heteroscedastic laws. A zero $\Pi c$ passes the test.

Let $\mathcal D_n$ be the cap around the initial estimator specified in Section~\ref{sec:tuning}, mapped to the original coordinates and intersected with
\[
\{[\beta]:\operatorname{angle}([\beta],[\widehat\beta])
                  \le \log(n+e)/\sqrt n\};
\]
$\mathcal D_n$ is empty when the point estimate fails the conditions of Section~\ref{sec:tuning}. The confidence set is the whole, possibly disconnected, set
\begin{equation}\label{eq:cs-region}
\mathcal C_{\rm CS}=\{[\beta]\in\mathcal D_n:
          F_{\rm CS}(\beta)\le F_{p-1,n-p-1;.95}\}.
\end{equation}

Its coverage is at least $.95-\Pp\{[\beta_0]\notin\mathcal D_n\}$. There is also a conditional check on the score variance. Set
\[
 \widehat K(\beta)=\frac{P_n\{(\Pi c)^2\}}{s^2}\,\frac{Z^TZ}{n-2}.
\]
At truth, $\E (\widehat K\mid x_\beta,Y)=n\Var(S\mid x_\beta,Y)$. When both use the same tuple at the estimate, $\widehat K-\widehat B=o_P(1)$ under the covariance rates of Theorem~\ref{thm:variance} (Supplement Section~\ref*{supp-sec:conditional-score-proofs}).

\begin{theorem}[Asymptotic shape of the confidence set]\label{thm:cs-efficient}
For the shrinking tuning grid that follows Theorem~\ref{thm:expansion} (Supplement~(\ref*{supp-inf:nonempty})) and in exact arithmetic, under the conditions of Theorems~\ref{thm:expansion} and~\ref{thm:variance}, the domain contains $[\beta_0]$ with probability tending to one. Every accepted component of \eqref{eq:cs-region} lies within $O_P(n^{-1/2})$ of the true line, and the set, centered at the point estimate and rescaled, converges to $\{v:v^T\mathcal I_{\rm raw}v\le\chi^2_{p-1,.95}\}$. Hence its coverage tends to $.95$ and its projective diameter is $O_P(n^{-1/2})$.
\end{theorem}
All shrinking-grid choices share the efficient score. At fixed positive smoothing, a separated inference rule selects a limit tuple whose influence determines the set's center and shape. Supplement Sections~\ref*{supp-up:selection} and~\ref*{supp-sec:conditional-score-proofs} give the proofs and treat ties.

\section{Information and inference after screening}
\label{sec:highdim}
When predictors outnumber observations, SI-KDR is fitted after screening, and the question of Section~\ref{sec:population} becomes whether the gain $G$ and the cost $\Lambda$ survive. If the screen uses an independent sample and retains the support, the retained model is again an independent-error single-index model with Gaussian design, so the comparison $\Lambda<G$, the case-weight covariance and the exact test carry over to the estimation sample; screening changes only the tangent geometry and the number of fitted coordinates.

Let $D_s,D_e$ be independent samples of sizes $n_s,n=n_e$, with $N=n_s+n_e$ and $n_s/N\to\pi\in(0,1)$. A screen using only $D_s$ and an independent seed returns $S\subseteq\{1,\ldots,P\}$, $q=|S|$, and SI-KDR is fitted to $D_e$ in $S$ with all preprocessing recomputed. Suppose $X\sim N_P(0,\Sigma_n)$ with $cI\preceq\Sigma_n\preceq CI$ and the link and error law satisfy Section~\ref{sec:population}. Let $E$ denote retention of
$S_0=\supp(\beta_0)$ in a model of admissible size $2\le q<n-1$ with
the uniform stage-two conditions below, and let $\Pp(E^c)\le\delta_{\rm scr}$.

\paragraph*{Dimension and smoothing.}
Write $v=\beta_0^T\Sigma_n\beta_0$. On the power grid of Supplement Section~\ref*{supp-sec:algorithm}, regression and density bandwidths have orders $n^{-1/6}$ and $n^{-1/15}$, $\lambda\asymp n^{-3/4}$ and $\rho=1/n$. With $r$ the Fourier order of Supplement Lemma~\ref*{supp-hdg:fourier}, assume $q^{r+1/2}=o(n^{9/20})$.

\begin{theorem}[Screened SI-KDR on the implemented smoothing grid]\label{thm:hd-estimation}
Condition on a screen containing $S_0$. Suppose $q=O(n^a)$, $0\le a<1/18$. Put $T=\beta_{0,S}^TX_S/\sqrt v$, let $W$ be the $q-1$ transverse coordinates of the population-whitened predictor, and write $\zeta_k=\kappa_k(T,Y^w)-T\E(T\kappa_k)$. Let $D_0$ map standardized to raw orthonormal tangents at the truth, as in Section~\ref{sec:population}. The Fourier initial estimator of Supplement Lemma~\ref*{supp-hdg:fourier} and finite projected ascent give $\dP(\widehat\beta_{S,k},\beta_{0,S})=O_P(\sqrt{q/n})$, uniformly over the preliminary and final tuples. In the raw chart,
\begin{equation}\label{eq:hd-expansion}
 \sqrt n\,\widehat\theta_{S,k}=\frac{D_0}{j(v,k)\sqrt n}\sum_{i\in D_e}W_i\zeta_{k,i}
 +\mathcal R_{n,k}.
\end{equation}
Here $j$ is the standardized population curvature. The vector remainder satisfies $\|\mathcal R_{n,k}\|=o_P(\sqrt q)$, and $\|\mathcal R_{n,k}\|=o_P(1)$ for $a<1/24$. With $b(v,k)=\E\zeta_k^2$, the full case-weight estimate at the preliminary or final fit is consistent in operator norm for $V_{S,k}=D_0D_0^T b(v,k)/j(v,k)^2$ in matching frames. The raw trace rule attains the minimum population trace per tangent coordinate up to $o_P(1)$. The conclusions hold at the selected index.
\end{theorem}

Slower smoothing permits faster growth of $q$. If all four parameters shrink as positive constants divided by $1+\log(n_e+e)$, line estimation and the normalized trace used for selection allow $a<2/3$, and operator consistency of the covariance estimate, fixed-contrast studentization and the absolute vector remainder hold for $a<1/2$, in each case together with $q^{r+1/2}=o(n_e^{9/20})$. For the cosine link ($r=1$) this covers $q\asymp n_e^{1/6}$ (Supplement Theorem~\ref*{supp-sb:dimension}; explicit bounds are in Supplement Sections~\ref*{supp-sb:relocated} and~\ref*{supp-sb:rates}).

\paragraph*{Error-density information after screening.}
Let $F$ span the raw orthonormal tangent, $\Omega_S=\Var(F^TX_S\mid\beta_{0,S}^TX_S)$, $A(v)=\E[m_0'(\sqrt v T)^2]$, and $I_\varepsilon$ denote raw error location information. The gain factor $G=\Var(\varepsilon)I_\varepsilon$ is unchanged by response standardization. Moving \eqref{eq:gain-loss} to the raw retained tangent with $D_0D_0^T=v\Omega_S^{-1}$ gives
\begin{align}
 \Lambda(v,k)&=vA(v)I_\varepsilon\,b(v,k)/j(v,k)^2\ge1,\nonumber\\
 V_{S,k}&=\Lambda(v,k)\,\frac{\Omega_S^{-1}}{A(v)I_\varepsilon}
       =\frac{\Lambda(v,k)}{G}V_{{\rm mean},S},\qquad
 V_{{\rm mean},S}=\frac{\Var(\varepsilon)}{A(v)}\Omega_S^{-1}.\label{eq:hd-gain}
\end{align}
Both bounds use $n_e$ observations. The scalar loss and population grid ordering are common to all correct screens; the traces depend on $\Omega_S^{-1}$. Since $\Lambda\to1$ uniformly on the spectral interval, non-Gaussian errors retain a strict gain at sufficiently small positive tuples.

\begin{corollary}[Independent screens and the support oracle]\label{cor:hd-oracle}
Given the screen, on $E$ the full influence, covariance, trace and candidate-inference results hold at the realized $q$. A retained-model error event of uniform probability $\epsilon_n$ therefore has unconditional probability at most $\epsilon_n+\delta_{\rm scr}$ when screen failure also counts as error. If the screen recovers $S_0$ except with probability $\delta_n^{=}$, its zero-embedded output agrees with the known-support SI-KDR on the same $D_e$, using the same rules and seed, with probability at least $1-\delta_n^{=}$. On $S_0$, with $|S_0|\ge2$, the leading covariance is $(\Lambda(v,k)/G)V_{{\rm mean},S_0}/n_e$ at a fixed positive-curvature tuple. A shrinking grid with common efficient influence and negligible remainders attains the independent-error efficiency bound on $S_0$ when $\delta_n^{=}\to0$, jointly at fixed or, for eligible contrasts, growing $|S_0|$.
\end{corollary}

Supplement Section~\ref*{supp-sb:interface} proves the corollary and verifies three screens. Top-$q$ selection by the biased distance correlation \citep{li2012distance} fails with probability at most
\begin{equation}\label{eq:hd-screen-error}
 \delta_{\rm DC}=\min\{1,C_0P[e^{-c_1n_s^{1-2(\nu_1+\nu_2)}}+n_s e^{-c_2 n_s^{\nu_2}}]\}
\end{equation}
when the weakest active squared distance correlation exceeds the $(q-|S_0|+1)$st largest inactive one by $4An_s^{-\nu_1}$, with $A>0$, $0\le\nu_1<1/2$ and $0<\nu_2<1/2-\nu_1$; taking $\nu_2=(1-2\nu_1)/3$ permits $\log P=o(n_s^{\nu_2})$. The SSIRvRP screen of \citet{zhang2026sparse} applies under an inverse-mean signal, and a least-squares Lasso screen \citep{sbNeykov2016} applies to correlated Gaussian predictors.

\begin{theorem}[Conditional inference after screening]\label{thm:hd-inference}
For $n>q+1$, on $E$ the candidate-only tuning rule gives an exact $F_{q-1,n-q-1}$ law at the true line conditional on $(D_s,X_S\beta_{0,S},Y_e)$ whenever its projected coefficient is nonzero. A zero coefficient is accepted. Keep the Fourier cap and all optimization and inference diagnostics. For $q\to\infty$ take the final raw angular radius $\min\{\pi/2,2\sqrt{\tr(\widehat V_{S,\widehat k})/n}\}$ about the fitted line. Let $\mathcal X_0$ denote exclusion of the truth by this domain, including global failures, and $\mathcal N_0$ a numerical false exclusion of a true accepted candidate within it. At level $\alpha$,
\begin{equation}\label{eq:hd-coverage}
 \Pp\{[\beta_0]\in\mathcal C^{\rm comp}\}
 \ge(1-\alpha)\Pp(E)-\Pp(E\cap\mathcal X_0)-\Pp(E\cap\mathcal X_0^c\cap\mathcal N_0).
\end{equation}
Under the growth conditions of Theorem~\ref{thm:hd-estimation}, conditionally on a successful screen the exact-arithmetic set has coverage $1-\alpha+o(1)$ and diameter $O_P(\sqrt{q/n})$. If $\delta_{\rm scr}\to0$ these conclusions hold after zero embedding.
\end{theorem}

Screening on each of two blocks of sizes $n_1,n_2$ and fitting SI-KDR on the other uses both samples for estimation: the leading line of $\sum_r(n_r/N)\widehat b_r\widehat b_r^T$ has the full-sample oracle's leading covariance $V_{S_0}/N$ at fixed $|S_0|$ when both screens recover $S_0$ with probability tending to one and the fits share the oracle's influence with negligible remainders, and intersecting the two full candidate inversions at level $\alpha/2$ gives coverage at least $1-\alpha-\delta_1-\delta_2$ (Supplement Section~\ref*{supp-sb:crossfit}).

\paragraph*{Tests for retained coordinates.}
On a correct screen, test $\beta_j=0$ by rerunning \mbox{SI-KDR}, including whitening, tuning and optimization, on $(X_{S\setminus j},Y)$. Let $\Pi_j$ remove the intercept and these $q-1$ predictors, and let $z_j$ be the cosine between $\Pi_jX_j$ and $\Pi_jc^{(-j)}$, where the restricted coefficient $c^{(-j)}$ in Supplement~(\ref*{supp-sbvar:c}) retains both density-center and link-refitting derivatives. Given the reduced data and screen, $\sqrt{n_e-q-1}\,z_j/\sqrt{1-z_j^2}$ has the $t_{n_e-q-1}$ law under the null for $n_e>q+1$ and nonzero coefficient projection; degenerate fits accept. Holm and Benjamini--Yekutieli
give respective FWER and FDR bounds $\alpha+(1-\alpha)\delta_{\rm scr}$.
For raw alternatives $\beta_j=\omega/\sqrt{n_e}+o(n_e^{-1/2})$
along the unit-norm path of Supplement Theorem~\ref*{supp-sbvar:power},
at a fixed tuple or one with a stabilized shift, the normal shift magnitude is
$\sqrt{G/\Lambda(v,k)}$ times that of the conditional-mean oracle
at the same sample size. Equivalently, the squared noncentrality of SI-KDR is $G/\Lambda(v,k)$ times that of the oracle, so $\Lambda<G$ gives higher local power at every nonzero $\omega$.

\section{Numerical results}\label{sec:experiments}

\subsection{Population Illustration of Finite-Smoothing Information Loss}
The following example uses population integrals only, without sampling, to separate the loss from smoothing the link from the loss from smoothing the residual density. Take $p=3$, $\Sigma_{jk}=.5^{|j-k|}$ and $\beta_0=(1,1,0)^T/\sqrt2$ in raw coordinates. We cross $\sin(2t)$ and $\cos(2t)$ with normal, asymmetric two-component, or symmetric three-component standardized errors (Supplement Section~\ref*{supp-sec:sim-details}). SI-KDR-R fits the same protected criterion using the reference grids and rule of Supplement Section~\ref*{supp-sec:variants}.

Figure~\ref{fig:pop-loss} crosses two regression bandwidths with two density bandwidths, holding the ridge and floor fixed. For the sine link with multimodal errors, the population variance ratio $V_k/\mathcal I^{-1}$ equals $13.245$ at the SI-KDR-R bandwidths. Reducing $h$ alone brings it to $11.763$, reducing $g$ alone to $2.562$, and reducing both to $1.847$. Most of the information in the modes is recovered by narrowing the regression bandwidth; the cosine link behaves similarly, and all 24 combinations are in Supplement Section~\ref*{supp-sec:sim-details}.

\begin{figure}[t]
\centering\includegraphics[width=\linewidth]{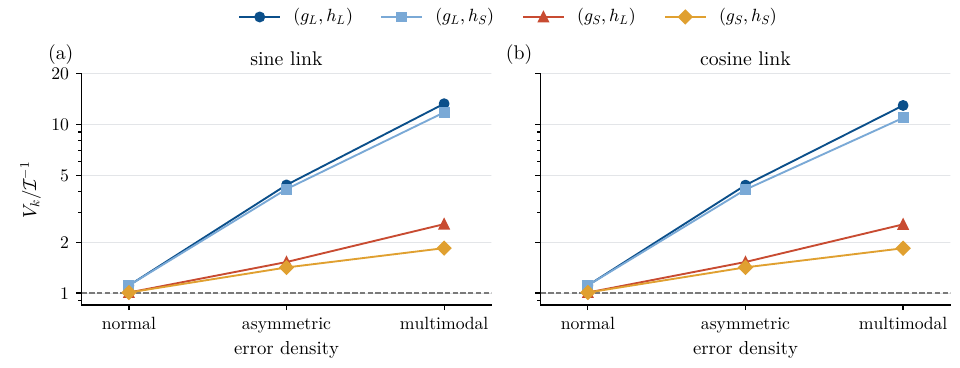}
\caption{Population illustration. Population variance of the SI-KDR influence function relative to the efficiency bound, $V_k/\mathcal I^{-1}$ on a log scale, for the regression bandwidths $g_L=.4312$, $g_S=.2156$ and the density bandwidths $h_L=.3232$, $h_S=.0970$, with $\lambda=.000113578$ and $\rho=1/600$; $(g_L,h_L)$ is the SI-KDR-R tuple at $n=600$. The dashed line marks the bound; the influence is $J_k^{-1}\eta_k$ with $\eta_k$ from Proposition~\ref{prop:influence}.}\label{fig:pop-loss}
\end{figure}

\subsection{Simulation design}
Each of the four primary cells (two links crossed with two non-Gaussian error laws) has 500 samples of size $n=600$, and SI-KDR, SI-KDR-R, mean MAVE and csMAVE are fitted to every sample. For the cosine link with asymmetric errors, larger samples add the Ma--Zhu estimator: 300 samples at $n=1600$ and 50 at $n=3200$. Confidence sets (Section~\ref{sec:cs-inference}) are evaluated on $500$ further samples per primary cell. At 500 samples the Monte Carlo standard error (MCSE) of a nominal coverage is about $0.97$ percentage points.

\subsection{Accuracy of the finite-smoothing expansion}
\label{sec:expansion-diagnostic}
\begin{figure}[t]
\centering
\includegraphics[width=\linewidth]{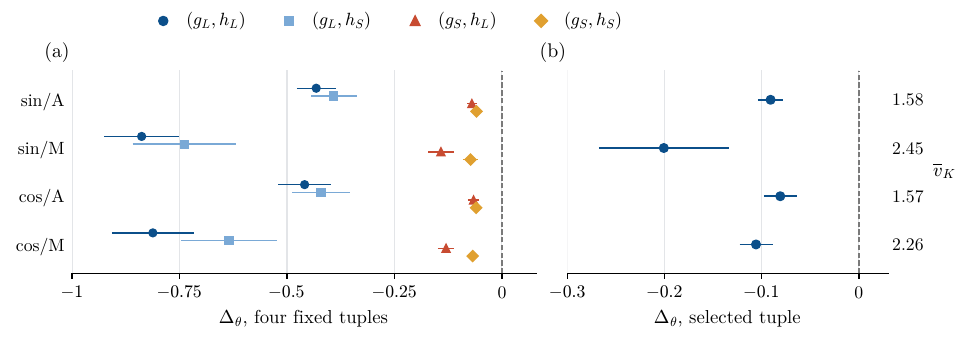}
\caption{Finite-smoothing versus efficient-limit approximation of the estimator at $n=600$: mean of $\Delta_\theta$ over 100 samples per cell, with bars of one paired MCSE. (a)~The four fixed tuples of Figure~\ref{fig:pop-loss}. (b)~The tuple selected by the variance-trace rule; the right margin gives the mean population variance ratio $\overline v_K$ of the selected tuples. A and M denote asymmetric and multimodal errors; the values are in Supplement Tables~\ref*{supp-tab:fixed-expansion} and~\ref*{supp-tab:expansion}.}
\label{fig:expansion-diagnostic}
\end{figure}

The expansion of Theorem~\ref{thm:expansion} predicts the estimator by $J_k^{-1}\mathbb G_k$, where $\mathbb G_k=n^{-1/2}\sum_i\eta_k(O_i)$, whereas the efficient limit predicts it by $\mathcal I^{-1}\mathbb G_{\rm eff}$ with $\mathbb G_{\rm eff}=n^{-1/2}\sum_iS_{{\rm eff},i}$. On the first 100 samples of each primary cell, we compare the two through
\[
\Delta_\theta=\|\sqrt n\widehat\theta-J_k^{-1}\mathbb G_k\|^2
             -\|\sqrt n\widehat\theta-\mathcal I^{-1}\mathbb G_{\rm eff}\|^2,
\]
which is negative when the finite-smoothing prediction is the closer of the two. In Figure~\ref{fig:expansion-diagnostic}, the mean of $\Delta_\theta$ is negative in all 16 combinations of the four fixed tuples with the four cells and, after tuning, in all four cells. The mean population variance ratio $\overline v_K$ of the selected tuples lies between $1.57$ and $2.45$, so finite smoothing is still visible at the bandwidths chosen by the data, and accounting for it improves the approximation of the estimator.

\subsection{Point estimation and coverage}
\begin{figure}[t]
\centering
\includegraphics[width=\linewidth]{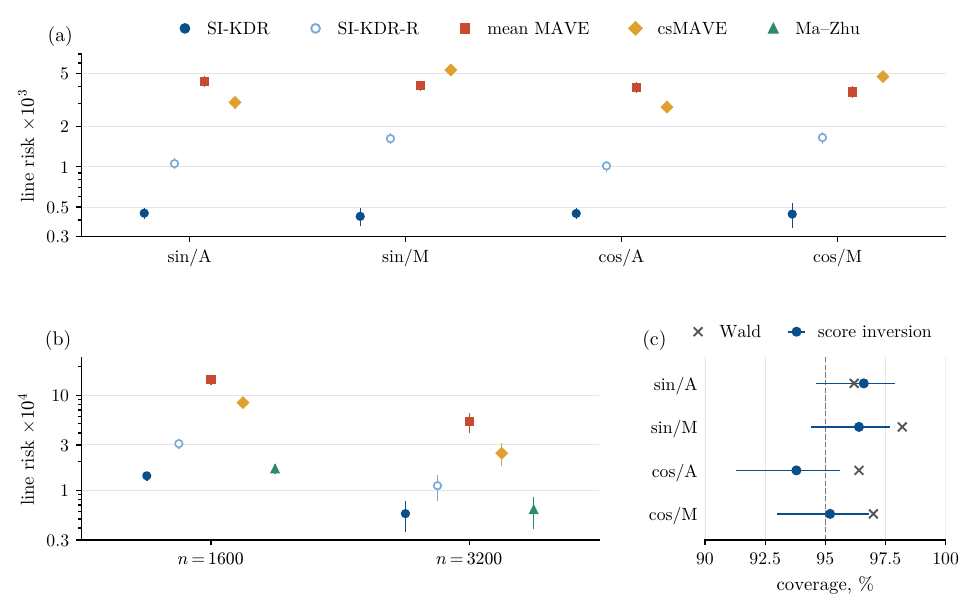}
\caption{Primary designs. (a)~Mean line risk at $n=600$, 500 samples per cell. (b)~Cosine link with asymmetric errors: 300 samples at $n=1600$ and 50 at $n=3200$. (c)~Coverage of the score-inversion sets with 95\% Wilson intervals, and Wald coverage, on 500 further samples per cell. Bars in (a) and (b) are $\pm1.96$ MCSEs. A and M denote asymmetric and multimodal errors. Values: Supplement Tables~\ref*{supp-tab:confirmation}, \ref*{supp-tab:large-risk} and~\ref*{supp-tab:conditional-confirmation}.}
\label{fig:point}
\end{figure}

Figure~\ref{fig:point}(a) compares the four methods on the same 500 samples in each primary cell. In every cell, variance-trace tuning lowers the risk of SI-KDR by $56.0$--$73.8\%$ relative to SI-KDR-R, and SI-KDR has risk $87.8$--$89.7\%$ below mean MAVE and $84.0$--$92.0\%$ below csMAVE.

In the larger samples of Figure~\ref{fig:point}(b), which add the Ma--Zhu efficient-score estimator (Supplement Section~\ref*{supp-sec:sim-details}), SI-KDR again has the lowest mean risk at both $n=1600$ and $n=3200$. At $n=1600$ SI-KDR's risk lies $15.4\%$ below Ma--Zhu's (paired difference $-2.58\times10^{-5}$, MCSE $0.93\times10^{-5}$), and $54\%$, $90\%$ and $83\%$ below SI-KDR-R, mean MAVE and csMAVE.

Score-inversion coverage is consistent with the nominal level in all four cells (Figure~\ref{fig:point}(c)): coverage lies between $93.8\%$ and $96.6\%$, and every Wilson interval contains $95\%$.

\subsection{Moderate and high dimensions}\label{sec:hd-numerical}
\paragraph*{Comparison with competing estimators.}
Fourteen settings compare SI-KDR with mean MAVE, csMAVE and Ma--Zhu (Supplement Section~\ref*{supp-sec:hd-protocol}). Eight fit $p=10$ or $20$ predictors directly at $n=600$; six screen $P=200$ or $1000$ predictors on $n_s=200$ observations and fit $q=10$ or $20$ retained coordinates on $n_e=400$, with SSIRvRP-BIC \citep{zhang2026sparse} also fitted to all 600 observations. SI-KDR, started from the mean-MAVE direction (the expansion needs only a $\sqrt n$-consistent start), had the lowest mean line risk in all fourteen settings: $35$--$89\%$ below mean MAVE in the direct settings and $22$--$95\%$ below every competitor after screening. Whenever the screen retained the support, the score-inversion set covered the truth in $94$--$96\%$ of the samples. The gain from the error density persists after screening.

\begin{figure}[!htbp]\centering
\includegraphics[width=\linewidth]{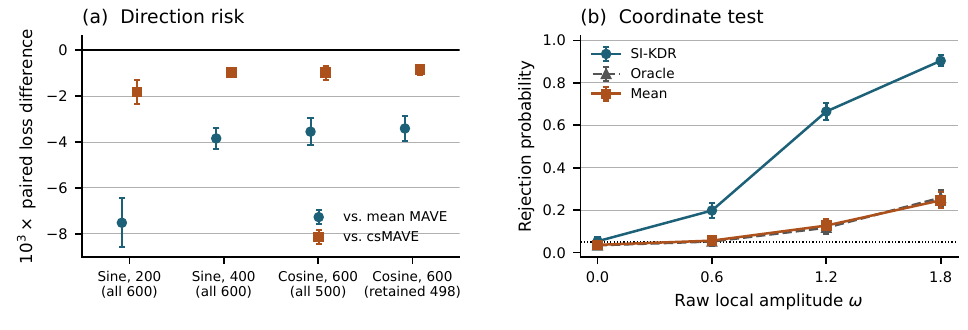}
\caption{(a) Paired SI-KDR-minus-competitor line loss on the same rows and screened columns, $P=1000$, $q=2$. Labels give the link, $n_e$ and the number of samples (all, or those retaining the support). Negative differences favor SI-KDR. (b) Rejection probability of the coordinate test for the forced target in correctly retained models; Mean is the feasible restricted mean test and Oracle the conditional-mean oracle. Bars show 1.96 MCSEs.}\label{fig:paired-HD}
\end{figure}

\paragraph*{Ambient dimension.}
With two active coordinates and the sine link, $N=600$ or $1200$ observations were split into $n_s=2N/3$ and $n_e=N/3$, and $q=2$ coordinates were retained from $P=1000$, $5000$ and $10{,}000$. Every screen retained the support, so the fits did not change with $P$; SI-KDR's mean line loss was $79$--$85\%$ below mean MAVE's and $49$--$60\%$ below csMAVE's. With the cosine link, $P=1000$ and $(n_s,n_e)=(2400,600)$, the screen recovered the support in 498 of 500 samples, SI-KDR's mean loss was $51\%$ and $22\%$ below those of mean MAVE and csMAVE, and on recovery SI-KDR coincided with the known-support fit (Figure~\ref{fig:paired-HD}a). Here all 20 fixed grid tuples have $\Lambda<G$ (Supplement Section~\ref*{supp-sb:cosine-new}).

\paragraph*{Pooling and growing dimension.}
Pooling the two cross-fitted directions from folds of 200 and 400 rows reduced the mean line loss of the single split from $.02908$ to $.01542$. On nested screens with $n_e=300$, the loss scaled by $n_e/(q-1)$ on correct screens was $.275$--$.320$ for $q$ from $5$ to $40$ on both schedules, and a studentized fixed contrast was accepted in $95.0$--$96.0\%$ of eligible fits on the logarithmic schedule (Supplement Sections~\ref*{supp-sb:x4-results} and~\ref*{supp-sb:x6-results}). Pooling recovers information spent on screening, and the scaled loss agrees with the $\sqrt{q/n}$ rate of Section~\ref{sec:highdim}.

\paragraph*{Coordinate tests.}
With $P=1000$, $q=4$ and $n_e=400$, a fixed rule kept the tested coordinate in the screen. At amplitudes $\omega=0,.6,1.2,1.8$ SI-KDR rejected in $5.3$, $19.8$, $66.5$ and $90.4\%$ of correctly retained samples, against $3.4$, $5.2$, $11.7$ and $25.8\%$ for the oracle (Figure~\ref{fig:paired-HD}b). Its mean signed statistic was about $2.6$ times the oracle's at nonzero amplitudes, and its single-null error rates were $3.6$--$5.0\%$ at level $5\%$ (Supplement Section~\ref*{supp-sbvar:x5}). As Section~\ref{sec:highdim} predicts when $\Lambda<G$, the error density raises coordinate-test power.
\FloatBarrier

\subsection{Public benchmark}
We predict heating load for 768 simulated building configurations \citep{uci_energy_2012}; roof area is determined by wall and surface area, leaving seven inputs. SI-KDR, \mbox{SI-KDR-R} and mean MAVE each reduce the inputs to one index and predict from it by one Gaussian kernel regression rule, on the same 30 splits into 576 training and 192 test cases, with all fitting on the training cases. On this discrete, non-Gaussian design, \mbox{SI-KDR}'s mean normalized test error was $13.6\%$ lower than SI-KDR-R's and $15.4\%$ lower than mean MAVE's (Table~\ref{tab:energy}).
\begin{table}[htbp]
\centering\small
\caption{Heating-load prediction on 30 random splits of the Energy Efficiency benchmark. NMSE is the test mean squared error divided by the training response variance; parentheses contain standard deviations over the splits.}
\label{tab:energy}
\begin{tabular}{lrrr}
\toprule
Method & NMSE (SD) & SI-KDR minus method (SD) & Relative reduction \\
\midrule
SI-KDR & .04821 (.00866) & --- & --- \\
SI-KDR-R & .05582 (.00562) & $-.00762$ (.00867) & $13.6\%$ \\
mean MAVE & .05699 (.00532) & $-.00879$ (.00941) & $15.4\%$ \\
\bottomrule
\end{tabular}
\end{table}

\FloatBarrier
\section{Discussion}\label{sec-conc}

The gain $G$ is the factor by which an independent error density can lower the conditional-mean bound, and $\Lambda_k$ the factor by which a given pair of bandwidths inflates the efficient variance. Realizing the gain requires a link fit that separates signal from error and a density fit that resolves the residuals; the case-weight covariance trace assesses both on the scale of direction error. Screening changes which coordinates are estimated and tested. Adding a noise coordinate to a correct screen changes the angular uncertainty but not the scalar law behind the smoothing loss, so correct screens share the population bandwidth ordering and $G/\Lambda_k$.

\FloatBarrier
\section{Disclosure statement}\label{disclosure-statement}

The authors report no competing interests to declare.

Grammarly (Grammarly, Inc.) was used for grammatical editing. The authors are responsible for the content.

\section{Data Availability Statement}\label{data-availability-statement}

The Energy Efficiency data are available from the UCI Machine Learning Repository \citep{uci_energy_2012}. The benchmark splits and predictions, the simulation code and records, and a reproduction guide are available from the authors.

\paragraph*{Supplementary material.}\label{supplementary-material}
The supplement gives the proofs, algorithms and further results.

  \bibliography{references}

@article{ichimura1993semiparametric,
  title={Semiparametric least squares ({SLS}) and weighted {SLS} estimation of single-index models},
  author={Ichimura, Hidehiko},
  journal={Journal of Econometrics},
  volume={58},
  number={1-2},
  pages={71--120},
  year={1993},
  publisher={Elsevier},
  doi={10.1016/0304-4076(93)90114-K}
}

@article{xia2002adaptive,
  title={An adaptive estimation of dimension reduction space},
  author={Xia, Yingcun and Tong, Howell and Li, Wai Keung and Zhu, Li-Xing},
  journal={Journal of the Royal Statistical Society Series B: Statistical Methodology},
  volume={64},
  number={3},
  pages={363--410},
  year={2002},
  publisher={Oxford University Press},
  doi={10.1111/1467-9868.03411}
}

@article{delecroix2003efficient,
  title={Efficient estimation in conditional single-index regression},
  author={Delecroix, Michel and H{\"a}rdle, Wolfgang and Hristache, Marian},
  journal={Journal of Multivariate Analysis},
  volume={86},
  number={2},
  pages={213--226},
  year={2003},
  publisher={Elsevier},
  doi={10.1016/S0047-259X(02)00046-5}
}

@article{chen2020kernel,
  title={Kernel density regression},
  author={Chen, Xin and Ma, Xuejun and Zhou, Wang},
  journal={Journal of Statistical Planning and Inference},
  volume={205},
  pages={318--329},
  year={2020},
  publisher={Elsevier},
  doi={10.1016/j.jspi.2019.09.001}
}

@article{kuchibhotla2023semiparametric,
  title={Semiparametric efficiency in convexity constrained single-index model},
  author={Kuchibhotla, Arun K and Patra, Rohit K and Sen, Bodhisattva},
  journal={Journal of the American Statistical Association},
  volume={118},
  number={541},
  pages={272--286},
  year={2023},
  publisher={Taylor \& Francis},
  doi={10.1080/01621459.2021.1927741}
}

@article{yuan2007semiparametric,
  title={Semiparametric regression with kernel error model},
  author={Yuan, Ao and De Gooijer, Jan G},
  journal={Scandinavian Journal of Statistics},
  volume={34},
  number={4},
  pages={841--869},
  year={2007},
  publisher={Wiley Online Library},
  doi={10.1111/j.1467-9469.2006.00531.x}
}

@article{ma2013efficient,
  title={Efficient estimation in sufficient dimension reduction},
  author={Ma, Yanyuan and Zhu, Liping},
  journal={The Annals of Statistics},
  volume={41},
  number={1},
  pages={250--268},
  year={2013},
  doi={10.1214/12-AOS1072}
}

@article{zhu2006fourier,
  title={Fourier methods for estimating the central subspace and the central mean subspace in regression},
  author={Zhu, Yu and Zeng, Peng},
  journal={Journal of the American Statistical Association},
  volume={101},
  number={476},
  pages={1638--1651},
  year={2006},
  publisher={Taylor \& Francis},
  doi={10.1198/016214506000000140}
}

@article{linton2008estimation,
  title={Estimation of a semiparametric transformation model},
  author={Linton, Oliver and Sperlich, Stefan and Van Keilegom, Ingrid},
  journal={The Annals of Statistics},
  volume={36},
  number={2},
  pages={686--718},
  year={2008},
  doi={10.1214/009053607000000848}
}

@article{delecroix2006semiparametric,
  title={On semiparametric {M}-estimation in single-index regression},
  author={Delecroix, Michel and Hristache, Marian and Patilea, Valentin},
  journal={Journal of Statistical Planning and Inference},
  volume={136},
  number={3},
  pages={730--769},
  year={2006},
  publisher={Elsevier},
  doi={10.1016/j.jspi.2004.09.006}
}

@article{li2017new,
  title={A new minimum contrast approach for inference in single-index models},
  author={Li, Weiyu and Patilea, Valentin},
  journal={Journal of Multivariate Analysis},
  volume={158},
  pages={47--59},
  year={2017},
  publisher={Elsevier},
  doi={10.1016/j.jmva.2017.03.009}
}

@article{ichimura2022influence,
  title={The influence function of semiparametric estimators},
  author={Ichimura, Hidehiko and Newey, Whitney K},
  journal={Quantitative Economics},
  volume={13},
  number={1},
  pages={29--61},
  year={2022},
  publisher={Wiley Online Library},
  doi={10.3982/QE826}
}

@article{perry2010rotation,
  title={A rotation test to verify latent structure},
  author={Perry, Patrick O and Owen, Art B},
  journal={Journal of Machine Learning Research},
  volume={11},
  pages={603--624},
  year={2010}
}

@misc{uci_energy_2012,
  author = {Tsanas, Athanasios and Xifara, Angeliki},
  title = {Energy Efficiency},
  year = {2012},
  howpublished = {UCI Machine Learning Repository},
  doi = {10.24432/C51307},
  note = {Dataset; CC BY 4.0}
}

@manual{MAVEPackage2025,
  author = {Xia, Yingcun and Hang, Weiqiang},
  title = {{MAVE}: Methods for Dimension Reduction},
  year = {2025},
  note = {R package version 1.3.12},
  url = {https://CRAN.R-project.org/package=MAVE}
}

@incollection{vandervaart2002semiparametric,
  title={Semiparametric statistics},
  author={van der Vaart, Aad W},
  booktitle={Lectures on Probability Theory and Statistics (Saint-Flour, 1999)},
  series={Lecture Notes in Mathematics},
  volume={1781},
  pages={331--457},
  year={2002},
  publisher={Springer}
}

@article{zhang2026sparse,
  author={Zhang, Jia and Wu, Runxiong and Chen, Xin},
  title={Optimal Sparse Sliced Inverse Regression via Random Projection},
  journal={Journal of Computational and Graphical Statistics},
  volume={35},
  number={2},
  pages={782--794},
  year={2026},
  doi={10.1080/10618600.2025.2571161}
}

@article{li2012distance,
  author = {Li, Runze and Zhong, Wei and Zhu, Liping},
  title = {Feature Screening via Distance Correlation Learning},
  journal = {Journal of the American Statistical Association},
  year = {2012},
  volume = {107},
  number = {499},
  pages = {1129--1139},
  doi = {10.1080/01621459.2012.695654}
}

@article{szekely2007distance,
  author = {Sz{\'e}kely, G{\'a}bor J. and Rizzo, Maria L. and Bakirov, Nail K.},
  title = {Measuring and Testing Dependence by Correlation of Distances},
  journal = {The Annals of Statistics},
  year = {2007},
  volume = {35},
  number = {6},
  pages = {2769--2794},
  doi = {10.1214/009053607000000505}
}

@article{sbNeykov2016,
  author = {Neykov, Matey and Liu, Jun S. and Cai, Tianxi},
  title = {{L1}-Regularized Least Squares for Support Recovery of High Dimensional
    Single Index Models with Gaussian Designs},
  journal = {Journal of Machine Learning Research},
  volume = {17},
  number = {87},
  pages = {1--37},
  year = {2016},
  url = {https://jmlr.org/papers/v17/16-006.html}
}

@misc{sbZhang2025supp,
  author = {Zhang, Jia and Wu, Runxiong and Chen, Xin},
  title = {Supplementary Material for {Optimal Sparse Sliced Inverse
    Regression via Random Projection}},
  year = {2025},
  note = {Taylor \& Francis, Figshare, version 1},
  doi = {10.6084/m9.figshare.30324843.v1},
  url = {https://doi.org/10.6084/m9.figshare.30324843.v1}
}

@article{he2000increasing,
  author = {He, Xuming and Shao, Qi-Man},
  title = {On Parameters of Increasing Dimensions},
  journal = {Journal of Multivariate Analysis},
  year = {2000},
  volume = {73},
  number = {1},
  pages = {120--135},
  doi = {10.1006/jmva.1999.1873}
}

@article{sbCook2004,
 author={Cook, R. Dennis},
 title={Testing predictor contributions in sufficient dimension reduction},
 journal={The Annals of Statistics},
 volume={32}, number={3}, pages={1062--1092}, year={2004},
 doi={10.1214/009053604000000292}
}

@article{sbLiu2022,
 author={Liu, Molei and Katsevich, Eugene and Janson, Lucas and Ramdas, Aaditya},
 title={Fast and powerful conditional randomization testing via distillation},
 journal={Biometrika}, volume={109}, number={2}, pages={277--293}, year={2022},
 doi={10.1093/biomet/asab039}
}

@article{sbHolm1979,
 author={Holm, Sture},
 title={A simple sequentially rejective multiple test procedure},
 journal={Scandinavian Journal of Statistics}, volume={6}, number={2},
 pages={65--70}, year={1979}, url={https://www.jstor.org/stable/4615733}
}

@article{sbBY2001,
 author={Benjamini, Yoav and Yekutieli, Daniel},
 title={The control of the false discovery rate in multiple testing under dependency},
 journal={The Annals of Statistics}, volume={29}, number={4}, pages={1165--1188},
 year={2001}, doi={10.1214/aos/1013699998}
}

@article{x23Huo2016,
  author = {Huo, Xiaoming and Sz{\'e}kely, G{\'a}bor J.},
  title = {Fast Computing for Distance Covariance},
  journal = {Technometrics}, year = {2016}, volume = {58}, number = {4},
  pages = {435--447}, doi = {10.1080/00401706.2015.1054435}
}

@article{x23Huo2017,
  author = {Huo, Xiaoming and Sz{\'e}kely, G{\'a}bor J.},
  title = {Clarifications for {\textquotedblleft Fast Computing for Distance Covariance\textquotedblright} by {Xiaoming Huo} and {G{\'a}bor J. Sz{\'e}kely}},
  journal = {Technometrics}, year = {2017}, volume = {59}, number = {1},
  pages = {134--135}, doi = {10.1080/00401706.2016.1258011}
}

@article{x1Lin2019,
  author={Lin, Qian and Zhao, Zhigen and Liu, Jun S.},
  title={Sparse Sliced Inverse Regression via Lasso},
  journal={Journal of the American Statistical Association},
  year={2019}, volume={114}, number={528}, pages={1726--1739},
  doi={10.1080/01621459.2018.1520115}
}

@article{fan2008sure,
  title={Sure independence screening for ultrahigh dimensional feature space},
  author={Fan, Jianqing and Lv, Jinchi},
  journal={Journal of the Royal Statistical Society Series B: Statistical Methodology},
  volume={70},
  number={5},
  pages={849--911},
  year={2008},
  doi={10.1111/j.1467-9868.2008.00674.x}
}

@article{lin2018consistency,
  title={On consistency and sparsity for sliced inverse regression in high dimensions},
  author={Lin, Qian and Zhao, Zhigen and Liu, Jun S.},
  journal={The Annals of Statistics},
  volume={46},
  number={2},
  pages={580--610},
  year={2018},
  doi={10.1214/17-AOS1561}
}

@article{radchenko2015high,
  title={High dimensional single index models},
  author={Radchenko, Peter},
  journal={Journal of Multivariate Analysis},
  volume={139},
  pages={266--282},
  year={2015},
  doi={10.1016/j.jmva.2015.02.007}
}

\end{document}


\def\spacingset#1{\renewcommand{\baselinestretch}%
{#1}\small\normalsize} \spacingset{1}


\if1\anon
{
  \title{\bf Supplement to ``Finite-Smoothing Efficiency and Inference in Single-Index Models''}
  \author{%
  \large
  Jiayang Xu\textsuperscript{1,}\thanks{Jiayang Xu, Yusen Jia and Yangzhou Chen contributed equally to this work.}\quad
  Yusen Jia\textsuperscript{2,}\footnotemark[1]\quad
  Yangzhou Chen\textsuperscript{1,}\footnotemark[1]\quad
  Xin Chen\textsuperscript{1,}\thanks{Corresponding author. Email: \texttt{chenx8@sustech.edu.cn}.}\\[0.8em]
  {\normalsize \textsuperscript{1}Department of Statistics and Data Science,}\\
  {\normalsize Southern University of Science and Technology,}\\
  {\normalsize Shenzhen, Guangdong 518055, China}\\[0.5em]
  {\normalsize \textsuperscript{2}Thrust of Financial Technology,}\\
  {\normalsize The Hong Kong University of Science and Technology (Guangzhou),}\\
  {\normalsize Guangzhou, Guangdong 511453, China}
}
\date{}
\maketitle
} \fi

\if0\anon
{
  \bigskip
  \bigskip
  \bigskip
  \begin{center}
    {\LARGE\bf Supplement to ``Finite-Smoothing Efficiency and Inference in Single-Index Models''}
\end{center}
  \medskip
} \fi

\bigskip
\noindent\textbf{Guide to the proofs.}
Screening $P$ predictors on $n_s$ observations leaves $q$ coordinates for fitting on $n=n_e$ independent rows ($N=n_s+n_e$). Section~\ref{sb:interface} gives the transfer and oracle coupling.

\noindent\begin{tabularx}{\linewidth}{@{}L{.25\linewidth}L{.25\linewidth}>{\raggedright\arraybackslash}X@{}}
\toprule
\textbf{Main result} & \textbf{Supplement proof} & \textbf{Conditions or use} \\
\midrule
Population information (Section~\ref*{main-sec:population})
& Propositions~\ref{inf:population}--\ref{inf:loss}
& At positive curvature, the complete influence determines the smoothing cost. \\
\addlinespace
Fitted direction and raw covariance (Theorems~\ref*{main-thm:expansion}--\ref*{main-thm:variance})
& Theorems~\ref{dir:theorem}, \ref{inf:caseweights} and~\ref{up:finite-selection}
& Case weights include refitting and preprocessing. At tied positive-grid minima use the joint expansion. \\
\addlinespace
Power grid (Theorem~\ref*{main-thm:hd-estimation})
& Theorem~\ref{sb:dimension}, using Lemmas~\ref{hdg:fourier}--\ref{hdg:ascent}
& For $q=O(n^a)$, estimation, trace, sandwich and studentization allow $a<1/18$; the absolute vector remainder requires $a<1/24$. \\
\addlinespace
Logarithmic grid (Section~\ref*{main-sec:highdim})
& Theorem~\ref{sb:dimension}
& Estimation and algebraic trace allow $a<2/3$; operator consistency, covariance availability, studentization and absolute vector remainder hold for $a<1/2$. \\
\addlinespace
Information after screening (main eq.~\ref*{main-eq:hd-gain})
& Proposition~\ref{hdg:loss} and Corollary~\ref{sb:oracle}
& On a correct screen $\Lambda<G$ gives a strict gain over the mean bound at the same $n_e$. Geometry enters through $\Omega_S$. \\
\addlinespace
Candidate test and localized set (Theorems~\ref*{main-thm:cs-pivot}--\ref*{main-thm:cs-efficient} and~\ref*{main-thm:hd-inference})
& Theorems~\ref{cs:pivot}, \ref{cs:efficient} and~\ref{hdg:coverage}
& For $n>q+1$ the nondegenerate pivot is exact at truth; zero accepts. Domain and numerical exclusions affect the computed set. \\
\addlinespace
Two folds and retained-coordinate inference
& Propositions~\ref{sb:pool}, \ref{sb:intersection}, \ref{sbvar:pivot} and Corollary~\ref{sbvar:wide-power}
& Pooling recovers point efficiency under a common influence. Coordinate power depends on retention and restricted tuning. \\
\bottomrule
\end{tabularx}

Both grids require $q^{r+1/2}=o(n^{9/20})$ for the four-matrix Fourier start, hence $q=o(n^{3/10})$ for cosine ($r=1$). Studentization requires a fixed unit raw contrast with tangent variance bounded away from zero. Theorem~\ref{hdg:sample} treats other positive grids.

\noindent\textbf{Numerics.} Sections~\ref{sec:variants}--\ref{sec:sim-details} cover fixed dimensions and the Energy data; Sections~\ref{sec:hd-protocol} and~\ref{sb:newexperiments} cover the moderate- and high-dimensional experiments.

\newpage
\paragraph*{Notation.} Except in Section~\ref{gauss-section}, where $h$ denotes the regression bandwidth, $h=h_1$ is the density bandwidth, $g=h_2$ the regression bandwidth, $\rho$ the density floor and $\lambda$ the regression ridge. In Section~\ref{inf:theory}, $s_j$ bounds the stochastic jets about the population smoothed profile, whereas the bounds $u_j$ of the preceding sections also include the approximation of the conditional mean. $P_n f=n^{-1}\sum_{i=1}^n f(O_i)$ denotes an empirical average. The ambient dimension $P$ may depend on $n$. The upright grid labels P and L in Section~\ref{sb:rates} denote the power and logarithmic smoothing schedules, respectively; its italic $L=1+\log(n+e)$ is the numerical logarithmic factor. $O_i=(X_i,Y_i)$ denotes the $i$th observation, and $Z$ and $z$ denote a generic observation in population moments. $J$, $B$, and $V$ denote the curvature, the score covariance, and the covariance coefficient of the root-$n$ linear approximation (its contribution on the estimator scale is $V/n$), and $\widehat K$ estimates the covariance of the normalized candidate-wise score.

\section{Empirical-integral bounds for the residual log-likelihood}\label{supp:operator}
\label{op-section}

The following bounds hold for every data set and control residual-fit errors through unweighted empirical norms. They account for all observations, both endpoints of each residual pair, and the density floor. Let $n\ge2$, $h>0$, $\rho>0$ and $x\in\mathbb R^n$. Define
\begin{align}
 F(x)&=\frac1n\sum_{i=1}^n\log\{d_i(x)+\rho\},\qquad
 d_i(x)=\frac1{n-1}\sum_{j\ne i}\phi_h(x_i-x_j),\notag\\
 b&=(n-1)\rho,\qquad
 P_{ij}(x)=\frac{\mathbf1\{i\ne j\}\phi_h(x_i-x_j)}
 {b+\sum_{k\ne i}\phi_h(x_i-x_k)},\notag\\
 P_{i0}(x)&=\frac b{b+\sum_{k\ne i}\phi_h(x_i-x_k)}.
 \label{op-definition}
\end{align}
Here, the index $j=0$ denotes a constant floor atom; with this atom included, the weights in each row sum to one. Write $\mathbb E_i$ for expectation under row $i$ of these weights, $z_{ij}=(x_i-x_j)/h$, and $\Delta_{ij}u=u_i-u_j$ for $j\ne0$. All derivative expressions involving $z$ or $\Delta$ are defined to be zero at the floor atom, which carries no observation or direction coordinate. For a vector $u$, put $\|u\|_{p,n}=(n^{-1}\sum_i|u_i|^p)^{1/p}$, and let $P_n$ denote averaging over the row index $i$.

\subsection{Derivatives and row and column weights}

The coefficient $a=n\nabla F$ satisfies
\begin{align}
 a_i(x)
 &=\frac1{n-1}\sum_{j\ne i}\phi_h'(x_i-x_j)
 \left\{\frac1{d_i(x)+\rho}+\frac1{d_j(x)+\rho}\right\}\notag\\
 &=-h^{-1}\sum_{j\ne i}z_{ij}(P_{ij}+P_{ji}).
 \label{op-full-score}
\end{align}
The second reciprocal arises because $x_i$ also enters the density estimates centered at the other observations. In particular $\sum_i a_i(x)=0$. For directions $u,v,w$, define $L_u=-h^{-1}z\Delta u$ and $L_{uv}=-h^{-2}\Delta u\Delta v$, with the floor convention above. Differentiating the normalized weights gives
\begin{align}
 DF_x[u]&=P_n\mathbb E_iL_u,\notag\\
 D^2F_x[u,v]&=P_n\{\mathbb E_iL_{uv}
                    +\operatorname{Cov}_i(L_u,L_v)\},
 \label{op-second-identity}\\
 D^3F_x[u,v,w]
 &=P_n\big\{\operatorname{Cov}_i(L_{uv},L_w)
       +\operatorname{Cov}_i(L_{uw},L_v)
       +\operatorname{Cov}_i(L_{vw},L_u)\notag\\
 &\hspace{38mm}+\operatorname{Cum}_{3,i}(L_u,L_v,L_w)\big\}.
 \label{op-third-identity}
\end{align}
For any three row variables, the third cumulant is
\begin{align*}
 \operatorname{Cum}_{3,i}(A,B,C)
 &=\mathbb E_iABC-\mathbb E_iAB\,\mathbb E_iC
   -\mathbb E_iAC\,\mathbb E_iB
   -\mathbb E_iBC\,\mathbb E_iA\notag\\
 &\quad+2\mathbb E_iA\,\mathbb E_iB\,\mathbb E_iC.
\end{align*}
Because the third derivative of the logarithm of a single Gaussian kernel vanishes, \eqref{op-third-identity} contains no further term. These formulas also hold under ties, since the positive floor makes $F$ smooth on all of $\mathbb R^n$.

\begin{lemma}[Deterministic incoming and outgoing moments]
\label{op-moment-lemma}
Set
\[
 \mathcal L=1+\log n+\log_+\{(hb)^{-1}\}.
\]
For each fixed real $q\ge0$, there is a constant $C_q$, independent of $n,h,\rho,x$, such that
\begin{equation}
 \max_i\sum_{j\ne i}P_{ij}|z_{ij}|^q
 +\max_i\sum_{j\ne i}P_{ji}|z_{ij}|^q
 \le C_q\mathcal L^{1+q/2}.
 \label{op-moments}
\end{equation}
\end{lemma}
\begin{proof}
Sort the observations, breaking ties by any fixed rule. If the ranks of $i$ and $j$ differ by $r\ge1$, the denominator of row $j$ includes $r$ observations, including $i$ but excluding $j$, at distance from $x_j$ no greater than $|x_i-x_j|$. Monotonicity of the Gaussian kernel therefore gives $P_{ji}\le1/r$, also under ties. Thus, outgoing row mass is at most one, and incoming column mass is at most $2H_{n-1}\le2(1+\log n)$.

Also $P_{ij}\le(\sqrt{2\pi}hb)^{-1}\exp(-z_{ij}^2/2)$. Choose
\[
 R_q^2=2q+4\log\{e+n/(hb)\}.
\]
For $|z|\le R_q$, we use the row and column mass bounds. For $|z|>R_q$, the function $t^q\exp(-t^2/2)$ is decreasing, and each row or column has at most $n$ terms, so the far contribution is at most
\[
 C\frac n{hb}R_q^q e^{-R_q^2/2}\le C_qR_q^q.
\]
Since $R_q\le C_q\sqrt{\mathcal L}$, combining the near and far bounds proves \eqref{op-moments}.
\end{proof}

For $\rho=c/n$ and $h=c_1n^{-a}$, with fixed $a>0$ and the positive multipliers bounded above and below, $\mathcal L=O\{\log(n+e)\}$. We keep $\mathcal L$ in the following bounds to display the contribution of the floor.

\subsection{Empirical-norm bounds}

For finite $p\ge1$, let
\[
 A_p(u)_i=(\mathbb E_i|\Delta u|^p)^{1/p},\qquad
 Z_p(u)_i=(\mathbb E_i|z\Delta u|^p)^{1/p}.
\]
The inequality $|u_i-u_j|^p\le2^{p-1}(|u_i|^p+|u_j|^p)$ and \eqref{op-moments} give
\begin{equation}
 \|A_p(u)\|_{p,n}\le C_p\mathcal L^{1/p}\|u\|_{p,n},\qquad
 \|Z_p(u)\|_{p,n}\le C_p\mathcal L^{1/2+1/p}\|u\|_{p,n}.
 \label{op-endpoint-norms}
\end{equation}
The column bound is needed here because the second endpoint is averaged with incoming rather than outgoing weights.

\begin{lemma}[Bilinear, trilinear and score bounds]
\label{op-derivative-lemma}
For arbitrary $x,y,d,u,v\in\mathbb R^n$,
\begin{align}
 |D^2F_x[u,v]|&\le Ch^{-2}\mathcal L^2
                  \|u\|_{2,n}\|v\|_{2,n},
 \label{op-bilinear}\\
 \|a(x)-a(y)\|_{2,n}&\le Ch^{-2}\mathcal L^2\|x-y\|_{2,n},
 \label{op-score-transfer}\\
 |D^3F_x[d,u,v]|&\le Ch^{-3}\mathcal L^{5/2}
             \|d\|_{2,n}\|u\|_{4,n}\|v\|_{4,n},
 \label{op-trilinear}\\
 \|a(x)\|_\infty&\le Ch^{-1}\mathcal L^{3/2},\qquad
 |DF_x[w]|\le Ch^{-1}\mathcal L^{3/2}\|w\|_{1,n}.
 \label{op-linear}
\end{align}
The trilinear inequality also holds with finite exponents $p_1,p_2,p_3$ satisfying $\sum_kp_k^{-1}=1$, replacing its three norms accordingly.
\end{lemma}
\begin{proof}
In \eqref{op-second-identity}, the contribution of $\mathbb E_i|\Delta u\Delta v|$ is bounded by $\|A_2(u)\|_{2,n}\|A_2(v)\|_{2,n}$. The two covariance contributions are each bounded by $\|Z_2(u)\|_{2,n}\|Z_2(v)\|_{2,n}$, by rowwise Cauchy--Schwarz followed by Cauchy--Schwarz over rows. Equation \eqref{op-endpoint-norms} now gives \eqref{op-bilinear}.

Under the empirical inner product, $D^2F_x[u,v]=n^{-1}u^{\mathsf T}Da(x)v$, so duality in \eqref{op-bilinear} gives $\|Da(x)v\|_{2,n}\le Ch^{-2}\mathcal L^2\|v\|_{2,n}$. The same bound holds at every point of the segment from $x$ to $y$, and integrating along this segment gives \eqref{op-score-transfer}.

For \eqref{op-trilinear}, apply H\"older's inequality first under each completed row probability and then over the row index, with exponents $(2,4,4)$. After factoring out $h^{-3}$, each covariance in \eqref{op-third-identity} contains three endpoint differences and one factor $z$. Both of its terms are bounded by two $A_p$ factors and one $Z_p$ factor. When two differences are inside one row expectation, the constant function $1$ takes up the unused H\"older exponent; this costs nothing because the completed row measure has unit mass. By \eqref{op-endpoint-norms}, the total logarithmic exponent of these terms is $1/2+(1/2+1/4+1/4)=3/2$.

Each of the five third-cumulant terms is bounded by three $Z_p$ factors. For example,
\[
 \mathbb E_i|z\Delta d\,z\Delta u|\le Z_2(d)_iZ_4(u)_i,
 \qquad \mathbb E_i|z\Delta v|\le Z_4(v)_i.
\]
Outer H\"older then bounds
\[
 P_n\mathbb E_i|z\Delta d\,z\Delta u|\mathbb E_i|z\Delta v|
\]
by the product of the three outer norms. The terms with a single-row expectation and those with a three-row expectation obey the same bound. Their total logarithmic exponent is $3/2+(1/2+1/4+1/4)=5/2$, which proves \eqref{op-trilinear}; the proof with general conjugate exponents is identical. Finally, \eqref{op-full-score} and \eqref{op-moments} with $q=1$ give \eqref{op-linear}.
\end{proof}

\subsection{Hessian transfer from empirical residual jets}

Let $r(\theta),e(\theta)$ be twice continuously differentiable maps from a fixed-dimensional parameter domain to $\mathbb R^n$. At the parameter point under consideration, let $J_{r,i},J_{e,i}$ be their row gradients and $K_{r,i},K_{e,i}$ their parameter Hessians. Define
\begin{align*}
 u_0&=\|r-e\|_{2,n},&
 u_1&=\{P_n\|J_{r,i}-J_{e,i}\|^2\}^{1/2},\\
 u_2&=\{P_n\|K_{r,i}-K_{e,i}\|_{\mathrm{op}}^2\}^{1/2},&
 B_{1,2}&=\{P_n\|J_{e,i}\|^2\}^{1/2},\\
 B_{1,4}&=\{P_n\|J_{e,i}\|^4\}^{1/4},&
 B_{2,2}&=\{P_n\|K_{e,i}\|_{\mathrm{op}}^2\}^{1/2}.
\end{align*}
All errors are unweighted empirical $L^2$ norms. Frobenius norms may replace matrix operator norms, yielding a stronger sufficient condition.

\begin{theorem}[Empirical-integral Hessian transfer]
\label{op-hessian-theorem}
With the preceding definitions,
\begin{align}
 &\|\nabla_\theta^2F(r(\theta))-
                   \nabla_\theta^2F(e(\theta))\|_{\mathrm{op}}
 \notag\\
 &\quad\le C\big[
 h^{-3}\mathcal L^{5/2}u_0B_{1,4}^2
 +h^{-2}\mathcal L^2u_1(2B_{1,2}+u_1)\notag\\
 &\hspace{19mm}+h^{-2}\mathcal L^2u_0B_{2,2}
 +h^{-1}\mathcal L^{3/2}u_2\big].
 \label{op-hessian-transfer}
\end{align}
The fitted jets enter the bound only through the empirical $L^2$ errors $u_0$, $u_1$ and $u_2$.
\end{theorem}
\begin{proof}
Test the matrix difference on unit parameter directions $s,t$. Set $v=J_rs$, $u=J_rt$, $v_e=J_es$, $u_e=J_et$, $w_i=s^{\mathsf T}K_{r,i}t$ and $w_{e,i}=s^{\mathsf T}K_{e,i}t$. The chain rule gives
\[
 s^{\mathsf T}\nabla_\theta^2F(r)t=D^2F_r[v,u]+DF_r[w].
\]
Split the first difference as
\begin{align*}
 D^2F_r[v,u]-D^2F_e[v_e,u_e]
 &=D^2F_r[v-v_e,u]+D^2F_r[v_e,u-u_e]\\
 &\quad+\int_0^1D^3F_{e+\tau(r-e)}[r-e,v_e,u_e]\,d\tau.
\end{align*}
The first two terms are bounded by $Ch^{-2}\mathcal L^2u_1(2B_{1,2}+u_1)$ using \eqref{op-bilinear}. Since only the oracle first jets enter the third derivative, \eqref{op-trilinear} bounds its integral by $Ch^{-3}\mathcal L^{5/2}u_0B_{1,4}^2$.

For the other chain-rule term,
\[
 DF_r[w]-DF_e[w_e]
 =DF_r[w-w_e]+\int_0^1D^2F_{e+\tau(r-e)}[r-e,w_e]\,d\tau.
\]
Since $\|w-w_e\|_{1,n}\le\|w-w_e\|_{2,n}\le u_2$, \eqref{op-linear} bounds the first part by $Ch^{-1}\mathcal L^{3/2}u_2$, and \eqref{op-bilinear} bounds the second by $Ch^{-2}\mathcal L^2u_0B_{2,2}$. Taking the supremum over $s,t$ gives \eqref{op-hessian-transfer}.
\end{proof}

Because the bound holds pointwise in the parameter, taking the suprema of its inputs over a local chart gives a uniform transfer directly. If the oracle moment norms are uniformly $O_{\mathbb P}(1)$, $u_1=o_{\mathbb P}(1)$ uniformly, and $h$ is bounded above, it yields the simplified fitted-to-oracle remainder
\begin{equation}
 R_{H,\mathrm{fit}}=
 O_{\mathbb P}\!\left[\mathcal L^{5/2}
       \left\{\frac{u_0}{h^3}+\frac{u_1}{h^2}+\frac{u_2}{h}\right\}\right].
 \label{op-hessian-rate}
\end{equation}
For example, a residual error vector with $k$ nonzero entries of unit magnitude has empirical norm $\sqrt{k/n}$, although its largest entry is one. The zeroth-jet term in \eqref{op-hessian-rate} accommodates such an error whenever $\mathcal L^{5/2}\sqrt{k/n}/h^3\to0$.

\section{Mixture-density and training-score bounds}\label{sec:core-error}
Here $f$ is a fixed finite Gaussian mixture with positive weights and variances, $\psi=f'/f$, $I_f=E_f\psi^2$, $h$ is the density bandwidth, and $\ell_n=\log(n+e)$. Constants depend on the fixed law. The empirical coefficient $a=n\nabla F$ includes contributions from both density endpoints, and all kernel row and column bounds are taken from Lemma~\ref{op-moment-lemma}.
\subsection{Gaussian-mixture envelopes}

Put $g_h=\ph_h*f$, $0<h\le1$. Component-wise Gaussian calculation gives
\begin{equation}\label{gm:mixture-envelope}
 f\le Cg_h,\qquad
 \frac{|f^{(j)}(e)|}{f(e)}+\frac{|g_h^{(j)}(e)|}{g_h(e)}
 \le C_j(1+|e|^j),\qquad j=1,2,3,4.
\end{equation}
For the first inequality, each component satisfies
\[
 \frac{\ph_{\sigma_l}(e-\zeta_l)}
 {\ph_{\sqrt{\sigma_l^2+h^2}}(e-\zeta_l)}
 \le\sqrt{1+h^2/\sigma_l^2}\le C.
\]
Summing over components preserves the inequality. Each Gaussian derivative ratio is a polynomial with coefficients bounded uniformly over the fixed means and variances, and a mixture derivative ratio is bounded by the largest absolute component ratio. Both $f$ and $g_h$ have uniform Gaussian tails.

For $0<\rho\le1/2$, define
\[
 A_\rho(h)=\int\frac{f(e)g_h(e)}{\{g_h(e)+\rho\}^2}\,de.
\]
Then
\begin{equation}\label{gm:tail-integrals}
 A_\rho(h)\le C\sqrt{\log(e/\rho)},\qquad
 \Pp\{g_h(\eps)<t\}\le Ct\sqrt{\log(e/t)},\quad0<t<1/2.
\end{equation}
To bound $A_\rho$, we use $f\le Cg_h$. The integrand is bounded on $[-R,R]$, and outside this interval it is at most $C\rho^{-2}g_h^2$. With $R=C_0\sqrt{\log(e/\rho)}$ and $C_0$ sufficiently large, the Gaussian tail integral is bounded. For the probability bound, integrate $f\le Cg_h<Ct$ over $[-R,R]$ and choose $R=C_0\sqrt{\log(e/t)}$ so that the remaining probability is at most $t^2$.

The first log-derivative bound also gives, for $|s|\le\eta$,
\begin{equation}\label{gm:shift-ratio}
 e^{-C\eta(1+|e|+\eta)}\le\frac{g_h(e+s)}{g_h(e)}
 \le e^{C\eta(1+|e|+\eta)}.
\end{equation}
Thus, polynomially vanishing shifts preserve constant-factor comparisons on $|e|\le R_n=C_0\sqrt{\ell_n}$. Increasing $C_0$ makes $\Pp(\max_i|\eps_i|>R_n)$ decay at any prescribed fixed polynomial rate.

\subsection{Population score and curvature}

Define
\begin{equation}\label{gm:population-score}
 s_{h,\rho}(e)=\frac{g_h'(e)}{g_h(e)+\rho},\qquad
 a_{h,\rho}(e)=s_{h,\rho}(e)
 +\left[\ph_h'*\frac f{g_h+\rho}\right](e).
\end{equation}
As $h,\rho\downarrow0$,
\begin{equation}\label{gm:population-score-limit}
 \E_fa_{h,\rho}=0,\qquad |a_{h,\rho}(e)|\le C(1+|e|),\qquad
 \norm{a_{h,\rho}-\psi}_{L^2(f)}\longrightarrow0.
\end{equation}
\begin{equation}
 J_{h,\rho}:=\E_f\left[s_{h,\rho}^2-
                  \frac{2g_h''}{g_h+\rho}\right]
 =I_f-\norm{\psi-s_{h,\rho}}_{L^2(f)}^2
 \longrightarrow I_f.\label{gm:curvature-identity}
\end{equation}
To prove these statements, set $r=f/(g_h+\rho)$. Equation~\eqref{gm:mixture-envelope} implies $r\le C$ and
\[
 |r'(u)|\le(f/g_h)\{|\psi(u)|+|g_h'(u)/g_h(u)|\}
 \le C(1+|u|).
\]
Integration by parts gives $\phi_h' * r = \phi_h * r'$, where $r'$ converges uniformly to 0 on compact sets. By these estimates and dominated convergence, the convolution term $E_Z[r'(e-hZ)]$ converges to 0 pointwise. Since both this convolution term and $s_{h,\rho}$ are bounded in absolute value by a function of the form $C(1+|e|)$, dominated convergence gives an $L^2$ limit and uniform upper bounds for $E a^2$ and $E(a^2\eps^2)$. By Fubini's theorem and oddness, the expectation of this convolution term is $-\int fg_h'/(g_h+\rho)$, so the two expectations cancel. Finally, $s'=g_h''/(g_h+\rho)-s^2$ and integration by parts give $J=2E\psi s-Es^2$; the boundary terms vanish by Gaussian tails.
\subsection{Empirical score error at training observation points}
In this section $h=c_h n^{-a}$ and $\rho=c_\rho/n$, where $a>0$, and $c_h$ and $c_\rho$ are fixed positive constants. Hence the logarithmic factor $\mathcal L$ in \eqref{op-moments} is $O(\log n)$. All constants below are uniform over a fixed, finite grid of positive multipliers.

For the iid error vector, write $e_i=\eps_i$, $g_i=g_h(e_i)$ and
\[
 d_i=\frac1{n-1}\sum_{j\ne i}\ph_h(e_i-e_j),\quad
 d_i'=\frac1{n-1}\sum_{j\ne i}\ph_h'(e_i-e_j),\quad
 R_i=\frac{g_i-d_i}{g_i+\rho}.
\]
The Gaussian inequality
\begin{equation}\label{gm:kernel-square}
 |\ph_h^{(j)}(u)|^2\le C_jh^{-2j-1}\ph_h(u)
\end{equation}
follows by bounding a squared Hermite polynomial times $\ph$. Conditionally on $\eps_i$, the leave-one-out summands have means $g_i$ and $g_i'$, and therefore
\begin{equation}\label{gm:relative-risk}
 E P_nR_i^2\le\frac{CA_\rho(h)}{nh},\qquad
 E P_n\left\{\frac{d_i'-g_i'}{g_i+\rho}\right\}^2
 \le\frac{CA_\rho(h)}{nh^3}.
\end{equation}
The outgoing score ratio satisfies
\begin{equation}\label{gm:ratio-identity}
 \frac{d_i'}{d_i+\rho}-\frac{g_i'}{g_i+\rho}
 =\frac{d_i'-g_i'}{g_i+\rho}
 +\frac{d_i'}{d_i+\rho}R_i.
\end{equation}
By \eqref{op-moments}, the outgoing ratio is bounded by $Ch^{-1}\ell_n^C$. Markov's inequality and \eqref{gm:relative-risk} therefore bound the empirical $L^2$ error in \eqref{gm:ratio-identity} by $\Op\{\ell_n^C/(\sqrt n h^{3/2})\}$.

When the incoming reciprocals are replaced by their population counterparts, the difference is $-h^{-2}\sum_j(e_i-e_j)P_{ji}R_j$. The nonnegative matrix $B_{ij}=h^{-2}|e_i-e_j|P_{ji}$ has maximum row and column sums at most $Ch^{-1}\ell_n^C$, by \eqref{op-moments}. Hence $\norm B_{2\to2}\le Ch^{-1}\ell_n^C$, and its action on $R$ has the same empirical $L^2$ rate.

With population denominators, the incoming empirical term has kernel $K_e(u)=\ph_h'(e-u)/(g_h(u)+\rho)$, whose conditional leave-one-out mean at $e_i$ is the convolution in \eqref{gm:population-score}. Integrating its conditional variance and exchanging the integrals gives
\begin{align}
 &E P_n\left|\frac1{n-1}\sum_{j\ne i}K_{e_i}(e_j)
                         -E K_{e_i}(\eps)\right|^2\nonumber\\
 &\quad\le\frac1{n-1}\iint
 \frac{f(e)f(u)|\ph_h'(e-u)|^2}{\{g_h(u)+\rho\}^2}\,du\,de
 \le\frac{CA_\rho(h)}{nh^3}.\label{gm:incoming-risk}
\end{align}
With $\norm{v}_n^2=n^{-1}\sum_i v_i^2$, these comparisons give
\begin{equation}\label{gm:iid-score-risk}
 \norm{a(\eps)-a_{h,\rho}(\eps)}_n
 =\Op\{\ell_n^C/(\sqrt n h^{3/2})\}.
\end{equation}
The expectation and Markov bounds apply regardless of the dependence among the leave-one-out errors.

\section{Integrated Gaussian regression jets and separation from uniform fitting}
\label{gauss-section}

This section gives first-stage bounds for the ridge-protected Nadaraya--Watson (NW) smoother. In this section only, $h$ denotes the regression bandwidth, which is called $g$ in the paper.

\subsection{Estimator, assumptions, and population calculation}
\label{gauss-setup}

Put $K_h(u)=h^{-1}\phi(u/h)$, $N=n-1$, and $s_i(\theta)=\beta(\theta)^\top X_i$, where $\beta(\theta)=(\beta_0+U\theta)/(1+\norm{\theta}^2)^{1/2}$. Define
\begin{align}
 D_i(\theta)&=N^{-1}\sum_{j\ne i}K_h(s_i-s_j),&
 N_i(\theta)&=N^{-1}\sum_{j\ne i}K_h(s_i-s_j)Y_j,\notag\\
 \widehat m_i^\lambda(\theta)&=\frac{N_i(\theta)}{D_i(\theta)+\lambda},&
 w_{ij}^\lambda(\theta)&=
 \frac{\ind\{j\ne i\}K_h(s_i-s_j)}
 {\sum_{k\ne i}K_h(s_i-s_k)+N\lambda}.
 \label{gauss-ridge}
\end{align}
Here $\lambda$ is measured in projected-density units and is held fixed in parameter derivatives.

Assume that $p$ is fixed, $X\sim N_p(\mu,\Sigma)$ with fixed positive definite $\Sigma$, and $Y=m(\beta_0^\top X)+\eps$, where $m\in C^6(\R)$ has bounded derivatives of orders zero through six. The error is centered, independent of $X$, and sub-Gaussian; every fixed finite Gaussian mixture with zero mean satisfies this requirement. Let
\begin{equation}
 h=c_hn^{-b},\qquad \lambda=c_\lambda n^{-\tau},\qquad
 b>0,\quad \tau>0,\quad b+\tau<1,
 \label{gauss-powers}
\end{equation}
with fixed positive multipliers, and the bounds are uniform over a fixed finite grid of such multipliers. Write $\ell=\log(n+e)$ and take $\mathcal B_n=\{\theta:\norm{\theta}\le r_n\}$, where $r_n=c_rn^{-d}$ for any fixed $d>0$. The radius may exceed $h$; the Gaussian density comparison below uses only $r_n\ell\to0$. Constants may depend on the fixed generating law, $p$, the positive multipliers, and $d$.

Let $g_\theta(x)=\E\{m(\beta_0^\top X')\mid \beta(\theta)^\top X'=\beta(\theta)^\top x\}$. For $X\sim N(0,I)$, set $s=\beta(\theta)^\top x$ and $c_\theta=\beta(\theta)^\top\beta_0$. If $q_h=\phi_{\sqrt{1+h^2}}$, the ordinary population kernel ratio and its ideal profile are given by
\begin{align}
 g_{\theta,h}(x)
 &=\E m\left\{\frac{c_\theta s}{1+h^2}
       +\sqrt{1-\frac{c_\theta^2}{1+h^2}}Z\right\},\notag\\
 g_\theta(x)
 &=\E m\{c_\theta s+\sqrt{1-c_\theta^2}Z\}.
 \label{gauss-population-formula}
\end{align}
For a general Gaussian design, both lines average $m$ under the Gaussian conditional law of $\beta_0^\top X'$ given the noisy index $\beta(\theta)^\top X'+hZ$, with $h=0$ in the second line, and the projected variances stay bounded above and below on a fixed chart neighborhood.

\begin{lemma}\label{gauss-population-lemma}
For $j=0,1,2$, the population ridged ratio
\[
 g_{\theta,h,\lambda}(x)=
 \frac{q_{\theta,h}(\beta(\theta)^\top x)}
 {q_{\theta,h}(\beta(\theta)^\top x)+\lambda}
 g_{\theta,h}(x)
\]
satisfies
\begin{align}
 \norm{\partial_\theta^j(g_{\theta,h}-g_\theta)(x)}
 &\le Ch^2(1+\norm{x})^C,\label{gauss-G1}\\
 \norm{\partial_\theta^j(g_{\theta,h,\lambda}-g_\theta)(x)}
 &\le C(1+\norm{x})^C
 \left\{h^2+\frac{\lambda}
 {q_{\theta,h}(\beta(\theta)^\top x)+\lambda}\right\}.
 \label{gauss-G2}
\end{align}
The polynomial in the second line may be taken to have degree at most four for the ridge part, and at most three for the ordinary smoothing part.
\end{lemma}

\begin{proof}
For $F(a,v)=\E m(a+\sqrt vZ)$, use the heat-semigroup identity $\partial_a^k\partial_v^lF =2^{-l}\E m^{(k+2l)}(a+\sqrt vZ)$, interpreted continuously at $v=0$. This removes the apparent square-root singularity in \eqref{gauss-population-formula}. The Gaussian conditional mean and variance, and their chart derivatives through order two, change by $O(h^2)(1+\norm{x})$ and $O(h^2)$ when $h$ is set to zero. The $C^6$ assumption therefore gives \eqref{gauss-G1}, in which the largest power of $x$ is three. Logarithmic derivatives of the Gaussian density have polynomial growth, so through order two the derivatives of $\lambda/(q_{\theta,h}+\lambda)$ are bounded by $C(1+\norm{x})^4\lambda/(q_{\theta,h}+\lambda)$. The product rule now gives \eqref{gauss-G2}.
\end{proof}

Put $T_i=\beta_0^\top X_i$. On $\max_i\norm{X_i}\le C_0\sqrt\ell$, uniformly over $\mathcal B_n$,
\begin{equation}
 C^{-1}\le
 \frac{q_{\theta,h}(s_i(\theta))}{q_{0,h}(T_i)}\le C.
 \label{gauss-G3}
\end{equation}
Indeed, the logarithm of this ratio is bounded by $Cr_n(1+\norm{X_i}^2)=o(1)$. By Gaussian tails, the complement of the sample event has probability $O(n^{-A})$ for any prescribed fixed $A$ after increasing $C_0$. Thus, a single function of $T_i$ dominates all chart values.

If $q$ is the normal density of $T$ and $q_h=q*K_h$, then $q\le Cq_h$ uniformly for $h\le1$. For every fixed $k$ and $v>1$, Gaussian tail integration gives
\begin{equation}
 \E\left[(1+|T|)^k
       \left\{\frac{\lambda}{q_h(T)+\lambda}\right\}^v\right]
 \le C\lambda\ell^C.
 \label{gauss-G4}
\end{equation}
The same order, with a possibly larger logarithmic factor, holds for $v=1$. To prove these bounds, split at $q_h=\lambda$, whose boundary lies at $O(\sqrt{\log(1/\lambda)})$. On the central set, integrate $\lambda^v q_h^{1-v}$, and outside it use the Gaussian tail probability. The standard Gaussian tail bound, obtained by integrating $ue^{-u^2/2}$, controls each boundary term. Equations \eqref{gauss-G2}--\eqref{gauss-G4} and one application of Markov's inequality to an empirical average imply
\begin{equation}
 \sup_{\theta\in\mathcal B_n}
 \norm{\partial_\theta^j(g_{\theta,h,\lambda}-g_\theta)(X)}_n
 =O_P\{\ell^C(h^2+\sqrt\lambda)\},\qquad j\le2,
 \label{gauss-G5}
\end{equation}
where $\norm{v}_n^2=n^{-1}\sum_i\norm{v_i}^2$.

\subsection{Uniform concentration before empirical integration}
\label{gauss-concentration}

For a deterministic evaluation point $x$, write $D(\theta,x)=\E K_h\{\beta(\theta)^\top(x-X)\}$ and $N(\theta,x)=\E[K_h\{\beta(\theta)^\top(x-X)\}Y]$. Use hats for their empirical versions, initially including all $n$ observations. Uniformly for $\norm{x}\le C_0\sqrt\ell$ and $\theta\in\mathcal B_n$, with probability $1-O(n^{-A})$,
\begin{align}
 &\norm{\partial^j(\widehat D-D)}+
       \norm{\partial^j(\widehat N-N)}\notag\\
 &\quad\le C\ell^5
 \left\{\sqrt{\frac{q_{\theta,h}(\beta(\theta)^\top x)}
                         {nh^{2j+1}}}
                +\frac1{nh^{j+1}}\right\},\qquad j=0,1,2.
 \label{gauss-G6}
\end{align}
Here and below the constant can depend on $A$.

To prove \eqref{gauss-G6}, note first that a chart derivative of the kernel is a finite sum of $K_h^{(l)}$, $l\le j$, times products of derivatives of $\beta(\theta)^\top(x-X)$. Conditional on $u=\beta(\theta)^\top X$, the remaining Gaussian coordinates have moments bounded by polynomials in $(\norm{x},|u|)$. The inequality $|K_h^{(l)}(v)|^2\le C_lh^{-2l-1}K_h(v)$ follows by bounding a squared Hermite polynomial times a Gaussian. Conditional Gaussian integration therefore bounds the second moment of each summand by $C\ell^3h^{-2j-1}q_{\theta,h}(\beta(\theta)^\top x)$. The same bound holds with the response factor, because $m$ is bounded and the independent error has fixed variance. Polynomial factors in $u$ are integrated under the Gaussian product density.

Fix $A>0$ and truncate all input coordinates and errors at $C_1\sqrt\ell$. Choose $C_0,C_1$ so that all omitted expectations through kernel derivative order three and the joint sample-tail probability are at most $n^{-A-20(1+b+\tau)}$; sub-Gaussian tails permit this choice. The derivative envelopes for $j\le2$ are then at most $C\ell^2h^{-j-1}$. We use a rectangular mesh of spacing $n^{-M}$ in the fixed $2p-1$ dimensional $(\theta,x)$ domain, with an integer
\begin{equation}
 M>20(1+b+\tau+A).
 \label{gauss-mesh}
\end{equation}
Bernstein's inequality with threshold $C_{A,M,p}\ell$, followed by a union bound over the mesh, gives \eqref{gauss-G6} there with failure probability $O(n^{-A})$.

For interpolation, differentiate the empirical and population deviations once more in $\theta$ or $x$. The numerator envelope is bounded by $C\ell^3h^{-4}$. Normalization by $q+\lambda$, including its derivative, costs at most $C\ell^2\lambda^{-2}$. Thus $C\ell^5n^{4b+2\tau}$ bounds the normalized Lipschitz constant. By \eqref{gauss-mesh}, the mesh error is smaller than $n^{-2}$, and it is therefore absorbed by the normalized linear Bernstein term. The Gaussian densities at neighboring grid points are comparable since their logarithmic derivatives are $O(\ell)$ on the evaluation window. Deleting observation $i$ changes a derivative by at most $C\ell^2/(nh^{j+1})$, which is already included in \eqref{gauss-G6}. Therefore, the inequality holds for all leave-one-out rows after evaluation at $X_i$. Independence is used only for the bounds at deterministic evaluation points.

For $j=0$, the largest relative error after division by $q+\lambda$ is at most $C\ell^5\{(nh\lambda)^{-1/2}+(nh\lambda)^{-1}\}=o(1)$. Hence, simultaneously over all rows and chart values,
\begin{equation}
 D_i(\theta)+\lambda\ge
 c\{q_{\theta,h}(s_i(\theta))+\lambda\}.
 \label{gauss-G7}
\end{equation}

\subsection{Influence geometry and integrated jet bounds}
\label{gauss-jets}

Set $P_{ij}=w_{ij}^\lambda$. Its row masses are at most one. Since $P_{ij}\le w_{ij}^0$ entrywise, its column masses are bounded by the column bound for the leave-one-out Gaussian NW weights without ridge, established in the proof of Lemma~\ref{op-moment-lemma}. For every fixed integer $k\ge0$,
\begin{equation}
 \max_i\sum_j P_{ij}|s_i-s_j|^k
 +\max_j\sum_i P_{ij}|s_i-s_j|^k\le Ch^k\ell^C.
 \label{gauss-G8}
\end{equation}
For the near part, split at $|s_i-s_j|\le Ah\sqrt\ell$ and use the row and column masses. For the far part use $P_{ij}\le C(n\lambda h)^{-1} \exp\{-(s_i-s_j)^2/(2h^2)\}$, sum at most $n$ terms, and choose fixed $A$ large enough to dominate the fixed powers $b,\tau$. The argument is pathwise.

On the common Gaussian sample event, chart derivatives of $s_i-s_j$ are $O(\sqrt\ell)$. Quotient differentiation and \eqref{gauss-G8} give
\begin{equation}
 \max_i\frac{\norm{\partial_\theta^kD_i}}{D_i+\lambda}
 \le C\ell^C h^{-k},\qquad k=1,2.
 \label{gauss-G9}
\end{equation}
Write $\widehat m-g_\lambda= \{(\widehat N-N)-g_\lambda(\widehat D-D)\}/(\widehat D+\lambda)$. Differentiate this identity up to order $j$. Each lower-order error of order $l$ is multiplied by at most $C\ell^4h^{-(j-l)}$. Equations \eqref{gauss-G3}, \eqref{gauss-G6}, \eqref{gauss-G7}, and \eqref{gauss-G9} therefore yield
\begin{align}
 &\sup_{\theta\in\mathcal B_n}
 \norm{\partial^j\{\widehat m_i-g_{\theta,h,\lambda}(X_i)\}}
 \notag\\
 &\quad\le C\ell^9
 \frac{\sqrt{q_{0,h}(T_i)/(nh^{2j+1})}+1/(nh^{j+1})}
      {q_{0,h}(T_i)+\lambda},\qquad j\le2.
 \label{gauss-G10}
\end{align}
The Gaussian integrals needed for empirical averaging are
\begin{equation}
 \E\frac{q_h(T)}{\{q_h(T)+\lambda\}^2}\le C\ell^C,
 \qquad
 \E\frac1{\{q_h(T)+\lambda\}^2}\le C\lambda^{-1}\ell^C.
 \label{gauss-G11}
\end{equation}
For example, the first integral is bounded by $C\int q_h^2/(q_h+\lambda)^2$, and splitting at $q_h=\lambda$ gives $O(\sqrt{\log(1/\lambda)})$. The second follows from $C\int q_h/(q_h+\lambda)^2$ and the same split.

\begin{lemma}[Integrated ridge jets]\label{gauss-integrated-jets}
Under \eqref{gauss-powers}, uniformly over $\mathcal B_n$,
\begin{align}
 &\norm{\partial_\theta^j(\widehat m-g_\theta)(X)}_n\notag\\
 &\quad=O_P\left[\ell^{20}\left\{h^2+\sqrt\lambda+
 (nh^{2j+1})^{-1/2}+\frac1{nh^{j+1}\sqrt\lambda}\right\}\right]
 \notag\\
 &\quad=O_P\left[\ell^{20}\{h^2+\sqrt\lambda+
                    (nh^{2j+1})^{-1/2}\}\right],\quad j\le2.
 \label{gauss-G12}
\end{align}
More precisely, for every fixed $A>0$ there are $C,n_0$ such that, for $n\ge n_0$, the probability that any of the three suprema exceeds $C\ell^{20}\{h^2+\sqrt\lambda+(nh^{2j+1})^{-1/2}\}$ is at most $Cn^{-A}+C\ell^{-2}$.
\end{lemma}

\begin{proof}
Square \eqref{gauss-G10}, average, and apply Markov's inequality once to the common functions in \eqref{gauss-G11}. Apply the same operation to the squared ridge-bias bound in \eqref{gauss-G2}, using \eqref{gauss-G4}. These common empirical functions do not depend on $\theta$, because \eqref{gauss-G3} was proved first. Their expectations, including the polynomial factors in the bias, are bounded by the displayed powers of $\lambda$ times at most $\ell^5$. Markov's inequality at level $\ell^2$ gives failure probability $O(\ell^{-2})$. Square roots add at most $\ell^{7/2}$ to the factor $\ell^9$, so the exponent $20$ suffices. Finally, the ratio of the linear Bernstein term to the square-root term is $(nh\lambda)^{-1/2}=o(1)$. The polynomial sample and concentration events contribute $O(n^{-A})$.
\end{proof}

For a fixed leave-one-out training set, define
\[
 \widehat m_{-i}(t)=
 \frac{N^{-1}\sum_{j\ne i}K_h(t-T_j)Y_j}
 {N^{-1}\sum_{j\ne i}K_h(t-T_j)+\lambda}.
\]
The preceding proof applies to $K_h^{(l)}(t-T_j)$, $l=0,1$. Its squared moment is at most $Ch^{-2l-1}q_h(t)$ and its truncated envelope is at most $C\ell^2h^{-l-1}$. The same quotient calculation and Gaussian integrals give
\begin{equation}
 \norm{A-m'(T)}_n
 =O_P\{\ell^{20}[h^2+\sqrt\lambda+(nh^3)^{-1/2}]\},\qquad
 A_i=\left.\partial_t\widehat m_{-i}(t)\right|_{t=T_i}.
 \label{gauss-G13}
\end{equation}
For the bandwidth operator $B_h=h\partial_h$, with $\lambda$ fixed, $B_hK_h(d)=\{(d/h)^2-1\}K_h(d)$. Its envelope and integrated squared order are both those of an ordinary zeroth-order kernel, namely $h^{-1}$ in each case. Thus, the same concentration proof gives the $j=0$ rate for $B_h$ derivatives. The population term $B_hg_h$ is $O(h^2)(1+|t|)^C$, and ridge differentiation adds at most a polynomial times $\lambda/(q_h+\lambda)$.

\subsection{The conditional tangent mean under a Gaussian design}
\label{gauss-conditional-mean}

In raw coordinates, write $V=U^\top X=\mu_V(T)+W$ with $\mu_V(t)=Bt+b_0$. The conditional covariance of $W$ is constant. At $\theta=0$ put
\[
 q_{ij}=\frac{\ind\{i\ne j\}K_h'(T_i-T_j)}
 {\sum_{k\ne i}K_h(T_i-T_k)+N\lambda},\qquad
 D_i^\mu=\sum_jq_{ij}\{\mu_V(T_i)-\mu_V(T_j)\}(Y_j-\widehat m_i).
\]
The ridge leaves the coefficient decomposition of the score and the conditional $W$-centering argument as in the case $\lambda=0$; in the noncentered term it adds the explicit row-mass correction displayed below. Since $K_h'(d)d=-(h\partial_hK_h(d)+K_h(d))$, quotient algebra gives
\begin{equation}
 D_i^\mu=-B\left\{h\partial_h\widehat m_i(h)
             +\frac{\lambda\widehat m_i(h)}{D_i+\lambda}\right\}.
 \label{gauss-G14}
\end{equation}
The ridge term $\lambda\widehat m_i(h)/(D_i+\lambda)$ is absent for an NW fit without a ridge, where the rows sum to 1. The population counterpart of $D_i^\mu$ has empirical $L^2$ order $O_P\{\ell^C(h^2+\sqrt\lambda)\}$ by the preceding bandwidth calculation and \eqref{gauss-G4}.

Condition on all $T_i$ and decompose $D^\mu=b^\mu+G\eps$, where $b^\mu$ uses the noiseless responses $m(T_j)$. Define
\begin{equation}
 l_{ij}=q_{ij}B(T_i-T_j),\qquad s_i=\sum_jl_{ij},\qquad
 G_{ij}=l_{ij}-s_iw_{ij}^\lambda,\qquad G_{ii}=0.
 \label{gauss-G15}
\end{equation}
The diagonal vanishes because the leave-one-out fit omits observation $i$. The zeroth-order concentration proof, applied also to $h\partial_hK_h$, and \eqref{gauss-G14} give
\begin{equation}
 \norm{b^\mu}_n
 =O_P\{\ell^C[h^2+\sqrt\lambda+(nh)^{-1/2}]\}.
 \label{gauss-G16}
\end{equation}

Let $Q=(q_{ij})$, and let $E_D$ denote the denominator event \eqref{gauss-G7}. On $E_D$, exchangeability and conditional integration give
\begin{align*}
 \E\left[\ind_{E_D}\sum_{i\ne j}q_{ij}^2\right]
 &\le CN^{-2}\sum_{i\ne j}
 \E\frac{K_h'(T_i-T_j)^2}{\{q_h(T_i)+\lambda\}^2}\\
 &\le Ch^{-3}\E\frac{q_h(T)}{\{q_h(T)+\lambda\}^2}
 \le Ch^{-3}\ell^C.
\end{align*}
The factor $N^{-2}$ cancels the order $n^2$ ordered pairs. For $G$, \eqref{gauss-G8} bounds $|s_i|$ by $C\ell^C$, and $|l_{ij}|\le C\{(T_i-T_j)/h\}^2w_{ij}^\lambda$, so that $|G_{ij}|^2\le C\ell^C[1+\{(T_i-T_j)/h\}^4] (w_{ij}^\lambda)^2$. The corresponding polynomially weighted squared kernel has integrated order $h^{-1}$, and the same calculation gives $\E[\ind_{E_D}\norm{G}_F^2]\le C\ell^C/h$. Markov's inequality, the probability of the denominator event, and the pathwise row and column bounds give
\begin{equation}
 \norm{G}_F^2=O_P(\ell^C/h),\qquad
 \norm{Q}_F^2=O_P(\ell^C/h^3),\qquad
 \norm{|Q|}_{\mathrm{op}}\le C\ell^C/h.
 \label{gauss-G17}
\end{equation}
These bounds suffice for averaging conditional variances.

For example, let $z_i=a_{h_1,\rho}(\eps_i)$ with $\E z_i=0$ and uniformly bounded $\E z_i^2$ and $\E z_i^2\eps_i^2$, as shown in Section~\ref{sec:core-error}. Then $n^{-1/2}\sum_i z_i b_i^\mu=o_P(1)$ whenever the bound in \eqref{gauss-G16} vanishes. In $n^{-1/2}\sum_{i,j}z_iG_{ij}\eps_j$, only identical or reversed ordered pairs contribute to the conditional variance, because $G_{ii}=0$, $\E z_i=0$, and $\E\eps_i=0$, and this variance is at most $C\norm{G}_F^2/n=o_P(1)$. Also $\norm{G\eps}_n=O_P\{\ell^C(nh)^{-1/2}\}$, so the same $u_0$ order controls $\norm{D^\mu}_n$. Replacing $z$ by the empirical coefficient $a(r)$ costs $\sqrt n\norm{a(r)-z}_n\norm{D^\mu}_n$.

\subsection{Separation from uniform fit consistency}
\label{gauss-separation}

Take $T_i\sim N(0,1)$, independent nondegenerate Gaussian-mixture errors, and $m=\sin$ or $m=\cos$. The covariates in this example may be correlated; for instance, take $p=2$, $\beta_0=(1,0)^\top$ and $\Sigma=\left(\begin{smallmatrix}1&1/2\\1/2&1\end{smallmatrix}\right)$. Only the marginal distribution of $T$ is used below.

First, consider the leave-one-out NW fit without a ridge. If $I_1,I_2,I_3$ index the three largest projected observations and $h=n^{-b}$, then
\begin{equation}
 w^0_{I_1,I_2}\longrightarrow_P1,\qquad
 \widehat m^0_{I_1}-m(T_{I_1})=\eps_{I_2}+o_P(1).
 \label{gauss-G18}
\end{equation}
To see this, put $a_n=n^{-b/2}$ and choose $\gamma\in(\max\{0,1-b/4\},1)$. The number of projected observations above $t_n=\sqrt{2\gamma\log n}$ diverges in probability. The expected number of pairs above $t_n$ whose separation is at most $a_n$ is bounded by
\[
 Cn^2a_n\int_{t_n-a_n}^{\infty}\phi(t)^2dt
 \le Cn^{2-2\gamma-b/2}/\sqrt{\log n}\longrightarrow0.
\]
Hence, with probability tending to 1, both spacings between the three largest order statistics exceed $a_n$; on this event, the kernel weight ratios for all points except the nearest neighbor are at most $\exp(-n^b/2)$, and the sum of these ratios over all $n$ indices goes to 0. By the Gaussian tail integral, the two largest observations fall in the interval $[u_n-2\log\log(n)/u_n, u_n+2\log\log(n)/u_n]$ (where $u_n=\sqrt{2\log n}$), with probability tending to 1, so their difference tends to 0. The Lipschitz continuity of $m$ and $\max_i|\eps_i|=O_P(\sqrt{\ell})$ then give \eqref{gauss-G18}. Since the sorted index depends only on $T$, $\eps_{I_2}$ has the nondegenerate error distribution. Hence the maximum fitting error of the NW fit without a ridge does not converge to 0.

For the ridge-protected estimator, the separation is stronger. Let $0<\tau<1-b$, $t_n=\sqrt{2\tau\log n}$, and select a deterministic interval $I_n\subset[t_n+2\pi,t_n+6\pi]$ of a fixed positive length on which $m(t)\ge1/2$. Such an interval exists for both sine and cosine. Uniformly there,
\[
 q_h(t)/\lambda\le C\exp\{-c\sqrt{\log n}\}\longrightarrow0,
 \qquad
 \Pp(T\in I_n)\ge cn^{-\tau}\exp\{-C\sqrt{\log n}\}.
\]
The convolution does not change these orders because $h^2\log n\to0$. Hence
\begin{equation}
 \Pp\{T_i\notin I_n\text{ for all }i\}
 \le\exp\{-n\Pp(T\in I_n)\}\longrightarrow0.
 \label{gauss-G19}
\end{equation}
The population ridge fit is uniformly $o(1)$ on $I_n$, whereas \eqref{gauss-G10} with $j=0$ bounds the empirical discrepancy at all sample points in $I_n$ by $C\ell^9\{(nh\lambda)^{-1/2}+(nh\lambda)^{-1}\}=o_P(1)$. Thus
\begin{equation}
 \Pp\left\{\max_i
 |\widehat m_i^\lambda(0)-m(T_i)|\ge\tfrac14\right\}
 \longrightarrow1.
 \label{gauss-G20}
\end{equation}
At the same time, \eqref{gauss-G12} shows that the empirical $L^2$ norms of the jet errors $\partial_\theta^j(\widehat m-g_\theta)$ of orders zero, one, and two all converge to zero when $b<1/5$ and $b+\tau<1$. Finally, $\max_i|T_i|\to\infty$ in probability, so the projected density admits no common positive lower bound at the observations.

\section{Gaussian oracle curvature on a common tail set}\label{sec:oracle}
This section concerns ideal conditional-mean residuals; Theorem~\ref{op-hessian-theorem} connects them to the fitted residuals. Let $X\sim N_p(\mu,\Sigma)$, with fixed positive-definite $\Sigma$, and let $m$ and its first six derivatives be bounded and continuous. The independent error has a fixed, finite Gaussian mixture density $f$ with positive variances. Use $\beta(\theta)=(\beta_0+U\theta)/(1+\|\theta\|^2)^{1/2}$ and put $g_\theta(x)=\E[m(\beta_0^TX')\mid\beta(\theta)^TX'=\beta(\theta)^Tx]$, $\delta_\theta(x)=m(\beta_0^Tx)-g_\theta(x)$, and $e_i(\theta)=\epsilon_i+\delta_\theta(X_i)$. Throughout $h=c_1n^{-a}$, $\rho=c/n$, $\ell=\log(n+e)$, $r_n=C_rn^{-d}$, $d>0$, and all multipliers are fixed positive constants.

\subsection{Profile derivatives and population information}
Set $s_\theta^2=\beta(\theta)^T\Sigma\beta(\theta)$, $c_\theta=\beta_0^T\Sigma\beta(\theta)$, $v_\theta=\beta_0^T\Sigma\beta_0-c_\theta^2/s_\theta^2$, and
\[
 u_\theta(x)=\beta_0^T\mu+
 \frac{c_\theta}{s_\theta^2}\beta(\theta)^T(x-\mu).
\]
The conditional Gaussian formula gives $g_\theta(x)=P_{v_\theta}m(u_\theta(x))$, where $P_vm(u)=\E m(u+\sqrt v Z)$ and $Z\sim N(0,1)$. The heat identities
\[
 \partial_u^j\partial_v^kP_vm=2^{-k}P_vm^{(j+2k)},\qquad j+2k\le6,
\]
follow by Gaussian integration by parts for $v>0$ and continuous extension at zero. They remove the apparent square-root singularity of $v_\theta$ at zero. Chain rules therefore give, on a fixed small chart,
\begin{equation}\label{eq:oracle-jets}
 |g_\theta(x)|\le\|m\|_\infty,\qquad
 \|\partial_\theta^jg_\theta(x)\|\le C_j(1+\|x\|^j),\quad j\le3.
\end{equation}
In particular, $|\delta_\theta|\le2\|m\|_\infty$ everywhere and $|\delta_\theta(x)|\le Cr_n(1+\|x\|)$ on the shrinking chart. The oracle first-jet empirical $L^4$ and second-jet empirical $L^2$ norms are uniformly $O_P(1)$ by Gaussian moments. With $T=\beta_0^TX$, $V=U^TX$, $W=V-\E(V\mid T)$, differentiation gives
\[
 \delta'_0(X)=-m'(T)W,\quad
 \E\delta'_0=\E\delta''_0=0,\quad
 \E\delta'_0\delta_0^{\prime T}=M.
\]
For the first identity, $v'_0=0$ and $u'_0=V-\E(V\mid T)$; the zero means follow by differentiating $\E\delta_\theta=0$, justified by \eqref{eq:oracle-jets}.

Write $g_h=f*\phi_h$ and
\[
 D_\theta(x,e)=\E_{X'}g_h\{e+\delta_\theta(x)-\delta_\theta(X')\},
 \qquad Q_{h,\rho}(\theta)=\E\log\{D_\theta(X,\epsilon)+\rho\}.
\]
Gaussian-mixture component calculations imply, uniformly $h\le1$, $|g_h^{(j)}(e)|/g_h(e)\le C_j(1+|e|^j)$ for $j\le3$ and
\begin{equation}\label{eq:oracle-ratio}
 e^{-C(1+|e|)}\le g_h(e+s)/g_h(e)\le e^{C(1+|e|)},
 \qquad |s|\le4\|m\|_\infty.
\end{equation}
Equations \eqref{eq:oracle-jets}--\eqref{eq:oracle-ratio} bound the first three relative derivatives of $D_\theta$ by a fixed polynomial in $\|x\|$ times $\exp\{C(1+|e|)\}$. This envelope has all the needed moments. Thus, for $H_\theta=\partial_\theta^2\log(D_\theta+\rho)$,
\begin{equation}\label{eq:oracle-outer}
 \sup_{\|\theta\|\le r_n}\|(P_n-P)H_\theta\|
 =O_P(n^{-1/2}+r_n),\qquad
 \sup_{\|\theta\|\le r_n}\|Q''_{h,\rho}(\theta)-Q''_{h,\rho}(0)\|=O(r_n).
\end{equation}
Indeed, apply Chebyshev's inequality at zero, and bound the remaining difference by $r_n$ times the empirical and population envelopes of $\partial_\theta H_\theta$. If $s=g_h'/(g_h+\rho)$, differentiation gives
\begin{equation}\label{eq:oracle-info}
 Q'_{h,\rho}(0)=0,\quad Q''_{h,\rho}(0)=-J_{h,\rho}M,
 \quad J_{h,\rho}=\E_f\{s^2-2g_h''/(g_h+\rho)\}
 =I_f-\|s-\psi\|_{L^2(f)}^2.
\end{equation}
For the last equality use $s'=g_h''/(g_h+\rho)-s^2$ and $\E_fs'=-\E_f\psi s$. Gaussian tails justify the boundary term.

\subsection{Uniform row replacement}
Choose $2a<\kappa<1-5a$ and $t_n=n^{-\kappa}$. The good evaluation set $\{g_h(\epsilon_i)\ge t_n\}$ is independent of $\theta$. Since $f\le Cg_h$, splitting at $B\sqrt{\log(e/t)}$ gives
\begin{equation}\label{eq:oracle-lowdensity}
 \Pp\{g_h(\epsilon)<t\}\le Ct\sqrt{\log(e/t)}.
\end{equation}
On the central interval, its probability is at most $Ct$ times the interval length; choose $B$ so that the outside Gaussian tail is at most $t^2$.

For any $A>0$, let $R_X=C_X\sqrt{\ell}$ and $R_e=C_e\sqrt{\ell}$. Outside an event of probability $O(n^{-A})$, all observed predictors and errors fall within these balls. We truncate the population integral over $X'$ to a larger logarithmic ball whose tail moments are $O(n^{-B})$. Within these balls, $|\delta_\theta(x)-\delta_\theta(x')|\le Cr_n\sqrt{\ell}$. By the log-density derivative bounds and $r_n\ell\to0$, $g_h(e+\delta_\theta(x)-\delta_\theta(x'))\asymp g_h(e)$ uniformly. For the finite mixture, there is a constant $C_*$ such that $g_h(e)\ge n^{-C_*}$ for all $|e|\le R_e+1$. We choose $B$ larger than $C_*+3a+\kappa$ and the mesh-interpolation exponents, plus $A+10$. After normalization, the tail errors, including those for the bad-density rows treated below, are then negligible. In particular, $D_\theta(x,e)\asymp g_h(e)$ on these logarithmic balls containing the observed data.

A total $j$th parameter derivative, $j\le2$, of a kernel summand is a finite sum of $\phi_h^{(l)}$, $l\le j$, times profile jets bounded by $\ell^C$ on these balls. The elementary bound $|\phi_h^{(l)}(u)|^2\le C_lh^{-2l-1}\phi_h(u)$ and integration over $\epsilon'$ give variance at most $C\ell^Ch^{-2j-1}g_h(e)$ and envelope $C\ell^Ch^{-j-1}$. Bernstein on a deterministic polynomial net in $(\theta,x,e)$ yields normalized errors
\begin{equation}\label{eq:oracle-bernstein}
 v_j=C\ell^C\left\{(nh^{2j+1}t_n)^{-1/2}
                    +(nh^{j+1}t_n)^{-1}\right\},\qquad j=0,1,2.
\end{equation}
The independent units are the pairs $(X_j,\epsilon_j)$. Establish the bounds for their common empirical measure before evaluation at $(X_i,\epsilon_i)$. Deleting one observation contributes at most the second term in \eqref{eq:oracle-bernstein}. Third-order parameter derivatives, first-order evaluation derivatives, the $C^6$ profile formula, and the polynomial lower bounds yield polynomial interpolation constants. Use a fine enough polynomial net on the enlarged set $g_h(e)\ge t_n/2$; the bound $|(\log g_h)'|\le C\sqrt\ell$ makes it cover $g_h(e)\ge t_n$. Its log-cardinality is $O(\ell)$, and the exceptional probability in Bernstein's inequality can be made $O(n^{-A})$.

Since $v_0=o(1)$, good empirical and population denominators are comparable. Expanding $D''/(D+\rho)-D'D'^T/(D+\rho)^2$ shows the good-row replacement error is
\begin{equation}\label{eq:oracle-good}
 O_P\left[\ell^C\{(nh^5t_n)^{-1/2}+(nh^3t_n)^{-1}\}\right].
\end{equation}
For every bad row, the row-weight bounds of Lemma~\ref{op-moment-lemma} and the logarithmically bounded oracle jets give an empirical row-Hessian bound $Ch^{-2}\ell^C$ uniformly in $\theta$. The population row Hessian is bounded by $\ell^C$ on the same sample balls, using the near-shift comparison and the chosen negligible inner tail. By \eqref{eq:oracle-lowdensity}, $P_n\ind\{g_h(\epsilon)<t_n\}=O_P(t_n\ell^{1/2})$. Thus, the total bad-row error is $O_P(t_nh^{-2}\ell^C)$.

\begin{proposition}[Gaussian oracle curvature]\label{prop:oracle}
Under the conditions of this section, if $0<a<1/7$ and $2a<\kappa<1-5a$, then
\begin{align}\label{eq:oracle-final}
 &\sup_{\|\theta\|\le r_n}
   \|\partial_\theta^2F(e(\theta))+I_fM\|\nonumber\\
 &\quad=O_P\left[\ell^C\{(nh^5t_n)^{-1/2}+(nh^3t_n)^{-1}
               +t_nh^{-2}+n^{-1/2}+r_n\}\right]
       +O\{\|s-\psi\|_{L^2(f)}^2\}=o_P(1).
\end{align}
\end{proposition}
The proposition follows by combining \eqref{eq:oracle-outer}, \eqref{eq:oracle-info}, and the two row replacements. The quantitative mixture comparison in Lemma~\ref{lem:mixture-rate} gives a rate for the last term.

\section{Score expansion with integrated fitting errors}\label{sec:score}
Observations $(X_i,\eps_i)$ are iid, with mean-zero error independent of the design. This section expands the score relative to the efficient limit; the finite-smoothing expansion of the paper, which uses the population smoothed profile as its reference, is developed in Section~\ref{inf:theory}. This section treats the score computed on the fitting sample, using the derivative bounds of Section~\ref{supp:operator} and the Gaussian regression and oracle-curvature bounds of Sections~\ref{gauss-section} and~\ref{sec:oracle}.

\subsection{Coefficients and empirical primitives}
Write $h>0$, $g>0$, $\rho>0$, $\lambda\ge0$, $N=n-1$, and use normalized Gaussian kernels. The Gaussian results use $\lambda>0$. At the true direction let $T_i=\beta_0^TX_i$, $V_i=U^TX_i$, and define
\begin{align}
 D_i&=\sum_{j\ne i}\phi_g(T_i-T_j)+N\lambda,&
 H_{ij}&=\frac{\ind\{i\ne j\}\phi_g(T_i-T_j)}{D_i},\nonumber\\
 \widehat m_i&=\sum_jH_{ij}Y_j,&
 Q_{ij}&=\frac{\ind\{i\ne j\}\phi_g'(T_i-T_j)}{D_i},\nonumber\\
 A_i&=\sum_jQ_{ij}(Y_j-\widehat m_i),&r_i&=Y_i-\widehat m_i.
 \label{eq:nw-coeff}
\end{align}
The ridge is in projected-density units. The quotient rule in the sphere chart gives
\[
 \partial_\theta\widehat m_i(0)
 =A_iV_i-\sum_jQ_{ij}(Y_j-\widehat m_i)V_j.
\]
For the residual log likelihood $F$ define $a(x)=n\nabla_xF(x)$. Its coefficient, including incoming density centers, is
\begin{align}
 a_i(x)&=\frac1N\sum_{j\ne i}\phi_h'(x_i-x_j)
 \left\{\frac1{d_i(x)+\rho}+\frac1{d_j(x)+\rho}\right\},\nonumber\\
 \nabla_\theta L_n(0)&=\frac1n\sum_iV_i c_i,\qquad
 c_i=-a_i(r)A_i+\sum_{k\ne i}a_k(r)Q_{ki}(Y_i-\widehat m_k).
 \label{eq:full-score}
\end{align}
Let $\mu_i=\E(V_i\mid T_i)$ and $W_i=V_i-\mu_i$. Exchanging the incoming indices gives the conditional-mean identity
\begin{equation}
 \frac1{\sqrt n}\sum_i\mu_i c_i
 =-\frac1{\sqrt n}\sum_i a_i(r)D_i^\mu,
 \qquad D_i^\mu=\sum_jQ_{ij}(\mu_i-\mu_j)(Y_j-\widehat m_i).
 \label{eq:mu-exact}
\end{equation}

Write $\|v\|_{n,p}=(n^{-1}\sum_i|v_i|^p)^{1/p}$, and let
\[
 u_0=\|\widehat m-m(T)\|_{n,2},\quad
 \alpha=\|A-m'(T)\|_{n,2},\quad
 B_Q=\||Q|\|_{2\to2},\quad F_Q=\|Q\|_F/\sqrt n.
\]
These are empirical fit and influence quantities. Put
\begin{align*}
 \ell_{ij}&=Q_{ij}(\mu_i-\mu_j),&s_i&=\sum_j\ell_{ij},\\
 b_i&=\sum_j\ell_{ij}\{m(T_j)-(Hm)_i\},&G_{ij}&=\ell_{ij}-s_iH_{ij}.
\end{align*}
Then $D^\mu=b+G\eps$ and $G_{ii}=0$, including the case of ridge weights that sum to less than one. Set
\[
 u_\mu=\|D^\mu\|_{n,2},\qquad b_\mu=\|b\|_{n,2},\qquad
 F_\mu=\|G\|_F/\sqrt n.
\]
The zero diagonal follows from the leave-one-out construction.

\subsection{A quantitative fixed-mixture score lemma}
Let $f$ be a fixed finite Gaussian mixture with positive component variances, $g_h=\phi_h*f$, $\psi=f'/f$, and
\[
 z_i=a_{h,\rho}(\eps_i),\qquad
 a_{h,\rho}(e)=\frac{g_h'(e)}{g_h(e)+\rho}
   +\left[\phi_h'*\frac f{g_h+\rho}\right](e).
\]
The training-error calculation in \eqref{gm:iid-score-risk} yields, for $\rho=c/n$ and polynomial $h$, $\|a(\eps)-z\|_{n,2}=\Op(\log^C(n)/(\sqrt n h^{3/2}))$. The proof conditions on each leave-one-out row and controls the incoming operator by its row and column sums. We also have $\E z=0$, $|z(e)|\le C(1+|e|)$, and uniformly bounded $\E z^2$ and $\E(z^2\eps^2)$.

\begin{lemma}[Quantitative density-score approximation]\label{lem:mixture-rate}
For all sufficiently small $h,\rho>0$,
\begin{equation}
 \|a_{h,\rho}-\psi\|_{L^2(f)}
 +\left\|\frac{g_h'}{g_h+\rho}-\psi\right\|_{L^2(f)}
 \le C\{h^2+\sqrt\rho\log^C(e/\rho)\}.
 \label{eq:mixture-rate}
\end{equation}
\end{lemma}
\begin{proof}
For each mixture component, the convolved-to-original density ratio is
\[
 \frac{\phi_{\sqrt{\sigma^2+h^2}}(e-\zeta)}{\phi_\sigma(e-\zeta)}
 =\frac\sigma{\sqrt{\sigma^2+h^2}}
 \exp\left\{\frac{h^2(e-\zeta)^2}{2\sigma^2(\sigma^2+h^2)}\right\}.
\]
The mean-value theorem in $h^2$, finite summation, and $f/g_h\le C$ imply
\[
 \frac{|g_h-f|}{f}\le Ch^2(1+e^2)e^{Ch^2e^2},\qquad
 \frac{|g_h'-f'|}{f}\le Ch^2(1+|e|^3)e^{Ch^2e^2}.
\]
Therefore $g_h'/g_h-\psi$ and $(f/g_h)'$ have bounds $Ch^2$ times a fixed polynomial and $e^{Ch^2e^2}$. Gaussian convolution preserves this form after enlarging the constants. For sufficiently small $h$, its squared integral under the fixed mixture is uniformly finite. Integration by parts gives $\phi_h'*(f/g_h)=\phi_h*(f/g_h)'$, which proves the $O(h^2)$ bound for $\rho=0$.

Put $s_0=g_h'/g_h$ and $r_0=f/g_h$. The two floor differences are
\[
 \frac{g_h'}{g_h+\rho}-s_0=-\frac{s_0\rho}{g_h+\rho},\qquad
 \left(\frac f{g_h+\rho}\right)'-r_0'
 =-\frac{r_0'\rho}{g_h+\rho}
   +\frac{r_0\rho g_h'}{(g_h+\rho)^2}.
\]
Their absolute values are bounded by $C(1+|e|)\rho/(g_h(e)+\rho)$ using the Gaussian-mixture derivative ratios. On $|e|\le R=C_0\sqrt{\log(e/\rho)}$, the relevant weighted square integrals are $O(\rho\log^C(e/\rho))$, since $f\le Cg_h$ and $g_h\rho^2/(g_h+\rho)^2\le\rho/4$. Outside this interval, Gaussian tails and $\rho/(g_h+\rho)\le1$ give $O(\rho^2)$ after increasing $C_0$. For the convolution of the incoming derivative difference $b$, Jensen and Fubini give
\[
 \int f(e)\{(\phi_h*b)(e)\}^2\,de
 \le\int g_h(u)b(u)^2\,du.
\]
The same bound holds under $g_h$, proving the result. All constants may depend on the fixed mixture.
\end{proof}

\subsection{Score expansion}
Lemma~\ref{op-derivative-lemma} gives, for every residual array,
\[
 \|a(x)\|_\infty\le Ch^{-1}\log^{3/2}(n),\qquad
 \|a(x)-a(y)\|_{n,2}\le Ch^{-2}\log^2(n)\|x-y\|_{n,2}.
\]
Hence, with $E_n=\|a(r)-z\|_{n,2}$,
\begin{equation}
 E_n=\Op\left[\log^C(n)
       \left\{\frac{u_0}{h^2}+\frac1{\sqrt{nh^3}}\right\}\right].
 \label{eq:E}
\end{equation}
\begin{proposition}[Score expansion]\label{prop:score}
Suppose $m$ and $m'$ are bounded, $\eps$ is independent of the design with the mixture properties above, and conditional on all $T_i$, the $W_i$ are independent, centered, with bounded conditional covariance. For each fixed deterministic tuning tuple,
\begin{equation}
 \sqrt n\nabla L_n(0)
 =\frac1{\sqrt n}\sum_i[-m'(T_i)W_i\psi(\eps_i)]+\Op(R_S),
 \label{eq:score-transfer}
\end{equation}
where, writing $\xi=\|a_{h,\rho}-\psi\|_{L^2(f)}$,
\begin{equation}
 \begin{split}
 R_S={}&\xi+\log^C(n)\left\{
 E_n(1+B_Q)+\frac\alpha h+B_Q u_0+F_Q\right.\\
 &\left.\hspace{36mm}+b_\mu+F_\mu+\sqrt n E_nu_\mu\right\}.
 \end{split}
 \label{eq:complete-rem}
\end{equation}
The primitives may be replaced by stochastic bounds.
\end{proposition}
\begin{proof}
Since $m$ is bounded and errors are sub-Gaussian, $\max_i|Y_i|+\max_i|z_i|=\Op(\sqrt{\log n})$. The outgoing difference is
\[
 -a_i(r)A_i+m_i'z_i
 =-(a_i(r)-z_i)m_i'-a_i(r)(A_i-m_i'),
\]
with empirical norm bounded by $CE_n+C\log^C(n)\alpha/h$. Replacing $a_k(r)$ by $z_k$ in the incoming term costs at most $\log^C(n)B_QE_n$: its absolute matrix is entrywise bounded by $2\max_i|Y_i|\,|Q|^T$, because $H$ has row mass at most one. Replacing $\widehat m_k$ by $m(T_k)$ then costs at most $\log^C(n)B_Qu_0$. The remaining coefficient is
\[
 C_i^0=\sum_{k\ne i}Q_{ki}z_k\{Y_i-m(T_k)\}.
\]
Conditional on all $T$ and $\eps_i$ its summands are independent and centered. Integrating $\eps_i$ and averaging gives $\E(\|C^0\|_{n,2}^2\mid T)\le C\|Q\|_F^2/n$. Thus
\[
 \|c+m'z\|_{n,2}
 =\Op\left[\log^C(n)
 \{E_n(1+B_Q)+\alpha/h+B_Qu_0+F_Q\}\right].
\]
Conditional on all $T$ and errors, the covariance of $n^{-1/2}\sum_iW_i(c_i+m_i'z_i)$ is bounded by a constant times the square of this norm. The conditional Chebyshev inequality bounds the centered contribution.

For the mean contribution, use
\[
 \frac1{\sqrt n}\sum_i z_iD_i^\mu
 =\frac1{\sqrt n}\sum_i z_i b_i
  +\frac1{\sqrt n}\sum_{i\ne j}z_iG_{ij}\eps_j.
\]
Conditional on $T$, the first term has second moment at most $Cb_\mu^2$. In the quadratic term only identical and reversed ordered pairs survive, because $G_{ii}=0$, $\E z=\E\eps=0$. Its second moment is at most
\[
 \{\E z^2\E\eps^2+(\E z\eps)^2\}\|G\|_F^2/n.
\]
Restoring $a(r)$ in place of $z$ costs at most $\sqrt n E_nu_\mu$ by empirical Cauchy--Schwarz. Finally the variance of the centered tangent sum with $z_i-\psi(\eps_i)$ is bounded by $C\xi^2$. This accounts for every term of \eqref{eq:complete-rem}.
\end{proof}

Each tuning tuple is fixed before conditional centering is applied. A selector from a finite grid is handled by a simultaneous event once these expansions are established.

\subsection{Rates under Gaussian design}
The Gaussian ridge-jet lemma supplies, with fixed polynomial tuning and up to logarithmic factors,
\begin{align}
 u_j&=g^2+\sqrt\lambda+(ng^{2j+1})^{-1/2},\quad j=0,1,2,\nonumber\\
 \alpha&=\Op(\log^C(n)u_1),\quad B_Q\le\log^C(n)/g,\quad
 F_Q=\Op\{\log^C(n)/\sqrt{ng^3}\},\nonumber\\
 u_\mu+b_\mu&=\Op(\log^C(n)u_0),\qquad
 F_\mu=\Op\{\log^C(n)/\sqrt{ng}\}.
 \label{eq:gaussian-primitives}
\end{align}
For raw Gaussian coordinates $\mu(t)=Bt+b$ the ridge contributes a correction. With $q_g$ the smoothed projected density and $m_{g,\lambda}$ the population ridged regression, the population expression is
\begin{equation}
 D_{\rm pop}^\mu(t)
 =-B\left\{g\partial_gm_{g,\lambda}(t)
       +\frac{\lambda m_{g,\lambda}(t)}{q_g(t)+\lambda}\right\}.
 \label{eq:ridge-mu}
\end{equation}
Its empirical integrated norm has the stated $g^2+\sqrt\lambda$ bias order. The diagonal of the noise matrix is again zero.

Put $E=\log^C(n)\{u_0/h^2+(nh^3)^{-1/2}\}$. Substitution gives
\begin{align}
 R_S&=\log^C(n)\left\{h^2+\sqrt\rho+E/g+u_1/h+u_0/g
 +(ng^3)^{-1/2}\right.\nonumber\\[-2pt]
 &\hspace{39mm}\left.+u_0+(ng)^{-1/2}+\sqrt n Eu_0\right\},
 \label{eq:rs-expanded}\\
 R_{H,\mathrm{fit}}&=\log^C(n)\{u_0/h^3+u_1/h^2+u_2/h\}.
 \label{eq:rh-fit}
\end{align}
The second display follows from Theorem~\ref{op-hessian-theorem}, applied
with the empirical $L^2$ norms of the fitted jets and the $L^4$ and $L^2$
norms of the oracle jets.

For the oracle Hessian use the common error-tail threshold $t=n^{-\kappa}$.
Proposition~\ref{prop:oracle}, with the identity $J_{h,\rho}=I_f-\|g_h'/ (g_h+\rho)-\psi\|_{L^2(f)}^2$, gives
\begin{align}
 R_H={}&R_{H,\mathrm{fit}}+\log^C(n)\left\{
 (nh^5t)^{-1/2}+(nh^3t)^{-1}+t/h^2\right.\nonumber\\
 &\hspace{47mm}\left.+n^{-1/2}+r_n+h^4+\rho\right\}.
 \label{eq:rh-total}
\end{align}
The threshold set depends only on the iid errors and the deterministic
bandwidth. Its omitted-row fraction has order $t\log^C(n)$.

Take the exponents
\begin{equation}
 a=\frac1{30},\qquad b=\frac4{25},\qquad
 \tau=\frac7{10},\qquad\kappa=\frac15,
 \quad(h,g,\lambda,t)=(n^{-a},n^{-b},n^{-\tau},n^{-\kappa}).
 \label{eq:exponents}
\end{equation}
The resulting polynomial orders are listed below; they hold for any fixed
positive multipliers.
\begin{center}
\begin{tabular}{@{}lll@{}}
\toprule
Quantity & Polynomial order, before logarithms & Role\\
\midrule
$u_0$ & $n^{-8/25}$ & residual error\\
$u_1$ & $n^{-13/50}$ & first-jet error\\
$u_2$ & $n^{-1/10}$ & second-jet error\\
$E/g$ & $n^{-7/75}$ & incoming score restoration\\
$\sqrt n Eu_0$ & $n^{-11/150}$ & conditional-mean restoration\\
$h^2$ & $n^{-1/15}$ & density-score approximation\\
$u_0/h^3$ & $n^{-11/50}$ & residual curvature transfer\\
$u_1/h^2$ & $n^{-29/150}$ & first-jet curvature transfer\\
$u_2/h$ & $n^{-1/15}$ & second-jet curvature transfer\\
$t/h^2$ & $n^{-2/15}$ & low-density rows\\
$(nh^5t)^{-1/2}$ & $n^{-19/60}$ & oracle derivative fluctuation\\
$(nh^3t)^{-1}$ & $n^{-7/10}$ & oracle Bernstein term\\
\bottomrule
\end{tabular}
\end{center}
Also $ng\lambda=n^{7/50}$, so the denominator event \eqref{gauss-G7} holds.
The total score rate is $\Op(\log^C(n)n^{-1/15})$ and the total curvature
rate is $\Op[\log^C(n)\{n^{-1/15}+r_n\}]$.

\subsection{Assembly and separation from uniform fitting}
For $X\sim N_p(\mu,\Sigma)$ with fixed positive definite $\Sigma$,
bounded nonconstant $C^6$ link and its first six derivatives, and independent fixed mean-zero Gaussian-mixture error, the Gaussian primitive and oracle-curvature results supply every input of the preceding argument.
Thus the displayed score expansion holds at $\theta=0$, and the Hessian bound holds uniformly on $\|\theta\|\le C_rn^{-d}$, $0<d<1/2$.

The information matrix
$\mathcal I=I_f\E[m'(T)^2WW^T]$ is positive definite: the Gaussian
conditional tangent covariance is a positive definite Schur complement,
$\E m'(T)^2>0$ for a nonconstant smooth link, and $I_f>0$.
On an event with probability tending to one, the Hessian is therefore negative definite throughout this ball, while the score at zero is $\Op(n^{-1/2})$.
Taylor expansion makes the objective on the boundary lower than at zero.
Its maximum over the closed ball is interior and unique. The score equation
gives its efficient influence expansion, and Section~\ref{sec:algorithm} identifies the computed estimate with this root.

Finally, a finite union ensures that the expansions are simultaneous over any fixed finite list of deterministic positive tuning multipliers. Their leading influence sum is common. Any measurable training-data selector from this list inherits the result on that simultaneous event.

\section{The Fourier initial estimator}\label{supp:fourier}
\label{fourier:appendix}

The Fourier central-subspace estimator below is due to \citet{zhu2006fourier}, equations (24)--(25), Proposition~6 and Theorem~1. We give a self-contained proof for the additive Gaussian model, allowing arbitrary measurable nonconstant links and arbitrary independent errors.

\subsection{Definition, rank, and exhaustiveness}
\label{fourier:definition}

Suppose $X\sim N(0,I_d)$, $\|\beta_0\|=1$, and $Y=m(\beta_0^\top X)+\eps$, where $m$ is real-valued, measurable, and nonconstant almost surely, and $\eps$ is any real-valued random variable independent of $X$. Fix $a,b>0$. For $z=(x,y)$, $z'=(x',y')$, and $\Delta=x-x'$, define
\begin{align}
 J(z,z')&=e^{-a\|\Delta\|^2/2-b(y-y')^2/2}
 \{aI_d+(x+a\Delta)(x'-a\Delta)^\top\},\label{fourier:kernel}\\
 \widehat M_n&=n^{-2}\sum_{i,j=1}^n J(Z_i,Z_j),\qquad
 M=\E J(Z,Z'),\label{fourier:matrix}
\end{align}
where $Z,Z'$ are independent in the population expectation. The sample sum runs over all ordered pairs, including the diagonal. Since $J(z',z)=J(z,z')^\top$, $\widehat M_n$ is real symmetric. Select a unit eigenvector for its largest algebraic eigenvalue, with measurable deterministic tie-breaking. The constants $a,b$ are fixed tuning constants of the initial estimator and are unrelated to the SI-KDR bandwidths.

\begin{proposition}[Exhaustiveness in the additive Gaussian model]
\label{fourier:rankprop}
For every model just specified,
\begin{equation}
 M=\lambda_d\beta_0\beta_0^\top,\qquad\lambda_d>0.
 \label{fourier:rank}
\end{equation}
For a fixed link and error law and fixed $a,b$,
\begin{equation}
 \lambda_d=(1+2a)^{-(d-1)/2}\lambda_1.
 \label{fourier:gapdimension}
\end{equation}
\end{proposition}
\begin{proof}
Let
\[
 \Phi(\omega,t)=\E[(X-i\omega)e^{i\omega^\top X+itY}],
 \qquad \omega\sim N(0,aI_d),\quad t\sim N(0,b),
\]
where the last two variables are integration variables. Gaussian integration gives
\begin{equation}
 M=\operatorname{Re}\iint\Phi(\omega,t)\Phi(\omega,t)^*
 \,dW_a(\omega)\,dW_b(t).
 \label{fourier:fourieridentity}
\end{equation}
Indeed, with $c(\Delta)=e^{-a\|\Delta\|^2/2}$,
\[
 \E_\omega e^{i\omega^\top\Delta}=c(\Delta),\quad
 \E_\omega[i\omega e^{i\omega^\top\Delta}]=-a\Delta c(\Delta),\quad
 \E_\omega[\omega\omega^\top e^{i\omega^\top\Delta}]
 =(aI-a^2\Delta\Delta^\top)c(\Delta).
\]
Expanding $(x-i\omega)(x'+i\omega)^\top$ yields $aI+xx'^\top-ax\Delta^\top+a\Delta x'^\top-a^2\Delta\Delta^\top$, which is the bracket in \eqref{fourier:kernel}. The $t$ integral gives the response factor. Fubini is justified by Gaussian moments and modulus-one response exponentials.

Write $X=\beta_0T+UV$, with $V$ standard Gaussian independent of $(T,Y)$. The identity $\E[(V-i\omega_\perp)e^{i\omega_\perp^\top V}]=0$ shows that $\Phi$ is a scalar multiple of $\beta_0$. Thus $M$ has the stated rank-one form with $\lambda_d\ge0$. Suppose $\lambda_d=0$. The integrand in \eqref{fourier:fourieridentity} is nonnegative, and the Gaussian weights are strictly positive. Continuity of $\Phi$, justified by the locally integrable envelope $\|X\|+\|\omega\|$, implies $\Phi(\omega,t)=0$ for all $\omega,t$. For fixed $t$, put
\[
 h_t(u)=\E(e^{itY}\mid T=u),\qquad
 G_t(v)=\E[h_t(T)e^{ivT}].
\]
Since $|h_t|\le1$, the function $G_t$ is differentiable. At $\omega=v\beta_0$ the $\beta_0$-coordinate of $\Phi$ equals
\[
 \E[(T-iv)h_t(T)e^{ivT}]=-i\{G_t'(v)+vG_t(v)\}.
\]
Hence $G_t'(v)+vG_t(v)=0$ and $G_t(v)=G_t(0)e^{-v^2/2}$. Uniqueness of Fourier transforms of finite complex measures gives $h_t(T)=G_t(0)$ almost surely. Taking a countable dense set of $t$ and then using conditional characteristic-function continuity proves $Y\perp T$.

But $h_t(u)=\phi_\eps(t)e^{itm(u)}$, and $\phi_\eps$ is nonzero on a neighborhood of zero. On a common probability-one set for a sequence of nonzero $t$ tending to zero, independence therefore implies $e^{it\{m(u)-m(v)\}}=1$ for almost every pair $(u,v)$. Any fixed finite difference must then be zero: eventually its absolute product with $t$ is less than $2\pi$. Thus $m$ is constant almost surely, a contradiction.

Finally the perpendicular Gaussian component of $\Phi$ contributes $e^{-\|\omega_\perp\|^2/2}$. Squaring and integrating over $N(0,aI_{d-1})$ yields $(1+2a)^{-(d-1)/2}$, proving \eqref{fourier:gapdimension}.
\end{proof}

\subsection{Rate, computation, and fixed-dimensional whitening}
\label{fourier:rate}
For unit vectors $a$ and $b$, $\dP(a,b)=\min\{\|a-b\|,\|a+b\|\}=\{2-2|a^Tb|\}^{1/2}$ denotes the projective distance.

\begin{proposition}[Rate of the initial estimator]\label{fourier:rateprop}
For fixed $a,b>0$,
\begin{equation}
 \E\|\widehat M_n-M\|_F^2\le C_a d^2/n,
 \qquad
 \dP(\widetilde\beta,\beta_0)
 =\Op\{d/(\lambda_d\sqrt n)\}.
 \label{fourier:ratebound}
\end{equation}
In fixed dimension, this is root-$n$ consistency. For a fixed scalar model, $d=o(\log n)$ gives the rate $n^{-1/2+o(1)}$.
\end{proposition}
\begin{proof}
The bounded response factor gives
\[
 \|J(Z,Z')\|_F\le a\sqrt d+C_a(\|X\|^2+\|X'\|^2),
 \qquad \E\|J(Z,Z')\|_F^2\le C_a d^2.
\]
The diagonal $J(Z,Z)=aI+XX^\top$ has the same second-moment order. Two off-diagonal summands with disjoint observation pairs are independent; at most $Cn^3$ ordered pairs of summands share an observation. Cauchy--Schwarz gives variance at most $C_ad^2/n$ for their centered average. The diagonal average and off-diagonal normalization bias have squared-norm contributions of at most $C_ad^2/n^2$. This proves the matrix bound.

On $\|\widehat M_n-M\|_{\rm op}\le\lambda_d/4$, the largest sample eigenvalue is at least $3\lambda_d/4$. Project its eigenvector equation onto $\beta_0^\perp$ to get
\[
 \|P_{\beta_0^\perp}\widetilde\beta\|
 \le \frac{4\|\widehat M_n-M\|_{\rm op}}{3\lambda_d}.
\]
Since $\dP(\widetilde\beta,\beta_0)\le \sqrt2\|P_{\beta_0^\perp}\widetilde\beta\|$, Markov's inequality and the matrix bound give the asserted stochastic rate; the complement of the displayed event has probability at most $C_ad^2/(n\lambda_d^2)$. Equation~\eqref{fourier:gapdimension} gives the rate in growing dimension.
\end{proof}

Evaluating $\widehat M_n$ costs $O(n^2d^2)$, followed by $O(d^3)$ for the eigendecomposition, and uses $O(d^2+nd)$ memory when pairs are streamed.

For $X\sim N(\mu,\Sigma)$ in fixed dimension, write $A_0=\Sigma^{-1/2}$ and $Z=A_0(X-\mu)$. The standardized direction is
\[
 \gamma_0=\frac{A_0^{-\top}\beta_0}{\|A_0^{-\top}\beta_0\|}.
\]
Compute the initial estimator on $\widehat Z_i=\widehat A(X_i-\widehat\mu)$, using the fitting sample mean and covariance and $\widehat A=\widehat\Sigma^{-1/2}$; if the sample covariance is singular, a fixed direction is used. Map its top eigenvector back by
\begin{equation}
 \widetilde\beta_{\rm raw}
 =\frac{\widehat A^\top\widehat\gamma}
 {\|\widehat A^\top\widehat\gamma\|}.
 \label{fourier:rawpilot}
\end{equation}
This estimator is also $\sqrt{n}$-consistent. Indeed, Gaussian moments give $\|\widehat\mu-\mu\|+\|\widehat A-A_0\|=O_P(n^{-1/2})$. On a compact neighborhood of the true affine transformation, the derivative of \eqref{fourier:kernel} in the two predictor arguments is bounded by $C(1+\|X_i\|+\|X_j\|)^3$; differentiation in the affine parameters adds at most one power. The response factor does not depend on the transformation and is bounded by one. By Gaussian moments and the law of large numbers, the average of this fourth-power envelope over all pairs is $O_P(1)$. The mean-value theorem gives
\begin{equation}
 \|\widehat M_n(\widehat Z,Y)-\widehat M_n(Z,Y)\|_F
 =O_P(n^{-1/2}).
 \label{fourier:plugin}
\end{equation}
The conclusion follows by combining \eqref{fourier:plugin}, \eqref{fourier:ratebound}, and the smoothness of the projective map \eqref{fourier:rawpilot}.

\section{Same-sample whitening and response standardization}\label{sec:preprocessing}
Both transformations are estimated on the fitting sample. The bounds below apply whenever their remainders vanish. Numerical exponents refer to the scaling $h\asymp n^{-1/30},g\asymp n^{-4/25},\lambda\asymp n^{-7/10}$; the rates for the tuning used in the paper and the finite-smoothing correction are given in Section~\ref{inf:theory}. We first prove bounds that hold uniformly over deterministic transformations near the population ones and then substitute the estimates.

\subsection{Model and moving-target identities}
Write $T=\beta_0^TX$, $v=\beta_0^T\Sigma\beta_0$, and let $U_0$ be a raw orthonormal tangent frame. The Gaussian conditional decomposition is
\begin{equation}
 X=\mu+k(T-\mu_T)+W,\quad k=\Sigma\beta_0/v,\quad
 \Gamma=\Cov(W)=\Sigma-\Sigma\beta_0\beta_0^T\Sigma/v.
 \label{eq:rp-decomp}
\end{equation}
Here $\beta_0^TW=0$, and $W$ is Gaussian and independent of $(T,\eps)$. The raw tangent information is $I_0=I_f\E[m_0'(T)^2]U_0^T\Gamma U_0$. In the analysis put $Y^0=(Y-\mu_Y)/\sigma_Y=m(T)+e$, where $\mu_Y=\E Y$, $\sigma_Y^2=\Var Y>0$, $m=(m_0-\mu_Y)/\sigma_Y$, and $e=\eps/\sigma_Y$. Its error score $\psi_0$ satisfies
\begin{equation}
 \psi_0(e)m'(T)=\psi_f(\eps)m_0'(T).
 \label{eq:rp-response-invariant}
\end{equation}
The estimates, using divisor $n$ for sample variances, are
\begin{equation}
 \widehat A=\widehat\Sigma^{-1/2},\quad
 \widehat Z_i=\widehat A(X_i-\overline X),\quad
 \widehat Y_i=\frac{Y_i-\overline Y}{\widehat\sigma_Y}
       =\widehat aY_i^0+\widehat b,
 \quad
 \widehat a=\frac{\sigma_Y}{\widehat\sigma_Y},\quad
 \widehat b=\frac{\mu_Y-\overline Y}{\widehat\sigma_Y}.
 \label{eq:rp-estimates}
\end{equation}
If a sample variance is zero or the sample covariance is singular, the estimate is a fixed direction and the confidence region is empty. Finite fourth moments and the smooth inverse-square-root map give
\[
 \|\widehat A-A_0\|+|\widehat a-1|+|\widehat b|=\Op(n^{-1/2}),
 \qquad A_0=\Sigma^{-1/2}.
\]
Indeed, sample second moments are root-$n$, sample-mean corrections are $\Op(n^{-1})$, and both population variances are positive. Bernstein applied to their sub-exponential envelopes places the estimates in a deterministic box of radius $\delta_n=\log(n+e)/\sqrt n$ with probability tending to one. For the numerical exponents the working-unit tuning parameters are $h=c_1n^{-1/30}$, $g=c_2n^{-4/25}$, $\lambda=c_\lambda n^{-7/10}$, and $\rho=c_\rho/n$, held fixed during fitting.

For deterministic nonsingular $A$ near $A_0$ define
\[
 \gamma_0(A)=\frac{A^{-T}\beta_0}{\|A^{-T}\beta_0\|},\qquad
 s(A)=\frac1{\|A^{-T}\beta_0\|}.
\]
Then $\gamma_0(A)^TAX=s(A)T$. Choose a smooth frame $U_A$ by projecting a fixed frame onto $\gamma_0(A)^\perp$ and normalizing its Gram matrix. Every derivative is bounded on this neighborhood; other frames only rotate the coordinates. Predictor centering cancels from all index differences. For response $Y^{a,b}=aY^0+b$, the normalized kernel identity gives
\begin{equation}
 \phi_g(s\Delta T)=s^{-1}\phi_{g/s}(\Delta T),\qquad
 \widehat m_{\rm work}
  =\widehat m_{\rm raw}(T,Y^{a,b};g/s,\lambda s).
 \label{eq:rp-ridge-scale}
\end{equation}
Thus the raw effective ridge is $\lambda s$. Let $c(s,a,b)$ be the coefficient in \eqref{eq:full-score} for this scalar raw family, with $h,\rho$ unchanged. Since $\phi_g'(s\Delta T)=s^{-2}\phi_{g/s}'(\Delta T)$, its working coefficient is $c(s,a,b)/s$, and
\begin{equation}
 \sqrt n\nabla_\gamma L_{A,a,b}(\gamma_0(A))
 =\frac1s U_A^TA\frac1{\sqrt n}\sum_iX_ic_i(s,a,b).
 \label{eq:rp-score-exact}
\end{equation}
The coefficients sum to zero, so $\overline X$ also cancels here. We use this identity only in the true direction, since away from it, the induced raw bandwidth and ridge vary with the direction. All working derivatives are taken at fixed working $h,g,\lambda,\rho$. For example, $F(r/t;h,\rho)=F(r;th,\rho/t)+\log t$ for $t>0$, and below the response family is differentiated with the working parameters held fixed.

\subsection{A modulus for the coefficient and a conditional Gaussian process}
Set $t=(s,a,b)$, $t_0=(v^{-1/2},1,0)$, and
\[
 \mathcal K_n=\{|s-v^{-1/2}|,|a-1|,|b|\le C\delta_n\}.
\]
Conditional on all $(T_i,e_i)$, the coefficient array $c(t)$ is deterministic and independent of the conditional Gaussian array $W$. On the event $\max_i|T_i|+\max_i|Y_i^0|\le C\sqrt{\log n}$, the weight and residual derivative bounds imply
\begin{equation}
 \sup_{t\in\mathcal K_n}\max_{j\le3}
       \|\partial_{t_j}c(t)\|_{n,2}
 \le K_n^*:=\log^C(n)h^{-2}g^{-2}.
 \label{eq:rp-modulus}
\end{equation}
To see this, vary the extended direction $s\beta_0$ at fixed working $g,\lambda$. With $H,Q,A$ as in \eqref{eq:nw-coeff} and $B_{ki}=Q_{ki}(Y_i-\widehat m_k)$, its coefficient is $c_{\rm ext}=-\operatorname{diag}(A)a(r)+B^Ta(r)$. Gaussian row and column displacement moments, quotient differentiation, and the logarithmic $T,Y$ envelope give
\begin{align*}
 \|A\|_\infty+\||Q|\|_{2\to2}+\||B|\|_{2\to2}
   &\le\log^C(n)/g,\\
 \|\partial_sA\|_\infty+
 \||\partial_sQ|\|_{2\to2}+\||\partial_sB|\|_{2\to2}
   &\le\log^C(n)/g^2,\\
 \|\partial_sr\|_{n,2}&\le\log^C(n)/g.
\end{align*}
Use $\phi_g'=-\Delta\phi_g/g^2$ and $\phi_g''=(\Delta^2/g^4-g^{-2})\phi_g$ for these bounds; the ridge is completed by its zero-response atom. The residual differential inequality then gives $\|\partial_sa(r)\|_{n,2}\le\log^C(n)/(h^2g)$ and $\|a(r)\|_\infty\le\log^C(n)/h$. Product differentiation proves \eqref{eq:rp-modulus} for $s$, since $c=s c_{\rm ext}$. For the response parameters,
\begin{equation}
 r^{a,b}=a r^{1,0}+b(1-H\boldsymbol1).
 \label{eq:rp-ridge-deficit}
\end{equation}
Hence $\|\partial_ar\|_\infty\le2\max|Y^0|$ and $\|\partial_br\|_\infty\le1$. The response derivatives of $A,B$ cost at most $\log^C(n)/g$, proving the stronger coefficient bound $\log^C(n)/(h^2g)$ for these parameters. Because the ridge rows sum to less than one, the translation $b$ enters the residuals in \eqref{eq:rp-ridge-deficit} even for a centered response.

Consider the centered conditional Gaussian process
\[
 G_n(t)=\frac1{\sqrt n}\sum_iW_i\{c_i(t)-c_i(t_0)\}.
\]
Its fixed-dimensional components have canonical distance at most $CK_n^*\|t-t'\|$, by \eqref{eq:rp-modulus}, and $G_n(t_0)=0$. Successive dyadic grids of the three-dimensional box have at most $C2^{3l}$ parent increments at level $l$, each of standard deviation at most $CK_n^*\delta_n2^{-l}$. The elementary Gaussian bound $\E\max_{j\le M}|Z_j|\le\sigma\sqrt{2\log(2M)}$ holds under arbitrary dependence. Summing $CK_n^*\delta_n2^{-l}\sqrt{l+1}$ gives
\begin{equation}
 \E\{\sup_{t\in\mathcal K_n}\|G_n(t)\|\mid T,e\}
 \le CK_n^*\delta_n,
 \qquad
 \sup_{t\in\mathcal K_n}\|G_n(t)\|
 =\Op\{\log^C(n)n^{-17/150}\}.
 \label{eq:rp-chaining}
\end{equation}
Finite sums of smooth coefficients have continuous paths, justifying the dyadic limit. Equivalently, increment thresholds proportional to $2^{-l}\sqrt{l+u+1}$ give a conditional exceptional probability at most $Ce^{-u}$. The exponent follows from $n^{-1/2}h^{-2}g^{-2}=n^{-17/150}$. The pointwise coefficient bound of Section~\ref{sec:score} is needed only at $t_0$.

\subsection{The affine mean and simultaneous score expansion}
At $A_0$, $U_{A_0}^TA_0k=0$, since $A_0k$ is parallel to $\gamma_0(A_0)$. Smoothness yields $\|U_A^TAk/s\|\le C\|A-A_0\|$. Define
\[
 D_i^T(t)=\sum_jQ_{{\rm raw},ij}(T_i-T_j)
                   (Y_j^{a,b}-\widehat m_i^{a,b}).
\]
Exchanging the incoming indices gives
\begin{equation}
 \frac1{\sqrt n}\sum_i(T_i-\mu_T)c_i(t)
 =-\frac1{\sqrt n}\sum_i a_i(r(t))D_i^T(t).
 \label{eq:rp-mean-exact}
\end{equation}
At $t_0$ the Gaussian ridge lemma gives $\|D^T(t_0)\|_{n,2}=\Op(\log^C(n)u_0)$, $u_0=g^2+\sqrt\lambda+(ng)^{-1/2}$. Differentiating $Q$, the logarithmic index displacement, and $\widehat m$ gives $\sup_t\max_j\|\partial_{t_j}D^T(t)\|_{n,2}\le\log^C(n)g^{-2}$. Thus $\sup_t\|D^T(t)\|_{n,2} =\Op\{\log^C(n)(u_0+\delta_ng^{-2})\}$. Together with $\|a(r)\|_\infty\le\log^C(n)/h$, the affine contribution to \eqref{eq:rp-score-exact} is
\begin{equation}
 \Op\{\log^C(n)\sqrt n\delta_nh^{-1}
                  (u_0+\delta_ng^{-2})\}
 =\Op\{\log^C(n)(n^{-43/150}+n^{-11/75})\}=o_P(1).
 \label{eq:rp-mean-rate}
\end{equation}

At $t_0$, the coefficient bound of Section~\ref{sec:score} gives $\|c(t_0)+m'(T)\psi_0(e)\|_{n,2} =\Op\{\log^C(n)n^{-1/15}\}$. Its conditional-$W$ sum has the same order by the Gaussian variance identity. Put
\[
 J_A=s^{-1}U_0^TA^TU_A,\qquad
 S_{0i}=-\psi_0(e_i)m'(T_i)U_0^TW_i.
\]
Combining the $t_0$ bound, \eqref{eq:rp-chaining}, and \eqref{eq:rp-mean-rate} gives simultaneously on the deterministic box
\begin{equation}
 \sqrt n\nabla L_{A,a,b}(\gamma_0(A))
 =J_A^T\frac1{\sqrt n}\sum_iS_{0i}
 +\Op\{\log^C(n)(n^{-1/15}+n^{-17/150}+n^{-11/75})\}.
 \label{eq:rp-score}
\end{equation}
It therefore also holds at the estimated transformations. Response effects are controlled by \eqref{eq:rp-ridge-deficit}, and \eqref{eq:rp-response-invariant} identifies the original-unit efficient score.

\subsection{Uniform curvature over the transformation parameters}
Use the working chart $\gamma_A(\theta)=(\gamma_0(A)+U_A\theta)/(1+\|\theta\|^2)^{1/2}$, $\|\theta\|\le r_n=C_rn^{-d}$, $0<d<1/2$. Index the empirical classes by the deterministic finite-dimensional tuple $(A,a,b,\theta,x)$. The matching ideal residual is $e_{{\rm oracle},i}=a[Y_i^0-\E\{Y^0\mid \gamma_A(\theta)^TAX_i\}]$; at $\theta=0$ it equals $ae_i$. The chart derivatives of this ideal residual are subtracted below. For each fixed tuple, these are functions of the iid observations. Eigenvalues of $A\Sigma A^T$ remain between fixed positive constants. Conditional Gaussian formulas and their first two chart derivatives are uniform, and response dependence is affine. The ideal first-jet $L^4$ and second-jet $L^2$ empirical norms are bounded in probability uniformly; their first three chart derivatives have polynomial envelopes in $\|X\|$.

For a common bound on the denominators put $l=A^T\gamma_A(\theta)$ and $l_0=A_0^T\gamma_0(A_0)=v^{-1/2}\beta_0$. The smoothed projected density at its row is
\[
 q_{l,g}(l^TX)=
 \frac{\exp[-\{l^T(X-\mu)\}^2/\{2(l^T\Sigma l+g^2)\}]}
      {\sqrt{2\pi(l^T\Sigma l+g^2)}}.
\]
Its log derivative in $l$ is bounded by $C(1+\|X\|^2)$ on this neighborhood. Since $\|l-l_0\|\le C(\delta_n+r_n)$, the log density ratio is $O((\delta_n+r_n)(1+\|X\|^2))=o(1)$ on the common Gaussian maximum event. Hence, one baseline-index function dominates all integrated inverse-denominator bounds. The Gaussian ridge-jet Bernstein proof applies to a polynomial net in this enlarged tuple, whose log cardinality is still $O(\log n)$. Variances are $qg^{-(2j+1)}$ up to logarithmic factors; interpolation uses $q+\lambda \ge\lambda$; the zeroth relative denominator error requires $ng\lambda\gg\log^C n$. Averaging the common envelope once gives
\begin{equation}
 \sup_{A,a,b,\theta}
 \|\partial_\theta^j(r-e_{\rm oracle})\|_{n,2}
 =\Op\left[\log^C(n)
       \{g^2+\sqrt\lambda+(ng^{2j+1})^{-1/2}\}\right],\quad j\le2.
 \label{eq:rp-jets}
\end{equation}

For oracle curvature use the one fixed good-error set $\{g_{h,0}(e_i)\ge n^{-\kappa}\}$, $\kappa=1/5$, $g_{h,0}=\phi_h*f_0$. Under response scaling, its truth density is $a^{-1}g_{h/a,0}(e_i)$, whose log ratio to $g_{h,0}(e_i)$ is $O(\delta_n(1+e_i^2))$ on the common error maximum event. Profile shifts are $O(r_n(1+\|X\|))$. Truncate the inner Gaussian integrals on larger logarithmic balls with arbitrarily small polynomial remainders, then apply the oracle Bernstein net argument, including $A,a,b$, before substituting evaluation rows. The relative derivative errors are again $(nh^{2j+1}t)^{-1/2}+(nh^{j+1}t)^{-1}$, $t=n^{-\kappa}$. The bad evaluation set is common, with a fraction $\Op(t\log^C n)$. At the truth, the information is $J_A^TI_0J_A$, independent of $a,b$. The mixture information identity gives uniform discrepancy $O(h^4+\rho\log^C n)$, and integrable population third derivatives give the $O(r_n)$ variation. Applying Theorem~\ref{op-hessian-theorem} pathwise to \eqref{eq:rp-jets} proves
\begin{equation}
 \sup_{A,a,b,\|\theta\|\le r_n}
 \|\nabla^2_\theta L_{A,a,b}(\theta)+J_A^TI_0J_A\|
 =\Op\{\log^C(n)(n^{-1/15}+r_n)\}.
 \label{eq:rp-hessian}
\end{equation}
The additional parameters change only the net dimension and the constants.

\subsection{Initial estimator, optimization and the region in original coordinates}
The influence of $\widehat A$ on the initial Fourier-based estimate is discussed in Section~\ref{supp:fourier}. Response centering cancels in pairwise response differences. For scales in a fixed range near one, differentiating the Gaussian response kernel in $a$ gives a factor proportional to $a(\Delta Y^0)^2$ times the kernel, which is uniformly bounded because $u^2e^{-cu^2}$ is bounded. For $A$ in a compact set of transformations, the remaining pairwise matrix factors of the Fourier matrices have a common polynomial envelope with finite Gaussian moments. Taking expectations, the pair-averaged derivative is $\Op(1)$. Hence $\widehat a-1=\Op(n^{-1/2})$ induces a matrix perturbation of order $\Op(n^{-1/2})$. For a fixed distribution of standardized responses, the positive population eigengap of the mean matrix (a finite average with positive weights) and the $\Op(n^{-1/2})$ sample perturbation give an $\Op(n^{-1/2})$ error for the estimated leading eigenspace; in working coordinates, its distance from $\gamma_0(\widehat A)$ is $\Op(n^{-1/2})$. For each point of a finite grid, \eqref{eq:rp-score}--\eqref{eq:rp-hessian} first hold uniformly over the transformations; since the grid is finite, they extend to any data-dependent selection from it. The contraction argument of Section~\ref{sec:ascent} gives an optimization error of order $o_P(n^{-1/2})$.

The raw map $B(A,\gamma)=A^T\gamma/\|A^T\gamma\|$ satisfies
\begin{equation}
 B(A,\gamma_0(A))=\beta_0\quad\hbox{for every nonsingular }A,
 \qquad
 D_\gamma B(A,\gamma_0(A))U_A
  =s^{-1}(I-\beta_0\beta_0^T)A^TU_A.
 \label{eq:rp-target-cancel}
\end{equation}
The raw tangent coordinates of this derivative are $J_A$, invertible between the two tangent spaces. Taylor expansion at the random target, keeping the realized $A$ in the identity, gives
\begin{align}
 \sqrt n\widehat\theta_{\rm raw}
 &=J_A(J_A^TI_0J_A)^{-1}J_A^T
        \frac1{\sqrt n}\sum_iS_{0i}+o_P(1)\nonumber\\
 &=I_0^{-1}\frac1{\sqrt n}\sum_iS_{0i}+o_P(1).
 \label{eq:rp-efficient}
\end{align}
The last equality holds because $J_A$ is invertible. The preprocessing effect on the moving target is canceled by \eqref{eq:rp-target-cancel}.

\subsection{Joint expansion of the training transformations}
Use the notation of \eqref{eq:rp-estimates}, and put $x_i=X_i-\mu$, $y_i=Y_i-\mu_Y$, $C=\Sigma^{1/2}$, $A_0=C^{-1}$,
\[
 E_i=x_ix_i^T-\Sigma,\qquad q_i=y_i^2-\sigma_Y^2.
\]
For a symmetric matrix $E$, let $\mathcal L(E)$ solve the Sylvester equation
\begin{equation}
 C\mathcal L(E)+\mathcal L(E)C=E,
 \qquad
 \mathcal D_A[E]=-A_0\mathcal L(E)A_0.
 \label{eq:joint-frechet}
\end{equation}
This solution is unique: in an eigenbasis of $C$ its $(j,k)$ entry is $E_{jk}/(c_j+c_k)$, where every $c_j>0$. Differentiating $C^2=\Sigma$ and $A_0=C^{-1}$ shows that $\mathcal D_A$ is the bounded Fr\'echet derivative of the inverse square-root map at $\Sigma$.

The divisor-$n$ definitions give the following expansions jointly:
\begin{align}
 \sqrt n(\overline X-\mu)
   &=n^{-1/2}\sum_i x_i,\nonumber\\
 \sqrt n(\widehat\Sigma-\Sigma)
   &=n^{-1/2}\sum_i E_i+o_P(1),\nonumber\\
 \sqrt n(\widehat A-A_0)
   &=n^{-1/2}\sum_i\mathcal D_A[E_i]+o_P(1),
 \label{eq:joint-x}\\
 \sqrt n(\overline Y-\mu_Y)
   &=n^{-1/2}\sum_i y_i,\nonumber\\
 \sqrt n(\widehat\sigma_Y-\sigma_Y)
   &=\frac{1}{2\sigma_Y}n^{-1/2}\sum_iq_i+o_P(1),\nonumber\\
 \sqrt n(\widehat a-1)
   &=-\frac{1}{2\sigma_Y^2}n^{-1/2}\sum_iq_i+o_P(1),\nonumber\\
 \sqrt n\widehat b
   &=-\frac{1}{\sigma_Y}n^{-1/2}\sum_i y_i+o_P(1).
 \label{eq:joint-y}
\end{align}
Indeed, centering the empirical second moments at the sample mean changes them only by a second-order sample-mean error of order $O_P(n^{-1})$, negligible at the root-$n$ scale. Assuming the population variance is positive, the remaining assertions follow from the multivariate delta method. Under a Gaussian design, a bounded link and a fixed error mixture, all terms have finite second moments. The multivariate central limit theorem therefore applies jointly to these quantities and the efficient-score sum.

\subsection{The two derivatives of the raw-direction map}
\label{sec:joint-preprocessing}
Put $\gamma_*=\gamma_0(A_0)$, $s_0=\|A_0^{-T}\beta_0\|^{-1}$, $P_0=I-\beta_0\beta_0^T$. Differentiating the normalized inverse map gives
\begin{equation}
\mathcal G[E]=D\gamma_0(A_0)[E]
=-(I-\gamma_*\gamma_*^T)A_0^{-T}E^T\gamma_*.
\label{eq:joint-target-derivative}
\end{equation}
For $B(A,\gamma)=A^T\gamma/\|A^T\gamma\|$ the two partial derivatives are
\[
D_AB[E]=s_0^{-1}P_0E^T\gamma_*,\qquad
D_\gamma B[v]=s_0^{-1}P_0A_0^Tv.
\]
Their sum along the moving target is
\begin{equation}
D_AB[E]+D_\gamma B[\mathcal G[E]]=0.
\label{eq:joint-cancellation}
\end{equation}
Indeed, the component along $\gamma_*$ vanishes under $P_0A_0^T$, and the remaining term cancels $D_AB[E]$. This is the derivative of \eqref{eq:rp-target-cancel}: to first order, the estimated preprocessing leaves the target fixed.

\section{Initialization and the algorithm}
\label{sec:algorithm}
The four Fourier matrices of Section~\ref{fourier:appendix}, with spatial constants $1/p, 1, 3,$ and $9$, response constant one and equal weights, are computed from the training-standardized observations. The initial estimate is the eigenvector of the largest eigenvalue; if this eigenvalue is not simple, the direction $e_1$ is used. In this step, the whitening inside the initial estimator is the identity, since the data are already whitened with sample statistics. Section~\ref{sec:preprocessing} controls the effect of response scaling and establishes a rate of order $n^{-1/2}$ relative to $\gamma_0(\widehat A)$. If the sample transformations are degenerate or non-finite, the estimate is $e_1$ and the confidence region is empty.

\subsection{Bandwidth selection}
At $n=600$ the density anchors are $600^{-1/30}\{.08,.12,.18,.27,.4\}$ and the regression anchors are $600^{-4/25}\{.4,.6,.8,1.2\}$. Multiply these by $(n/600)^{-1/15}$ and $(n/600)^{-1/6}$, respectively, and use $\lambda_n=.01\,600^{-7/10}(n/600)^{-3/4}$, $\rho_n=1/n$. A preliminary SI-KDR fit uses the same density anchors and regression anchors $600^{-4/25}\{1.2,1.4,1.6\}$, rescaled by the same factor as the regression anchors above, with the same $\lambda_n$ and $\rho_n$. The preliminary fit uses the paper's training standardization and proceeds as follows.
\begin{enumerate}
\item At the initial Fourier direction, select the regression bandwidth from its grid by minimizing the leave-one-out mean squared prediction error of the protected Nadaraya--Watson fit for the standardized response.
\item With the regression bandwidth fixed, select the density bandwidth by maximizing $L_n$ at the same direction, that is, the mean log leave-one-out residual density, with the positive offset $\rho_n$ added to each density estimate.
\item In both steps, ties are broken by taking the first grid point.
\item With these values fixed, maximize $L_n$ from the initial Fourier direction by the projected ascent of Section~\ref{sec:ascent}, with the same cap, step size, iteration limit and stopping tolerance as the final fit. The resulting direction, mapped back to the original coordinates, is the preliminary direction.
\end{enumerate}
At the preliminary direction, the positive-curvature tuple minimizing the trace of the case-weight covariance in original tangent coordinates is chosen among the twenty candidates. Ties go to the first tuple in grid order, with the density bandwidth as the outer index. If no tuple has positive curvature, the tuple is selected at the preliminary direction by the leave-one-out squared regression error on the training sample and then by the leave-one-out log-density criterion. If the preliminary estimate is not finite, the direction is $e_1$ and the confidence region is empty.

The selected tuple is then fixed and the procedure is rerun from the Fourier initial estimate. The case-weight derivative and its covariance are defined in \eqref{inf:ijdefinition}, and Theorem~\ref{inf:caseweights} proves their consistency. The limiting distribution is common to all grid points and efficient. Inference for a candidate direction uses a separate selection criterion that depends only on the candidate index and the response (Section~\ref{sec:conditional-score-proofs}).

\subsection{Projected ascent}\label{sec:ascent}
Use any measurable orthonormal $U_p\perp\widetilde\gamma$, $\gamma_p(z)=(\widetilde\gamma+U_pz)/(1+\|z\|^2)^{1/2}$ and $R_n=\min(1.5,2n^{-1/20})$. Write $F_n(z)=L_n\{\gamma_p(z)\}$ for the criterion in this chart, with the tuning tuple held fixed. Begin at $z=0$, use $J_n=\lceil\log^3(n+e)\rceil$ iterations, initial step $\eta_n=1/\log(n+e)$, and Euclidean projection onto the cap, with the zero vector fixed. Each line search uses at most 80 trials with successive step halving, and the step $d=z_{\rm new}-z$ is taken when
\[
 F_n(z+d)\ge F_n(z)+\tfrac12\nabla F_n(z)^Td.
\]
Iteration stops when the gradient norm is at most $n^{-1}$. A trial step with a non-finite objective value is discarded and the step size halved; if the resulting function value or gradient is non-finite, the estimate is $e_1$ and the confidence region is empty.

On a working chart of radius $16n^{-1/20}$ around the true direction, the score and Hessian bounds of Theorem~\ref{inf:rates}, with the preprocessing bounds of Section~\ref{sec:preprocessing}, give uniform negative curvature and a local maximum at distance of order $n^{-1/2}$ from the truth. Root-$n$ consistency of the initial estimate implies that the cap lies inside this chart and, with probability approaching one, contains the maximizer in its interior. The chart transitions have uniformly bounded second derivatives, and their first-derivative singular values tend to one. The additional Hessian term, the true-chart gradient times the chart curvature, has norm of order $O_P(R_n)$. Thus the eigenvalues of the Hessian in the working chart lie in the interval $[-C, -c]$ for some fixed constants $0 < c < C$.

For $\eta_n\le1/C$, projection gives $\nabla F_n(z)^Td\ge\|d\|^2/\eta_n$, and Taylor's lower bound shows that the first Armijo trial satisfies the sufficient-increase condition. The unprojected gradient map has derivative eigenvalues in $[0,1-c\eta_n]$, and projection is nonexpansive and fixes the interior maximizer $z_*$. Therefore
\[
 \|z_{J_n}-z_*\|\le2R_n(1-c\eta_n)^{J_n}
 \le2R_n\exp\{-c\log^2(n+e)\}=o(n^{-q})
\]
for every fixed $q>0$. A stop by the gradient criterion before $J_n$ iterations gives $\|\widehat z-z_*\|\le n^{-1}/c$. Both imply $o_P(n^{-1/2})$ error. The contraction also eventually brings the score below the stopping tolerance, so the exceptional outcomes above have a probability tending to zero.

\subsection{Inference conditions and the confidence region}\label{sec:inference-conditions}
The inference conditions are an interior solution, finiteness of the initial estimate and both fits, positive $\widehat J$ and $\widehat B$, positive observed working curvature, and a working-score norm at most $n^{-1/2}/\log(n+e)$. Otherwise, the confidence region is empty. The confidence set is \eqref{cs:local-domain}; the Wald set of Section~\ref{sec:wald} is given for comparison. The Jacobian of the raw map in \eqref{eq:rp-target-cancel} transfers the covariance in both directions; the additional Hessian term away from a stationary point is described in Theorem~\ref{dir:theorem}. The point expansion and the regularity follow from Theorem~\ref{inf:rates} and Section~\ref{sec:core-efficiency}.

Computing the Fourier matrices requires $O(n^2p^2)$ operations; computing each score function requires $O(n^2p)$; the optimization requires $O(n^2p\log^3 n)$ (with a limit of 80 trials per one-dimensional search); and computing the analytical Hessian requires $O(n^2p^2)$. Runtimes are reported in Section~\ref{sec:sim-details}.
\subsection{Wald region}\label{sec:wald}
When the inference conditions of Section~\ref{sec:inference-conditions} hold, map the covariance to the fitted working chart. Let $d=p-1$, let $\widehat U$ be an orthonormal frame for $\widehat\gamma^\perp$ in working coordinates, and let $F$ be the raw tangent frame in which $\widehat V$ was computed. With $D=(I-\widehat\beta\widehat\beta^T)\widehat A^T\widehat U/\|\widehat A^T\widehat\gamma\|$ and $G=F^TD$, put $\widehat V_w=G^{-1}\widehat V G^{-T}$. The Wald region is
\begin{equation}\label{core:reference-wald}
\mathcal C_n=\left\{\left[\frac{\widehat A^T(\widehat\gamma+\widehat Uu)}
 {\|\widehat A^T(\widehat\gamma+\widehat Uu)\|}\right]:
 n u^T\widehat V_w^{-1}u\le\chi^2_{d,.95}\right\}.
\end{equation}

Otherwise, the Wald region is empty.

\section{Efficient information and regularity}\label{sec:core-efficiency}
The efficiency bound concerns the model in which the link, the density of the independent mean-zero errors, and the covariate distribution are all unknown. Let $T=\beta_0^TX$, $V=U_0^TX$, $\mu(T)=E(V\mid T)$, $W=V-\mu(T)$, and $M=E\{m'(T)^2WW^T\}$. For tuning sequences tending to zero, this applies to the expansion in Theorem~\ref{inf:rates} and to the estimate computed by the algorithm of Section~\ref{sec:algorithm}. If instead the smoothing parameter is held at a fixed positive value, the score influence function in tangent coordinates is $\eta_k$ and the parameter influence function is $J_k^{-1}\eta_k$; Proposition~\ref{inf:loss} gives the resulting information loss. For the information argument we assume that $|\psi(e)|\le C(1+|e|)$ and $\int\{(\sqrt f)'\}^2=I_f/4<\infty$. Hence $\sqrt f \in H^1$, and the translation operator is differentiable in $L^2$. Since the predictors are Gaussian, the link derivative is bounded and smooth, and link perturbations are smooth and bounded, the regression path is differentiable in quadratic mean (DQM). The squared derivative in the location parameter is bounded above by a constant multiple of $(1+\|X\|^2)I_f$ and is integrable. Truncation and dominated convergence extend the $L^2$ derivative of the translation operator to these stochastic shifts. Bounded multiplicative perturbations of the error density or the design law give the corresponding nuisance scores. Let $t$ denote the derivative of the index, $a(T)$ the derivative of the link function, $b(\eps)$ the score for the error, and $d(X)$ the score for the design. The path score is
\[
s=-\psi(\eps)\{m'(T)V^Tt+a(T)\}+b(\eps)+d(X),
\qquad Eb=E(\eps b)=Ed=0.
\]
Writing $V=\mu(T)+W$, projection onto the link nuisance tangent space removes the term $-\psi m'\mu(T)$. The removed term lies in the closure of the link tangent space, because smooth compactly supported link perturbations approximate the square-integrable function $m'\mu$ (truncate in $T$, then approximate by smooth compactly supported functions). Gaussian second moments justify this truncation in the $I_f$-weighted $L^2$ norm. The remaining term, $S_{\rm eff} = -\psi m'W$, is orthogonal to the link score (since $E(W\mid T)=0$), the error score (since $E(m'W)=0$), and the design score (since $E\psi=0$). Its covariance is $I_fM$, and $\varphi = (I_fM)^{-1}S_{\rm eff}$ satisfies $E(\varphi s) = t$. Here $S_{\rm eff}$ is the efficient score of \citet{ma2013efficient}. For a shrinking tuning sequence satisfying the conditions of Theorem~\ref{inf:rates}, the SI-KDR estimator is asymptotically linear under the underlying law, with influence function $\varphi$. Along DQM paths, local asymptotic normality and contiguity carry the remainder terms and localization events over to the measures $P_{u/\sqrt{n}}$. By a joint iid central limit theorem for $(\varphi, s)$ and Le Cam's third lemma, the normalized limit is shifted by $uE(\varphi s) = ut$. Centering the path at the local chart parameter removes this shift, leaving covariance $(I_fM)^{-1}$. This establishes regularity for models with unconstrained designs \citep{vandervaart2002semiparametric}.

By Gaussian tails, integration by parts gives $E[\eps\psi(\eps)] = -1$. The Cauchy--Schwarz inequality implies $I_f\Var(\eps) \ge 1$. Equality holds if and only if $\psi(e) = -e/\Var(\eps)$ almost everywhere, that is, for a Gaussian density. In the model restricting only the conditional mean, evaluated at a law with constant conditional error variance, the lower bound is $\Var(\eps)M^{-1}$.

\section{Finite-smoothing bias in raw coordinates}\label{supp:rawbias}
\label{sec:bias-diagnostic}

This section computes the population score of SI-KDR in the original, unwhitened coordinates. For a correlated design ($k\ne0$ in \eqref{eq:bias-decomposition}), finite regression smoothing can produce a nonzero term in the score, and whitening removes it. The fourth-order expansion below holds for a fixed positive density bandwidth and floor.

\subsection{Gaussian population derivative}
Let $X\sim N(0,\Sigma)$, $T=\beta_0^TX$, $V=U^TX$, where $U^T\beta_0=0$ and $U^TU=I$. Write
\begin{equation}
 v=\Var(T),\qquad k=U^T\Sigma\beta_0,\qquad
 V=(k/v)T+W,\qquad W\perp T.
 \label{eq:bias-decomposition}
\end{equation}
Let $Y=m(T)+\epsilon$, with independent centered error having a fixed finite Gaussian-mixture density $f$. Fix the regression bandwidth $g>0$, density bandwidth $h>0$, and density floor $\rho>0$. Initially, take regression ridge zero. The population Gaussian NW fit is
\begin{equation}
 M_{\beta,g}(s)=\E\{m(T)\mid\beta^TX+gZ=s\},\qquad
 Z\sim N(0,1),
 \label{eq:bias-population-nw}
\end{equation}
where the auxiliary Gaussian is independent. Put
\begin{equation}
 D=v+g^2,\qquad \alpha=v/D,\qquad
 A_j(t)=\E m^{(j)}\{\alpha t+\sqrt{\alpha g^2}Z\}.
 \label{eq:bias-A}
\end{equation}
At the truth, the fitted value is $A_0(T)$. The first-stage derivative, including the movement of the evaluation index, in the normalized chart $\beta(\theta)=(\beta_0+U\theta)/(1+\|\theta\|^2)^{1/2}$ is
\begin{equation}
 \left.\partial_\theta
 M_{\beta(\theta),g}\{\beta(\theta)^TX\}\right|_0
 =\alpha A_1(T)W+\frac{kg^2}{D^2}
                   \{2TA_1(T)-vA_2(T)\}.
 \label{eq:bias-first-derivative}
\end{equation}
Indeed, for $c_\beta=\Cov(T,\beta^TX)$ and $v_\beta=\Var(\beta^TX)$ the Gaussian conditional mean and variance in \eqref{eq:bias-population-nw} are $u=c_\beta s/(v_\beta+g^2)$ and $\zeta=v-c_\beta^2/(v_\beta+g^2)$. At the moving evaluation row,
\[
 \dot u=\alpha W+2kg^2T/D^2,
 \qquad \dot\zeta=-2vkg^2/D^2.
\]
Gaussian differentiation gives $d\E m(u+\sqrt\zeta Z)=A_1\,du+(A_2/2)\,d\zeta$, proving \eqref{eq:bias-first-derivative} without a smoothing expansion.

\subsection{The residual coefficient and the fourth-order term}
Set $R=\epsilon+\delta_g(T)$, $\delta_g=m-A_0$, and let $p$ be its density. For $K_h=\phi_h$ and $d=K_h*p$, consider
\[
 \mathcal F(p)=\int p(r)\log\{(K_h*p)(r)+\rho\}\,dr.
\]
Differentiating both the outer evaluation and all moving density centers under a residual perturbation $\dot R$ gives
\begin{equation}
 \dot{\mathcal F}=\E\{a_p(R)\dot R\},\qquad
 a_p(r)=\frac{d'(r)}{d(r)+\rho}
        +K_h'*\left\{\frac{p}{d+\rho}\right\}(r).
 \label{eq:bias-complete-coefficient}
\end{equation}
The convolution term follows from $K_h'(s-r)=-K_h'(r-s)$ for a moving center. Translation invariance, or an interchange of the two integrals, gives $\E_p a_p(R)=0$. Since $R$ is a function of $(T,\epsilon)$ independent of $W$, the finite-smoothing score is
\begin{equation}
 S_{g,h,\rho}
 =-\frac{kg^2}{D^2}
   \E\left[a_p\{\epsilon+\delta_g(T)\}
                  \{2TA_1(T)-vA_2(T)\}\right].
 \label{eq:bias-exact-score}
\end{equation}
It is parallel to $k$ and vanishes for every $g,h,\rho$ when $k=0$. In particular, population whitening removes this population score. Whitening estimated from the same data is treated in Section~\ref{sec:preprocessing}.

Suppose $m$ and its first six derivatives are bounded. Define
\begin{equation}
 b(t)=\frac{tm'(t)}v-\frac{m''(t)}2.
 \label{eq:bias-b}
\end{equation}
Expansion of the Gaussian conditional mean yields, for every fixed finite $q$, $\delta_g(T)=g^2b(T)+O_{L^q}(g^4)$ and $2TA_1-vA_2=2vb(T)+O_{L^q}(g^2)$. All polynomial factors in $T$ have finite moments. For $d_0=K_h*f$, $\psi=f'/f$, put
\begin{equation}
 J_{h,\rho}=I_f-\int f
       \left\{\frac{d_0'}{d_0+\rho}-\psi\right\}^2.
 \label{eq:bias-J}
\end{equation}
Integration by parts, justified by Gaussian-mixture tails, gives
\begin{equation}
 \E_f a_f'
 =-2\int f'\frac{d_0'}{d_0+\rho}
     +\int f\left(\frac{d_0'}{d_0+\rho}\right)^2
 =-J_{h,\rho}.
 \label{eq:bias-J-identity}
\end{equation}
The incoming term contributes $\int d_0''f/(d_0+\rho)$; integration by parts supplies the second copy of the cross term in \eqref{eq:bias-J-identity}.

With $h,\rho$ fixed positive, the residual density satisfies $p=f-g^2(\E b)f'+O(g^4)$ in the required weighted integral norms. Translation equivariance of \eqref{eq:bias-complete-coefficient} then gives
\begin{equation}
 \E_\epsilon a_p\{\epsilon+\delta_g(T)\}
 =-J_{h,\rho}g^2\{b(T)-\E b(T)\}+O_{L^2(T)}(g^4).
 \label{eq:bias-centered-expansion}
\end{equation}
To justify the remainder, expand the shifted Gaussian-mixture density twice and then the two quotients in \eqref{eq:bias-complete-coefficient}. Their denominators are at least the fixed positive $\rho$; Gaussian kernel derivatives are bounded. Density derivatives have integrable Gaussian envelopes even under the bounded shift $\delta_g$. The displayed $L^q$ remainders thus give the weighted $L^2(T)$ remainder. The linear density perturbation is the common translation $g^2\E b$, which subtracts $\E b$ in \eqref{eq:bias-centered-expansion}. Substitution into \eqref{eq:bias-exact-score} proves
\begin{equation}
 S_{g,h,\rho}
 =\frac{2J_{h,\rho}}v\,k g^4\Var\{b(T)\}+O(g^6).
 \label{eq:bias-fourth-order}
\end{equation}
The mixture-score lemma of Section~\ref{sec:core-error} gives $J_{h,\rho}\to I_f$ as $h,\rho\to0$, so $J_{h,\rho}>0$ for small $h,\rho$.

When $J_{h,\rho}>0$, the leading score points toward increasing raw index variance. Define $H_{g,h,\rho}$ as the negative population Hessian at truth. In the regime where this matrix is positive definite, the first-order local parameter bias is $H_{g,h,\rho}^{-1}S_{g,h,\rho}$, with the finite-smoothing information in place of the efficient information. For the simulation design of Section~\ref*{main-sec:experiments} of the paper, $v=1.5$ and $\Sigma\beta_0-v\beta_0=(0,0,.75/\sqrt2)^T$. An order-$g^2$ translation term cancels because the score includes the incoming derivative.

\subsection{The ridge correction}
Now let the regression ridge $\lambda$ be positive. Define
\begin{gather*}
 q_g(t)=\phi_{\sqrt{v+g^2}}(t),\qquad
 s_g(t)=\frac{q_g(t)}{q_g(t)+\lambda},\\
 M_g(t)=s_g(t)A_0(t),\qquad R=\epsilon+m(T)-M_g(T).
\end{gather*}
The affine conditional-mean identity yields
\[
 D_\mu(T)=-\frac{k}v\left\{g\partial_gM_g(T)
                   +\frac{\lambda M_g(T)}{q_g(T)+\lambda}\right\}.
\]
Hence, with $p$ now the density of the ridge-fit residual,
\begin{equation}
 S_{g,h,\rho,\lambda}
   =\frac{k}v\E\{a_p(R)F_g(T)\},\qquad
 F_g=g\partial_gM_g+(1-s_g)M_g.
 \label{eq:bias-ridge-score}
\end{equation}
Here $g$ is differentiated holding $\lambda$ fixed. A computation gives
\begin{align}
 g\partial_gM_g
 ={}&s_g\left\{-\frac{2vg^2T}{D^2}A_1
                     +\frac{v^2g^2}{D^2}A_2\right\}
       \nonumber\\
 &+s_g(1-s_g)A_0\frac{g^2(T^2-D)}{D^2}.
 \label{eq:bias-ridge-derivative}
\end{align}
The term $(1-s_g)M_g$ in \eqref{eq:bias-ridge-score} arises because the ridge weights sum to less than one. These equations show that the population score also vanishes for $k=0$ with the ridge; the expansion \eqref{eq:bias-fourth-order} concerns zero ridge.

This finite-smoothing score motivates the training standardization used in the paper.

\section{Finite smoothing, influence and covariance estimation}
\label{inf:theory}
This section uses the empirical-integral bounds and Gaussian jet bounds of Sections~\ref{supp:operator}--\ref{sec:oracle}.

\subsection{The population target}
\looseness=-1 The proof uses population-standardized units: $X=(T,W)$, $T\sim N(0,1)$, $W\sim N(0,I_d)$ with $d=p-1$, independent of $(T,\eps)$, $Y=m(T)+\eps$, $\E Y=0$, $\Var Y=1$. The link is a nonconstant bounded $C^6$ function with bounded derivatives, and the independent centered error has a fixed finite Gaussian mixture density with positive weights and component variances. The nonsingular tangent map from Section~\ref{sec:preprocessing} carries the conclusions back to the original units.

We distinguish the conditional-mean profile $m_\theta(t)=\E[Y\mid\beta(\theta)^TX=t]$, the population smoothed profile
\begin{align}
 q_{\theta,g}(t)&=\E K_g(t-\beta(\theta)^TX),\notag\\
 M_{\theta,g,\lambda}(t)&=
 \frac{\E\{K_g(t-\beta(\theta)^TX)Y\}}{q_{\theta,g}(t)+\lambda},
 \label{inf:profile}
\end{align}
and the leave-one-out profile computed from the sample, whose denominator is $\sum_{j\ne i}K_g(T_i-T_j)+(n-1)\lambda$. Let $R_\theta=Y-M_{\theta,g,\lambda}(\beta(\theta)^TX)$, with density $p_\theta$. The population target is
\begin{equation}
 Q(\theta)=\int p_\theta(r)\log\{(K_h*p_\theta)(r)+\rho\}\,dr.
 \label{inf:target}
\end{equation}
The four smoothing parameters are held fixed during differentiation. Put $b=\nabla Q(0)$ and $J=-\nabla^2Q(0)$.

At truth define $\alpha=(1+g^2)^{-1}$, $q(t)=\phi_{\sqrt{1+g^2}}(t)$, $s(t)=q(t)/(q(t)+\lambda)$, and
\begin{align}
 A_j(t)&=\E m^{(j)}\{\alpha t+\sqrt{\alpha g^2}Z\},&
 M(t)&=s(t)A_0(t),&\delta(t)&=m(t)-M(t).\label{inf:gaussianprofile}
\end{align}
For $R=\eps+\delta(T)$, $p=\mathcal L(R)$, $d=K_h*p$, the residual coefficient is
\begin{equation}
 a(r)=\frac{d'(r)}{d(r)+\rho}
       +\left[K_h'*\left\{\frac p{d+\rho}\right\}\right](r).
 \label{inf:fulla}
\end{equation}
Exchanging the arguments of the odd kernel derivative gives $\E a(R)=0$.

\begin{proposition}[Population score and curvature]\label{inf:population}
For every positive $g,h,\lambda,\rho$, $b=0$ and $J=jI_d$, where
\begin{align}
 N(T)&=-M''(T)+TM'(T)+s(T)\alpha\{TA_1(T)-A_2(T)\},\notag\\
 j&=-\E\{a'(R)M'(T)^2+a(R)N(T)\}.\label{inf:curvature}
\end{align}
\end{proposition}
\begin{proof}
Along a unit tangent use $\beta(\theta)=\cos(\theta)e_1+\sin(\theta)u$. The projected variance stays one, so $q_{\theta,g}=q$. The conditional Gaussian argument at fixed $t$ has mean $\alpha\cos(\theta)t$ and variance $1-\alpha\cos(\theta)^2$. Their first derivatives vanish at zero; their second derivatives are $-\alpha t$ and $2\alpha$. Thus the fixed-argument profile second derivative is $s(t)\alpha\{-tA_1(t)+A_2(t)\}$. The evaluation index has derivatives $W_u,-T$, giving
\begin{align*}
 \dot R&=-M'(T)W_u,\\
 \ddot R&=-M''(T)W_u^2+TM'(T)+s(T)\alpha\{TA_1(T)-A_2(T)\}.
\end{align*}
Conditional centering gives $\dot p=0$ and $\E\{a(R)\dot R\}=0$. The second derivative is $\E\{a'(R)\dot R^2+a(R)\ddot R\}$; varying the density coefficient produces zero because $\dot p=0$. Integrate $W_u^2$; rotational invariance gives equal diagonal curvatures and zero mixed terms. Differentiation under the integral is justified by polynomial integrable envelopes: the residual law is a bounded-location mixture of fixed Gaussian error components, and the population ridge profile has polynomial index derivatives.
\end{proof}
As in Section~\ref{supp:rawbias}, without whitening, the score equals $(k/v)\E[a(R)\{g\partial_gM+(1-s)M\}]$, which is generally nonzero; with fixed density smoothing and no ridge, its leading term is of order $g^4$.

\subsection{First-order influence}
Define
\begin{align}
 \kappa(t,y)={}&-a\{y-M(t)\}M'(t)\notag\\
 &+\E\left[a(R')\frac{K_g'(T'-t)\{y-M(T')\}}{q(T')+\lambda}\right],
 \label{inf:kappa}\\
 C={}&\E\{T\kappa(T,Y)\},&
 \widetilde\kappa(t,y)={}&\kappa(t,y)-Ct.\label{inf:whiteningprojection}
\end{align}
The first term carries both roles of the density through $a$; the second arises from refitting the NW smoother.

\begin{proposition}[Influence function with training preprocessing]
\label{inf:influence}
At fixed population preprocessing, the Gateaux derivative of the direction score in direction $\delta_z-P$ is $W_z\kappa(T_z,Y_z)$. Including training predictor and response moments gives, at the true direction in working coordinates,
\begin{equation}
 \eta(z)=W_z\widetilde\kappa(T_z,Y_z),\qquad
 B=\Var\{\eta(Z)\}=\E\{\widetilde\kappa(T,Y)^2\}I_d.
 \label{inf:eta}
\end{equation}
It is mean zero with finite moments. The parameter influence is $J^{-1}\eta$ followed by the nonsingular raw tangent transformation.
\end{proposition}
\begin{proof}
For $z=(t_z,w_z,y_z)$ the NW variation is
\begin{equation}
 H_z(t)=\frac{K_g(t-t_z)\{y_z-M(t)\}-\lambda M(t)}{q(t)+\lambda}.
 \label{inf:nwvariation}
\end{equation}
The ridge term arises from the denominator because the ridge weights do not sum to one. The directional derivative of this variation is
\[
 H_z'(t)w-\frac{K_g'(t-t_z)\{y_z-M(t)\}}{q(t)+\lambda}w_z.
\]
The outer-measure variation is $-a(R_z)M'(t_z)w_z$. Variations of $a$ and $R$ multiplied by $\dot R=-M'(T)W$ depend only on $(T,\eps,t_z,y_z)$ and integrate to zero against centered independent $W$. The same holds for the $w$ part above; the remaining $w_z$ part is the incoming term of \eqref{inf:kappa}.

Predictor centering cancels from pair differences. At the isotropic design, the covariance influence is $zz^T-I$ and inverse-square-root influence is $-(zz^T-I)/2$. Combining this with the moving frame at $\gamma_0(A)=A^{-T}e_1/\norm{A^{-T}e_1}$ gives tangent-index covariance derivative $\Pi(A+A^T-2I)e_1$, hence $-W_zT_z$. The affine-mean population identity adds $-W_zT_zC_0$, with
\begin{equation}
 C_0=\E[a(R)\{g\partial_gM(T)+(1-s(T))M(T)\}].\label{inf:cidentity}
\end{equation}
To see that $C_0=C$, vary the Gaussian covariance with derivative $e_1u^T+ue_1^T$. Its score is $TW_u$. The Gateaux formula above gives $\E(T\kappa)u$, and the affine-mean identity gives $C_0u$ for the same functional. Hence $C_0=C$, which also fixes the sign of the correction.

For every fixed response shift and positive scale, the rotational argument gives an identically zero population direction score, ridge term included, so its derivatives in these two parameters vanish. Thus, the estimated response mean and scale contribute no first-order term in the population. Finally, the identity $B(A,\gamma_0(A))=\beta_0$ cancels the moving coordinate target on returning to raw units.
\end{proof}

\subsection{Information loss}
Let $S_{\rm eff}=-Wm'(T)\psi(\eps)$, $\psi=f'/f$, and $I_{\rm eff}=i_{\rm eff}I_d$, $i_{\rm eff}=I_f\E m'^2$, the efficient score and information in the independent-error model with unrestricted design.
\begin{proposition}[Information loss]\label{inf:loss}
For $j>0$,
\begin{align}
 \E(\eta S_{\rm eff}^T)&=J,\label{inf:crossidentity}\\
 J^{-1}BJ^{-T}-I_{\rm eff}^{-1}
 &=\E\big[(J^{-1}\eta-I_{\rm eff}^{-1}S_{\rm eff})
          (J^{-1}\eta-I_{\rm eff}^{-1}S_{\rm eff})^T\big]\succeq0.
 \label{inf:lossidentity}
\end{align}
Generally, $B\ne J$ at finite smoothing.
\end{proposition}
\begin{proof}
Rotate the true direction while keeping the standard Gaussian design, link, and error law. The model score is $S_{\rm eff}$. The population criterion score is zero at that model's true direction for every rotation. Differentiating gives $\E(\eta S_{\rm eff}^T)+\nabla^2Q=0$. Expanding the square gives the result. The $-CT$ term is orthogonal to $S_{\rm eff}$ since $\E\psi(\eps)=0$. Nonsingular raw tangent transformations preserve the identity.
\end{proof}
In the limiting case $g=\lambda=0$, $\kappa=-m'a_{f,h,\rho}(\eps)$, $C=0$, and $j=J_{h,\rho}\E m'^2$, with
\[
 J_{h,\rho}=I_f-\left\|\frac{(K_h*f)'}{K_h*f+\rho}-\psi\right\|_{L^2(f)}^2.
\]
Density smoothing alone, therefore, yields a sandwich covariance that generally differs from the inverse curvature, and finite link smoothing and its refit enter through \eqref{inf:kappa}--\eqref{inf:curvature}. With fixed, nonzero smoothing, the comparison is made within the Gaussian submodel; efficiency in the unrestricted-design model is recovered as the bandwidths shrink.

\subsection{Population bounds}
Put $v=g^2+\sqrt\lambda$, $\xi=h^2+\sqrt\rho\log^C(e/\rho)$ and $\mathcal L=1+\log_+\{1/(h\rho)\}$. Constants depend only on the fixed model and multipliers in a fixed compact grid. The exponent $C$ of the logarithm may change from line to line; only logarithmic factors are absorbed in this way, never a polynomial factor of an inverse bandwidth, ridge, or floor.
\begin{lemma}[Continuum integrated bound]\label{inf:transport}
For coupled residuals with continuous interpolating laws,
\begin{equation}
 \norm{a_{\mathcal L(U)}(U)-a_{\mathcal L(V)}(V)}_2
 \le Ch^{-2}\mathcal L^2\norm{U-V}_2.\label{inf:transportbound}
\end{equation}
Consequently
\begin{align}
 d_\kappa:=\norm{\kappa+m'\psi}_2
 &\le\log^C n\{\xi+v/(gh^2)+v/h\},\label{inf:kappabound}\\
 \norm{\widetilde\kappa+m'\psi}_2&\le2d_\kappa,&
 |j-i_{\rm eff}|&\le\log^C n\{v/h^3+h^4+\rho\}.
 \label{inf:jbound}
\end{align}
\end{lemma}
\begin{proof}
Fix $r$, let $d(s)=\int K_h(s-z)dP(z)$ and on $s\ge r$ put $u(s)=P([r,s])$. Monotonicity gives $d(s)\ge K_h(s-r)u(s)$, hence
\[
 \int_{s\ge r}\frac{K_h(s-r)}{d(s)+\rho}dP(s)
 \le\int_0^1\min\{u^{-1},C/(h\rho)\}\,du\le C\mathcal L.
\]
The other side is identical; outgoing mass is at most one. Splitting standardized distances at $C\sqrt{\mathcal L}$ and using the Gaussian kernel divided by $\rho$ for the remainder yields the same displacement moments as Lemma~\ref{op-moment-lemma}. The proof of Lemma~\ref{op-derivative-lemma} then applies with probability integrals and gives \eqref{inf:transportbound}. For $U=\eps+\delta(T)$, every intermediate law is continuous.

The bias portion of the Gaussian jet calculation gives $\norm\delta_2+\norm{M'-m'}_2\le\log^C(e/\lambda)v$, also with fixed polynomial index weights. Smoothing contributes $g^2$; integrating the squared ridge term and its derivatives gives $C\lambda\log^C(e/\lambda)$. The normalized derivative-NW operator has $L^2(P_T)$ norm $\log^C(e/\lambda)/g$: use displacement bounds for rows, $q_0/q_g\le C$ and $\int|K_g'|=C/g$ for columns, and Schur. Bounded $M$ and conditional response second moments handle $y-M(T')$.

In the incoming integral, subtract $a_{f,h,\rho}(\eps')$, whose contribution is zero. The transport and derivative-operator bounds apply. The outgoing term additionally uses the derivative bias and the continuum bound $\norm a_\infty\le Ch^{-1}\mathcal L^{3/2}$. The mixture inequality of Lemma~\ref{lem:mixture-rate} $\norm{a_{f,h,\rho}-\psi}_2\le C\xi$ proves \eqref{inf:kappabound}. Gaussian tails permit polynomial index and response weights at a logarithmic cost. Since $\E(Tm'\psi)=0$, $|C|\le d_\kappa$. The continuum Hessian transfer applied to population jet biases has bound $\log^C n\,v/h^3$ for $h\le1$; the oracle information discrepancy is $O(h^4+\rho\log^C n)$. This proves \eqref{inf:jbound}, and eventual positive curvature follows from $i_{\rm eff}>0$.
\end{proof}

\subsection{Sample expansion of the score and Hessian}
\label{dir:section}
We use the population profile, residual law, curvature, and influence function of \eqref{inf:profile}--\eqref{inf:eta}, now indexed by $k=(g,h,\lambda,\rho)$. In this subsection $d=K_h*p$, $a_p$ denotes \eqref{inf:fulla}, and $M=sA_0$. The model conditions are those stated at the start of this section, and all direction and weight derivatives hold the tuning parameters fixed. In particular, the second jet of the population residual includes $s\alpha(TA_1-A_2)$, which is a population term rather than a stochastic fitting error.
\paragraph*{Rates.} Put
\begin{gather*}
 \ell_n=1+\log n+\log_+(1/g)+\log_+(1/h)
                  +\log_+(1/\lambda)+\log_+(1/\rho),\\
 \sigma_j=(ng^{2j+1})^{-1/2},\quad
 \omega_j=(ng^{j+1}\sqrt\lambda)^{-1},\quad s_j=\sigma_j+\omega_j.
\end{gather*}
Assume eventually $n^{-A}\le g,h,\lambda,\rho\le1$ for fixed finite $A$; fixed positive endpoints are allowed. Let $r_n=c_rn^{-\nu_r}$, $0<\nu_r<1/2$, and $t_H=n^{-\nu_H}$, $\nu_H>0$. Define
\begin{align}
 D_S={}&\ell_n^{32}\left[
 \frac{s_0/h^2+(nh^3)^{-1/2}}g+\frac{s_1}h+
 \frac{s_0}g+\sigma_1+\frac1{\sqrt n\,g h^2}\right],\label{dir:DS}\\
 D_H(r_n)={}&\ell_n^{32}\left[
 \frac{s_0}{h^3}+\frac{s_1}{h^2}+\frac{s_2}h+
 (nh^5t_H)^{-1/2}+(nh^3t_H)^{-1}+\frac{t_H}{h^2}
 +n^{-1/2}+r_n\right].\label{dir:DH}
\end{align}
Here $\sigma_j$ is the stochastic fluctuation and $\omega_j$ is the linear term arising from the Bernstein inequality and the leave-one-out construction. The condition $ng\lambda/\ell_n^{12}\to\infty$ absorbs $\omega_j$ into $\sigma_j$. The ridge enters polynomially only through the factor $\lambda^{-1/2}$ in $\omega_j$, and the floor enters only through $\ell_n$. The population smoothing bias, $d_\kappa$, and the discrepancy $|j_k-i_{\rm eff}|$ are population quantities and do not enter the sample remainders.

\begin{theorem}[Finite-smoothing expansion]\label{dir:theorem}\label{inf:rates}
Under the primitive model above, suppose $ng\lambda/\ell_n^{12}\to\infty$ and $D_S+D_H(r_n)\to0$. For the same-sample leave-one-out (LOO) NW criterion, including all divisor-$n$ training moments and the return to raw directions,
\begin{align}
 \sqrt n\nabla L_{n,k}(0)
 &=n^{-1/2}\sum_i\eta_k(O_i)+O_P(D_S),\label{dir:score}\\
 \sup_{\|\theta\|\le r_n}\|-\nabla^2L_{n,k}(\theta)-J_k\|
 &=O_P\{D_H(r_n)\}.\label{dir:hessian}
\end{align}
These statements hold simultaneously over a fixed finite deterministic grid. If $\inf_nj_k>0$, an interior approximate root with score tolerance $\tau_n=o(n^{-1/2})$ satisfies
\begin{equation}
 \sqrt n\widehat\theta_k=J_k^{-1}n^{-1/2}\sum_i\eta_k(O_i)
 +O_P\{D_S+D_H(r_n)+\sqrt n\tau_n\}.\label{dir:estimate}
\end{equation}
The finite Armijo algorithm, starting from the Fourier initial estimator, produces such a root under the conditions verified in Section~\ref{dir:curvature}. For general fixed positive-definite predictor covariance, let $D_0$ map standardized tangent parameters to a raw orthonormal tangent. Scores transform by $D_0^{-T}$, curvatures by $D_0^{-T}J_kD_0^{-1}$ and parameter covariances by $D_0V_kD_0^T$. Away from a root, the Hessian additionally has the gradient times the second derivative of the coordinate map; uniformly on the local ball, it is $O_P(n^{-1/2}+r_n)$, and it is $O_P(n^{-1/2})$ at the truth or a root-$n$ evaluation point. The leading asymptotic parameter covariance is $V_k/n$.
\end{theorem}

\subsection{Population envelopes and positive curvature}
\label{dir:popbounds}
The law of $R_\theta$ is a bounded-location Gaussian mixture: its component variances are the fixed error variances and its means lie in a fixed bounded interval, since $|m|,|M_{P,\theta}|\le\|m\|_\infty$. Componentwise calculation gives, uniformly over every such mixing law,
\begin{align}
 p&\le Cd,\qquad |p^{(j)}(r)|/p(r)+|d^{(j)}(r)|/d(r)
 \le C_j(1+|r|^j),\quad j\le4,\notag\\
 \int\frac{pd}{(d+\rho)^2}&\le C\sqrt{\log(e/\rho)},\qquad
 P\{d(R)<t\}\le Ct\sqrt{\log(e/t)}.\label{dir:mixture}
\end{align}
The integral bounds follow by splitting at a sufficiently large multiple of $\sqrt{\log(e/\rho)}$ or $\sqrt{\log(e/t)}$; the central integrands are bounded using $p\le Cd$ and the outside integrals use uniform Gaussian tails. Differentiating $p/(d+\rho)$ and moving $K_h'$ to that quotient gives
\begin{equation}
 |a_p^{(j)}(r)|\le C_j(1+|r|^{j+1}),\quad j=0,1,2,\label{dir:aenv}
\end{equation}
uniformly in all four parameters in $(0,1]$, with constants free of $\rho^{-1}$. The Gaussian profile formula similarly gives $|M^{(j)}(t)|\le C_j(1+|t|^j)$ for $j\le3$; mixed direction and preprocessing derivatives through order three have degree at most six.

Let $b_k(t)=E_\varepsilon a_p\{\varepsilon+m(t)-M(t)\}$ and $u_g(t)=\phi(t)/(q_g(t)+\lambda)$. Then $|u_g|\le\sqrt2$, $u_g'=u_g(-t+\alpha ts)$, $|b_k|\le C$ and $|b_k'|\le C(1+|t|)$. An integration by parts, with vanishing Gaussian boundary term, yields
\begin{equation}
 \kappa_k(t,y)=-a_p\{y-M(t)\}M'(t)
 -\int K_g(u-t)\partial_u[u_g(u)b_k(u)\{y-M(u)\}]\,du.\label{dir:kappaibp}
\end{equation}
Thus $|\kappa_k(t,y)|\le C(1+|t|)(1+|y|)$ and all finite moments of $\eta_k$ are bounded uniformly. Also $E\kappa_k=0$: integrating its incoming term over $(t,y)$ uses $E[K_g'(u-T)\{Y-M(u)\}]=(q_g(u)+\lambda)M'(u)$ and cancels its outgoing mean. These facts make an inverse-$g$ moment bound unnecessary and justify empirical centering in the covariance connection.

If all four parameters tend to zero, $M^{(j)}(t)\to m^{(j)}(t)$ pointwise for $j\le2$. On compact residual intervals $p^{(j)}\to f^{(j)}$ and $a_p^{(j)}\to\psi^{(j)}$, $j\le1$; the uniform envelopes extend convergence to weighted moments. In \eqref{dir:kappaibp}, $b_k(t)\to E\psi=0$ and $b_k'(t)=\{m'(t)-M'(t)\}Ea_p'(R\mid T=t)\to0$. Substitution $u=t+gZ$ and domination give
\begin{equation}
 \eta_k\to-Wm'(T)\psi(\varepsilon)\text{ in }L^2,\qquad
 j_k\to i_{\rm eff}=E[m'(T)^2]I_f>0.\label{dir:poplimit}
\end{equation}
This population limit holds whatever the relative rates at which the four parameters tend to zero. It provides eventual identification and the common limiting central limit theorem, and it does not enter the sample remainders.

\subsection{Stochastic bounds for the fitted residual jets}
\label{dir:jets}
For the LOO fitted residual $r_\theta$ and the population-smoothed residual $R_\theta$,
\begin{equation}
 \sup_\theta\|\partial_\theta^j(r_\theta-R_\theta)\|_{n,2}
 =O_P(\ell_n^{20}s_j),\quad j=0,1,2,\qquad
 \|A-M'(T)\|_{n,2}=O_P(\ell_n^{20}s_1).\label{dir:jetsrate}
\end{equation}
These bounds hold uniformly on a deterministic preprocessing box of radius $n^{-1/2}\ell_n$. They follow from the stochastic part of the Gaussian ridge-jet proof in Section~\ref{gauss-section}: use \eqref{gauss-G10} before adding the population bias in \eqref{gauss-G5}. The common density comparison \eqref{gauss-G3} extends to the box since its log ratio is $O\{(r_n+n^{-1/2}\ell_n)(1+\|X_i\|^2)\}=o(1)$ on the Gaussian maximum event. The fixed-dimensional mesh in \eqref{gauss-G6} is enlarged to include these scalar and matrix parameters, and its logarithmic cardinality is still $O(\ell_n)$; derivatives and interpolation are polynomial because $g,h,\lambda,\rho\ge n^{-A}$. The zeroth relative denominator error is $C\ell_n^5\{(ng\lambda)^{-1/2}+(ng\lambda)^{-1}\}=o(1)$. The row bound \eqref{gauss-G10} is $\ell_n^9$ times $\{\sqrt{q_g(T_i)/(ng^{2j+1})}+1/(ng^{j+1})\}/(q_g(T_i)+\lambda)$. Square and average this common envelope using \eqref{gauss-G11}: the square-root term gives $\sigma_j$ and the linear leave-one-out term gives $\omega_j$. Markov's inequality at level $\ell_n^2$ and the fixed polynomial index weights add at most $\ell_n^{7/2}$, so the power $20$ in \eqref{dir:jetsrate} suffices. The evaluation-index version of the same argument gives the bound for $A-M'$. No term $g^2+\sqrt\lambda$ appears, because the centering is the population-smoothed profile itself.

\subsection{Score coefficients and sample preprocessing}\label{dir:coef}
At truth write $D_i=(n-1)^{-1}\sum_{j\ne i}K_g(T_i-T_j)$ and
\[
 Q_{ij}=\frac{1_{i\ne j}K_g'(T_i-T_j)}{(n-1)(D_i+\lambda)},\quad
 A_i=\sum_jQ_{ij}(Y_j-\widehat M_i),\quad
 c_i=-a_i(r)A_i+\sum_{j\ne i}a_j(r)Q_{ji}(Y_i-\widehat M_j).
\]
Differentiation gives $\nabla L_n(0)=P_n(Wc)$; given $(T_i,Y_i)$, all coefficients are fixed while $W_i$ are independent centered Gaussians. Apply the iid training-density calculation to $R_i$, with law $p$. For $e_i=\{d(R_i)-d_i(R)\}/\{d(R_i)+\rho\}$, conditional LOO variances and \eqref{dir:mixture} give
\[
 E\|e\|_{n,2}^2\le C\ell_n^{1/2}/(nh),\qquad
 E P_n\{(d_i'(R)-d'(R_i))/(d(R_i)+\rho)\}^2
 \le C\ell_n^{1/2}/(nh^3).
\]
The outgoing ratio identity replaces its empirical denominator; incoming reciprocal replacement is a normalized Gaussian derivative matrix applied to $e$, with row and column sums at most $Ch^{-1}\ell_n^{3/2}$. After insertion of population denominators, the incoming conditional variance is at most
\[
 \frac1{n-1}\iint\frac{p(r)p(u)|K_h'(r-u)|^2}{(d(u)+\rho)^2}\,dr\,du
 \le C\ell_n^{1/2}/(nh^3).
\]
Taking a single expectation of the empirical squared norm accounts for the dependence between LOO errors. Markov's inequality at level $\ell_n^2$ and the deterministic residual transport bound give
\begin{equation}
 E_a:=\|a_n(r)-a_p(R)\|_{n,2}
 =O_P\{\ell_n^{22}s_0/h^2+\ell_n^3(nh^3)^{-1/2}\}.\label{dir:arisk}
\end{equation}
This bound includes both the residual evaluation and the derivatives with respect to the kernel centers.

The ordered Gaussian row and column displacement bounds give $\||Q|\|_{2\to2}\le Cg^{-1}\ell_n^{3/2}$. On the maximum event, $\max(|Y_i|+|a_p(R_i)|+|M'(T_i)|)\le C\ell_n^{1/2}$ and $\|a_n(r)\|_\infty\le Ch^{-1}\ell_n^{3/2}$. Replacing outgoing fitted factors costs $C\ell_n^{1/2}E_a+Ch^{-1}\ell_n^{43/2}s_1$; replacing $a_n(r)$ and $\widehat M$ in the incoming term costs $Cg^{-1}\ell_n^2E_a+Cg^{-1}\ell_n^{22}s_0$. For $e_j^g=\{q_g(T_j)-D_j\}/\{q_g(T_j)+\lambda\}$ the conditional variance calculation yields $\|e^g\|_{n,2}=O_P(\ell_n^2\sigma_0)$. Since $Q_{ji}-\bar Q_{ji}=Q_{ji}e_j^g$ with population-denominator $\bar Q$, replacing these denominators costs $O_P(g^{-1}\ell_n^{9/2}\sigma_0)$.

The remaining incoming array is the LOO average of
\[
 F_{t,y}(T',Y')=
 a_p(R')K_g'(T'-t)\{y-M(T')\}/\{q_g(T')+\lambda\}.
\]
Conditional on the evaluation $(T_i,Y_i)$ its mean is the incoming term of \eqref{inf:kappa}. Conditional response moments are uniformly bounded by \eqref{dir:aenv} and bounded $m,M$. Its integrated conditional variance is at most
\[
 \frac Cn\iint\frac{\phi(t)\phi(u)K_g'(u-t)^2}{(q_g(u)+\lambda)^2}\,dt\,du
 \le\frac C{ng^3}\int\frac{\phi(u)q_g(u)}{(q_g(u)+\lambda)^2}\,du
 \le C\ell_n^{1/2}/(ng^3).
\]
This bound accounts for the shared sample centers and the NW refit. Combining gives
\begin{equation}
 \|c-\kappa_k(T,Y)\|_{n,2}
 =O_P\left\{\ell_n^{24}\left[
 \frac{s_0/h^2+(nh^3)^{-1/2}}g+\frac{s_1}h+
 \frac{s_0}g+\sigma_1\right]\right\}.\label{dir:crisk}
\end{equation}

For training preprocessing, let $s_T^2=P_n(T-\bar T)^2$, $t_i=(T_i-\bar T)/s_T$, standardize $Y$ by its sample mean and standard deviation, and let $c_i^*$ denote the unprojected coefficient calculated from these standardized data. With $C_n=P_n(tc^*)$, raw directional standardization gives
\begin{equation}
 \nabla L_n(0)=s_T^{-1}P_n[W\{c^*-tC_n\}].\label{dir:exactpre}
\end{equation}
Indeed $\partial t_i=s_T^{-1}\{W_i-\bar W-t_iP_n(tW)\}$ and $\sum_i c_i^*=0$ by index-translation invariance. Thus, the whitening correction enters with a minus sign, and the coefficient $c^*-tC_n$ depends only on $(T,Y)$.

For deterministic index scale $s$, response scale $a$ and response shift $b$ near $(1,1,0)$, normalized Gaussian locality gives
\begin{equation}
 \|\partial_{s,a,b}c\|_{n,2}\le C\ell_n^7/(gh^2).\label{dir:scalarLip}
\end{equation}
In detail $s\partial_sK_g(sv)=-(sv/g)^2K_g(sv)$, so index scaling introduces no extra inverse $g$. The fitted-value scale derivative is at most $C\ell_n^{5/2}$, the derivative-NW matrix scale derivative has operator norm at most $C\ell_n^4/g$, and $A$ has scale derivative at most $C\ell_n^{9/2}/g$. The bound $\|D_ra\|_{2\to2}\le Ch^{-2}\ell_n^2$ and the product rule for both terms of $c$ give \eqref{dir:scalarLip}. Response shift gives $\partial_b r_i=1-\sum_jH_{ij}$, which is bounded by one. All sample scalar increments are $O_P(n^{-1/2})$. The uniform moment bound above gives $(P_n-P)T\kappa_k=O_P(n^{-1/2})$; hence $C_n-C_k$ is the coefficient error plus $O_P(n^{-1/2})$. The coefficient in \eqref{dir:exactpre} differs in empirical norm from $\kappa_k-C_kT$ by $O_P(D_S)$. The conditional Gaussian variance equals $I_d$ times the empirical squared coefficient error, which proves \eqref{dir:score}.

\subsection{Hessian approximation and convergence of the algorithm}\label{dir:curvature}
Apply the deterministic empirical-integral Hessian transfer to the sample residual jets and the population-smoothed jets. Polynomial population envelopes give uniformly bounded first-jet $L^4$ and second-jet $L^2$ empirical norms. Equation \eqref{dir:jetsrate} yields the fitting term $O_P\{\ell_n^{23}(s_0/h^3+s_1/h^2+s_2/h)\}$; the quadratic first-jet error is absorbed since $D_H\to0$ implies $\ell_n^{20}s_1=o(1)$. The centering jets include the fixed-argument population NW derivative.

For density row replacement, use the common good set $\{d(R_i)\ge t_H\}$. For a deterministic chart or preprocessing parameter $v$ put $D_v(z)=E K_h\{R_v(z)-R_v(Z')\}$. On logarithmic predictor and error balls, the residual shifts and their first three jets are at most $C\ell_n^3$. Gaussian log-density derivative bounds imply $D_v(z)\asymp d(R_0(z))$: the log ratio is $O\{(r_n+n^{-1/2}\ell_n)\ell_n^4\}=o(1)$. Truncate the inner Gaussian predictor integral on a larger logarithmic ball; its omitted polynomial moments are $n^{-B}$ for arbitrarily large fixed $B$, whereas $d(R_0(z))\ge n^{-B_0}$ on the observed ball. Thus, the normalized inner tails are negligible. The argument accommodates the dependence of $R$ on $T$.

A total $j$th derivative of a kernel summand has variance at most $C\ell_n^{12}d(R_0(z))h^{-2j-1}$ and envelope $C\ell_n^6h^{-j-1}$ for $j\le2$. Bernstein's inequality on the finite-dimensional polynomial mesh, before evaluation at the data, gives normalized good-row errors at most
\[
 C\ell_n^8\{(nh^{2j+1}t_H)^{-1/2}+(nh^{j+1}t_H)^{-1}\}.
\]
LOO deletion costs the linear term. Third parameter derivatives suffice for interpolation, so six link derivatives are enough. The population first and second relative derivatives on these balls are at most $C\ell_n^{7/2}$ and $C\ell_n^7$. Expand $D''/(D+\rho)-D'D'^T/(D+\rho)^2$. The good-row replacement is bounded by $C\ell_n^{16}\{(nh^5t_H)^{-1/2}+(nh^3t_H)^{-1}\}$; $D_H\to0$ makes the relative errors small. For every bad row, the normalized Gaussian displacement moments yield a sample row-Hessian bound of $Ch^{-2}\ell_n^8$, and the population counterpart is at most $C\ell_n^7$. The bad-row fraction is $O_P(t_H\ell_n^{1/2})$ by \eqref{dir:mixture}, so their total contribution to the objective is $O_P(t_Hh^{-2}\ell_n^{17/2})$.

The population row Hessian and its third mixed direction and preprocessing derivatives have a common finite-second-moment envelope of the form $C(1+\|X\|+|\varepsilon|)^C e^{C|\varepsilon|}$. Residual shifts are bounded; Gaussian derivative ratios are polynomial, and the ratio of component densities with bounded relative location shift is at most $e^{C(1+|\varepsilon|)}$. Integrating polynomial inner predictor jets preserves this envelope. Chebyshev at the anchor and a mean-value bound therefore give outer empirical error $O_P(n^{-1/2}+r_n)$ and population curvature continuity $O(r_n)$. The deterministic preprocessing box adds $O(n^{-1/2}\ell_n)$ to population curvature. The mesh and the common projected-density envelope make every stochastic bound uniform on that box before the random preprocessing value is substituted. These terms are bounded by \eqref{dir:DH}, which proves \eqref{dir:hessian}.

Integrate the Hessian between zero and an interior approximate root. Its uniform positive lower eigenvalue and the bounded score variance first give root-$n$ consistency, then \eqref{dir:estimate}. For the algorithm of Section~\ref{sec:algorithm}, the contraction proof given there applies with the score and curvature above. First, apply the bound on the radius-$16n^{-1/20}$ chart containing the optimization cap. The root-$n$ Fourier start, positive limiting curvature, and the shrinking step $1/\log(n+e)$ give acceptance of the first Armijo trial and contraction by $1-c/\log(n+e)$. The $\lceil\log^3(n+e)\rceil$ iterations give error $O_P(n^{-1/20})e^{-c\log^2n}$, or an earlier $n^{-1}$ score stop. After root-$n$ localization, the expansion may use a smaller $r_n\gg n^{-1/2}$, with the same optimization cap and stopping rule.

\subsection{Admissible smoothing rates and an example}\label{dir:ranges}
For fixed positive $k$ with $j_k>0$, any $0<\nu_H<1$ gives vanishing remainders. Uniform moments yield $\sqrt n\widehat\theta_k\Rightarrow N_d(0,V_k)$. A nondegenerate instance with positive curvature is given below. For polynomially shrinking parameters with exponents $(h,g,\lambda,\rho)\asymp(n^{-\nu_h},n^{-\nu_g},n^{-\nu_\lambda},n^{-\nu_\rho})$, sufficient primitive inequalities are
\begin{equation}
 \nu_g+\nu_\lambda<1,\quad \nu_g+6\nu_h<1,\quad3\nu_g+4\nu_h<1,\quad5\nu_g+2\nu_h<1,
 \qquad2\nu_h<\nu_H<1-5\nu_h.\label{dir:region}
\end{equation}
All exponents are positive and fixed, and $\nu_\rho$ is unrestricted. Positive curvature follows from \eqref{dir:poplimit}. The practical grid uses fixed positive multipliers and
\begin{equation}
 h\asymp n^{-1/15},\quad g\asymp n^{-1/6},\quad
 \lambda\asymp n^{-3/4},\quad\rho\asymp n^{-1},\quad
 t_H=n^{-1/5},\quad t_D=n^{-2/3}.\label{inf:nonempty}
\end{equation}
These parameters satisfy the inequalities and give
\[
 D_S=O(\ell_n^{32}n^{-7/60}),\qquad
 D_H(r_n)=O\{\ell_n^{32}(n^{-1/60}+r_n)\}.
\]
Another sufficient choice is $(\nu_h,\nu_g,\nu_\lambda,\nu_\rho,\nu_H)=(1/10,1/20,1/2,1,3/10)$, for which $D_S=O(\ell_n^{32}n^{-9/40})$ and $D_H(r_n)=O\{\ell_n^{32}(n^{-1/10}+r_n)\}$.

For an example with positive curvature and strictly positive information loss, use $m(t)=\sin t$ and Gaussian error variance $s=(1+e^{-2})/2$, so $\operatorname{Var}Y=1$. Begin with the density-only limit $g=\lambda=\rho=0$ and fixed $h>0$. For $u=h^2$, $D=s^2+su+u^2$, $b=u/D$ and $K=(s+u)/\sqrt D$, Gaussian integration gives
\begin{equation}
 a_h(r)=-r/(s+u)-Kb\,r e^{-br^2/2},\qquad
 j_h^{\rm den}=(s+2u)/(s+u)^2>0.\label{dir:normal}
\end{equation}
Put $A_m=E m'(T)^2$. The direction curvature is $A_mj_h^{\rm den}$. Since $a_h(r)/r$ is nonconstant for every $u>0$, its parameter influence is not the efficient influence, and the information projection gives a strictly positive loss. In closed form,
\[
 E a_h(\varepsilon)^2=s/(s+u)^2+
 \frac{2Kbs}{s+u}(1+sb)^{-3/2}+K^2b^2s(1+2sb)^{-3/2}.
\]
The uniform envelope argument gives continuity as $g,\lambda,\rho\downarrow0$ while $h$ is fixed. Therefore, within the bounded nonmonotone primitive model, an open set of tuples with all four parameters strictly positive has $j_k>0$, $B_k>0$, and a strictly positive loss, thereby verifying the fixed-endpoint conditions.

Expanding \eqref{dir:normal} in $u=h^2$ gives, in weighted $L^2$,
\begin{align}
 a_h(r)&=-r/s+h^4r(r^2-s)/(2s^4)+O_{L^2}(h^6),\notag\\
 (j_h^{\rm den})^{-1}&=s+h^4/s+O(h^6),\notag\\
 (A_mj_h^{\rm den})^{-1}\{-Wm'a_h(\varepsilon)\}
 -\mathcal I^{-1}S_{\rm eff}
 &=-\frac{h^4Wm'(T)\varepsilon(\varepsilon^2-3s)}{2A_ms^3}
   +O_{L^2}(h^6).\label{dir:normalif}
\end{align}
Since $E\varepsilon^2(\varepsilon^2-3s)^2=6s^3$, the information loss is $3h^8/(2A_ms^3)+o(h^8)$ per tangent coordinate. Now take the positive parameters
\begin{equation}
 h=n^{-1/100},\quad g=n^{-2/25},\quad\lambda=n^{-4/5},\quad\rho=n^{-1},
 \quad t_H=n^{-1/3},\quad r_n=n^{-1/5}.\label{dir:resolvable}
\end{equation}
The continuum integrated comparison, applied to the density-only finite-$h$ limit instead of the efficient score, gives with $v=g^2+\sqrt\lambda$
\begin{align*}
 \|\eta_k+Wm'a_{h,0}(\varepsilon)\|_2
 &\le C\ell_n^{32}\{v/(gh^2)+v/h+\sqrt\rho\}=o(h^4),\\
 |j_k-A_mj_h^{\rm den}|
 &\le C\ell_n^{32}\{v/h^3+\rho\}=o(h^4).
\end{align*}
As for the logarithmic factors, the weighted population profile-jet biases in \eqref{gauss-G2}--\eqref{gauss-G4} are bounded by $C\ell_n^5v$ through order two; continuum transport contributes $\ell_n^2/h^2$, the derivative-NW Schur bound at most $C\ell_n^2/g$, and truncating fixed polynomial response and index weights adds at most $\ell_n^4$. Thus, the power $32$ also dominates these population-comparison factors. The regression profiles are compared first at a common floor. To remove the Gaussian density floor afterwards, write $l_0=-r/(s+h^2)$ and $\tau=\rho/(d+\rho)$; then $J_{h,\rho}-J_{h,0}=E[2l_0(l_0-\psi)\tau-l_0^2\tau^2]$. The first-power integral $E[(1+R^2)\tau]=O(\rho\ell_n^{3/2})$ follows from $f\le Cd$ on a logarithmic interval and Gaussian tails; therefore the floor changes the curvature by $O(\rho\ell_n^{3/2})$, which improves on the Cauchy--Schwarz bound $O(\sqrt\rho)$. Moving $K_h'$ to the quotient derivative similarly gives $\|a_{h,\rho}-a_{h,0}\|_2 =O(\sqrt\rho\ell_n^2)$. The largest polynomial orders are $n^{-3/50}$ and $n^{-13/100}$, both smaller than $h^4=n^{-1/25}$. Thus, the population influence covariance of SI-KDR with positive ridge and floor satisfies
\begin{equation}
 V_k-\mathcal I^{-1}=\frac{3h^8}{2A_ms^3}I_d+o(h^8).\label{dir:loss}
\end{equation}
Meanwhile $D_S=O(\ell_n^{32}n^{-9/25})$ and $D_H(n^{-1/5})=O(\ell_n^{32}n^{-1/5})$, both $o(h^8)$. The optimization cap is proportional to $n^{-1/20}$; the smaller radius here is used only after root-$n$ localization. The score tolerance $\tau_n=n^{-1}$ gives $\sqrt n\tau_n=o(h^8)$. For the case-weight covariance, the Frobenius bounds in the proof of Theorem~\ref{inf:caseweights} with $t_D=n^{-4/5}$ are $O(\ell_n^{32}n^{-31/100})$ before the coefficient, curvature, and localization errors are added, hence also $o(h^8)$. This gives a nondegenerate example in which the finite-smoothing loss is asymptotically resolvable.

The stochastic integrated jet bounds and the expansion above require only integrated, rather than uniform, control of the fitted profile.

\subsection{Case-weight derivatives}
\label{up:caseweights}
All quantities in this subsection use one raw tangent chart, and the tuning constants are held fixed in each weight derivative. Put $\bar w_i=w_i/\sum_jw_j$, $\ell_i=1-\bar w_i$, and use
\begin{gather*}
 m_i(w)=\frac{\sum_{j\ne i}\bar w_jK_g(t_i-t_j)y_j(w)}
 {\sum_{j\ne i}\bar w_jK_g(t_i-t_j)+\ell_i\lambda},\\
 d_i(w)=\ell_i^{-1}\sum_{j\ne i}\bar w_jK_h(r_i-r_j),\qquad
 L_w=\sum_i \bar w_i\log\{d_i(w)+\rho\}.
\end{gather*}
The weighted directional standardization is
\begin{equation}\label{inf:directionalstandardization}
t_i=(\beta^TX_i-P_w\beta^TX)/s_T(w).
\end{equation}
The response is $y_i(w)=(Y_i-P_wY)/s_Y(w)$, with divisor-one weighted variances. This scalar index coincides with the index obtained by weighted whitening followed by the normalized inverse map, so it carries the derivative of the weighted covariance. Let $S_w=\nabla_\theta L_w$, and define
\begin{equation}\label{inf:ijdefinition}
 \widehat\eta_i=n\partial_{w_i}S_w|_1,\quad
 \widehat J=-\partial_\theta S_1,\quad
 \widehat B=P_n\widehat\eta_i\widehat\eta_i^T,\quad
 \widehat V=\widehat J^{-1}\widehat B\widehat J^{-T}.
\end{equation}
Weight homogeneity gives $\sum_i\widehat\eta_i=0$. The estimated leading asymptotic covariance of the parameter is $\widehat V/n$, whereas $\widehat B$ estimates the covariance of its normalized score.

We first record the algebra needed to differentiate the two leave-one-out masses $\ell_i\rho$ and $\ell_i\lambda$. Write $D_i=\sum_{j\ne i}w_j\{K_h(r_i-r_j)+\rho\}$ and, with zero diagonals, $P_{ij}=K_h(r_i-r_j)/D_i$, $S_{ij}=K_h'(r_i-r_j)/D_i$, $M_D=P+(\rho\ind\{i\ne j\}/D_i)_{ij}$. The weighted residual coefficient is $a_w=(S-S^T)w$. Hence, at unit weights and fixed residuals,
\begin{equation}\label{up:density-weight}
 \partial_w a|_r=S-S^T-\diag(S1)M_D+S^TM_D.
\end{equation}
For the regression analogues $H,Q,M_g=H+(\lambda/G_i)_{i\ne j}$, with $G_i=\sum_{j\ne i}w_j\{K_g(t_i-t_j)+\lambda\}$, let $A_i=\sum_jw_jQ_{ij}(y_j-m_i)$ and $B_{ki}=Q_{ki}(y_i-m_k)$. Holding the index and response arrays fixed,
\begin{align}
 D_m&=H\diag(y)-\diag(m)M_g,\notag\\
 D_A&=Q\diag(y)-\diag(m)Q-\diag(A)M_g-\diag(Q1)D_m,\notag\\
 D_a&=\partial_w a|_r-(D_ra)D_m,\notag\\
 D_c&=(B^T-\diag(A))D_a-\diag(a)D_A
       +B^T\diag(a)(I-M_g)-Q^T\diag(a)D_m.\label{up:weight-matrices}
\end{align}
Indeed $c_{w,i}=-a_{w,i}A_i+\sum_kw_ka_{w,k}B_{ki}$; differentiating this identity gives the last line. In particular $B^T\diag(a)$ is the derivative of the incoming observation's outer weight. The floor and ridge terms in $M_D$ and $M_g$ are the derivatives of $\ell_i\rho$ and $\ell_i\lambda$.

\subsection{Consistency of the case-weight covariance estimator}
Use $\ell_n,\sigma_j,\omega_j,s_j,D_S,D_H$ from Theorem~\ref{dir:theorem} and put
\begin{align}
 \Delta_D={}&\ell_n^{32}\left\{\frac{s_0}{gh^2}
 +\frac{(nh)^{-1/2}+\sqrt{t_D}}{gh}
 +(nh^4g^3)^{-1/2}+(ng^2h^4)^{-1/2}\right\},\label{up:direct-delta}\\
 q_D={}&D_S+\Delta_D+\ell_n^{32}\left\{
 (\sqrt n gh^2)^{-1}+(\sqrt n gh^3)^{-1}
 +a_n/(g^2h^2)\right\},\quad a_n=\log(n+e)/\sqrt n.\label{up:ij-q}
\end{align}
The term $s_0=\sigma_0+\omega_0$ consists of the stochastic fluctuation and the leave-one-out correction from the Bernstein bound; it is free of regression-smoothing bias.
\begin{theorem}[Covariance consistency]\label{inf:caseweights}\label{up:ij-direct}
Under Theorem~\ref{dir:theorem}, suppose $nht_D/\ell_n^{32}\to\infty$ and $q_D\to0$. Simultaneously over the finite grid and uniformly on $\|\theta\|\le a_n$ in matching tangent frames, hence at any root-$n$ consistent preliminary or final direction,
\[
 \|\widehat B-B_k\|=O_P(q_D+q_D^2+n^{-1/2}),\qquad
 \|\widehat J-J_k\|=O_P\{D_H(r_n)\}.
\]
If $\inf_k\lambda_{\min}(J_k)>0$, then $\widehat V-V_k=o_P(1)$. At fixed positive scales, the covariance error is $O_P(n^{-1/2})+O(\|\theta\|)$ uniformly locally, hence $O_P(n^{-1/2})$ at any root-$n$ consistent evaluation point; no threshold is needed. Here $B_k$, $J_k$ and $V_k$ are the score covariance, curvature and parameter covariance, and the leading asymptotic covariance of the estimator is $V_k/n$.
\end{theorem}
\begin{proof}
Anchor the density rows at the iid finite-smoothed residuals $R_i=Y_i-M(T_i)$. Their law is a bounded-location Gaussian mixture and $p\le C(K_h*p)$, since $\phi_\sigma(x)\le\sqrt{1+h^2/\sigma^2}\phi_{\sqrt{\sigma^2+h^2}}(x)$ componentwise. Splitting at $|r|=C\sqrt{\log(1/t_D)}$ gives $P\{(K_h*p)(R)<t_D\}\le Ct_D\sqrt{\log(1/t_D)}$; the exterior tail is $O(t_D^2)$. Conditional Bernstein bounds on the remaining rows and integration of $p/(K_h*p)$ consequently give
\[
 F_{P(R)}=O_P\{\ell_n^{1/4}((nh)^{-1/2}+\sqrt{t_D})\},\quad
 F_H=O_P(\ell_n^{1/4}\sigma_0),\quad
 F_Q=O_P(\ell_n^{1/4}\sigma_1).
\]
Here $F_M=\|M\|_F/\sqrt n$. The denominator event follows from the condition $ng\lambda/\ell_n^{12}\to\infty$ of Theorem~\ref{dir:theorem}. For transport, $dP_{ij}=P_{ij}(\ell_{ij}-E_i\ell)$, $\ell_{ij}=-(r_i-r_j)(e_i-e_j)/h^2$. Summing its square using $P_{ij}^2\le P_{ij}$ and the row and column second displacement moments gives $\|P(r+e)-P(r)\|_F/\sqrt n\le C\ell_n\|e\|_{n,2}/h$. Equation~\eqref{dir:jetsrate} gives $\|r-R\|_{n,2}=O_P(\ell_n^{20}s_0)$, so anchoring the density rows costs only the stochastic fitting error.

To track logarithmic factors, we bound the deterministic matrix factors in the weight identities \eqref{up:density-weight}--\eqref{up:weight-matrices} as follows:
\[
\begin{array}{c|c}
H,P,M_g,M_D\text{ in operator norm}&C\ell_n\\
Q,S\text{ in operator norm}&C\ell_n^{3/2}/g, C\ell_n^{3/2}/h\\
\|A\|_\infty,\|B\|_{\rm op}&C\ell_n^2/g\\
\|a\|_\infty,\|D_ra\|_{\rm op}&C\ell_n^{3/2}/h, C\ell_n^2/h^2\\
\|D_r^2a[e]\|_{\rm op}&C\ell_n^{5/2}\|e\|_\infty/h^3.
\end{array}
\]
Schur's inequality and the Gaussian displacement bound prove each line. Splitting standardized residual displacements at $C\sqrt{\ell_n}$ also gives $F_S\le C\ell_n^{1/2}F_P/h+O(n^{-1/2}/h)$; the polynomial lower bounds on the tuning parameters make the Gaussian exterior tail this small. Using $\|AB\|_F\le\min(\|A\|_{\rm op}\|B\|_F, \|A\|_F\|B\|_{\rm op})$ in every product of the derivative identities yields
\[
 \frac{\|D_wc\|_F}{\sqrt n}
 \le C\ell_n^5\left\{\frac{F_P}{gh}+\frac{F_H}{gh^2}
                    +\frac{F_Q}{h}+\frac1{\sqrt n gh^2}\right\}.
\]
The largest logarithmic product is $\ell_n^{5+1+20}$, bounded by $\ell_n^{32}$. The last term includes the derivatives of both leave-one-out masses.

For population truth coordinates $(T,W)$, weighted directional standardization gives
\[
 S_w=P_w(Wd_w),\quad d_w=s_T(w)^{-1}\{c_w-t_wP_w(t_wc_w)\},\quad
 \widehat\eta_j=W_jd_j-S_1+\sum_iW_i\partial_{w_j}d_i.
\]
All scalar coefficients depend only on $(T,Y)$. Conditionally, the empirical squared norm of the last term has mean $d\|D_wd\|_F^2/n$. Differentiating training moments adds rank-one matrices with scaled Frobenius norm at most $C\ell_n^7/(\sqrt n gh^2)$; normalized Gaussian scale locality prevents an additional inverse $g$. Equation~\eqref{dir:exactpre} and the coefficient bounds of Section~\ref{dir:coef} give $\|d-\widetilde\kappa_k\|_{n,2}=O_P(D_S)$.

For local evaluation at the estimated direction, rotate at a deterministic $\beta$ and couple $Y$ to $Y_\beta=m(T_\beta)+\varepsilon$. Their maximum difference is $Ca_n\max_i\|X_i\|$. The displayed matrix bounds, differentiated once in the response, give $\|D_yD_wc[e]\|_{\rm op}\le C\ell_n^6\|e\|_\infty/(gh^3)$. The leading coefficient has local modulus $C\ell_n^7a_n/(g^2h^2)$. The common Frobenius envelope is obtained as follows. Relative to population truth indices and residuals, on the common maximum and preprocessing event, the local index changes are at most $Ca_n\sqrt{\ell_n}$ and ideal residual changes at most $Ca_n\ell_n^3$ (the population mixed derivatives have degree at most six). Gaussian kernel log ratios are consequently bounded by $Ca_n\ell_n/g^2$ and $Ca_n\ell_n^4/h^2$, both $o(1)$ by $q_D\to0$. Thus, every local $H$ and ideal density matrix $P$ is entrywise bounded by a fixed multiple of its truth counterpart, including its denominator. For the derivative matrix, the identity $K_g'(u)=-uK_g(u)/g^2$ gives $F_{Q,\beta}\le C\{F_{Q,0}+a_n\sqrt{\ell_n}F_{H,0}/g^2\}$; the extra term is $o_P(\ell_n^{1/4}\sigma_1)$. The projected and ideal residual densities also have bounded ratios, so their inverse-density and bad-row averages have one common envelope. These bounds establish the shared Frobenius event before a polynomial directional net is taken, so no union bound over directions is needed. Conditional Gaussian tails add only $\ell_n^{1/2}$; a finer polynomial mesh is justified by the positive ridge and floor. This gives an empirical influence error $O_P(q_D)$, including the fitted location; recall that $a_n$ already contains one logarithm.

Finally \eqref{dir:kappaibp} gives $\sup_kE\|\eta_k\|^4<\infty$, so $\|P_n\eta_k\eta_k^T-B_k\|=O_P(n^{-1/2})$. The Cauchy--Schwarz inequality yields the covariance bound, and \eqref{dir:hessian}, together with matrix inversion, completes the argument. At fixed positive scales, Lemma~\ref{up:smooth-functional} gives all three root-$n$ expansions. At nonstationary preliminary locations, the raw Hessian includes its gradient-times-coordinate connection term.
\end{proof}

\subsection{Tuning parameter selection and confidence sets}
\label{up:selection}
Let $\mathcal L_k(P,\beta)$ be the population version of the weighted procedure. At a raw direction $\beta(b)$ use a smooth orthonormal frame $F(b)$ and the local criterion
\[
 \ell_k(P,b,v)=\mathcal L_k\left(P,
       \frac{\beta(b)+F(b)v}{\sqrt{1+\|v\|^2}}\right),\quad
 S_k=\partial_v\ell_k|_0,\quad J_k=-\partial_v^2\ell_k|_0.
\]
Put $\eta_k(z)=D_PS_k[\delta_z-P]$, $B_k=P\eta_k\eta_k^T$ and $V_k=J_k^{-1}B_kJ_k^{-T}$. This moving chart is the chart in which the raw trace-variance grid is computed; away from the root, a Hessian in one fixed truth-centered chart differs from it. Trace is invariant to orthonormal frame rotations, so this smooth frame gives the same criterion as the null-space frame of the algorithm.

\begin{lemma}[Positive-scale empirical expansion]\label{up:smooth-functional}
For a fixed positive grid under the primitive Gaussian-mixture model, $S_k,J_k,B_k$ and the two tuning criteria have first-order empirical measure expansions with remainder $O_P(n^{-1})$, locally uniformly in $b$. Their LOO versions have the same expansions. This also holds for $v_k=\tr V_k$ wherever $J_k$ is invertible.
\end{lemma}
\begin{proof}
All denominators have fixed positive lower bounds, and all fixed-order kernel, quotient, and moment derivatives have polynomial envelopes with every moment finite. Expand successively in the moments, the NW quotient, the residuals, the density, and the outer logarithm. The quotient rule used is
\[
 \frac{A+a}{D+d}=\frac AD+\frac{a-(A/D)d}{D}
       -\frac{d\{a-(A/D)d\}}{D^2}+O((|a|+|d|)^3).
\]
Conditionally on an evaluation observation, centering and conditional independence remove every index pattern in which some index appears only once, so the sixth moment of each centered primitive kernel mean is $O(n^{-3})$. Hence, by H{\"o}lder's inequality, the third-order remainder is $O_P(n^{-3/2})$. Expanding the product before taking the outer average handles the shared residual centers. All remaining second-order kernel terms are doubly centered. Examining the index pairs shows that off-diagonal second-order moments are $O(n^{-2})$, while the diagonal terms are $O_P(n^{-1})$. Similar algebra and differentiation give the expansion of the case-weight covariance. Sufficiently many location derivatives and finite-dimensional Sobolev bounds then give uniformity. These location derivatives act on a Gaussian kernel of fixed scale and a moment map, rather than on $m$. Given all required polynomial moments, a $C^6$ link suffices. The LOO correction term has the form $\sum_i \bar w_i^2b_i$, with a polynomial envelope. Summed over all LOO components, this correction term and the corresponding scaled case-weight derivative are $O_P(n^{-1})$ in the empirical $L^2$-norm. Smooth inversion completes the proof.
\end{proof}

We next describe the joint behavior of the trace-variance grid, including its preliminary direction. With $H_k(z,z')=D_P^2S_k[\delta_z-P,\delta_{z'}-P]$, define
\begin{align}
 \mathcal J_k(z)&=D_PJ_k[\delta_z-P],\notag\\
 \mathcal B_k(z)&=\eta_k(z)\eta_k(z)^T-B_k
 +P\{H_k(Z,z)\eta_k(Z)^T+\eta_k(Z)H_k(Z,z)^T\},\notag\\
 \zeta_k(z)&=\tr\{J_k^{-1}\mathcal B_k(z)J_k^{-T}
  -J_k^{-1}\mathcal J_k(z)V_k-V_k\mathcal J_k(z)^TJ_k^{-T}\},\quad
 a_k=D_b\tr V_k(P,0).\label{up:trace-influence}
\end{align}
The centering derivative in $\delta_Z-P$ contributes $-\eta_k(z)$, whose cross product integrates to zero because $P\eta_k=0$. All terms are derivatives of the protected population integrals and include preprocessing and the raw chart connection. Lemma~\ref{up:smooth-functional} gives
\begin{equation}\label{up:trace-expansion}
 \sqrt n(\widehat v_k-v_k)=n^{-1/2}\sum_i\zeta_k(O_i)
                    +a_k^T\sqrt n\widetilde\theta+o_P(1).
\end{equation}
The preliminary-direction term $a_k^T\sqrt n\widetilde\theta$ is also present when the preliminary fit itself selects a tied tuple.

\begin{theorem}[Positive-grid selection]\label{up:finite-selection}
Suppose grid curvatures are nonzero, at least one is positive, and the conditions of Theorem~\ref{dir:theorem} hold for every possible preliminary and final fit, with $\inf_nj_l>0$ for every preliminary tuple $l$ selected with nonvanishing probability. Let $\mathcal E=\{k:J_k\succ0\}$ and $\mathcal M=\arg\min_{k\in\mathcal E}\tr V_k$. A unique population minimizer $k_*$ is selected consistently and gives $\sqrt n\widehat\theta\Rightarrow N(0,V_{k_*})$. For tied minimizers, simultaneously
\[
 \mathbb G_{n,k}=n^{-1/2}\sum_i\eta_k(O_i),\quad
 \sqrt n\widehat\theta=J_{\widehat k}^{-1}\mathbb G_{n,\widehat k}+o_P(1),
 \quad \widehat V-V_{\widehat k}=o_P(1),\quad
 P(\widehat k\in\mathcal M)\to1.
\]
The score and trace-influence sums have a joint Gaussian limit. If the limiting contrasts in \eqref{up:trace-expansion} have a unique minimum $K$ almost surely, the selected limit is $J_K^{-1}\mathbb G_K$ and its Wald quadratic is $\mathbb G_K^TB_K^{-1}\mathbb G_K$, which is not automatically chi-square.
\end{theorem}
\begin{proof}
The uniform convergence of the Hessian and covariance estimators identifies $\mathcal E$. Hence the probability of falling back on the LOO selection rule, which is used when no tuple has positive estimated curvature, tends to zero. Since the smallest positive population trace gap separates the population minimizer from all other tuples, selection is consistent when this minimizer is unique. The simultaneous expansions of Theorem~\ref{dir:theorem} apply to every selector on a finite grid. A standard multivariate CLT applies to the influence functions, whose moments are finite. Under regular ties, apply the continuous argmin map to \eqref{up:trace-expansion}. If the preliminary tuple is separated, substitute $\sqrt n\widetilde\theta=J_l^{-1}\mathbb G_{n,l}+o_P(1)$ into that expression. If ties are present, select and substitute the corresponding expansion from the broader joint vector. By Proposition~\ref{fourier:rateprop}, the Fourier pilot is root-$n$ consistent. Since the population directional derivatives of the preliminary scalar tuning criteria are zero, their first-order contrasts evaluated at this pilot change by $o_P(1)$. Hence, under regular ties at the preliminary stage, their joint scalar influences determine the index selected by the sequential argmin and argmax processes. For degenerate contrasts, the stated simultaneous expansion holds across all joint subsequential limits of $(\widehat k,\mathbb G_n)$. The conditions concerning curvature, separation, and contrasts are population-integral conditions. By the population bounds above, tuples with small positive values satisfy the positive-curvature condition. For a shrinking grid, the common efficient influence yields both normal and Wald inference for any training selection.
\end{proof}

\begin{theorem}[Score confidence sets under positive smoothing]\label{up:fixed-shape}
Suppose that the uniform bounds of Theorem~\ref{dir:theorem} hold for all possible inference tuples. Assume that their $J_k$ and $B_k$ are positive definite, the initialization within the local cap is $o_P(r_n)$, and the finite optimization process satisfies its contraction bound. Let $k_I$ be the population inference tuple, which minimizes the population inference MSE over $g$ and then maximizes the log-score over $h$, and suppose both optima are separated. In raw root-$n$ coordinates centered at the truth, the local score set is approximated from inside and outside by
\begin{equation}\label{up:fixed-ellipse}
\Bigl\{v:\bigl(v-J_{k_I}^{-1}\mathbb G_{n,k_I}\bigr)^T V_{k_I}^{-1}
\bigl(v-J_{k_I}^{-1}\mathbb G_{n,k_I}\bigr)
\le\chi^2_{d,1-\alpha}\Bigr\}.
\end{equation}
The cutoff value for these inclusions can be shifted by any fixed $\epsilon > 0$. Under regular ties, the same holds, except that $K_I$ is a random index independent of the joint limiting score vector. Here regularity means that the minimum of the first-order MSE contrast at the population minimizers is unique with probability 1 and that, for each minimizing $g$, the maximum of the log-score contrast at its population maximizers is also unique with probability 1. In this setting, the coverage approaches $1-\alpha$, and the diameter is $O_P(n^{-1/2})$.
\end{theorem}

\begin{proof}
The candidate selector uses only $(x_\beta, Y)$, as described in Section~\ref{sec:conditional-score-proofs}. At the true parameter value, the directional derivatives of both scalar population criteria vanish by a Gaussian rotation. Their first-order measure-influence functions depend only on $(T, Y)$, and $E(\eta_k\mid T,Y)=0$ for $\eta_k = W\widetilde\kappa_k(T, Y)$; thus the Gaussian limits of their normalized sums are independent of those of the $\eta_k$. By the positive-scale lemma, criterion contrasts within a bounded root-$n$ ball differ from their true values by only $o_P(n^{-1/2})$. Hence, if the contrasts are separated or regular, a sequential selector within this ball remains fixed.

From the expansion of Theorem~\ref{dir:theorem} and the covariance connection, we obtain, uniformly over this ball, $\sqrt{n}S_k = \mathbb G_{n,k} - J_kv + o_P(1)$ and $K_{n,k} = B_k + o_P(1)$. We use the identities $nS_k^TK_{n,k}^{-1}S_k = (n-2)U$ and $dF = (n-d-2)U/(1-U)$. By positive curvature, every point in the score set lies within distance $O_P(n^{-1/2})$ of the true line. With probability tending to 1, both caps contain bounded root-$n$ balls. This determines the set's inclusions and diameter. By independence, the null quadratic form remains chi-square, even at a regular random inference index.
\end{proof}

For degenerate inference contrasts, the score set can be characterized through a candidate-dependent selector $k_n(v)$ in the expression
\[
\bigl(\mathbb G_{n,k_n(v)}-J_{k_n(v)}v\bigr)^T B_{k_n(v)}^{-1}
\bigl(\mathbb G_{n,k_n(v)}-J_{k_n(v)}v\bigr).
\]
Because of the uniform positive curvature, the diameter is $O_P(n^{-1/2})$, and the conditional test described in Section~\ref{sec:conditional-score-proofs} provides asymptotic coverage as $P(\beta_0\notin\mathcal D_n)$ and $P(N_{\beta_0})$ approach zero. For a finite $n$, the localized coverage bound is $1-\alpha-P(\beta_0\notin\mathcal D_n)-P(N_{\beta_0})$.
\section{Confidence sets by conditional score inversion}
\label{sec:conditional-score-proofs}

This section proves the results on the confidence set obtained by inverting a score test. The null distribution of the score is calibrated conditionally on the candidate index and the response. The Gaussian residual-subspace argument is classical \citep[pp.~606--608]{perry2010rotation}; we show that it applies to the SI-KDR coefficient with the preprocessing and the tuning rule, and we relate the resulting set to the efficient limit.

\subsection{The score in raw coordinates}

Let $X_i\sim N_p(\mu,\Sigma)$ independently, with $\Sigma\succ0$, and let $Y_i=m_0(\beta_0^TX_i)+\varepsilon_i$, where $\varepsilon_i\perp X_i$. The conditional calculation requires only $Y\perp X F_{\beta_0}\mid X\beta_0$; the bounded smooth link and the Gaussian-mixture error of the preceding sections enter through the efficient limit. Fix a candidate unit $\beta$ and a deterministic orthonormal tangent frame $F_\beta\in\R^{p\times d}$, $d=p-1$. Set
\begin{gather}
 x_\beta=X\beta,\qquad W=XF_\beta,\qquad
 s^2=P_n(x_\beta-P_nx_\beta)^2,\qquad
 t=(x_\beta-P_nx_\beta)/s,\notag\\
 \Pi=I_n-\boldsymbol1\boldsymbol1^T/n-tt^T/n,\qquad
 Z=\Pi W,\qquad \nu=\operatorname{rank}(\Pi)=n-2.
 \label{cs:projection}
\end{gather}
The response is standardized by its own training mean and standard deviation, as in the point estimator, and we write $y$ for the standardized vector; this transformation depends on $Y$ alone.

For a positive tuning tuple $k=(g,h,\lambda,\rho)$, consider the double leave-one-out criterion with ridge at $t$. With zero diagonals define
\begin{align}
 H_{ij}&=\frac{\phi_g(t_i-t_j)}{
       \sum_{l\ne i}\phi_g(t_i-t_l)+(n-1)\lambda},&
 Q_{ij}&=\frac{\phi_g'(t_i-t_j)}{
       \sum_{l\ne i}\phi_g(t_i-t_l)+(n-1)\lambda},\notag\\
 \widehat m&=Hy,&r&=y-\widehat m,\notag\\
 A_i^{\rm NW}&=\sum_jQ_{ij}(y_j-\widehat m_i),&
 d_i&=(n-1)^{-1}\sum_{j\ne i}\phi_h(r_i-r_j),\notag\\
 a_i&=\frac1{n-1}\sum_{j\ne i}\phi_h'(r_i-r_j)
       \{(d_i+\rho)^{-1}+(d_j+\rho)^{-1}\},\notag\\
 (c_k)_i&=-a_iA_i^{\rm NW}+\sum_l a_lQ_{li}(y_i-\widehat m_l).
 \label{cs:complete-coefficient}
\end{align}
The superscript on $A^{\rm NW}$ distinguishes the slope vector from the studentization matrix below. Differentiation with respect to the index yields $n\partial L_k/\partial t_i=(c_k)_i$. To see this, differentiate each NW numerator and denominator, use $\partial(t_i-t_j)/\partial t_l=\ind\{i=l\}-\ind\{j=l\}$, and exchange the $i,j$ sums; the outgoing term is $-a_iA_i^{\rm NW}$, and after the exchange the incoming term is the second term in \eqref{cs:complete-coefficient}. Summing these two terms gives $\sum_i(c_k)_i=0$.

\begin{lemma}[Standardized gradient]\label{cs:gradient}
In the normalized raw chart $\beta(v)=(\beta+F_\beta v)/\sqrt{1+\norm v^2}$, with the tuning tuple $k$ held fixed,
\begin{equation}
 S_k(\beta):=\left.\nabla_vL_k\{\beta(v)\}\right|_{v=0}
       =\frac1{ns}Z^Tc_k=\frac1{ns}Z^T\Pi c_k.
 \label{cs:gradient-identity}
\end{equation}
\end{lemma}
\begin{proof}
The first chart derivative of $x_\beta$ is $W$. Differentiating its empirical centering and standard deviation gives
\[
 \dot t=\frac{W-\boldsymbol1P_nW}{s}
       -\frac{tP_n\{t(W-\boldsymbol1P_nW)\}}{s}
       =\Pi W/s.
\]
The result follows from the chain rule and \eqref{cs:complete-coefficient}, since the response moments do not depend on $\beta$. Training whitening followed by the raw direction map gives the scalar index $\beta^T(X_i-\bar X)/\sqrt{\beta^T\widehat\Sigma\beta}$, so the calculation includes the effect of estimating the covariance through the term $tP_n(tc_k)$.
\end{proof}

\subsection{Tuning at each candidate and the pivotal statistic}

At each candidate $\beta$, the inference tuple is selected using only $(x_\beta,Y)$ and fixed constants. The $g$ and $h$ grids and the values of $\lambda$ and $\rho$ are those of the point estimator. First, select $g$ based on the leave-one-out squared error of the ridge fit at $t$. At that $g$, select $h$ by the double leave-one-out log-density criterion. Ties are broken in favor of the first index. Then evaluate \eqref{cs:complete-coefficient}, holding the selected tuple fixed when the score is differentiated.

The point estimator selects its tuple by the variance-trace criterion, which involves $W$ through the Hessian and the covariance estimate; the inference selector above uses only $(x_\beta,Y)$, as the conditional argument requires. Write $c=c_{k(\beta)}$ and $u=\Pi c$. If $u\ne0$, put
\begin{equation}
 \bar u=u/\norm u,\qquad z=Z^T\bar u,\qquad
 A=Z^T(\Pi-\bar u\bar u^T)Z=Z^TZ-zz^T.
 \label{cs:studentization}
\end{equation}

\begin{theorem}[Conditional finite-sample calibration]\label{cs:pivot}
Suppose $p\ge2$, $n>p+1$, $s>0$, $s_Y>0$ and $\Sigma\succ0$. Under $H_\beta:Y\perp XF_\beta\mid X\beta$, conditionally on $(x_\beta,Y)$ and on $\norm u>0$,
\begin{equation}
 F_{\rm CS}(\beta)=\frac{n-d-2}{d}\,z^TA^{-1}z
       \ \sim\ F_{d,n-d-2}.
 \label{cs:F-law}
\end{equation}
Equivalently, with $U=z^T(Z^TZ)^{-1}z$,
\begin{equation}
 U\sim\operatorname{Beta}\{d/2,(n-d-2)/2\},\qquad
 F_{\rm CS}=\frac{n-d-2}{d}\frac{U}{1-U}.
 \label{cs:beta-law}
\end{equation}
The result holds with the selector above, which depends only on the candidate index and the response, for any independent error law on the event of well-defined standardization, any positive definite Gaussian design covariance, and positive bandwidths. Rate conditions are needed only for the asymptotic shape of the set in Theorem~\ref{cs:efficient}.
\end{theorem}
\begin{proof}
Completing the square in the joint Gaussian density of $(x_\beta,W)$ gives conditional row means affine in $x_\beta$ and independent row conditional covariances
\begin{equation}
 \Omega=F_\beta^T\Sigma F_\beta
 -F_\beta^T\Sigma\beta(\beta^T\Sigma\beta)^{-1}
       \beta^T\Sigma F_\beta\succ0.
 \label{cs:Omega}
\end{equation}
Projection by $\Pi$ removes those affine means. Under $H_\beta$, conditioning additionally on all responses leaves this conditional Gaussian law unchanged. Consequently
\begin{equation}
 Z\mid(x_\beta,Y)\sim MN_{n,d}(0,\Pi,\Omega).
 \label{cs:conditional-normal}
\end{equation}
The selector, $c$, and $\bar u$ are fixed under this conditioning. Complete $\bar u$ to an orthonormal basis $[\bar u,Q_2]$ of the range of $\Pi$. The matrix $Q_2$ has $n-3$ columns. Then $z=Z^T\bar u$ is $N_d(0,\Omega)$, while $Q_2^TZ$ has $n-3$ independent $N_d(0,\Omega)$ rows. Their cross covariance is $\bar u^TQ_2$ times $\Omega$, hence zero. Joint Gaussianity makes the two blocks independent. Thus $A=(Q_2^TZ)^T(Q_2^TZ)$ is Wishart with scale $\Omega$ and $n-3$ degrees of freedom, independently of $z$. Since $n-3\ge d$, it is invertible almost surely.

Whiten the columns by $\Omega^{-1/2}$, which does not change $U$. In the $\nu$-dimensional range of $\Pi$, the resulting $\nu\times d$ matrix has independent standard normal entries. Its $d$-dimensional column space is invariant under every orthogonal transformation of $\R^\nu$, since its density is proportional to $\exp(-\norm E_F^2/2)$. The squared projection of a fixed unit $\bar u$ onto this space therefore has the distribution of the sum of the first $d$ squared coordinates of a uniform point on the unit sphere in $\R^\nu$. A standard normal vector divided by its norm is such a point: its polar density is radial. It follows that
\[
 U\ \stackrel{d}=\
 \frac{\sum_{j=1}^dG_j^2}{\sum_{j=1}^{\nu}G_j^2},\qquad
 G_j\stackrel{\rm iid}{\sim}N(0,1).
\]
The numerator and complementary sum are independent chi-square variables with $d$ and $\nu-d$ degrees of freedom. Their scaled ratio gives the $F$ law. Finally Sherman--Morrison gives $z^TA^{-1}z=U/(1-U)$, and both formulas follow.
\end{proof}

The degrees of freedom reflect two removed observation directions: centering and the candidate index. The independent Wishart block has $n-3$ rows, and the degrees of freedom in the denominator are $n-d-2$. For $d=1$, the statistic is the ordinary partial-regression $t^2$ statistic with $n-3$ residual degrees of freedom.

If $u=0$, set $F_{\rm CS}=0$, so that the candidate belongs to the set. Under the continuous Gaussian-mixture model, a zero index or response scale and rank deficiency occur with probability zero. For a general error law, the event $s_Y=0$ may have positive probability and is treated separately. In computation, a candidate at which a vanishing scale, a rank deficiency, or a numerical breakdown occurs is excluded from the set, and the coverage bound below accounts for this event.

\subsection{The acceptance set and localization}

Define the acceptance set in projective space
\begin{equation}
 \mathcal C_{\rm full}
 =\{[\beta]:F_{\rm CS}(\beta)\le F_{d,n-d-2;1-\alpha}\}.
 \label{cs:full-set}
\end{equation}
It is a set of lines: changing $\beta$ to $-\beta$ changes the sign of $t$, $Q$ and $c$, leaves the residuals and selected bandwidths unchanged, and leaves the squared projection in \eqref{cs:beta-law} unchanged. The choice of the orthonormal tangent frame also leaves that projection invariant. Under the continuous mixture model, the scales are positive almost surely. At $\beta_0$, Theorem~\ref{cs:pivot} gives $\Pp\{[\beta_0]\in\mathcal C_{\rm full}\}\ge1-\alpha$, with equality if $u\ne0$ almost surely. Since a nonmonotone link can produce stationary directions far from $\beta_0$, the set is localized around the point estimate.

Let $\mathcal C_{{\rm init},n}$ be the cap around the Fourier initial estimate used in the optimization, mapped to raw lines. Let $\widehat\beta$ be the SI-KDR point estimate. Let $\mathcal E_n$ be the whole projective space when all inference conditions of Section~\ref{sec:inference-conditions} hold and the empty set otherwise. Define
\begin{align}
 a_n&=\log(n+e)/\sqrt n,\notag\\
 \mathcal D_n&=\mathcal E_n\cap\mathcal C_{{\rm init},n}
 \cap\{[\beta]:\operatorname{angle}([\beta],[\widehat\beta])\le a_n\},
 \qquad
 \mathcal C_{\rm CS}^{\rm math}=\mathcal D_n\cap\mathcal C_{\rm full}.
 \label{cs:local-domain}
\end{align}
The radius $a_n$ exceeds the root-$n$ rate of the point estimate.

Write $\mathcal C_{\rm CS}^{\rm comp}$ for the set evaluated in floating-point arithmetic, and let $N_\beta$ denote the event that the evaluation of the test at $\beta$ breaks down numerically; such candidates are excluded from $\mathcal C_{\rm CS}^{\rm comp}$. If the point estimate does not satisfy the inference conditions of Section~\ref{sec:inference-conditions}, or the initial estimate is not available, $\mathcal D_n=\varnothing$.

\begin{corollary}[Finite localized coverage]\label{cs:domain}
For the data-dependent domain $\mathcal D_n$,
\begin{align}
 \Pp\{[\beta_0]\in\mathcal C_{\rm CS}^{\rm math}\}
 &\ge1-\alpha-\Pp\{[\beta_0]\notin\mathcal D_n\},\notag\\
 \Pp\{[\beta_0]\in\mathcal C_{\rm CS}^{\rm comp}\}
 &\ge1-\alpha-\Pp\{[\beta_0]\notin\mathcal D_n\}-\Pp(N_{\beta_0}).
 \label{cs:finite-coverage}
\end{align}
\end{corollary}
\begin{proof}
The true line is excluded from $\mathcal C_{\rm CS}^{\rm math}$ only if the test rejects it or $[\beta_0]\notin\mathcal D_n$; it is excluded from $\mathcal C_{\rm CS}^{\rm comp}$ additionally on $N_{\beta_0}$. The union bound and Theorem~\ref{cs:pivot} give both inequalities.
\end{proof}
Under this model, $N_{\beta_0}$ is empty in exact arithmetic. For an error law under which the sample response scale vanishes with positive probability, this global preprocessing failure leaves the domain empty and is therefore already included in $\Pp\{[\beta_0]\notin\mathcal D_n\}$.

\subsection{Connection to the influence covariance}

Define
\begin{equation}
 K_{n,k}(\beta)=\frac{P_n\{(\Pi c_k)^2\}}{s^2}\frac{Z^TZ}{n-2}.
 \label{cs:K-definition}
\end{equation}
At the true candidate \eqref{cs:conditional-normal} implies
\begin{equation}
 n\Var\{S_k(\beta_0)\mid x_{\beta_0},Y\}
       =\frac{P_n\{(\Pi c_k)^2\}}{s^2}\Omega.
 \label{cs:conditional-covariance}
\end{equation}
The covariance estimate and the score share the matrix $Z$; in Theorem~\ref{cs:pivot} this dependence is handled by removing the direction $\bar u$, which yields the independent matrix $A$. In population-standardized coordinates, the coefficient is $\kappa-C_kT$, $C_k=\E(T\kappa)$. Its empirical counterpart is $\Pi c_k=c_k-P_nc_k-tP_n(tc_k)$, which includes the whitening contribution. In a raw orthonormal tangent, with $\sigma_T^2=\beta_0^T\Sigma\beta_0$,
\begin{equation}
 B_{n,k,\rm raw}=\frac{\Omega}{\sigma_T^2}
                      \E\{\kappa_k(T,Y)-C_kT\}^2.
 \label{cs:raw-B}
\end{equation}
This agrees with the transformation by $D_0$ in Section~\ref*{main-sec:population} of the paper.

\begin{lemma}[Uniform covariance connection]\label{cs:covariance}
Under the conditions of Theorems~\ref{dir:theorem} and~\ref{inf:caseweights}, including $D_S+D_H+q_D\to0$, for every fixed $C_0$,
\begin{equation}
 \max_{k\in\mathcal K}\sup_{\norm\theta\le C_0a_n}
 \norm{K_{n,k}\{\beta(\theta)\}-B_{n,k,\rm raw}}=o_P(1),
 \label{cs:uniform-K}
\end{equation}
after matching tangent coordinates. At the computed estimate, $K_{n,k}-\widehat B_k=o_P(1)$ for the case-weight covariance. These conclusions cover fixed positive and shrinking smoothing. If all four parameters shrink to zero, both additionally have the common efficient information limit $\mathcal I_{\rm raw}$.
\end{lemma}
\begin{proof}
Use a smooth local frame about $\beta_0$; its coordinate changes and inverses are bounded. Uniformly on the shrinking ball, $Z^TZ/(n-2)\to_P\Omega$ and $s^2\to_P\sigma_T^2$. Indeed, both are continuous rational functions of $\beta$ and the empirical first and second predictor moments, while $\beta^T\widehat\Sigma\beta$ is bounded away from zero with probability tending to one. Fixed dimension and finite Gaussian fourth moments give convergence of those empirical moments, and their population counterparts vary by $O(a_n)$ on the ball.

The fitted-coefficient step in Theorem~\ref{inf:caseweights} gives, in matching coordinates and with $c_k^*(\theta)=\Pi_\theta c_k(\theta)$,
\begin{equation}
 \max_k\sup_{\norm\theta\le C_0a_n}
 \norm{c_k^*(\theta)-c_k^*(0)}_{n,2}
 =O_P\Bigl(\max_k\ell_n^{32}a_ng_k^{-2}h_k^{-2}\Bigr)=o_P(1).
 \label{cs:coefficient-modulus}
\end{equation}
Here $g_k$ and $h_k$ are the two bandwidths of tuple $k$. The sample scale of the index enters only through a bounded, uniformly convergent factor. Under \eqref{inf:nonempty} the polynomial rate in \eqref{cs:coefficient-modulus} is $n^{-1/30}$; in the general range of these conditions the vanishing of each of the finitely many terms is part of the condition $q_D\to0$. The argument is that used for the leading term in Theorem~\ref{inf:caseweights}, with the finite-smoothed coefficient as the anchor. At the fixed truth, the empirical coefficient bound and preprocessing bound from the same proof give
\begin{equation}
 \max_k\norm{c_k^*(0)-
       \{\kappa_k(T_i,Y_i)-C_kT_i\}_{i=1}^n}_{n,2}=o_P(1).
 \label{cs:fixed-coefficient}
\end{equation}
The projection removes the empirical scalar mean and empirical index component. The scalar population mean is zero: differentiating $D(t)M(t)=\E\{\phi_g(t-T)Y\}$ with $D=q_g+\lambda$ gives
\[
 D(t)M'(t)=\E[\phi_g'(t-T)\{Y-M(t)\}].
\]
Integrating the incoming term of $\kappa$ therefore gives $\E\{a(R)M'(T)\}$, cancelling its outgoing term. The index projection is $C_kT$ by definition. Put $\widetilde\kappa_k=\kappa_k-C_kT$. The uniform fourth-moment bound from \eqref{dir:kappaibp} gives, simultaneously over the finite grid,
\[
 (P_n-P)\widetilde\kappa_k^2=O_P(n^{-1/2}),\qquad
 P_n\widetilde\kappa_k^2=O_P(1).
\]
Equations~\eqref{cs:fixed-coefficient} and \eqref{cs:coefficient-modulus}, followed by Cauchy--Schwarz, imply $\sup_\theta|P_n\{c_k^*(\theta)^2\}-P\widetilde\kappa_k^2|=o_P(1)$. Multiplication by the predictor covariance and scale limits establishes \eqref{cs:uniform-K} at $B_{n,k,\rm raw}$, with the finite-smoothed coefficient itself. For the assertion under shrinking smoothing, \eqref{dir:poplimit} then gives the common information limit.

Theorem~\ref{inf:caseweights} gives $\widehat B_k-B_{n,k,\rm raw}=o_P(1)$ at the computed estimate. Subtracting this from \eqref{cs:uniform-K} proves the last assertion.
\end{proof}

\subsection{Asymptotic shape of the confidence set}

\begin{theorem}[Asymptotic shape of the confidence set]
\label{cs:efficient}
Under the preceding conditions and the shrinking tuple \eqref{inf:nonempty}, with exact arithmetic for the computed estimate and the initial estimate, $\Pp\{[\beta_0]\in\mathcal D_n\}\to1$. All accepted points in $\mathcal D_n$ are at projective distance $O_P(n^{-1/2})$ from $\beta_0$. In a raw tangent chart, write the coordinate of the estimate as $\widehat\theta$ and let $\mathcal I=\mathcal I_{\rm raw}$. The rescaled acceptance set is asymptotically the efficient ellipsoid $\{v:v^T\mathcal I v\le\chi^2_{d,1-\alpha}\}$, in the inner and outer sense specified in \eqref{cs:set-inclusions}. In particular its projective diameter is $O_P(n^{-1/2})$ and its coverage tends to $1-\alpha$.
\end{theorem}
\begin{proof}
We treat the domain, localization, and the center and shape in turn. For the domain, Theorem~\ref{dir:theorem}, simultaneously over the finite grid, gives
\[
 \sqrt n\widehat\theta
 =\mathcal I^{-1}n^{-1/2}\sum_iS_{{\rm eff},i}+o_P(1).
\]
Consequently $\operatorname{angle}([\widehat\beta],[\beta_0])=O_P(n^{-1/2})$, so the angular cap with radius $a_n$ contains $\beta_0$ with probability tending to one. The initial-cap event and all inference conditions of Section~\ref{sec:inference-conditions} also hold with probability tending to one by Theorems~\ref{inf:rates} and~\ref{inf:caseweights} and the positivity of the limiting information. This proves the domain claim. The angular cap lies in a deterministic $C_0a_n$ ball about $\beta_0$ with probability tending to one, so Lemma~\ref{cs:covariance} applies throughout it. The error of the initial estimator is $o_P(r_n)$, and $r_n$ exceeds the root-$n$ scale. Thus bounded root-$n$ neighborhoods of $\beta_0$ are inside both domain caps with probability tending to one.

For localization, at any nondegenerate candidate, \eqref{cs:gradient-identity} and \eqref{cs:K-definition} give
\begin{equation}
 nS_k(\beta)^TK_{n,k}(\beta)^{-1}S_k(\beta)=(n-2)U,
 \qquad
 dF_{\rm CS}(\beta)=\frac{(n-d-2)U}{1-U}.
 \label{cs:quadratic-identities}
\end{equation}
Acceptance implies $U\le u_{n,\alpha}$, where
\[
 u_{n,\alpha}=\frac{dF_{d,n-d-2;1-\alpha}}
 {n-d-2+dF_{d,n-d-2;1-\alpha}}=O(n^{-1}).
\]
Lemma~\ref{cs:covariance} bounds the eigenvalues of $K_{n,k}$ above and away from zero uniformly on the domain, with probability tending to one. It follows that every accepted candidate obeys $\sqrt n\norm{S_k(\beta)}=O_P(1)$ uniformly over the finite grid. The uniform Hessian bound of Theorem~\ref{dir:theorem} gives negative curvature bounded away from zero in the common local chart. Taylor's integral identity then gives
\[
 \norm{\nabla_\theta L_k(\theta)}
 \ge c\norm\theta-\norm{\nabla_\theta L_k(0)}
\]
uniformly for some fixed $c>0$: take the inner product of the gradient difference with $\theta/\norm\theta$ and use the negative-definite integral Hessian. The score expansion gives $\max_k\norm{\nabla L_k(0)}=O_P(n^{-1/2})$. Smooth frame transformations preserve these bounds up to bounded constants. Hence every accepted point in $\mathcal D_n$ has $\norm\theta=O_P(n^{-1/2})$.

For the center and shape, let $M$ be bounded; the simultaneous common influence limit and uniform Hessian convergence imply
\begin{equation}
 \max_k\sup_{\norm v\le M}
 \norm{\sqrt n\nabla_\theta L_k(\widehat\theta+v/\sqrt n)
                         +\mathcal I v}=o_P(1).
 \label{cs:common-center}
\end{equation}
For the centering, note that uniformly over $k$, $\sqrt n\nabla L_k(0)=n^{-1/2}\sum_iS_{{\rm eff},i}+o_P(1)$, and $\mathcal I\sqrt n\widehat\theta$ equals the same sum plus $o_P(1)$, although the point selector involves $W$. Taylor's integral identity then gives \eqref{cs:common-center}. It holds for every grid point, and hence for the candidate-dependent inference selector. Lemma~\ref{cs:covariance} and \eqref{cs:quadratic-identities} now yield, uniformly on bounded $v$,
\begin{equation}
 dF_{\rm CS}\{\beta(\widehat\theta+v/\sqrt n)\}
       =v^T\mathcal I v+o_P(1).
 \label{cs:local-shape}
\end{equation}
Here $U=O_P(n^{-1})$, so replacing $(n-2)U$ by $(n-d-2)U/(1-U)$ costs $o_P(1)$. Also $dF_{d,n-d-2;1-\alpha}\to q_\alpha:=\chi^2_{d,1-\alpha}$: in the independent chi-square ratio, the denominator divided by its increasing degrees of freedom converges to one. For every $\epsilon>0$,
\begin{equation}
 \{v:v^T\mathcal I v\le q_\alpha-\epsilon\}
 \subseteq\sqrt n(\mathcal C_{\rm CS}^{\rm math}-\widehat\theta)
 \subseteq\{v:v^T\mathcal I v\le q_\alpha+\epsilon\}
 \label{cs:set-inclusions}
\end{equation}
with probability tending to one, in the following sense. First, as above, choose the localization radius $M$ so large that the probability of an accepted point outside this ball is arbitrarily small. Then use the uniform convergence within this ball. The interior ellipsoid lies within both caps with probability tending to 1. Letting this residual probability tend to zero gives the set inclusions, so the rescaled sets converge to the ellipsoid in the Hausdorff sense. By Lemma~\ref{cs:covariance}, the probability of degenerate coefficients or violated inference conditions tends to zero. The probability that the matrix is rank-deficient or that a scale is zero is exactly zero. Since $\sqrt n\widehat\theta\Rightarrow N(0,\mathcal I^{-1})$, the coverage and diameter claims follow from these inclusions.
\end{proof}

\section{Simulations with the reference bandwidth rule}\label{sec:variants}
This section reports simulations for SI-KDR-R, which uses the training standardization of the paper but selects the bandwidths by a reference rule in place of the variance-trace rule. The study has 2,600 simulated datasets with correlated Gaussian predictors, sine and cosine links and two non-Gaussian error densities, generated independently of the datasets of Section~\ref{sec:sim-details}.

In the units of the whitened predictors and standardized response, the reference grids, ridge and floor are
\[
 g\in n^{-4/25}\{1.2,1.4,1.6\},\qquad
 h\in n^{-1/30}\{.4,.6,.8\},\qquad
 \lambda=.01n^{-7/10},\quad\rho=n^{-1}.
\]
At the Fourier initial direction, $g$ minimizes the protected leave-one-out NW prediction mean squared error over its grid; with this $g$, $h$ maximizes the protected leave-one-out log density of the residuals over its grid. Ties go to the first grid point. The selected tuple is held fixed while the direction is optimized. Initialization, the protected criterion, projected ascent, and the handling of numerical exceptions are as in Section~\ref{sec:algorithm}; the grids there belong to the variance-trace rule. The rule selected $(g,h)=(.4312,.3232)$ at $n=600$, $(.3686,.3128)$ at $n=1600$ and $(.3299,.3056)$ at $n=3200$ in every sample.

\subsection{Data-generating laws}
At dimension three, let $\Sigma_{jk}=.5^{|j-k|}$, $\beta_0=(1,1,0)^T/\sqrt2$ and $T=\beta_0^TX$. The links are $\sin(2T)$ and $\cos(2T)$. Generate
\[
 E_a\sim .7N(-.9,.5^2)+.3N(2.1,.3^2),\qquad
 E_m\sim\tfrac13\{N(-1.5,.25^2)+N(0,.25^2)+N(1.5,.25^2)\}.
\]
Use $\epsilon=(E-\E E)/\sqrt{\Var(E)}$, so both error shapes have mean zero and variance one.

The $n=600$ cells have 500 replicates for each of two links and two errors. The $n=1600$ cosine/asymmetric cell has 500 replicates, and the $n=3200$ cosine/asymmetric cell has 100.

\subsection{Outcomes and coordinates}
The direction loss is $2-2|\widehat\beta^T\beta_0|$. With a fixed true tangent frame $U_0$, the normalized error is $\sqrt n\widehat\theta=\sqrt n U_0^T\widehat\beta/(\beta_0^T\widehat\beta)$. The estimated covariance in this chart uses the derivative
\[
 D_0(\widehat\beta)
 =\{U_0^T-\widehat\theta\beta_0^T\}/(\beta_0^T\widehat\beta),
\]
where $\widehat\theta$ is the unnormalized chart coordinate, applied to the inverse observed information (the negative Hessian of the SI-KDR-R working criterion) after mapping to the raw direction. Section~\ref{sec:sim-details} reports the separate study using the case-weight covariance.

\begin{table}[tbp]
\centering\footnotesize
\setlength{\tabcolsep}{3pt}
\caption{First coordinate of $\sqrt n\widehat\theta$ in the true chart: bias, empirical variance over all replicates, mean estimated variance, and the efficiency bound. The last two columns are ratios of the empirical variance to the estimated variance and to the bound.}
\label{tab:variance-1}
\vspace{3pt}
\begin{tabular}{rllrrrrrr}
\toprule
$n$ & Link & Error & Bias & Empirical & Estimated & Efficient & Emp./est. & Emp./eff. \\
\midrule
600 & Sine & Asymmetric & 0.038 & 0.397 & 0.491 & 0.085 & 0.809 & 4.680 \\
600 & Sine & Multimodal & 0.002 & 0.681 & 0.986 & 0.046 & 0.691 & 14.941 \\
600 & Cosine & Asymmetric & -0.007 & 0.364 & 0.483 & 0.085 & 0.754 & 4.290 \\
600 & Cosine & Multimodal & -0.027 & 0.677 & 0.977 & 0.046 & 0.693 & 14.844 \\
1600 & Cosine & Asymmetric & 0.015 & 0.281 & 0.324 & 0.085 & 0.867 & 3.318 \\
3200 & Cosine & Asymmetric & -0.028 & 0.238 & 0.259 & 0.085 & 0.920 & 2.810 \\
\bottomrule
\end{tabular}
\end{table}

\begin{table}[tbp]
\centering\footnotesize
\setlength{\tabcolsep}{3pt}
\caption{As Table~\ref{tab:variance-1}, for the second coordinate.}
\label{tab:variance-2}
\vspace{3pt}
\begin{tabular}{rllrrrrrr}
\toprule
$n$ & Link & Error & Bias & Empirical & Estimated & Efficient & Emp./est. & Emp./eff. \\
\midrule
600 & Sine & Asymmetric & -0.005 & 0.247 & 0.302 & 0.052 & 0.817 & 4.729 \\
600 & Sine & Multimodal & -0.039 & 0.434 & 0.602 & 0.028 & 0.720 & 15.454 \\
600 & Cosine & Asymmetric & 0.014 & 0.242 & 0.298 & 0.052 & 0.811 & 4.633 \\
600 & Cosine & Multimodal & -0.035 & 0.423 & 0.604 & 0.028 & 0.700 & 15.078 \\
1600 & Cosine & Asymmetric & -0.044 & 0.166 & 0.200 & 0.052 & 0.828 & 3.172 \\
3200 & Cosine & Asymmetric & -0.024 & 0.105 & 0.160 & 0.052 & 0.658 & 2.016 \\
\bottomrule
\end{tabular}
\end{table}

\begin{table}[tbp]
\centering\footnotesize
\setlength{\tabcolsep}{3pt}
\caption{Eigenvalues of $I_{\rm eff}^{1/2}\widehat{\operatorname{Cov}}(\sqrt n\widehat\theta)I_{\rm eff}^{1/2}$ in the true chart, using all point estimates. Parentheses give delete-one jackknife Monte Carlo standard errors. Both eigenvalues equal one for the efficient limiting distribution.}
\label{tab:covariance-eigen-mc}
\vspace{3pt}
\begin{tabular}{rllrrr}
\toprule
$n$ & Link & Error & Reps & Smaller eigenvalue (MCSE) & Larger eigenvalue (MCSE) \\
\midrule
600 & Sine & Asymmetric & 500 & 4.33 (0.27) & 4.92 (0.30) \\
600 & Sine & Multimodal & 500 & 13.94 (0.92) & 15.92 (1.07) \\
600 & Cosine & Asymmetric & 500 & 4.17 (0.27) & 4.66 (0.26) \\
600 & Cosine & Multimodal & 500 & 14.29 (1.01) & 15.34 (0.98) \\
1600 & Cosine & Asymmetric & 500 & 3.15 (0.20) & 3.32 (0.20) \\
3200 & Cosine & Asymmetric & 100 & 1.98 (0.29) & 2.81 (0.38) \\
\bottomrule
\end{tabular}
\end{table}

\FloatBarrier
\section{Details of the numerical studies}\label{sec:sim-details}
\begin{table}[t]
\centering\small
\caption{Population quantities for the four bandwidth combinations at $n=600$. Every tuple uses $\lambda=.01\,600^{-7/10}$ and $\rho=1/600$. The ratio is $i_{\rm eff}B/J^2$.}\label{tab:population}
\begin{tabular}{llrrrrr}
\toprule Link & Error & $g$ & $h$ & $J$ & $B$ & Ratio \\
\midrule
Sine & Normal & 0.431 & 0.323 & 1.906 & 1.345 & 1.110 \\
Sine & Normal & 0.431 & 0.097 & 1.936 & 1.388 & 1.110 \\
Sine & Normal & 0.216 & 0.323 & 2.785 & 2.607 & 1.008 \\
Sine & Normal & 0.216 & 0.097 & 2.838 & 2.706 & 1.008 \\
Sine & Asymmetric & 0.431 & 0.323 & 6.877 & 5.397 & 4.374 \\
Sine & Asymmetric & 0.431 & 0.097 & 9.060 & 8.848 & 4.132 \\
Sine & Asymmetric & 0.216 & 0.323 & 17.137 & 11.709 & 1.528 \\
Sine & Asymmetric & 0.216 & 0.097 & 29.637 & 32.514 & 1.419 \\
Sine & Multimodal & 0.431 & 0.323 & 3.427 & 2.183 & 13.245 \\
Sine & Multimodal & 0.431 & 0.097 & 4.734 & 3.699 & 11.763 \\
Sine & Multimodal & 0.216 & 0.323 & 13.510 & 6.562 & 2.562 \\
Sine & Multimodal & 0.216 & 0.097 & 45.493 & 53.632 & 1.847 \\
Cosine & Normal & 0.431 & 0.323 & 1.909 & 1.349 & 1.111 \\
Cosine & Normal & 0.431 & 0.097 & 1.939 & 1.392 & 1.111 \\
Cosine & Normal & 0.216 & 0.323 & 2.786 & 2.607 & 1.008 \\
Cosine & Normal & 0.216 & 0.097 & 2.839 & 2.706 & 1.008 \\
Cosine & Asymmetric & 0.431 & 0.323 & 6.935 & 5.475 & 4.363 \\
Cosine & Asymmetric & 0.431 & 0.097 & 9.137 & 8.941 & 4.105 \\
Cosine & Asymmetric & 0.216 & 0.323 & 17.165 & 11.755 & 1.529 \\
Cosine & Asymmetric & 0.216 & 0.097 & 29.666 & 32.638 & 1.421 \\
Cosine & Multimodal & 0.431 & 0.323 & 3.469 & 2.184 & 12.935 \\
Cosine & Multimodal & 0.431 & 0.097 & 4.918 & 3.705 & 10.916 \\
Cosine & Multimodal & 0.216 & 0.323 & 13.562 & 6.588 & 2.553 \\
Cosine & Multimodal & 0.216 & 0.097 & 45.595 & 53.786 & 1.844 \\
\bottomrule
\end{tabular}
\end{table}

\subsection{Data generation}
The low-dimensional point-estimation and inference studies of Section~\ref*{main-sec:experiments} of the paper use $p=3$, $X\sim N_3(0,\Sigma)$, $\Sigma_{jk}=.5^{|j-k|}$ and $\beta_0=(1,1,0)^T/\sqrt2$. The index in the original coordinates is $t=X^T\beta_0$, whose variance is $1.5$. The links are $\sin(2t)$ and $\cos(2t)$. Independent errors are standardized to mean zero and variance one. Before standardization their laws are
\[
.7N(-.9,.5^2)+.3N(2.1,.3^2),\qquad
\tfrac13\{N(-1.5,.25^2)+N(0,.25^2)+N(1.5,.25^2)\}.
\]
For weights $w_j$, means $\mu_j$ and standard deviations $s_j$, the generator subtracts $\sum_jw_j\mu_j$ and divides by the square root of $\sum_jw_j\{(\mu_j-\sum_lw_l\mu_l)^2+s_j^2\}$. The predictor and response standardizations are estimated from each generated dataset.

The four primary cells (sine, then cosine, each with asymmetric and multimodal errors) have 500 replicates at $n=600$. The cosine/asymmetric comparison at larger sample sizes uses 300 replicates at $n=1600$ and 50 at $n=3200$. The confidence-set study uses 2000 further independent datasets, 500 in each primary cell. All studies use the same SI-KDR point estimator, and the seeds are recorded with the code.

The line loss is $2-2|\widehat\beta^T\beta_0|$. Before biases are computed, the sign of each estimate is aligned with $\beta_0$. Table~\ref{tab:variance} reports the empirical and estimated variance diagonals in the gnomonic chart $U_0^T\widehat\beta/(\beta_0^T\widehat\beta)$, together with the eigenvalues of the information-standardized tangent covariance and their Monte Carlo standard errors. Table~\ref{tab:cost} reports the computing times at $n=600$. In the confidence-set study, every fit met the inference conditions of Section~\ref*{main-sec:tuning} of the paper, and the true direction lay in the initial cap in every replicate.
\begin{table}[htbp]
\centering\small
\caption{Bias in the original coordinates, chart variances and Monte Carlo standard errors.}\label{tab:variance}
\resizebox{\textwidth}{!}{%
\begin{tabular}{lllllll}
\toprule
Cell & $n$ & Bias (3 coords.) & Emp. chart diag. & Est. chart diag. & $\lambda_1$ (SE) & $\lambda_2$ (SE) \\
\midrule
Sin/A & 600 & -0.000208, -0.00011, 0.000579 & 0.164, 0.106 & 0.197, 0.124 & 1.77 (0.109) & 2.11 (0.145) \\
Sin/M & 600 & -0.000129, -0.000173, -0.000415 & 0.168, 0.0887 & 0.211, 0.124 & 3.12 (0.247) & 3.67 (0.355) \\
Cos/A & 600 & -0.000562, 0.000245, -0.000733 & 0.164, 0.104 & 0.195, 0.122 & 1.88 (0.121) & 2.12 (0.137) \\
Cos/M & 600 & 0.000182, -0.000495, 0.000764 & 0.176, 0.0904 & 0.195, 0.12 & 3.2 (0.261) & 3.85 (0.483) \\
\bottomrule
\end{tabular}}
\par\smallskip\parbox{\textwidth}{\footnotesize Chart variances are for $\sqrt n$ times the chart coordinate. The chart covariance uses all estimates; the eigenvalues are those of the information-standardized covariance of the tangent deviations in the original coordinates. Standard errors are delete-one jackknife estimates.}
\end{table}

\begin{table}[htbp]
\centering\small
\caption{Computing time in seconds at $n=600$ (Monte Carlo standard errors).}\label{tab:cost}
\resizebox{\textwidth}{!}{%
\begin{tabular}{llllllll}
\toprule
Cell & $n$ & Total fit & Prelim. & Selection & Final+var. & Variance & Overall \\
\midrule
Sin/A & 600 & 4.9 (0.0327) & 0.633 (0.00761) & 3.47 (0.0244) & 0.79 (0.011) & 0.0853 (0.000969) & 4.98 (0.033) \\
Sin/M & 600 & 5.11 (0.0345) & 0.598 (0.00828) & 3.61 (0.0243) & 0.902 (0.0154) & 0.0921 (0.00112) & 5.21 (0.0349) \\
Cos/A & 600 & 4.96 (0.0293) & 0.663 (0.00805) & 3.49 (0.0223) & 0.804 (0.011) & 0.0855 (0.000923) & 5.05 (0.0296) \\
Cos/M & 600 & 5.02 (0.0304) & 0.602 (0.00852) & 3.55 (0.0219) & 0.87 (0.0146) & 0.0888 (0.00102) & 5.11 (0.0307) \\
\bottomrule
\end{tabular}}
\par\smallskip\parbox{\textwidth}{\footnotesize Total fit includes the preliminary fit, the selection among twenty tuples, and the final fit with variance estimation; the Variance column is part of Final+var. The Overall column additionally includes data generation and the diagnostics computed after fitting.}
\end{table}

\subsection{Paired comparisons with other estimators}
Table~\ref{tab:confirmation} reports four methods in the four primary cells on the same 500 samples per cell. Mean MAVE and csMAVE are fitted with the R package MAVE, version 1.3.12 \citep{MAVEPackage2025}, with arguments max.dim $=1$ and screen $=p$. SI-KDR-R follows Section~\ref{sec:variants}. Every method receives the same training-whitened predictors and standardized response, and its direction is mapped back to the original coordinates before evaluation. The Monte Carlo standard error of a difference is the sample standard deviation of the paired losses divided by $\sqrt{500}$.

In all four cells, SI-KDR reduces risk by $87.8$--$89.7\%$ relative to mean MAVE and by $84.0$--$92.0\%$ relative to csMAVE. All four methods are fitted to the same datasets, which differ from those used in the SI-KDR-R study in Section~\ref{sec:variants}.
\begin{table}[t]
\centering\footnotesize
\setlength{\tabcolsep}{3pt}
\caption{Paired $n=600$ comparison on the same 500 samples per cell. Entries are mean line loss (Monte Carlo standard error), multiplied by $10^3$.}
\label{tab:confirmation}
\begin{tabular}{lrrrr}
\toprule
Cell & SI-KDR & mean MAVE & csMAVE & SI-KDR-R \\
\midrule
Sin/A & 0.450 (0.023) & 4.382 (0.221) & 3.042 (0.160) & 1.058 (0.051) \\
Sin/M & 0.426 (0.032) & 4.053 (0.192) & 5.313 (0.251) & 1.627 (0.078) \\
Cos/A & 0.448 (0.021) & 3.924 (0.193) & 2.801 (0.146) & 1.017 (0.050) \\
Cos/M & 0.443 (0.049) & 3.637 (0.186) & 4.738 (0.231) & 1.657 (0.091) \\
\bottomrule
\end{tabular}

\smallskip
\parbox{\linewidth}{\footnotesize A and M denote asymmetric and multimodal errors. All methods use the same data and training standardization, and all directions are mapped back to the original coordinates. No numerical failures occurred.}
\end{table}

The comparison at larger sample sizes adds a feasible efficient-score estimator based on \citet{ma2013efficient}. Our implementation starts from the mean-MAVE direction computed on the same data, uses the chart whose largest initial coordinate is fixed to 1, and refits the conditional nuisances at each candidate. The conditional mean uses NW regression, and the conditional density score uses double-kernel local-linear fitting. All kernels are Gaussian, with bandwidths $s_U n^{-1/5}$ for the conditional predictor mean, $s_U n^{-1/6}$ for the index in the density fit and $s_Y n^{-1/7}$ for the response, where $s_U$ and $s_Y$ are sample standard deviations with divisor $n-1$. A damped Newton solver uses at most 30 iterations, 16 backtracking steps, and a score tolerance of $10^{-7}$. A fit whose score is not resolved, or whose local fit is numerically singular, receives line loss two. The other methods are fitted as above.

Table~\ref{tab:large-risk} gives the mean risks and Table~\ref{tab:large-paired} all paired contrasts. All five methods returned valid directions on all 350 samples, and no SI-KDR, SI-KDR-R or Ma--Zhu fit reached its iteration limit. At $n=1600$ the risk of SI-KDR is $15.39\%$ below that of the Ma--Zhu implementation, and the paired interval lies below zero.
\begin{table}[tbp]
\centering\small
\setlength{\tabcolsep}{4pt}
\caption{Cosine/asymmetric design at larger sample sizes: mean line risk with Monte Carlo standard error in parentheses, both multiplied by $10^4$; $N$ is the number of samples.}
\label{tab:large-risk}
\begin{tabular}{rrrrrrr}
\toprule
$n$ & $N$ & SI-KDR & SI-KDR-R & mean MAVE & csMAVE & Ma--Zhu \\
\midrule
1600 & 300 & 1.420 (0.079) & 3.089 (0.179) & 14.488 (0.940) & 8.364 (0.555) & 1.678 (0.096) \\
3200 & 50 & 0.568 (0.105) & 1.117 (0.177) & 5.249 (0.643) & 2.457 (0.339) & 0.626 (0.118) \\
\bottomrule
\end{tabular}
\end{table}

\begin{table}[tbp]
\centering\footnotesize
\setlength{\tabcolsep}{4pt}
\caption{Paired comparisons for the cosine link with asymmetric errors at $n=1600$ and $n=3200$. The risk ratio is SI-KDR/reference and $\Delta$ is SI-KDR minus reference, so negative differences favor SI-KDR. Differences, MCSEs and interval endpoints are multiplied by $10^4$. Intervals are pointwise paired Student-$t$ intervals with $N-1$ degrees of freedom, without multiplicity adjustment.}
\label{tab:large-paired}
\begin{tabular}{llrrr}
\toprule
$n$ ($N$) & Reference & Risk ratio & $\Delta$ (MCSE) & 95\% interval \\
\midrule
1600 (300) & SI-KDR-R & 0.460 & $-1.670$ (0.140) & $[-1.946,-1.393]$ \\
1600 (300) & mean MAVE & 0.098 & $-13.069$ (0.929) & $[-14.897,-11.241]$ \\
1600 (300) & csMAVE & 0.170 & $-6.944$ (0.546) & $[-8.018,-5.870]$ \\
1600 (300) & Ma--Zhu & 0.846 & $-0.258$ (0.093) & $[-0.441,-0.075]$ \\
\midrule
3200 (50) & SI-KDR-R & 0.509 & $-0.549$ (0.126) & $[-0.802,-0.296]$ \\
3200 (50) & mean MAVE & 0.108 & $-4.680$ (0.666) & $[-6.020,-3.341]$ \\
3200 (50) & csMAVE & 0.231 & $-1.889$ (0.365) & $[-2.623,-1.155]$ \\
3200 (50) & Ma--Zhu & 0.908 & $-0.058$ (0.096) & $[-0.250,0.135]$ \\
\bottomrule
\end{tabular}
\end{table}

\subsection{Population and sample expansion diagnostics}
The population comparison of Table~\ref{tab:population} holds $\lambda$ and $\rho$ fixed and varies $g\in\{g_L,g_S\}$ and $h\in\{h_L,h_S\}$ for both links and all three error laws. The population profile and influence include the ridge, the floor, the incoming refit term, and the whitening correction. The quadrature uses $(2049,16385,100)$ index-grid, residual-grid and Gaussian-node counts; increasing these counts to $(4097,32769,180)$ changes $J$, $B$ and $V$ by at most $6.26\times10^{-6}$ in relative terms across the 24 combinations.

The sample diagnostic uses the first 100 point-study samples in each primary cell and fits all four tuples, giving 1600 fixed-tuple fits. The selected-tuple diagnostic uses the corresponding 400 fits. The true direction is used neither for fitting nor for tuning. For $\mathbb G_k=n^{-1/2}\sum_i\eta_k(O_i)$ and $\mathbb G_{\rm eff}=n^{-1/2}\sum_iS_{\rm eff}(O_i)$, we compare the parameter predictions $J_k^{-1}\mathbb G_k$ and $\mathcal I^{-1}\mathbb G_{\rm eff}$ through
\[
 \Delta_\theta=
 \|\sqrt n\widehat\theta-J_k^{-1}\mathbb G_k\|^2
 -\|\sqrt n\widehat\theta-\mathcal I^{-1}\mathbb G_{\rm eff}\|^2.
\]
Table~\ref{tab:fixed-expansion} reports paired means and Monte Carlo standard errors. All 16 mean differences are negative, so in every combination the finite-smoothing expansion is on average closer to the estimator than the efficient limit. Table~\ref{tab:expansion} gives the same comparison at the selected tuple, together with the population efficiency ratio $\overline v_K$ of the selected tuples, which lies between $1.57$ and $2.45$ in the primary cells.
\begin{table}[htbp]
\centering\small
\caption{Paired parameter approximation error, finite smoothing minus efficient limit: mean $\Delta_\theta$ with paired Monte Carlo standard error in parentheses, on the first 100 samples of each primary cell at $n=600$. All four tuples use the same ridge and floor; negative values favor the finite-smoothing prediction.}\label{tab:fixed-expansion}
{\setlength{\tabcolsep}{6pt}
\begin{tabular}{llrrrr}
\toprule
Link & Error & $g_L,h_L$ & $g_L,h_S$ & $g_S,h_L$ & $g_S,h_S$\\
\midrule
Sine & Asymmetric & -0.432 (0.045) & -0.391 (0.053) & -0.070 (0.011) & -0.059 (0.011)\\
Sine & Multimodal & -0.838 (0.087) & -0.739 (0.120) & -0.142 (0.030) & -0.073 (0.018)\\
Cosine & Asymmetric & -0.459 (0.061) & -0.421 (0.067) & -0.066 (0.013) & -0.060 (0.012)\\
Cosine & Multimodal & -0.812 (0.095) & -0.635 (0.111) & -0.130 (0.019) & -0.068 (0.013)\\
\bottomrule
\end{tabular}}
\end{table}
\begin{table}[htbp]
\centering\small
\caption{Parameter approximation error and population efficiency at the selected tuple; $\overline v_K=N^{-1}\sum_j\operatorname{tr}(\mathcal I V_{K_j})/d$.}\label{tab:expansion}
{\setlength{\tabcolsep}{6pt}
\begin{tabular}{lrrrr}
\toprule
Cell & $n$ & $N$ & $\Delta_{\theta}$ (MCSE) & $\overline v_K$\\
\midrule
Sine / asym. & 600 & 100 & -0.091 (0.013) & 1.584\\
Sine / multi. & 600 & 100 & -0.201 (0.067) & 2.448\\
Cosine / asym. & 600 & 100 & -0.081 (0.017) & 1.569\\
Cosine / multi. & 600 & 100 & -0.106 (0.017) & 2.263\\
\bottomrule
\end{tabular}}
\end{table}

\subsection{Coverage and area of the confidence sets}
A set covers the truth when the true direction satisfies the conditions that define it, including the inference conditions, the initial cap, and the angular ball. Table~\ref{tab:conditional-confirmation} reports the coverage with Wilson intervals, the Wald coverage on the same samples, and the area of the sets.

Area is measured in projective directions, in steradians, on the first 100 replicates of each of the four primary cells. Let $F$ be a tangent frame at the point estimate $\widehat\beta$, write $\beta(v)=(\widehat\beta+Fv)/\sqrt{1+\|v\|^2}$, and put $a_n=\log(n+e)/\sqrt n$, $A_n=2\pi(1-\cos a_n)$. The spherical Jacobian is $j(v)=(1+\|v\|^2)^{-3/2}$. The proposal mixes $.8$ times a two-dimensional Student-$t_4$ density with shape matrix $\chi^2_{2,.95}F^T\widehat{\operatorname{Cov}}(\widehat\beta)F$ and $.2$ times the uniform-area density on the entire angular ball. Its planar density is therefore
\[
q(v)=.8q_t(v)+.2\,j(v)A_n^{-1}
      1\{\arctan\|v\|\le a_n\}.
\]
For independent proposal draws, the area estimate is the mean of $1\{[\beta(v)]\in\mathcal C_{\rm CS}\}j(v)/q(v)$. Every component of the set lies in the support of the proposal, and the weights are at most $5A_n$. Each dataset uses $2048$ independent importance draws, for a total of $819{,}200$ across the 400 datasets.
\begin{table}[htbp]
\centering\small
\caption{Coverage and area of the score-inversion confidence sets. Wilson intervals summarize the Monte Carlo uncertainty of the estimated coverage, and the Wald coverage is computed on the same samples. Areas, in steradians and multiplied by $10^3$, use the first 100 samples per cell.}\label{tab:conditional-confirmation}
{\setlength{\tabcolsep}{4pt}
\begin{tabular}{lrrrrrr}
\toprule
Cell & $n$ & $N$ & \shortstack{Coverage \%\\(MCSE)} & Wilson 95\% & \shortstack{Wald\\coverage \%} & \shortstack{Area $\times10^3$\\(MCSE)} \\
\midrule
Sin/A & 600 & 500 & 96.6 (0.8) & [94.6, 97.9] & 96.2 & 5.59 (0.23) \\
Sin/M & 600 & 500 & 96.4 (0.8) & [94.4, 97.7] & 98.2 & 4.13 (0.18) \\
Cos/A & 600 & 500 & 93.8 (1.1) & [91.3, 95.6] & 96.4 & 5.25 (0.18) \\
Cos/M & 600 & 500 & 95.2 (1.0) & [93.0, 96.8] & 97.0 & 4.19 (0.30) \\
\bottomrule
\end{tabular}}
\end{table}

\subsection{Benchmark data}
The Energy Efficiency data \citep{uci_energy_2012} contain 768 simulated configurations, all with complete data. Heating load is the first of the two responses. Surface area equals wall area plus twice the roof area, so we drop roof area. In source order, the retained inputs are relative compactness, surface area, wall area, overall height, orientation, glazing area and glazing-area distribution. Orientation and glazing-area distribution retain their original numeric codes. We fixed the protocol on five preliminary splits and drew all 30 reported splits before fitting any of them. None of the five preliminary partitions recurs in the reported set.

On each split, we center and symmetrically whiten the predictors and standardize the response using the $n_{\rm tr}=576$ training observations, with divisor $n_{\rm tr}$ for the predictor covariance and response variance. SI-KDR-R and mean MAVE are fitted as above. For each fitted direction, the Gaussian NW prediction bandwidth minimizes training leave-one-out squared error on $s_T n_{\rm tr}^{-1/5}\{0.25,0.4,0.6,0.8,1,1.4,2,3\}$, where $s_T$ is the training index standard deviation with divisor $n_{\rm tr}-1$. Ties go to the first bandwidth. We apply the training transformation to the 192 test predictors and return predictions to the original heating-load scale. NMSE divides test mean squared error by training response variance with divisor $n_{\rm tr}-1$. The standard deviations in Table~\ref*{main-tab:energy} describe variation across overlapping splits of this one dataset.

\FloatBarrier

\section{Distance screening in the original coordinates}\label{sec:hd-screen-proofs}
Here $P=P_n$ denotes the ambient dimension, not the empirical-average operator used in earlier sections. The link $m_0$ and the independent, mean-zero finite Gaussian-mixture error law are fixed as $P$ and the sample sizes vary. Assume $\|\beta_0\|=1$, $s=|S_0|\le s_{\max}\le q$, $q=q_n\ge2$, $P\ge q$, and
\begin{equation}\label{hd:spectrum}
 c_\Sigma I_P\preceq\Sigma_n\preceq C_\Sigma I_P,
 \qquad 0<c_\Sigma\le C_\Sigma<\infty.
\end{equation}
The link has the boundedness and smoothness imposed in the fixed-dimensional theory. Only the independent screening sample $\mathcal D_s$ is used in this section. These conditions, along with the rank condition below, are used only for the screening extension.

\subsection{The statistic and its normalization}
For scalar random variables with the moments used below, write $\mathcal D(U,V)=\operatorname{dcov}^2(U,V)$. With independent copies of the paired observation,
\[
\mathcal D(U,V)=E|U-U'||V-V'|+E|U-U'|E|V-V'|
 -2E|U-U'||V-V''|.
\]
Thus $u_j=\mathcal D(X_j,Y)/\{\mathcal D(X_j,X_j)\mathcal D(Y,Y)\}^{1/2}$ is the \emph{squared} distance correlation. Finite first moments suffice for the characteristic-function definition and its independence characterization \citep{szekely2007distance}; all displayed moment products are finite here.

Let $N=n_s$, $a^{(j)}_{il}=|X_{ij}-X_{lj}|$, $b_{il}=|Y_i-Y_l|$, and double-center each distance matrix, including its diagonal:
\[
 A^{(j)}_{il}=a^{(j)}_{il}-\bar a^{(j)}_{i\cdot}-\bar a^{(j)}_{\cdot l}+\bar a^{(j)}_{\cdot\cdot},
 \qquad B_{il}=b_{il}-\bar b_{i\cdot}-\bar b_{\cdot l}+\bar b_{\cdot\cdot}.
\]
All row means use the divisor $N$, and the grand mean uses the divisor $N^2$. Define
\begin{equation}\label{hd:screen-stat}
 \widehat u_j=
 \frac{N^{-2}\sum_{i,l}A^{(j)}_{il}B_{il}}
 {\{(N^{-2}\sum_{i,l}(A^{(j)}_{il})^2)
       (N^{-2}\sum_{i,l}B_{il}^2)\}^{1/2}},
\end{equation}
with value zero if its denominator is zero. This is the biased, ordinary double-centered statistic. Expanding its numerator gives
\[
 N^{-2}\sum_{i,l}a^{(j)}_{il}b_{il}
 +\bar a^{(j)}_{\cdot\cdot}\bar b_{\cdot\cdot}
 -2N^{-3}\sum_{i,l,r}a^{(j)}_{ir}b_{lr}.
\]
It therefore agrees exactly with $\widehat\omega_j$ in Section~2.2 of \citet{li2012distance}, including the square and the diagonal convention. Order coordinates by decreasing $\widehat u_j$, breaking ties by increasing original coordinate number, and retain the first $q$. This defines $\widehat S$ even on exceptional samples.

\subsection{Visibility and a nonempty separated class}
\begin{lemma}[Gaussian marginal visibility]\label{hd:visibility}
Let $\tau^2=\beta_0^T\Sigma_n\beta_0$ and
\[
 r_j=\frac{(\Sigma_n\beta_0)_j}{\sqrt{\Sigma_{n,jj}}\,\tau}.
\]
Then $u_j>0$ if $r_j\ne0$, and $u_j=0$ if $r_j=0$. In particular, when $\Sigma_n=I_P$, $u_j>0$ exactly for $j\in S_0$. For a fixed nonconstant $m_0$ and fixed error law, the independent-design subclass with $|\beta_{0j}|\ge b_0>0$ on $S_0$ has a uniform positive active-score gap, provided it is nonempty.
\end{lemma}
\begin{proof}
Put $T=\beta_0^TX/\tau$ and $Z=X_j/\sqrt{\Sigma_{n,jj}}$. They are standard normal with correlation $r_j$. Expand the bounded function $m_0(\tau t)$ in the orthonormal Hermite basis: $m_0(\tau t)=\sum_{k\ge0}c_k H_k(t)$ in $L^2(\phi)$. Some $c_k$, $k\ge1$, is nonzero: otherwise continuity and the full support of the normal distribution would make $m_0$ constant. The Gaussian generating function gives $E\{H_k(T)\mid Z\}=r_j^kH_k(Z)$; taking $L^2$ limits yields
\[
 E(Y\mid Z)=\sum_{k\ge0}c_k r_j^k H_k(Z).
\]
When $r_j\ne0$ this conditional mean is nonconstant, so $Y$ and $Z$ are dependent. When $r_j=0$, Gaussian independence gives $Z\perp(T,\varepsilon)$ and hence $Z\perp Y$. The distance-correlation characterization gives the assertion.

For $\Sigma_n=I_P$, $\tau=1$ and $r_j=\beta_{0j}$. The law of $(X_j,Y)$ depends on $j,\beta_0,P$ only through $r_j$: couple it as
\[
 Z\sim N(0,1),\quad T=rZ+\sqrt{1-r^2}G,
 \quad Y=m_0(T)+\varepsilon,
\]
with independent $Z,G,\varepsilon$. All distance moments are continuous in $r\in[-1,1]$ by this coupling, boundedness of $m_0$, and uniform integrability. Their denominators are strictly positive. Thus $u(r)$ is continuous and positive on the compact set $b_0\le |r|\le1$, and has a positive minimum $\Delta_0$. Every inactive score is zero. For example, $s\le s_{\max}$, equal nonzero coefficients $1/\sqrt{s}$, and $0<b_0\le1/\sqrt{s_{\max}}$ give a nonempty subclass.
\end{proof}

\subsection{Uniform concentration and fixed-size retention}
\begin{proposition}[Instantiated DC-SIS concentration]\label{hd:screen-uniform}
Fix $A>0$, $0\le\nu_1<1/2$, $b_s=A n_s^{-\nu_1}$ and $0<\nu_2<1/2-\nu_1$. Under the fixed link and error assumptions and \eqref{hd:spectrum}, constants $C_0,c_1,c_2>0$ and $N_0<\infty$, independent of $P$ and the coordinate, exist such that for $n_s\ge N_0$,
\begin{align}
 &\Pp\left(\max_{j\le P}|\widehat u_j-u_j|>b_s\right)
 \le \delta_{\rm scr},\nonumber\\
 &\delta_{\rm scr}:=\min\left\{1,C_0P\left[
 e^{-c_1n_s^{1-2(\nu_1+\nu_2)}}+n_s e^{-c_2n_s^{\nu_2}}\right]\right\}.
 \label{hd:screen-delta}
\end{align}
The constants depend only on the stated model bounds and on $A,\nu_1,\nu_2$.
\end{proposition}
\begin{proof}
We apply the first assertion, equation~(2.6), of Theorem~1 of \citet{li2012distance}; its threshold-screening assertion is not used. The statistic matches by \eqref{hd:screen-stat}. Their condition (C1) requires uniform exponential moments of the \emph{squared} marginal predictors and response. These hold here. Indeed,
\[
 \sup_{n,j}E e^{tX_j^2}\le(1-2tC_\Sigma)^{-1/2},\qquad
 E e^{tY^2}\le e^{2t\|m_0\|_\infty^2}E e^{2t\varepsilon^2}.
\]
A fixed finite Gaussian mixture has the latter moment for all sufficiently small positive $t$, since every component variance is finite. Choose the same sufficiently small $t$ for all $n,j$. These estimates also uniformly bound the moment products in their proof, without requiring independence among predictor coordinates.

The normalization keeps the constants in that theorem uniform. Put $d_Z=\mathcal D(Z,Z)>0$ for $Z\sim N(0,1)$. Scaling gives $\mathcal D(X_j,X_j)=\Sigma_{n,jj}d_Z\ge c_\Sigma d_Z$. The response has the law $Y_\tau=m_0(\tau T)+\varepsilon$ with $\tau\in[\sqrt{c_\Sigma},\sqrt{C_\Sigma}]$. Its distance variance is positive for every $\tau$, since convolution with a positive-variance Gaussian-mixture error makes it nondegenerate. It is continuous in $\tau$, by dominated distance moments in the common coupling. Compactness therefore gives $\inf_\tau\mathcal D(Y_\tau,Y_\tau)>0$. The corresponding upper bounds are uniform as well.

The denominator enters the cited concentration proof as follows. For each of the three pairs $(X_j,Y)$, $(X_j,X_j)$ and $(Y,Y)$, its Appendix~B distance-covariance argument bounds the deviation at level $t$ by
\[
 C\{e^{-c t^2 n_s^{1-2\nu_2}}+n_s e^{-c'n_s^{\nu_2}}\}
 \quad\text{when }t\text{ is a fixed multiple of }n_s^{-\nu_1}.
\]
This argument decomposes an ordinary distance statistic into order-2 and order-3 $U$-statistics, plus terms with repeated indices. Truncation at a constant multiple of $n_s^{\nu_2}$ gives the stated bound, and the repeated-index biases are $O(n_s^{-1})$. Hence the bound applies to the biased statistic itself. If the two estimated distance variances stay within half of their common positive lower bound, the ratio map $(a,b,c)\mapsto a/\sqrt{bc}$ is uniformly Lipschitz on the domain given by the bounded population ranges and their sufficiently small neighborhoods. With $t$ a sufficiently small constant multiple of $b_s$, a union bound over these three deviations bounds the ratio deviation by $b_s$. The probability of the complementary event is bounded by the sum of the three deviation probabilities. This gives the uniform denominator condition. Finally, by taking the union over all $P$ coordinates, we obtain \eqref{hd:screen-delta}. All constants and the minimum sample size are determined by the uniform exponential-moment, denominator and moment bounds established above.
\end{proof}

For $r=q-s+1$, let $v_r$ be the $r$th largest inactive population score, with $v_r=-\infty$ if fewer than $r$ inactive coordinates exist. The latter case, under $P\ge q$, can occur only when $P=q$, for which every variable is retained. Put $u_*=\min_{j\in S_0}u_j$.
\begin{theorem}[Top-$q$ sure screening]\label{hd:screen-rank}
If $P>q$ and
\begin{equation}\label{hd:screen-gap}
 u_*-v_{q-s+1}\ge4b_s,
\end{equation}
then $|\widehat S|=q$ and
\[
 \left\{\max_{j\le P}|\widehat u_j-u_j|\le b_s\right\}
 \subseteq E_{\rm scr}:=\{S_0\subseteq\widehat S\},
 \qquad \Pp(E_{\rm scr}^c)\le\delta_{\rm scr}.
\]
For $P=q$, $E_{\rm scr}$ is certain and one may set $\delta_{\rm scr}=0$. For $q=s<P$, condition \eqref{hd:screen-gap} compares the weakest active coordinate to the strongest inactive coordinate, and gives exact support recovery.
\end{theorem}
\begin{proof}
On the uniform-error event, every active empirical score is at least $u_*-b_s$. Except for at most $q-s$ inactive coordinates, every inactive empirical score is at most $v_{q-s+1}+b_s\le u_*-3b_s$. Thus, any active coordinate can be preceded by at most $s-1$ other active coordinates and $q-s$ exceptional inactive coordinates, for a total of $q-1$. Its rank is at most $q$. The positive separation $2b_s$ from all other inactive coordinates makes the conclusion unaffected by ties among exceptional or active coordinates. The deterministic tie rule always returns exactly $q$ coordinates. Proposition~\ref{hd:screen-uniform} gives the failure bound.
\end{proof}

Taking $\nu_2=(1-2\nu_1)/3$ makes $\delta_{\rm scr}\to0$ when
\begin{equation}\label{hd:ambient-growth}
 \log P=o\{n_s^{(1-2\nu_1)/3}\}.
\end{equation}
This permits $P>n_{\rm tot}$ and suitable subexponential dimensions; retention at finite sample sizes is examined in Section~\ref{sec:hd-protocol}.

\section{Retained coordinates and the fixed-dimensional specialization}
\label{hd:inheritance-proofs}
The submodel and coordinate identity below hold for every admissible retained dimension $q$. The sharper bandwidth bounds and joint Gaussian approximation later in this section concern fixed $q$; Section~\ref{hdg:exp-section} tracks growing $q$. All empirical averages below use the estimation sample $D_e$ of size $n_e$. The symbol $P$ for ambient dimension is unrelated to the empirical-average notation $P_n$ in the earlier sections.

\subsection{A uniform model class and the retained submodel}
\label{hd:uniform-assumptions}
For the screening extension, fix one nonconstant bounded $C^6$ function $m_0$ with bounded derivatives through order six and one centered, nondegenerate finite Gaussian-mixture error law. These two objects do not vary with $n$ or the retained set. Let $\|\beta_0\|=1$, $|S_0|\le s_{\max}\le q$, and assume
\begin{equation}\label{hd:inherit-spectrum}
 cI_P\preceq\Sigma_n\preceq CI_P,\qquad 0<c<C<\infty.
\end{equation}
The screening rank-gap assumption is imposed separately, and all these conditions are used only for the extension. A fixed finite collection of such links and error laws is also allowed by taking the smallest of the finitely many positive constants established below.

Write $\mathcal S_n=\{S:S_0\subseteq S,\ |S|=q\}$ and let $\iota_S:\R^q\to\R^P$ insert zeros outside $S$. For a deterministic $S\in\mathcal S_n$, put $\Sigma_S=\Sigma_{n,SS}$, $\beta_S=\beta_{0,S}$ and
\[
 v_n=\beta_0^T\Sigma_n\beta_0=\beta_S^T\Sigma_S\beta_S\in[c,C],\qquad
 A_S=\Sigma_S^{-1/2},\qquad
 \gamma_S=\Sigma_S^{1/2}\beta_S/\sqrt{v_n}.
\]
Choose orthonormal frames $F_S\perp\beta_S$ and $U_S\perp\gamma_S$, each with $d=q-1$ columns. In population-standardized coordinates,
\begin{equation}\label{hd:scalar-family}
 T=\gamma_S^TA_SX_S,\quad W=U_S^TA_SX_S,
 \quad Y^w=m_v(T)+\varepsilon_v,
 \quad m_v(t)=\frac{m_0(\sqrt v\,t)-\mu_v}{\sigma_v},
\end{equation}
where $v=v_n$, $\mu_v=E m_0(\sqrt v\,T)$, $\sigma_v^2=\Var\{m_0(\sqrt v\,T)\}+\Var(\varepsilon)$, and $\varepsilon_v=\varepsilon/\sigma_v$. Here $T\sim N(0,1)$, $W\sim N(0,I_d)$ and $W\perp(T,Y^w)$. For fixed $n$, the scalar law in \eqref{hd:scalar-family} is the same for every $S\in\mathcal S_n$.

\begin{lemma}[Correct submodel and coordinate recovery]\label{hd:submodel}
Conditional on $D_s$, on $E_{\rm scr}=\{S_0\subseteq\widehat S\}$ the observations $(X_{i,S},Y_i)$, $i\in D_e$, with $S=\widehat S$, are iid from
\[
 Y=m_0(\beta_S^TX_S)+\varepsilon,\quad
 X_S\sim N_q(0,\Sigma_S),\quad \varepsilon\perp X_S,
 \quad \|\beta_S\|=1.
\]
Whitening and response standardization must be recomputed using only $D_e$. For any nonsingular realized whitening matrix $A$, the standardized target is $\gamma_S(A)=A^{-T}\beta_S/\|A^{-T}\beta_S\|$, and
\begin{equation}\label{hd:map-cancellation}
 \frac{A^T\gamma_S(A)}{\|A^T\gamma_S(A)\|}=\beta_S.
\end{equation}
The derivative from population-standardized tangent coordinates to raw orthonormal tangent coordinates is
\begin{equation}\label{hd:raw-map}
 D_{0,S}=\sqrt{v_n}\,F_S^T\Sigma_S^{-1/2}U_S,
 \quad \|D_{0,S}\|+\|D_{0,S}^{-1}\|\le2\sqrt{C/c}.
\end{equation}
Zero insertion preserves projective distance and maps the retained raw tangent space into $\{h\in\R^P:\supp(h)\subseteq S,\ h^T\beta_0=0\}$.
\end{lemma}
\begin{proof}
Independent splitting makes $D_e$ independent of the selected set. On screening success, every deleted coefficient is zero, so the displayed regression identity is exact. Principal-submatrix eigenvalues remain in $[c,C]$. Normalization gives \eqref{hd:map-cancellation}; differentiating the normalized map at $A=A_S$ gives \eqref{hd:raw-map}. Its inverse is $v_n^{-1/2}U_S^T\Sigma_S^{1/2}F_S$, as follows by differentiating the inverse normalized map. Both norms are at most $\sqrt{C/c}$. Finally, $\iota_S$ is a Euclidean isometry and preserves both signs of a direction. The centered sample covariance is positive definite almost surely for Gaussian data when $n_e-1\ge q$. The stricter condition $q<n_e-1$ is imposed for the positive residual degrees of freedom in the candidate $F$ test. A continuous nondegenerate error also makes the sample response variance positive almost surely. Neither step uses an ambient covariance estimate; thus $P>n_{\rm tot}$ is permitted. The lemma requires screening success.
\end{proof}

\subsection{Uniform constants, initialization, and stochastic bounds}
For a remainder $R_{n,S}$, write $R_{n,S}=O_{P,u}(a_n)$ if
\[
 \lim_{M\to\infty}\limsup_{n\to\infty}
 \sup_{S\in\mathcal S_n}
 P_{e,S}(\|R_{n,S}\|>Ma_n)=0,
\]
where $P_{e,S}$ is the law of the estimation sample in that retained submodel. The bound is uniform over the individual submodel probabilities and involves no event simultaneous over all subsets.

\begin{lemma}[Uniform inheritance]\label{hd:uniform-inheritance}\label{hd:uniform}
For fixed $q\ge2$, under \eqref{hd:inherit-spectrum} and the fixed link and error conditions above, the moment, Gaussian-tail, mixture-envelope, preprocessing, and Fourier initialization bounds in Sections~\ref{supp:fourier}, \ref{sec:preprocessing}, and \ref{inf:theory} have constants independent of $n$, $P$, and $S\in\mathcal S_n$. The efficient information in raw tangent coordinates has eigenvalues bounded away from zero and infinity. The four-matrix Fourier initial estimator, with spatial constants $1/q,1,3,9$ and response constant one, is uniformly root-$n_e$ consistent.

For a fixed finite grid of positive smoothing tuples, require additionally that its population standardized curvatures, including all possible preliminary and inference tuples, satisfy
\begin{equation}\label{hd:positive-curvature}
 \inf_{v\in[c,C],\,k}j_{v,k}>0.
\end{equation}
Such grids exist. For a fixed-cardinality shrinking grid, \eqref{hd:positive-curvature} holds eventually if every tuple tends to zero. Under the rate conditions of Theorem~\ref{dir:theorem} and Theorem~\ref{inf:caseweights}, with $n=n_e$, all their probability remainders are uniform in the preceding sense. Their simultaneous conclusions cover the fixed finite preliminary, final, and candidate-inference grids.
\end{lemma}
\begin{proof}
First, $\Var(\varepsilon)\le\sigma_v^2\le\Var(\varepsilon)+\|m_0\|_\infty^2$, and $\mu_v,\sigma_v$ depend continuously on $v\in[c,C]$. For $1\le j\le6$, $m_v^{(j)}$ is bounded uniformly by $C^{j/2}\|m_0^{(j)}\|_\infty/\sqrt{\Var(\varepsilon)}$, and $\|m_v\|_\infty\le2\|m_0\|_\infty/\sqrt{\Var(\varepsilon)}$. The standardized mixture has the same fixed component weights, bounded component means, and component variances in a common positive compact interval. Hence all fixed-order moments and exponential-square tail bounds used in the earlier proofs are uniform. The residual density remains a bounded-location mixture with these same variance bounds. The constants in \eqref{dir:mixture}, \eqref{dir:aenv}, and the Gaussian profile-derivative bounds can therefore be chosen uniformly. The integration-by-parts representation \eqref{dir:kappaibp} supplies a common polynomial envelope for $\kappa$ and its influence, rather than a divergent inverse-bandwidth bound.

For information, the standardized scalar coefficient equals
\begin{equation}\label{hd:info-coefficient}
 i(v)=v E\{m_0'(\sqrt v\,T)^2\}I_{\varepsilon},
 \qquad I_{\varepsilon}=E\!\left[\left\{\frac{f_\varepsilon'(\varepsilon)}{f_\varepsilon(\varepsilon)}\right\}^2\right].
\end{equation}
Here $f_\varepsilon$ is the error density in the original response units. The coefficient $i(v)$ is finite, positive, and continuous in $v$. Positivity follows because a nonconstant continuously differentiable function has derivative nonzero on some open interval and every Gaussian density is positive there; $I_\varepsilon>0$ for a nonconstant integrable mixture density. Compactness gives $0<i_-\le i(v)\le i_+<\infty$. The raw information is $D_{0,S}^{-T}\{i(v_n)I_d\}D_{0,S}^{-1}$, so \eqref{hd:raw-map} gives the claimed eigenvalue bounds.

For each of the four fixed Fourier spatial constants, Proposition~\ref{fourier:rankprop} applies to every scalar law \eqref{hd:scalar-family}, giving a strictly positive rank-one eigenvalue $\lambda_q(v)$. This eigenvalue is continuous in $v$: couple the laws through common $(T,W,\varepsilon)$ and use the polynomial predictor envelope in the proof of Proposition~\ref{fourier:rateprop}; the response factor is bounded by one. Thus $\min_{v\in[c,C]}\lambda_q(v)>0$, also for their positive average. The matrix mean-square bound, affine-map derivative envelope, and response-standardization perturbation in Section~\ref{sec:preprocessing} all have common constants. Eigenvector perturbation therefore gives uniformly $\dP(\widetilde\beta_S,\beta_S)=O_{P,u}(n_e^{-1/2})$; the argument uses that $q$ is fixed.

The population convergence \eqref{dir:poplimit} is uniform in $v$. To verify this, consider any sequence $v_l\in[c,C]$ and any tuning tuples tending to zero. A subsequence has $v_l\to v_*$. In the coupled scalar representation, the bounded $C^6$ link derivatives converge on every compact index interval, the Gaussian-mixture densities and their derivatives converge on compact residual intervals, and the common polynomial envelopes above control the complements by Gaussian tails. The proof of \eqref{dir:poplimit}, including substitution $u=t+gZ$ in \eqref{dir:kappaibp}, then gives the same weighted-moment convergence along this subsequence. A contradicting sequence for uniform convergence is therefore impossible. Uniform positive information implies eventual positive curvature. It also gives a nonempty sufficiently small neighborhood of positive fixed tuples satisfying \eqref{hd:positive-curvature}. The information-projection identity \eqref{inf:lossidentity} and the positive curvature gap also bound the score covariance away from zero, uniformly on any such grid.

For the sample results, uniformity is read off from their proofs. There, the deterministic empirical-integral inequalities are independent of the model; Gaussian conditional formulas use bounded covariance eigenvalues; and the Bernstein and moment inequalities use exactly the common envelopes just established. After conditioning on a fixed $S$, all predictor maxima involve only $q$ coordinates, and the preprocessing and local-direction nets have fixed dimension and log cardinality $O(\ell_{n_e})$. No maximum over the ambient $P$ coordinates is needed. Inverse-square-root Fr\'echet derivatives are bounded on the compact covariance set by the Sylvester formula \eqref{eq:joint-frechet}; the common lower bound on $\sigma_v$ controls response standardization. The proofs of \eqref{dir:jetsrate}, \eqref{dir:score}, \eqref{dir:hessian}, and Theorem~\ref{inf:caseweights} consequently have the same deterministic rates and exceptional-probability constants for each $S$. The Armijo argument uses only the common curvature bounds and the uniform pilot rate, so its localization and optimization failure probabilities also vanish uniformly. A union over a fixed number of tuning tuples is harmless.

Finally, conditional independence removes the need for a union over subsets. If every submodel's bad event has probability at most $r_n$, then
\[
 P(E_{\rm scr}\cap\mathrm{bad})
 =E\{1_{E_{\rm scr}}P_{e,\widehat S}(\mathrm{bad})\}
 \le\sup_{S\in\mathcal S_n}P_{e,S}(\mathrm{bad})\le r_n.
\]
In particular, no factor $\binom Pq$ arises.
\end{proof}

\subsection{Expansion, covariance, selection, and ambient error}
Let $D_{S,e}$ and $D_{H,e}$ denote the score and Hessian rates \eqref{dir:DS}--\eqref{dir:DH} with $n=n_e$, and let $q_{D,e}$ be \eqref{up:ij-q} with the same substitution. Here, the subscript $S$ in $D_{S,e}$ denotes the score. The auxiliary threshold is $t_H=n_e^{-\nu_H}$. In particular,
\[
 \sigma_{j,e}=(n_eg^{2j+1})^{-1/2},\quad
 \omega_{j,e}=(n_eg^{j+1}\sqrt\lambda)^{-1},\quad
 r_e=c_rn_e^{-\nu_r},\quad 0<\nu_r<1/2.
\]
All tuning scales, ridge, floor, caps, iteration counts and tolerances use $n_e$.

\begin{theorem}[Screened SI-KDR expansion and covariance]\label{hd:stage-two}
Assume Lemma~\ref{hd:uniform-inheritance}, $n_eg\lambda/\ell_{n_e}^{12}\to\infty$, $D_{S,e}+D_{H,e}\to0$, and $\tau_e=o(n_e^{-1/2})$. Conditional on $D_s$, on $E_{\rm scr}$, in a raw orthonormal tangent chart at $\beta_S$,
\begin{equation}\label{hd:expansion}
 \sqrt{n_e}\widehat\theta_{S,k}
 =J_{S,k}^{-1}\frac1{\sqrt{n_e}}\sum_{i\in D_e}\eta_{S,k}(O_{i,S})
 +O_{P,u}(D_{S,e}+D_{H,e}+\sqrt{n_e}\tau_e).
\end{equation}
The simultaneous expansion includes all preliminary and final tuples. In this display
\begin{align}\label{hd:influence-map}
 \eta_{S,k}&=D_{0,S}^{-T}W\{\kappa_{v_n,k}(T,Y^w)-C_{v_n,k}T\},\notag\\
 J_{S,k}&=D_{0,S}^{-T}(j_{v_n,k}I_d)D_{0,S}^{-1},\quad
 B_{S,k}=\Var(\eta_{S,k}),\quad
 V_{S,k}=J_{S,k}^{-1}B_{S,k}J_{S,k}^{-T}.
\end{align}
The functions $\kappa$ and $C$ are exactly \eqref{inf:kappa}--\eqref{inf:whiteningprojection}, evaluated in the retained scalar model. In particular, the incoming NW refit, both residual-density roles, and the second-stage training-whitening correction are retained.

If also $n_eht_D/\ell_{n_e}^{32}\to\infty$ and $q_{D,e}\to0$, the case-weight procedure of the main text, differentiating only the estimation sample and holding $S$ and the chosen tuple fixed, gives
\begin{align}\label{hd:covariance}
 \widehat B_{S,k}-B_{S,k}&=O_{P,u}(q_{D,e}+q_{D,e}^2+n_e^{-1/2}),\notag\\
 \widehat J_{S,k}-J_{S,k}&=O_{P,u}(D_{H,e}),\qquad
 \widehat V_{S,k}-V_{S,k}=o_{P,u}(1).
\end{align}
The parameter covariance approximation is $V_{S,k}/n_e$, estimated by $\widehat V_{S,k}/n_e$.

On a shrinking finite grid that satisfies these rates, every data-dependent point-estimation selector, including the variance-trace rule, has the same efficient expansion in its retained submodel. In matching tangent frames,
\begin{equation}\label{hd:studentized-limit}
 \widehat V_{S,\widehat k}^{-1/2}\sqrt{n_e}\widehat\theta_{S,\widehat k}
 \quad\hbox{has a uniformly valid }N_d(0,I_d)\hbox{ approximation},
\end{equation}
conditional on $D_s$ and screening success; the approximation is indexed by $S$.
\end{theorem}
\begin{proof}
Using uniform score and Hessian bounds at the true raw direction, a mean-value expansion is applied to an approximate score root. A uniform positive-curvature bound and Fourier/Armijo localization then give \eqref{hd:expansion}. Here $D_{S,e}$, $D_{H,e}$ and the optimization remainder are as in Theorem~\ref{dir:theorem}. The moving-target identity \eqref{hd:map-cancellation} cancels the target shift that would otherwise arise from reversing the estimated whitening. The first-order empirical whitening component of the score is the term $-CTW$. Differentiating the full estimation-sample empirical measure gives the case-weight equations of Section~\ref{up:caseweights}. Then \eqref{hd:covariance} follows from Lemma~\ref{hd:uniform-inheritance} and smooth matrix inversion. Since the calculations condition on the independent screening sample, the screening stage does not affect the derivative.

On a shrinking grid, given the uniform $L^2$ limit established above, the difference between the sum of the individual influences for each tuple and the efficient influence sum is $o_{P,u}(1)$. Since the conditional means are zero, this follows from the variance of the difference. The same applies to their curvatures and covariance terms. The efficient influences have uniformly bounded third moments, and their information is uniformly nondegenerate. Hence the fixed-dimensional triangular-array central limit theorem holds uniformly. This follows, for example, by bounding the Taylor remainder of the characteristic function by the third moment and using tightness and subsequence arguments. Uniform covariance consistency and Slutsky's argument give \eqref{hd:studentized-limit}. Since the tangent-frame changes between the true values and the computed roots are $o_{P,u}(1)$, the conclusion is unchanged.
\end{proof}

Finally, write $\widehat\beta^{\rm amb}=\iota_{\widehat S}\widehat\beta_{\widehat S}$, using any fixed unit fallback on an estimation failure. By isometry and uniform localization, for every $M>0$,
\begin{equation}\label{hd:ambient-risk}
 P\{\sqrt{n_e}\dP(\widehat\beta^{\rm amb},\beta_0)>M\}
 \le\delta_{\rm scr}
 +\sup_{S\in\mathcal S_n}P_{e,S}\{\sqrt{n_e}\dP(\widehat\beta_S,\beta_S)>M\}.
\end{equation}
The second term tends to zero as first $n\to\infty$ and then $M\to\infty$. Hence $\dP(\widehat\beta^{\rm amb},\beta_0)=O_P(n_e^{-1/2})$ when $\delta_{\rm scr}\to0$.

\section{An envelope bound for other positive smoothing grids}
\label{hdg:exp-section}
Write $n=n_e$, $d=q-1$, and condition on $D_s$ with $S_0\subseteq S$.
Use the fixed bounded $C^6$ link, fixed nondegenerate Gaussian mixture, and
retained spectral bounds in \eqref{hd:inherit-spectrum}. All probabilities
below are uniform over individual correct retained models, as in
Section~\ref{hd:inheritance-proofs}. The cardinality of the union of all
preliminary, point and inference grids is bounded. Each component of a tuple
is strictly positive and bounded above by a fixed constant. Define
\begin{align}
 e_n&=\sqrt{\log(n+1)/n},\qquad
 L_n=\left(2+\sqrt{\log(n+1)}+
 \max_{k}\{g^{-1},h^{-1},\lambda^{-1},\rho^{-1}\}\right)^{100},\notag\\
 H_n(u)&=qL_n(e_n+\sqrt q\,u),\qquad
 A_n(u)=\sqrt q\,L_n(e_n+\sqrt q\,u).\label{hdg:rates}
\end{align}
The powers give one sufficient range. They come from protected denominator
bounds throughout the sample, including its tails. Section~\ref{vtwo:section} treats the implemented shrinking grid by integration.

Use coordinates $x=(T,W)$ and response $Y^w$ in \eqref{hd:scalar-family}.
In the proof the chart coordinate $b$ represents $e_1+(0,b)$ up to
normalization. Its fitted scalar index under a probability $Q$ is exactly
\begin{equation}
 t_b(z)=\frac{T+b^TW-Q(T+b^TW)}{s_b},\qquad
 s_b^2=Q\{T+b^TW-Q(T+b^TW)\}^2 .\label{hdg:chart}
\end{equation}
Use $y=(Y^w-QY^w)/s_Y$ similarly. Thus population whitening here is a
coordinate system for analyzing the training-standardized criterion.
No population transformation enters its implementation. Denote by
$\widehat\eta_i(b)$ the case-weight derivative of the score in this chart,
by $\widehat B(b)$ its empirical second moment, and by $\widehat J(b)$
the negative Hessian. Write $f_k=\kappa_k-C_kT$ as in
\eqref{inf:kappa}--\eqref{inf:whiteningprojection}.

\begin{theorem}[Sample derivatives and the fitted direction]
\label{hdg:sample}
Suppose $q<n-1$, $R\sqrt q=O(1)$, $H_n(R)=o(1)$, and
$\inf_{v,k}j_{v,k}>0$ on the possible tuples. Uniformly over this grid,
\begin{align}
 \sqrt n\,S_{n,k}(0)&=n^{-1/2}\sum_i W_i f_k(T_i,Y_i^w)
               +O_P(\sqrt q\,L_ne_n),\label{hdg:score}\\
 \sup_{\|b\|\le u}\|\widehat J_k(b)-j_{v,k}I_d\|_{\rm op}
   &=O_P\{H_n(u)\},\notag\\
 \sup_{\|b\|\le u}\max_i\|\widehat\eta_{i,k}(b)-W_i f_k(T_i,Y_i^w)\|
   &=O_P\{A_n(u)\},\qquad 0\le u\le R.\label{hdg:matrix}
\end{align}
The same events give these bounds at all such radii. The hypotheses also
hold on a fixed multiple of $R$, with altered constants. Put $B_k=b_{v,k}I_d$
with $b_{v,k}=E f_k^2$, and $V_k=j_{v,k}^{-2}B_k$. In these coordinates
\begin{align}
 \sup_{\|b\|\le u}\|\widehat V_k(b)-V_k\|_{\rm op}
 &=O_P\{H_n(u)+q/\sqrt n+A_n(u)+A_n(u)^2\} .\label{hdg:sandwich}
\end{align}
In a raw orthonormal tangent frame these targets become
$D^{-T}W_i f_k$, $D^{-T}(j_{v,k}I_d)D^{-1}$,
$D^{-T}B_kD^{-1}$, and $D V_kD^T$, respectively, where $D=D_{0,S}$
is given by \eqref{hd:raw-map}. The score expansion and the Hessian and
sandwich bounds retain the displayed orders in matching raw frames.

Suppose a finite fit has score tolerance $\tau=o(n^{-1/2})$ in this cap,
$\sqrt{q/n}=o(R)$, and the initialization and ascent conditions below hold.
For $u_n=\sqrt{q/n}$ its fitted raw chart coordinate obeys
\begin{equation}
 \sqrt n\,\widehat\theta_k=D\,j_{v,k}^{-1}n^{-1/2}
    \sum_i W_i f_k(T_i,Y_i^w)+O_P\{\sqrt q\,L_ne_n+
      \sqrt q\,H_n(u_n)+\sqrt n\,\tau\}.\label{hdg:estimate}
\end{equation}
In particular, $\dP(\widehat\beta_k,\beta_S)=O_P(\sqrt{q/n})$.
The covariance coefficient of the linear term is $D V_kD^T$ and its
contribution on the estimator scale is $D V_kD^T/n$.
\end{theorem}

\begin{proof}
We give details of the empirical comparison since its dimension cost
differs for the score, curvature and the case-weight covariance. Write
$Q_n$ for the empirical law and $P$ for the retained population law.
For an observation $z$ with a deletion mark $s(z)$ define the row integral
\[
 {\cal A}_z F=\{Q_{z'} F(z,z')-s(z)F(z,z)\}/\{1-s(z)\}.
\]
On the empirical support take $s_i=\bar w_i=w_i/\sum_l w_l$. The entire
objective is the following sequence of scalar operations:
\begin{align*}
 D_g(z)&={\cal A}_z K_g(t-t')+\lambda,&
 M(z)&={\cal A}_z\{K_g(t-t')y'\}/D_g(z),\\
 r(z)&=y-M(z),& D_h(z)&={\cal A}_z K_h(r-r')+\rho,\qquad
 L=Q_z\log D_h(z).
\end{align*}
At $P$ put $s=0$. At unit empirical weights the unprojected row coefficient
obtained by differentiation and exchanging summation indices is
\begin{align*}
 a(z)&={\cal A}_z[K_h'(r-r')\{D_h(z)^{-1}+D_h(z')^{-1}\}],\\
 B_g(z,z')&=K_g'(t-t')\{y'-M(z)\}/D_g(z),\qquad
 A(z)={\cal A}_z B_g(z,z'),\\
 c(z)&=-a(z)A(z)+{\cal A}_z\{a(z')B_g(z',z)\},\\
 d(z)&=s_b^{-1}\{c(z)-t_b(z)Q(t_bc)\}.
\end{align*}
The common mark $s=1/n$ makes the same row operator valid on interchanging
the indices. The case-weight calculation below differentiates $L$ itself,
where every mark varies with its own normalized weight. At truth the score
is $Q_n(Wd)$ since $Q_nc=0$ by index translation. The population $d$ is
$f_k$ by \eqref{inf:kappa}. Both density roles and the incoming NW refit
appear in these equations. Differentiating $s_b$ in \eqref{hdg:chart}
gives the subtraction $-t_bQ(t_bc)$.

\emph{Envelopes.} Let $H=L_n^{1/100}$. With probability tending to one
all scalar predictor coordinates and responses have absolute value at most
$C_0\sqrt{\log(n+1)}$, the sample predictor covariance has eigenvalues
in $[1/2,2]$ in population-standardized units and the sample response
variance lies in $[1/2,2]$. For the covariance event, on a $1/4$-net of
at most $9^q$ unit directions the Gaussian mean-square tails are
$2\exp(-c n t^2)$ for $0<t<1$ by the chi-square moment generating
function. The net and the centered Gaussian sample mean give an operator
error $O_P(\sqrt{q/n})$ (also $O_P(n^{-1/2})$ when $q$ is bounded).
The response assertion follows from its uniform fourth moment. The
coordinate maximum bound uses at most $n(q+2)$ uniform sub-Gaussian
tails. On this event $|T+b^TW|\le C H$ throughout $\|b\|\le R$.
Thus the bound holds for the whole cap without a directional mesh.

The following table bounds all required coordinate derivatives. A pair
$(a,b_*)$ means that an order-$l$ mixed derivative is bounded by
$C_lH^{a+b_*l}$ for $0\le l\le4$. Allow any coordinate derivatives
of the chart and one mass derivative in direction $\delta_{z_i}-Q$.
At empirical rows a deletion parameter may also be varied. Write
$K^{(j)}$ for the $j$th derivative in its scalar argument, $j=0,1$.
\begin{center}
\begin{tabular}{lc@{\qquad}lc}
quantity & $(a,b_*)$ & quantity & $(a,b_*)$\\ \hline
$t,y$ & $(1,2)$ & $K_g^{(j)}(t-t')$ & $(1+j,4)$\\
$D_g^{-1}$ & $(1,6)$ & $M,r$ & $(3,6)$\\
$K_h^{(j)}(r-r')$ & $(1+j,10)$ & $D_h^{-1}$ & $(1,12)$\\
$\log D_h$ & $(1,12)$ & $a,\ B_g$ & $(3,12),\ (6,6)$\\
$c,\ d$ & $(9,12),\ (11,12)$ &&
\end{tabular}
\end{center}
Products add zeroth exponents and take the larger
slope. An inverse with denominator at least $c/H$ and input pair $(1,b_*)$
has pair $(1,b_*+2)$, by expanding its $l$th derivative into products
of $j\le l$ derivatives divided by the $(j+1)$st denominator power.
The means in the scale variances and their derivatives are bounded by
$CH^2$ and the variances have a constant lower bound. This gives $(1,2)$
for standardized values. The Gaussian derivative bound
$\sup_x|K_s^{(j)}(x)|\le C_js^{-1-j}$ makes composition with input
$(a,b_*)$ cost slope $1+a+b_*$. The ridge ratio multiplies $(2,4)$ by
$(1,6)$. The density denominator has pair $(1,10)$; its log derivatives
cost at most $12l$ powers for $l>0$. A row average preserves a pair:
its measure derivative is the difference of an evaluation and an integral,
and each derivative of $(1-s)^{-1}$ is bounded when $s\le1/n$.
These rules also bound the NW profile and density jets through order three.
The constants depend on the fixed number of differentiated coordinates,
not on $q$. In particular $H^{100}$ bounds every table entry needed here,
including one extra derivative in comparisons.

\emph{Empirical replacement at truth.} We give the argument also for a
mixed case-weight and coordinate derivative. First set all deletion marks
to zero. In a fixed coordinate entry at $b=0$, the row integrals and
all their differentiated integrands are built only from $(T,Y^w)$ and
at most two specified coordinates of $W$. A mass derivative marks one
fixed observation $z_i$; it replaces an integration by evaluation at $z_i$
minus integration and differentiates its integrand. The expression for an
entry has a fixed number of integration nodes and a fixed maximum number
$m$ of free row arguments, independent of $q,n$. Indeed each additional
free row is introduced by one of the displayed integrations or by the
single mass derivative, and products only juxtapose such arguments.

At any such node compare $Q_n F_n$ and $P F_P$ by writing their
difference as $Q_n(F_n-F_P)+(Q_n-P)F_P$. For fixed values of its free
sample rows, $F_P$ is deterministic and the other observations are iid.
Remove the at most $m$ coincident observations temporarily. Conditional
Hoeffding, followed by a union bound over at most $C n^m q^2$ index and
coordinate choices and the finite list of nodes, bounds the second term
by $C e_n$ times its envelope; restoring the excluded observations costs
at most $C m/n$ times this envelope. To apply this bounded-variable
argument, truncate the integration observation's at most four scalar
components at $C_0\sqrt{\log(n+1)}$; fixed free rows are already within
these bounds on the common maximum event. The population integrals obey
the same table bounds with $H$ enlarged by the sum of absolute scalar components
of their free arguments: products and integration preserve the indicated
powers since the fixed-dimensional Gaussian and mixture moments are
uniformly bounded. Population scale variances at truth are one. Thus
Cauchy--Schwarz and the sub-Gaussian tail bound make each truncated
integral error $O(n^{-3})$ after choosing $C_0$ sufficiently large. Here
$H=O(n)$ follows already from $H_n(R)=o(1)$; arbitrarily large fixed
polynomial tail precision is available. Conditional Hoeffding is applied
to the truncated independent rows before imposing the common maximum
event, so that event does not change their conditional law.

An error in an order-$l$ entry of pair $(a,b_*)$ is bounded
on this event by $C e_n H^{a+b_*(l+1)}$. This follows by induction with
the preceding decomposition at each integration. For products, telescope
one factor at a time. At a smooth scalar node interpolate linearly between
the two collections of input derivatives and differentiate the interpolant:
the one varied input uses the extra order in the exponent. For inverses
the interpolated input value lies between the two positive denominators,
so the same inverse-derivative bound applies. The standardized means and
second moments give errors $O(e_nH)$ and $O(e_nH^2)$ respectively;
interpolating the centered variances, whose endpoint lower bounds are
constant, proves the initial $(1,2)$ rule. This inductive comparison acts
only on population integrands when applying concentration. In particular,
no independence of fitted residuals or their centers is assumed.

\emph{Deletion and its derivative.} At equal weights turning a common mark
from zero to $1/n$ changes any indicated entry by $O(H^{100}/n)$ by
integration of its mark derivative. For all intermediate marks the
unnormalized retained mass is nonnegative at each empirical row, so
$D_g\ge\lambda$ and $D_h\ge\rho$. For a case derivative in direction
$\delta_{z_j}-Q_n$, with its empirical weights followed, the marks have
velocity $v_i=\ind\{i=j\}-1/n$. The part due solely to these mark
velocities in a scalar coordinate of $S$ is $O(H^{100}Q_n|v|)$.
To see this, differentiate the displayed row formula holding $Q_n$ fixed.
A mark differentiated at row $i$ introduces a factor $v_i$ times a
bounded row integrand. Following a product, quotient or kernel derivative
retains that factor; summing a child row replaces it by its row average.
The root averages the outer row as well. All averaging kernels have thus
been bounded by their envelopes in the table, and each remaining normalized row
sum has column sums at most $2$ (its weights are bounded by $2/n$).
Diagonal evaluations simply identify the parent and child rows. Each term
therefore integrates $|v|$ against weights bounded by a constant times
$Q_n$, with the remaining differentiated factors bounded by the table.
Since $Q_n|v|\le2/n$, this is $O(H^{100}/n)$, uniformly in $j$.
The other part is the derivative at fixed common mark, already covered by
the common-mark interpolation. The empirical contamination derivative
equals $n\partial_{w_j}$ at unit weights. This proves the comparison for
the original weight-dependent ridge and floor masses.

We have proved simultaneously at truth, for every scalar entry of the
Hessian and of every case influence, that its discrepancy from the
population functional evaluated at the same free observations is
$O_P(L_ne_n)$. The identical argument on the scalar row-coefficient
formula gives $\max_i|d_i-f_k(T_i,Y_i^w)|=O_P(L_ne_n)$ on an event
measurable in $(T_i,Y_i^w)_{i\le n}$ alone. Conditional on these data,
$n^{-1/2}\sum W_i(d_i-f_{k,i})$ is exactly centered Gaussian with covariance
$Q_n(d-f_k)^2 I_d$. This proves \eqref{hdg:score}. The population
Hessian and case influence are $-j_{v,k}I_d$ and $W_i f_{k,i}$ by
Propositions~\ref{inf:population} and \ref{inf:influence}.

\emph{Extension over the cap.} Each third chart derivative of $L$ and
each mixed derivative with two chart and one case derivative is bounded
by $C L_n$ on the entire cap by the table; here individual empirical
case marks can be differentiated directly since $|v_i|\le1$. Thus a
single entry changes by at most $CL_n\|b\|_1\le CL_n\sqrt q\,u$.
The operator norm of a $d\times d$ matrix is bounded by $d$ times its
largest absolute entry, and the vector norm costs $\sqrt d$. This proves
\eqref{hdg:matrix} without a union bound over local directions. It also
gives $\sup_{\|b\|\le u}\max_i|d_i(b)-d_i(0)|\le CL_n\sqrt q\,u$
for the scalar row coefficient above.

By \eqref{dir:kappaibp} the scalar influence has a uniformly bounded
fourth moment, including on shrinking grids. The exact factorization
$W\perp(T,Y^w)$ therefore gives $E\|W f_k\|^4\le Cq^2$. Independence of
rows and the squared Frobenius norm imply that the oracle empirical
second moment differs from $B_k$ by $O_P(q/\sqrt n)$ in operator norm.
This oracle matrix is used only for the proof. For any vector arrays $U,V$,
$\|Q_n UV^T\|_{\rm op}^2\le\|Q_n UU^T\|_{\rm op}\|Q_n VV^T\|_{\rm op}$.
Apply this inequality to the influence error array and the oracle array;
the latter has bounded empirical second-moment operator norm. We obtain
an error $O_P\{q/\sqrt n+A_n(u)+A_n(u)^2\}$ in $\widehat B(b)$.
Positive curvature and inverse perturbation prove \eqref{hdg:sandwich}.

The projective map from the fixed population coordinates to raw orthonormal
tangents and its inverse have uniformly bounded first three derivative
operator norms. This follows by differentiating $x\mapsto Ax/\|Ax\|$
with $\|A\|,\|A^{-1}\|\le C$ and the normalized affine chart. For moving
frames one may rotate the initial normal to the current one in their
spanned plane, using the smooth rotation $I+K+K^2/(1+u^Tv)$ for unit
normals $u,v$, $K=vu^T-uv^T$. Singular values stay bounded on the cap.
The raw Hessian also has a coordinate connection bounded by $C\|S(b)\|$,
which is $O_P(\sqrt{q/n}+u)$ by the curvature and truth score bounds.
The case-weight chart map is deterministic when the evaluation direction
and its tangent frame are held fixed in a weight derivative;
its matrix differs from its truth value by $O(u)$ and the corresponding
second-moment perturbation is $O_P(u)$. These contributions are absorbed
by \eqref{hdg:matrix}--\eqref{hdg:sandwich}. The fitted sample whitening
inverse cancels exactly on the raw target by \eqref{hd:map-cancellation}.

Finally, the conditional-Gaussian scalar factor and its second moment give
$\|S_{n,k}(0)\|=O_P(\sqrt{q/n})$. Positive empirical curvature throughout
the cap puts an approximate score root within $O_P(u_n+\tau)=O_P(u_n)$
of truth by integrating its Hessian along the segment. The finite ascent
argument below supplies that root. Substitution of this smaller radius in
\eqref{hdg:matrix} and inversion gives \eqref{hdg:estimate}; the chart
Taylor error $O_P(\sqrt n\,u_n^2)=O_P(q/\sqrt n)$ is absorbed. This last
step establishes a probability expansion and does not exchange limits
with expectation over fitting or screening failures.
\end{proof}

For deterministic tuples, if $\sqrt q\,H_n(\sqrt{q/n})+\sqrt n\,\tau=o(1)$,
\eqref{hdg:estimate} yields a studentized univariate normal limit for
each deterministic unit raw contrast (also for one fixed by $D_s$).
Indeed its linear term is conditionally normal given the scalar data and
its variance divided by its population value is $Q_nf_k^2/Ef_k^2\to_P1$.
The denominator is uniformly positive by \eqref{inf:crossidentity} and
the information upper bound, and the common fourth moment proves the
variance convergence. Equation~\eqref{hdg:sandwich} allows using the
raw estimated contrast variance. The uniform error in trace divided by
$d$ is bounded by the operator error. Consequently the trace selected on
a consistent preliminary fit attains the minimum population trace per
tangent dimension up to $o_P(1)$; a positive separation in normalized
trace excludes all nonminimizers with probability tending to one. With
a unique separated minimizer the contrast limit applies to the selected
fit. With positive-scale ties only the simultaneous, selected-index
expansion and covariance-functional consistency follow in general.
On a shrinking finite grid, scalar convergence in \eqref{dir:poplimit},
uniform in $v$ by the compactness argument of Section~\ref{hd:inheritance-proofs},
makes every contrast's linear term differ from its common efficient one
by $o_P(1)$: its mean square is bounded by a constant times the scalar
$L^2$ discrepancy squared. Taking the maximum over the finite grid thus
allows any training selector for this contrast. Replacing the whole
increasing-dimensional vector by its efficient version would cost a
further factor $\sqrt q$ in this bound.

One polynomial example uses $m_0(u)=\cos u$, $q=\max\{2,\lfloor
n^{1/100}\rfloor\}$, and four scales proportional to $n^{-1/5000}$ on
a finite grid. With $R=2n^{-1/20}$, $L_n=O(n^{1/50})$,
$q^{3/2}L_n R=O(n^{-3/200})$, and $\sqrt q H_n(u_n)=o(1)$.
The cosine admits $r=1$ in Lemma~\ref{hdg:fourier}, so the actual
Fourier start lies inside this basin with probability tending to one.

\subsection{The empirical Fourier start and finite ascent in a growing retained model}
\label{hdg:algorithm}
Throughout this subsection $n=n_e$ and $q=|S|$. Condition on an independently selected correct retained model. All constants below are uniform over such models under \eqref{hd:inherit-spectrum} and the fixed scalar family \eqref{hd:scalar-family}. Write $\gamma_*=\Sigma_S^{1/2}\beta_S/\sqrt v$, and let $\widetilde\gamma$ be the leading eigenvector of the equally weighted four training-standardized matrices in Section~\ref{sec:algorithm}.

\begin{lemma}[Fourier gap and training-standardized start]\label{hdg:fourier}
There is a finite nonnegative integer $r$, depending only on the scalar family, for which the population eigengap of the four-matrix average is at least $c_0q^{-r}$. For $q=o(n)$ the operator error of its training-standardized sample matrix, relative to the population-standardized average, is $O_{P,u}(\sqrt{q/n})$. Thus, if $q^r\sqrt{q/n}\to0$,
\begin{equation}\label{hdg:pilot}
 \dP(\widetilde\gamma,\gamma_S(\widehat A))+
 \dP\!\left(\frac{\widehat A^T\widetilde\gamma}{\|\widehat A^T\widetilde\gamma\|},\beta_S\right)
 =O_{P,u}(q^r\sqrt{q/n}).
\end{equation}
Here $\widehat A$ and the response scaling use the same $n$ observations as the pair sums. One may take $r=0$ for a strictly monotone link and $r=1$ for $m_0(u)=\cos u$.
\end{lemma}

\begin{proof}
Let $H_j$ be the probabilists' Hermite polynomials, $\nu=N(0,1)$ and
$A_{j,v}(t)=E[H_j(T)\exp(itY^w)]$. For each $v$ some $A_{j,v}$ with $j\ge1$ has positive $L^2(\nu)$ norm. In fact, simultaneous vanishing would give, by continuity in $t$ and completeness of the Gaussian polynomials, $E(e^{itY^w}\mid T)=E(e^{itY^w})$ for all rational $t$ on a common full-measure set. Continuity would imply independence, contradicting the nonconstant additive link as in Proposition~\ref{fourier:rankprop}. Coupling the scalar family through the same $(T,\varepsilon)$ proves continuity of each norm in $v$ by dominated convergence. A finite subcover of $[c,C]$ consequently gives $r<\infty$ with
\[
 \inf_{v\in[c,C]}\sum_{j=1}^{r+1}\|A_{j,v}\|_{L^2(\nu)}^2>0 .
\]
For a strictly monotone link $E[Tm_0(\sqrt v T)]\ne0$ by the two-independent-copies covariance identity. For the cosine $E[(T^2-1)\cos(\sqrt v T)]=-v e^{-v/2}$. Each nonzero moment gives a nonzero derivative at $t=0$ of the corresponding $A_{j,v}$ after response standardization.

Write $u=\gamma_*^T\omega$ for the longitudinal Fourier frequency. The scalar factor of the vector in \eqref{fourier:fourieridentity}, after removing the transverse Gaussian factor, is
\[
 E[(T-iu)e^{iuT+itY^w}]
 = e^{-u^2/2}\sum_{k\ge0} A_{k+1,v}(t)(iu)^k/k! .
\]
This follows by differentiating the Hermite generating function. Since $|A_{k+1,v}(t)|\le\sqrt{(k+1)!}$, the tail after $k=r$ is $O(|u|^{r+1})$ uniformly for $|u|\le1$ inside the sum. Put $u=\sqrt a\,z$ and restrict $|z|\le1$. The Gram matrix of $1,z,\ldots,z^r$ under $1_{[-1,1]}\,d\nu(z)$ is positive definite. Thus the $L^2(d\nu(z)d\nu(t))$ norm of the polynomial on this interval is at least
\[
 c_r\left\{\sum_{k=0}^r a^k\|A_{k+1,v}\|_{L^2(\nu)}^2/(k!)^2\right\}^{1/2}
 \ge c'_r a^{r/2}.
\]
The omitted tail has norm $O(a^{(r+1)/2})$, and $e^{-u^2/2}\ge e^{-1/2}$. Squaring gives a uniform longitudinal eigenvalue lower bound of order $a^r$ as $a\downarrow0$. At $a=1/q$ the transverse factor $(1+2/q)^{-(q-1)/2}$ is at least $e^{-1}$. Each of the four matrices is positive semidefinite in population and has the same range by Proposition~\ref{fourier:rankprop}. The $1/q$ term alone therefore supplies the claimed gap after multiplication by $1/4$. Any finitely many smaller dimensions are covered by continuity and positivity. Fixed spatial scales $1,3,9$ can have exponentially small transverse factors.

For the sampling bound first put $Z_i=\Sigma_S^{-1/2}X_{i,S}$ and use population-standardized responses. For $a\in\{1/q,1,3,9\}$, substitute empirical expectations for $\Phi$ in \eqref{fourier:fourieridentity}; the resulting Gram matrix equals the ordered-pair average including the diagonal. In $L^2(dW_a(\omega)d\nu(t))$ with Euclidean vector values,
\[
 E\|\widehat\Phi-\Phi\|_{L^2}^2
 \le n^{-1}E\!\int\|Z-i\omega\|^2dW_a(\omega)=q(1+a)/n .
\]
The population norm is uniformly bounded because only the longitudinal component survives and its modulus is at most $E|T|+|u|$. The Gram operator perturbation is at most $2\|\Phi\|_{L^2}\|\widehat\Phi-\Phi\|_{L^2}+\|\widehat\Phi-\Phi\|_{L^2}^2$, giving the asserted order before empirical standardization.

Write the fitted predictors as $BZ_i+b$, $B=\widehat A\Sigma_S^{1/2}$, and the fitted response as $a_yY_i^w+b_y$. Then
\[
 \|B-I\|_{\rm op}+\|b\|=O_{P,u}(\sqrt{q/n}),\qquad
 |a_y-1|=O_{P,u}(n^{-1/2}).
\]
Indeed, the sample mean has expected squared norm $q/n$ in standardized coordinates. For every fixed unit vector the uncentered quadratic sample moment has a scaled $\chi^2_n$ law. Its exponential tail $2\exp[-c n\min(t^2,t)]$ and a $1/4$-net of size $9^q$ yield an operator error $O_{P,u}(\sqrt{q/n}+q/n)$. Centering adds $O_{P,u}(q/n)$. On the fixed spectral interval the inverse-square-root map is operator Lipschitz, as follows by subtracting its resolvent integrals $M^{-1/2}=\pi^{-1}\int_0^\infty s^{-1/2}(M+sI)^{-1}ds$. The uniform fourth response moment gives the scalar scale bound.

This substitution does not require independence of $B,b,a_y$ from the pair sum. Interpolate linearly from $(I,0,1)$ to $(B,b,a_y)$ and call the predictors on this path $x_i$. The centered response translation cancels. For $\Delta=Z_i-Z_j$ the derivative along the path of the spatial exponential is bounded in modulus by $C\|B-I\|_{\rm op}$, since all path matrices are well conditioned and $s e^{-s/2}$ is bounded on $[0,\infty)$. The response exponential derivative is bounded by $C|a_y-1|$ for the same reason. Let $l_{ij}=(1+a)x_i-a x_j$ and $p_{ij}=(1+a)x_j-a x_i$, the two outer-product factors in \eqref{fourier:kernel}. For every unit $u$, simultaneously along the path,
\[
 n^{-2}\sum_{ij}(u^T l_{ij})^2
 \le (1+2a)^2\left\|n^{-1}\sum_i x_ix_i^T\right\|_{\rm op}
 =O_{P,u}(1),
\]
also for $p_{ij}$. For their path derivatives the same bound is $O_{P,u}((\|B-I\|_{\rm op}+\|b\|)^2)$ by the preceding sample moment bound. Cauchy--Schwarz over all pairs bounds each bilinear form of the differentiated matrix, including its $aI$ term, by $O_{P,u}(\|B-I\|_{\rm op}+\|b\|+|a_y-1|)$, uniformly over both unit vectors and all four scales. Integration along the path proves the trained-sample operator bound. Projecting the empirical leading eigenvector equation onto $\gamma_*^\perp$ as in Proposition~\ref{fourier:rateprop} gives the gap-divided rate. The moving target differs from $\gamma_*$ by $O_{P,u}(\sqrt{q/n})$. The normalized raw map is uniformly Lipschitz on this spectral event and maps the moving target exactly to $\beta_S$ by \eqref{hd:map-cancellation}. This proves \eqref{hdg:pilot}.
\end{proof}

\begin{lemma}[Actual finite optimization]\label{hdg:ascent}
Use the Fourier chart, projection, iteration budget and stopping rule of Section~\ref{sec:ascent} with $n=n_e$ and $R=\min(1.5,2n^{-1/20})$. Assume
\begin{equation}\label{hdg:basin}
 q^r\sqrt{q/n}=o(R).
\end{equation}
Suppose the score at the truth is $O_{P,u}(\sqrt{q/n})$ and the raw Hessian on $\|\theta\|\le KR$ has negative eigenvalues bounded away from zero and infinity in magnitude with probability tending to one, for each fixed $K$. These are the local score and Hessian conclusions of the growing-dimension expansion. Then uniformly over the finite grid and both fits the Fourier cap contains the interior local maximum. The finite iterate differs from this maximum by $O_{P,u}(n^{-1})$. Its distance from the true line is $O_{P,u}(\sqrt{q/n})$.
\end{lemma}
\begin{proof}
Select signs for the analysis to give an acute angle at the truth. The normalized raw direction for a working chart coordinate $z$ is proportional to $\widehat A^T(\widetilde\gamma+U_pz)$. Hence its raw chart coordinate is
\[
 \theta(z)=\frac{F_S^T\widehat A^T(\widetilde\gamma+U_pz)}
 {\beta_S^T\widehat A^T(\widetilde\gamma+U_pz)} .
\]
On the spectral event and \eqref{hdg:pilot} the denominator is uniformly bounded away from zero for $\|z\|\le R$. This affine quotient and its inverse (interchange coordinates and use $\widehat A^{-T}$) have dimension-independent bounds on their first and second operator derivatives on the caps. In particular the derivative's smallest singular value stays positive. The entire cap maps inside a raw truth-centered ball of radius $KR$ for some fixed $K$, since the pilot error is $o_P(R)$. The negative Hessian in the working chart is the congruence of the negative raw Hessian plus a term bounded by $C\sup_{\|\theta\|\le KR}\|\nabla L_n(\theta)\|=O_{P,u}(\sqrt{q/n}+R)$. It therefore has spectrum in $[c_1,C_1]$ with probability tending to one for fixed $0<c_1<C_1$.

In working coordinates the true line has coordinate $z_0=o_{P,u}(R)$ and score $O_{P,u}(\sqrt{q/n})$. The ball of radius $R/2$ about $z_0$ lies inside the cap with probability tending to one. The concave quadratic upper bound there is $F_n(z_0)+\|\nabla F_n(z_0)\|u-c_1u^2/2$ at distance $u$. The maximum over this closed ball lies in its interior, and the bound gives $\|z_*-z_0\|\le 2\|\nabla F_n(z_0)\|/c_1$. Strict concavity gives uniqueness on the entire cap. In particular all bounded multiples of $\sqrt{q/n}$ about the truth lie inside the raw image of the cap with probability tending to one.

At step size $\eta=1/\log(n+e)$, for the projected increment $d$ one has $\nabla F_n(z)^Td\ge \|d\|^2/\eta$. Once $\eta C_1\le1$, Taylor's lower bound accepts the first Armijo trial. The derivative of the unprojected gradient map is $I+\eta\nabla^2 F_n$ and has operator norm at most $1-c_1\eta$. Projection is nonexpansive and fixes the interior maximizer. After $\lceil\log^3(n+e)\rceil$ iterations the error is bounded by $2R\exp\{-c_1\log^2(n+e)\}$; the gradient norm is at most $C_1$ times that error. An earlier stop at score tolerance $1/n$ has distance at most $1/(nc_1)$ by strong concavity. Thus the working numerical score also meets the inference threshold $n^{-1/2}/\log(n+e)$ eventually. The data select the preliminary tuple at the Fourier direction and select the final tuple by the estimated trace at the preliminary fit. Both maximizations start at $z=0$, and the preceding event holds for every preliminary and final tuple simultaneously. Only the theoretical centers $z_0$ and $\theta=0$ use the unknown direction.
\end{proof}

For example $q=\max(2,\lfloor\log(n+e)\rfloor)$ satisfies \eqref{hdg:basin} for every fixed scalar family above. On this path $q^r\sqrt{q/n}/R=O(n^{-9/20}\log^{r+1/2}(n+e))$. More generally \eqref{hdg:basin} permits $q^{r+1/2}=o(n^{9/20})$ subject to the estimation and covariance restrictions. The earlier direct and screened numerical settings in Section~\ref{hd:protocol} used a mean-MAVE start. Lemmas~\ref{hdg:fourier} and~\ref{hdg:ascent} concern the four-matrix Fourier start used in the growing-dimension analysis.

\section{Information and bandwidth selection in the retained model}
\label{hdg:information}

Condition on a support-containing screen $S$, write $b_0=\beta_{0,S}$,
$\Sigma=\Sigma_S$, $v=b_0^T\Sigma b_0$, and $d=|S|-1$. Set
$\gamma=\Sigma^{1/2}b_0/\sqrt v$, and choose orthonormal tangent frames
$F\perp b_0$, $U\perp\gamma$. With $T=\gamma^T\Sigma^{-1/2}X_S$
and $W=U^T\Sigma^{-1/2}X_S$, $W\sim N_d(0,I_d)$ independently of
$(T,Y)$. Put $\sigma_v^2=\Var(Y)$, $Y^w=(Y-\E Y)/\sigma_v$,
$m_v(t)=\{m_0(\sqrt v\,t)-\E Y\}/\sigma_v$ and
$\varepsilon_v=\varepsilon/\sigma_v$. Thus $(T,Y^w)$ has a scalar
law indexed by $v$. Define
\[
 A(v)=\E m_0'(\sqrt v\,T)^2,\quad
 I_\varepsilon=\E\{(f_\varepsilon'/f_\varepsilon)(\varepsilon)^2\},
 \quad i(v)=vA(v)I_\varepsilon,\quad
 G=\Var(\varepsilon)I_\varepsilon .
\]
Here and throughout this section $f_\varepsilon$ is the density in raw
response units. For a positive tuple $k=(g,h,\lambda,\rho)$, use
$j=j(v,k)$ from \eqref{inf:curvature} and
$\zeta_k=\kappa_k(T,Y^w)-T\E\{T\kappa_k(T,Y^w)\}$ from
\eqref{inf:kappa}--\eqref{inf:whiteningprojection}.
Write $b=b(v,k)=\E\zeta_k^2$.

\begin{proposition}[Loss and gain after screening]\label{hdg:loss}
Let $j>0$. If $\Omega_S=\Var(F^TX_S\mid b_0^TX_S)$, the raw
tangent derivative $D=\sqrt v F^T\Sigma^{-1/2}U$ and the covariance
coefficient $V_{S,k}$ of the leading root-$n_e$ term satisfy
\begin{align}
 DD^T&=v\Omega_S^{-1},&
 V_{S,k}&=v(b/j^2)\Omega_S^{-1},\label{hdg:geometry}\\
 \mathcal I_S^{-1}&=\{A(v)I_\varepsilon\}^{-1}\Omega_S^{-1},&
 V_{S,k}&=\Lambda(v,k)\mathcal I_S^{-1}
 =\{\Lambda(v,k)/G\}V_{{\rm mean},S},\label{hdg:gain}
\end{align}
where $\Lambda(v,k)=i(v)b/j^2\ge1$ and
$V_{{\rm mean},S}=\Var(\varepsilon)A(v)^{-1}\Omega_S^{-1}$ is the
conditional-mean efficiency bound evaluated at this homoscedastic law.
Thus the same-sample-size covariance comparison with the mean bound is
strictly favorable in every nonzero retained contrast exactly when
$\Lambda(v,k)<G$. The leading covariance approximation for the
unscaled estimator is $V_{S,k}/n_e$, the covariance of its linear
approximation. A mean-risk expansion additionally requires moment
control of the remainder and of failed fits.
\end{proposition}

\begin{proof}
The normalized map from a whitened direction $a$ to the raw direction
is $a\mapsto\Sigma^{-1/2}a/\|\Sigma^{-1/2}a\|$. Its tangent derivative
at $\gamma$ is $D$. Since $UU^T=I-\gamma\gamma^T$ and $F^Tb_0=0$,
$DD^T=vF^T\Sigma^{-1}F$. Block inversion of
$[b_0,F]^T\Sigma[b_0,F]$ gives $F^T\Sigma^{-1}F=\Omega_S^{-1}$,
where the Schur complement is
\[
 \Omega_S=F^T\Sigma F-F^T\Sigma b_0b_0^T\Sigma F/v .
\]
Consequently $cI_d\preceq\Omega_S\preceq CI_d$ under the assumed
spectral bounds, for every $d$. In standardized coordinates the
complete parameter influence is $W\zeta_k/j$. Its raw covariance is
$D(b/j^2)D^T$. In raw units the efficient score is
$-\psi_\varepsilon(\varepsilon)m_0'(b_0^TX_S)
\{F^TX_S-\E(F^TX_S\mid b_0^TX_S)\}$, where
$\psi_\varepsilon=f_\varepsilon'/f_\varepsilon$.
The Gaussian residual tangent is independent of the index and of the
error. This gives $\mathcal I_S=A(v)I_\varepsilon\Omega_S$.
The conditional-mean bound of Section~\ref{sec:core-efficiency}
uses $1/\Var(\varepsilon)$ in place of $I_\varepsilon$.
These calculations establish both covariance equalities.

Write $\psi_v$ for the score of $\varepsilon_v$ and let
$\zeta_*=-m_v'(T)\psi_v(\varepsilon_v)$. Response scales cancel,
so $\E\zeta_*^2=i(v)$. Along each unit standardized tangent $w$,
rotation of the model direction leaves the population SI-KDR score at its
own truth zero. Differentiating this identity as in
\eqref{inf:crossidentity} gives
$\E\{(w^TW)^2\zeta_k\zeta_*\}=j$, hence
$\E\zeta_k\zeta_*=j$. Therefore
\begin{equation}
 \Lambda(v,k)-1=i(v)\E\{\zeta_k/j-\zeta_*/i(v)\}^2 .
 \label{hdg:loss-square}
\end{equation}
The independent-error and mean bounds concern the same retained
coordinates, design law, and $n_e$ observations.
\end{proof}

\begin{corollary}[Uniform preservation of the non-Gaussian gain]
\label{hdg:uniform-gain}
For a fixed nonconstant bounded $C^6$ link with bounded derivatives,
a fixed centered nondegenerate finite Gaussian-mixture error, and
$v\in[c,C]$, $\Lambda(v,k)\to1$ uniformly as $\|k\|_\infty\to0$.
If the error density is non-Gaussian, there exists $a>0$ such that
every $0<g,h,\lambda,\rho\le a$ has positive curvature and
$\sup_{v\in[c,C]}\Lambda(v,k)<G$. This includes fixed positive
grids. For a given $n$, every $S$ containing the true support has
the same $v$ and hence the same $\Lambda$.
\end{corollary}

\begin{proof}
Continuity and positivity give $\inf_{v\in[c,C]}i(v)>0$: a
nonconstant differentiable link has derivative nonzero on an open
interval, and all the index laws give it positive probability. The
mixture has finite positive location information.
For completeness, take any sequence $v_l\to v_*$ in this compact
interval and $k_l\to0$. Couple the scalar experiments through common
$(T,\varepsilon)$. The response scales belong to a fixed positive
compact interval. The standardized links and their derivatives
converge locally uniformly through order two, and the fixed component
means and variances after scaling lie in compact sets, the variances
bounded away from zero. The bounded-location mixture and Gaussian
profile bounds \eqref{dir:mixture}--\eqref{dir:aenv} then give common
polynomial envelopes for $M,M',M''$ and the residual coefficients
$a_p,a_p'$. On compact sets $M,M',M''$ converge to the limiting
standardized link and its derivatives. The convolutions defining
$a_p,a_p'$ converge there to $\psi_{v_*},\psi_{v_*}'$ by Gaussian
approximation to the identity and local positivity of the mixture
density. In the incoming term of \eqref{dir:kappaibp}, the factors
$\E_\varepsilon a_p(\varepsilon_v+m_v-M)$ and their first index
derivatives tend to zero, since $\E\psi_{v_*}=0$ and $M'\to m_{v_*}'$.
Substituting $u=t+gZ$ in that integral yields a common polynomial
envelope in $(T,Y^w,Z)$. Gaussian tails allow dominated convergence in
squared mean. Thus $\|\zeta_{k_l}-\zeta_{*,v_l}\|_2\to0$; the same
bounds in \eqref{inf:curvature}, and $-\E\psi_{v_*}'=I_{\varepsilon_{v_*}}$,
give $j(v_l,k_l)-i(v_l)\to0$. The subsequence criterion gives both
convergences uniformly over $[c,C]$, without a dimension factor.
Equation~\eqref{hdg:loss-square} gives the uniform statement for
$\Lambda$. Integration by parts gives $\E(\varepsilon
\psi_\varepsilon)=-1$, so $G\ge1$, with equality only when
$\psi_\varepsilon(e)=-e/\Var(\varepsilon)$ almost everywhere.
The positive mixture density then solves the Gaussian density
differential equation. For non-Gaussian errors take, for example,
$\sup_v|\Lambda-1|<(G-1)/2$. Finally, zero omitted coefficients give
$v=\beta_0^T\Sigma_n\beta_0$ for every successful retained model.
\end{proof}

\begin{corollary}[Trace selection]\label{hdg:tuning}
For the same deterministic positive-curvature grid $\mathcal K$,
population raw traces rank its tuples identically across all
support-containing retained sets at the same $n$. Suppose $\inf_{n,k}j(v_n,k)>0$ and the raw
case-weight covariances at the preliminary estimator obey
$\Delta_n=\max_{k\in\mathcal K}
\|\widehat V_{S,k}-V_{S,k}\|_{\rm op}=o_P(1)$ in matching frames and
the estimated curvature is positive throughout the grid with probability
tending to one. With $T_{S,k}=d^{-1}\tr V_{S,k}$, the actual
trace selector satisfies
\[
 0\le T_{S,\widehat k}-\min_k T_{S,k}\le2\Delta_n
\]
on that event. This bounds the excess trace per tangent coordinate.
\end{corollary}
\begin{proof}
$T_{S,k}=\Lambda(v,k)d^{-1}\tr\mathcal I_S^{-1}$ with common
strictly positive multiplier bounded above and below. Also
$|d^{-1}\tr(\widehat V_{S,k}-V_{S,k})|\le\Delta_n$. Comparing the
empirical minimum with a population minimum proves the claim. On a
fixed positive grid a unique minimizer is selected with probability
tending to one if its normalized gap is bounded away from zero
(or dominates $\Delta_n$ in probability). The same separation from
the complement suffices to select within a tied minimizing set.
Near a crossing as $v=v_n$ varies, the excess-trace bound still applies.
At an exact tie the selectors and their score vectors can be dependent.

For a deterministic or consistently selected tuple and a deterministic
unit raw contrast $a$, the one-observation influence
$a^TDW\zeta_k/j$ has variance $a^TV_{S,k}a$ bounded above and away
from zero. Its standardized third absolute moment is uniformly bounded
by the scalar envelopes above, bounded singular values of $D$ and the
one-dimensional Gaussian moment. The Lyapunov ratio is $O(n_e^{-1/2})$.
For an $o_P(1)$ root-$n_e$ contrast remainder, this gives
\[
 \frac{a^T\sqrt{n_e}\,\widehat\theta_k}{(a^TV_{S,k}a)^{1/2}}
 \ \Longrightarrow\ N(0,1),
\]
also for unit contrasts depending deterministically on the retained
model and $n$. On a fixed-cardinality shrinking grid, replacing
$\zeta_k/j$ by $\zeta_*/i(v)$ changes each such leading contrast
by $o_P(1)$ in squared-mean norm, uniformly by the preceding proof.
A maximum over the grid gives the same limit for a data-selected
tuple. For ties on a fixed positive grid the simultaneous
tuple-specific linear representations remain available; a single
selected normal limit requires further control of the joint selector
and scores. These are contrast statements for the growing model.
\end{proof}

\section{Conditional inference with a growing retained dimension}
\label{hdg:inference}

Write $n=n_e$, $q=|S|$, $d=q-1$ and condition on $D_s$ with
$S_0\subseteq S$. All following operations use the estimation rows
alone. For a fixed raw unit candidate $b$ and an orthonormal tangent frame
$F_b$ determined by $(S,b)$, let $x=X_Sb$, $W_b=X_SF_b$, $s_b^2=P_n(x-P_nx)^2$,
$t=(x-P_nx)/s_b$,
$\Pi=I-\boldsymbol1\boldsymbol1^T/n-tt^T/n$, and $Z=\Pi W_b$.
The candidate rule first minimizes protected LOO regression squared
error over $g$, then maximizes protected LOO residual log score over $h$,
with predetermined ridge, floor and tie breaking. Write $c$ for
\eqref{cs:complete-coefficient} at the chosen tuple. Both the tuple
and $c$ are functions of $(D_s,x,Y_e)$. Holding the chosen tuple fixed,
the raw tangent score equals $Z^T\Pi c/(ns_b)$ by differentiating the
empirically standardized index. For $u=\Pi c\ne0$ put
\[
 \bar u=u/\|u\|,\qquad z=Z^T\bar u,\qquad A=Z^TZ-zz^T .
\]
If $u=0$ the candidate passes the test.

\begin{proposition}[Calibration at a retained candidate]\label{hdg:pivot}
For each $n>q+1$, if $X_S$ is Gaussian with positive definite covariance
and $Y\perp X_S^TF_b\mid b^TX_S$ (using column notation for a single
row), then conditionally on $(D_s,x,Y_e)$ with positive index and
response scales and $u\ne0$,
\begin{equation}
 F(b)=\frac{n-d-2}{d}\,z^TA^{-1}z\ \sim\ F_{d,n-d-2}.
 \label{hdg:F}
\end{equation}
\end{proposition}
\begin{proof}
The Gaussian conditional means of the tangent rows are affine in $x$
and their common conditional covariance is the positive Schur
complement $\Omega_b$. The null conditional independence leaves the
row law unchanged given $Y_e$. Thus $Z$ has conditional row and column
covariance factors $\Pi$ and $\Omega_b$. In the $n-2$ dimensional range
of $\Pi$, complete the conditionally fixed $\bar u$ to an orthonormal basis.
The first Gaussian row gives $z\sim N_d(0,\Omega_b)$, and the remaining
$n-3$ independent Gaussian rows give
$A\sim\operatorname{Wishart}_d(\Omega_b,n-3)$ independently. In
particular $A$ is invertible almost surely, including $n=q+2$.
Equivalently, after column whitening, $U=z^T(Z^TZ)^{-1}z$ is the
squared projection of $\bar u$ onto a uniformly oriented $d$-subspace
of $\R^{n-2}$. Rotational invariance gives
$U\stackrel{\mathcal L}=\chi_d^2/(\chi_d^2+\chi_{n-d-2}^2)$ for
independent chi squares, while rank-one inversion gives
$z^TA^{-1}z=U/(1-U)$. This proves \eqref{hdg:F} at every admissible
dimension.
\end{proof}

\paragraph*{Local domain.}
For $d=d_n\to\infty$, use the actual trace-tuned point estimate
$\widehat b$, its final raw tangent covariance coefficient
$\widehat V=\widehat V_{S,\widehat k}$, and the raw angular radius
\begin{equation}
 a_{n,d}=\min\{\pi/2,\ 2\sqrt{\tr(\widehat V)/n}\}.
 \label{hdg:radius}
\end{equation}
Here $\widehat V$ is the case-weight estimate for the root-$n$ term,
evaluated at the final fit. The domain $\mathcal D$ is the intersection
of the mapped Fourier optimizer cap with the angular ball of radius
$a_{n,d}$ about $[\widehat b]$. A global preprocessing, initialization,
optimization or inference-diagnostic failure makes the domain empty,
as in Section~\ref{sec:inference-conditions}. The point estimate,
domain and radius can use the tangent design. Only the candidate
test statistic is conditioned on its scalar inputs. For fixed $q$
the radius $\log(n+e)/\sqrt n$ and its shape result are
given in Section~\ref{sec:conditional-score-proofs}.
The mathematical set $\mathcal C^{\rm math}$ zero-embeds the part of $\mathcal D$ that passes \eqref{hdg:F} at level $\alpha$,
including zero-coefficient candidates. An implicit membership
evaluation suffices to test inclusion of any specified line.

\begin{theorem}[Localization, coverage and diameter]\label{hdg:coverage}
Suppose $d\to\infty$, $d/n\to0$, and use the growing-dimensional
estimation and covariance conditions for the preliminary, final and
inference grids. On screen success, uniformly in the individual retained
models, the true line belongs to $\mathcal D$ with conditional
probability $1-o(1)$. Its mathematical coverage conditional on $D_s$ is
$1-\alpha+o(1)$, and the projective diameter is $O_P(\sqrt{d/n})$.
For every finite $n$ with the pivot degrees of freedom above, let
$E=\{S_0\subseteq S\}$, $\mathcal X_0=\{[b_0]\notin\mathcal D\}$ on $E$, and
let $\mathcal N_0$ denote numerical false exclusion of a candidate that truly
belongs to the localized set. Then
\begin{align}
 \Pp\{[\beta_0]\in\mathcal C^{\rm comp}\}
 \ \ge\ &(1-\alpha)\Pp(E)-\Pp(E\cap\mathcal X_0)
 -\Pp(E\cap\mathcal X_0^c\cap\mathcal N_0).
 \label{hdg:coverage-bound}
\end{align}
For the mathematical set, omit the final probability. If
$\Pp(E^c)\le\delta_{\rm scr}\to0$, its unconditional coverage tends
to $1-\alpha$, with the same diameter rate. The same coverage conclusion
holds for a computed subset when its false-exclusion term vanishes.
\end{theorem}

\begin{proof}
The sole dimension concentration needed to prove inclusion in the
observable angular ball is at the truth. For each point tuple,
the leading root-$n$ raw tangent displacement is
\[
 H_k=\frac{D}{j_k\sqrt n}\sum_{i=1}^n W_i\zeta_{k,i},
 \quad V_{S,k}=\frac{b_k}{j_k^2}DD^T,
 \quad b_k=\E\zeta_{k,i}^2 .
\]
Here $D$ is the population tangent derivative in
\eqref{hdg:geometry}, and the standard Gaussian tangent rows $W_i$
are independent of the entire collection of scalar rows
$(T_i,Y_i^w)$. The uniform lower curvature and information bounds and
\eqref{hdg:loss-square} bound $b_k$ away from zero. The scalar
fourth-moment bound gives $B_{n,k}=P_n\zeta_k^2/b_k
=1+O_P(n^{-1/2})$. Conditional on the scalar rows,
$H_k\sim N_d(0,B_{n,k}V_{S,k})$, so
\begin{align*}
 \E(\|H_k\|^2\mid T_{1:n},Y^w_{1:n})
 &=B_{n,k}\tr V_{S,k},\\
 \Var(\|H_k\|^2\mid T_{1:n},Y^w_{1:n})
 &=2B_{n,k}^2\tr V_{S,k}^2 .
\end{align*}
The eigenvalues of every $V_{S,k}$ lie in a common positive compact
interval. In particular $\tr(V_{S,k}^2)/(\tr V_{S,k})^2=O(1/d)$,
and $\|H_k\|^2/\tr V_{S,k}=1+O_P(d^{-1/2}+n^{-1/2})$.
All these bounds hold simultaneously over the finite grid. This
simultaneity allows data-dependent point tuning, including ties.

The uniform expansion has
$\|\sqrt n\,\widehat\theta_k-H_k\|=o_P(\sqrt d)$, in the normalized
raw chart $(b_0+F\theta)/\sqrt{1+\|\theta\|^2}$. The corresponding
angle to the truth is $\arctan\|\theta\|$, and the angles tend to
zero since $d/n\to0$. Operator consistency at the final fit implies
$|\tr\widehat V_{S,k}-\tr V_{S,k}|=o_P(d)$. Consequently
\[
 \frac{\sqrt n\,\operatorname{angle}([\widehat b],[b_0])}
 {\sqrt{\tr\widehat V_{S,\widehat k}}}\ \longrightarrow_P\ 1 .
\]
Any fixed multiplier greater than one in \eqref{hdg:radius}
therefore suffices. The localized Fourier pilot and the finite ascent
result put the truth inside the mapped optimization cap and guarantee
the global diagnostics with probability tending to one. In particular,
empirical positive curvature and positive case-weight covariance
follow from operator consistency and the population spectral bounds.
This proves the domain inclusion. The angular triangle inequality and
$\dP(b,b')=2\sin\{\operatorname{angle}([b],[b'])/2\}$ bound the
diameter of every component together by $2a_{n,d}=O_P(\sqrt{d/n})$.
Zero embedding preserves this distance.

At truth, $\Pi c_k=0$ implies that the exact raw score of tuple $k$
vanishes. Its root-$n$ expansion has leading term
$D^{-T}n^{-1/2}\sum W_i\zeta_{k,i}$ and norm error $o_P(\sqrt d)$.
Apply the same conditional Gaussian second-moment calculation to
that leading vector, with raw covariance
$B_{S,k}=b_kD^{-T}D^{-1}$. Its norm divided by
$\sqrt{\tr B_{S,k}}$ converges in probability to one and
$\tr B_{S,k}\asymp d$. Simultaneity over the inference grid gives
$\Pp(u=0\mid D_s)=o(1)$ also for the candidate-selected tuple.
Exact conditional
calibration integrated over the scalar inputs yields full-test
acceptance probability $1-\alpha+\alpha\Pp(u=0\mid D_s)$.
The raw index and response scales are positive almost surely by the
model. Losing $o(1)$ to the domain gives the asserted conditional
coverage.

For finite $n$, rejection on $E$ has probability at most
$\alpha\Pp(E)$ by the conditional pivot; domain exclusion is charged
to $E\cap\mathcal X_0$. A failed solve at an accepted candidate, an incorrect
numerical zero-coefficient decision that excludes it, or a set
representation that misses it belongs to the last event in
\eqref{hdg:coverage-bound}. This proves the inequality without
conditioning on passage of the point-estimation diagnostics. If the
screen misses an active coordinate, the embedded set cannot contain
the true line. Integrating the conditional results over $D_s$ gives
the final claims.
\end{proof}

\section{Growing retained dimension on the implemented grid}\label{vtwo:section}
This section proves a first sufficient power-grid bound. Theorem~\ref{sb:dimension} sharpens it and treats the separate logarithmic schedule.
\subsection{A scalar coefficient for the training score}
\label{v2:truth-score-section}
Write $n=n_e$, $d=q-1$, and condition on $D_s$ with $S_0\subseteq S$.
In the population coordinates of \eqref{hd:scalar-family}, put
$\mathcal A=\sigma(D_s,T_{1:n},Y^w_{1:n})$ and let $W$ have rows $W_i^T$.
Its entries are iid standard normal given $\mathcal A$. For each deterministic
tuple let $\zeta_k=\kappa_k(T,Y^w)-T\E(T\kappa_k)$,
$b_k=\E\zeta_k^2$, and use $D_{S,k}$ from \eqref{dir:DS}.

\begin{lemma}[Score at the true direction]\label{v2:truth-score}
Assume $n^{-A}\le g,h,\lambda,\rho\le1$ for a fixed $A$,
$ng\lambda/\ell_n^{12}\to\infty$, and the retained scalar model in
\eqref{hd:scalar-family}. Let $c_k^*$ be the complete coefficient in
\eqref{eq:full-score} calculated after training standardization,
$t=(T-\bar T)/s_T$, $s_T^2=P_n(T-\bar T)^2$, and
$\Pi=I-\boldsymbol1\boldsymbol1^T/n-tt^T/n$. Then
\begin{align}
 d_k&=s_T^{-1}\Pi c_k^*,&
 \|d_k-\zeta_k\|_{n,2}&=O_{P,u}(D_{S,k}),\label{v2:scalar-score}\\
 \sqrt n S_{n,k}(0)&=n^{-1/2}W^T\zeta_k+E_k,&
 E_k\mid\mathcal A&\sim N_d(0,\|d_k-\zeta_k\|_{n,2}^2I_d).
 \label{v2:gaussian-score}
\end{align}
Here $\zeta_k$ on the right denotes the vector of its observed values.
Thus $\|E_k\|=O_{P,u}(\sqrt d\,D_{S,k})$, whereas
$a^TE_k=O_{P,u}(D_{S,k})$ for each unit $a$ fixed given $D_s$.
These bounds hold simultaneously for finitely many tuples.
\end{lemma}
\begin{proof}
At fixed scalar inputs the unprojected coefficient is
$-a_iA_i+\sum_{j\ne i}a_jQ_{ji}(y_i-\widehat M_j)$, with both
denominators in $a_i$ as in \eqref{eq:full-score}. In particular
$P_nc_k^*=0$ by exchanging the same ordered pairs. Differentiating
\eqref{hdg:chart} at zero gives the matrix $\Pi W/s_T$. This proves the
exact identity $S_{n,k}(0)=W^Td_k/n$, including the derivative of the
training covariance. Both scalar smoothers use the original $n-1$
other observations. Neither computation uses a perpendicular coordinate.

For the scalar comparison, first use population-standardized observations.
The outgoing fit and incoming refit estimates of \eqref{dir:crisk} give
\[
 O_{P,u}\!\left(\ell_n^{24}\left[
 \{s_0/h^2+(nh^3)^{-1/2}\}/g+s_1/h+s_0/g+\sigma_1\right]\right)
\]
for the empirical $L^2$ error of the unprojected coefficient from $\kappa_k$.
The proof conditions on one evaluation row for the incoming integral,
then takes one expectation of the empirical squared error. The ridge-jet
and residual-density estimates there involve only $(T,Y^w)$. Their
constants are uniform in $v\in[c,C]$ by the scalar envelope argument
of Section~\ref{hd:uniform-assumptions}. In particular \eqref{dir:kappaibp}
gives $|\kappa_k(t,y)|\le C(1+|t|)(1+|y|)$, uniformly in the scales
and in $v$. Training index scaling and response scaling and centering add
$O_{P,u}(\ell_n^7/(\sqrt n\,g h^2))$ by \eqref{dir:scalarLip}.
The projection $\Pi$ has norm one, $\|t-T\|_{n,2}+|s_T-1|=O_{P,u}(n^{-1/2})$,
$\E\kappa_k=0$, and $(P_n-P)(\kappa_k,T\kappa_k)=O_{P,u}(n^{-1/2})$.
Substitution into $s_T^{-1}\Pi c_k^*$ proves \eqref{v2:scalar-score}.
All random coefficients used so far are $\mathcal A$-measurable, hence
\eqref{v2:gaussian-score} follows, including the degenerate case. A finite
union takes place after the bounds for deterministic tuples.
\end{proof}

For \eqref{inf:nonempty} the six terms in \eqref{dir:DS}, grouping
$s_0=\sigma_0+\omega_0$ and $s_1=\sigma_1+\omega_1$, have exponents
$-7/60,-7/30,-11/60,-1/4,-1/4,-1/5$ respectively, apart from logarithms.
Here $\omega_j/\sigma_j=(ng\lambda)^{-1/2}=O(n^{-1/24})$.
Thus a sufficient condition for an absolute $o_P(1)$ Euclidean truth-score
error on this grid is $\sqrt q\,\ell_n^{32}n^{-7/60}\to0$. Consistency
of an estimator and covariance at its fitted location also require the
curvature, initialization, and case-weight bounds of Propositions~\ref{vtwo:hessian} and \ref{vtwo:cov}.

The projection also controls the zero-coefficient branch of
the candidate test. The identity $\E(\zeta_k\zeta_*)=j(v,k)$ and
$\E\zeta_*^2=i(v)$ imply $b_k\ge j(v,k)^2/i(v)$. On a grid of
uniformly positive curvatures, \eqref{v2:scalar-score} and
$P_n\zeta_k^2=b_k+O_{P,u}(n^{-1/2})$ imply
$\min_k\|\Pi c_k^*\|_{n,2}$ is bounded away from zero with probability
tending to one whenever $\max_kD_{S,k}\to0$. This conclusion applies to
the scalar candidate selector by simultaneous control of its grid. Exact
conditional calibration at nonzero coefficients is stated in
Proposition~\ref{hdg:pivot}.

\subsection{Conditional Gaussian matrix and trace bounds}
\begin{lemma}[Conditionally weighted Gaussian Gram matrix]\label{vtwo:weightedgram}
Let $W$ have $n$ rows in $\mathbb R^d$ that are independent $N_d(0,I_d)$
given a scalar-data sigma field $\mathcal A$. For real, possibly signed,
$\mathcal A$-measurable weights $a_1,\ldots,a_n$, put
$G_a=n^{-1}W^T\diag(a)W$, $\bar a=P_n a$, and $m_a=\max_i|a_i|$.
For every $x>0$, with conditional probability at least $1-2e^{-x}$,
\begin{equation}
 \|G_a-\bar a I_d\|_{\rm op}\le
 C\left\{\sqrt{(d+x)P_n a^2/n}+(d+x)m_a/n\right\}.
 \label{vtwo:weightedgram-op}
\end{equation}
For any symmetric matrix $A$ fixed given $\mathcal A$,
\begin{equation}
 E\!\left[\frac{\tr\{A(G_a-\bar a I_d)\}}d\,\middle|\,\mathcal A\right]=0,
 \qquad
 \Var\!\left[\frac{\tr\{A(G_a-\bar a I_d)\}}d\,\middle|\,\mathcal A\right]
 =\frac{2 P_n a^2\,\tr(A^2)}{n d^2}.
 \label{vtwo:weightedgram-tr}
\end{equation}
If $a_i$ is a rowwise function of the iid scalar observations, has a
uniformly bounded second moment (in particular a bounded fourth moment
suffices), and $m_a=O_P(L^s)$ with $L=1+\log n$ for fixed $s$, the operator error from
$(Ea_i)I_d$ is $O_P(\sqrt{d/n}+L^s d/n)$.
For $\|A\|_{\rm op}\le C$ the normalized trace error from this target is
$O_P(n^{-1/2}+(nd)^{-1/2})$.
\end{lemma}
\begin{proof}
Condition on $\mathcal A$. For a fixed unit $v$, the moment generating
function of $n v^T(G_a-\bar a I_d)v$ at $t$ is
$\prod_i e^{-ta_i}(1-2ta_i)^{-1/2}$. Its logarithm is bounded by
$t^2\sum_i a_i^2/(1-2|t|m_a)$ for $2|t|m_a<1$, by expanding the
logarithm. Applying this bound to both signs and optimizing $t$ gives
a probability at most $2e^{-z}$ for the absolute value to exceed
$2\sqrt{z\sum_i a_i^2}+2zm_a$. The all-zero case is immediate.
A $1/4$-net of the unit sphere has at most $9^d$ points and controls
a symmetric matrix's operator norm by twice its maximum absolute
quadratic form on the net. Set $z=x+d\log 9$ to obtain
\eqref{vtwo:weightedgram-op}. For \eqref{vtwo:weightedgram-tr}, sum the
independent centered quadratic forms $a_i(W_i^TAW_i-\tr A)/(nd)$
and use $\Var(W_i^TAW_i)=2\tr(A^2)$ by orthogonal diagonalization.
In the iid case $\bar a-Ea_i=O_P(n^{-1/2})$ and $P_n a^2=O_P(1)$.
The scalar mean error in the normalized trace is multiplied by
$\tr(A)/d$, which has absolute value at most $\|A\|_{\rm op}$.
\end{proof}

\subsection{Integrated curvature on the Fourier cap}
\label{v2:curvature}

Use the scalar model \eqref{hd:scalar-family}, write $Z=(T,W)$,
$d=q-1$, $n=n_e$, and $\ell=\log(n+e)$. For a vector of row values
write $\|v\|_{p,n}=(P_n|v|^p)^{1/p}$. All constants are uniform in
$v\in[c,C]$ and the finitely many bandwidth multipliers. In the
population-standardized truth chart $b\in\mathbb R^d$, let $L_n(b)$
denote the actual criterion after training standardization. The direction
in these coordinates is proportional to $(1,b)$. Write $j_{v,k}$ for the
finite-tuple curvature in \eqref{inf:curvature}.

\begin{proposition}[Practical-grid curvature and finite optimization]
\label{vtwo:hessian}
Take $g\asymp n^{-1/6}$, $h\asymp n^{-1/15}$,
$\lambda\asymp n^{-3/4}$ and $\rho\asymp n^{-1}$. For any
fixed $\delta>0$, suppose $2\le q\le n^{1/30-\delta}$. For fixed $K$
and $R=\min(1.5,2n^{-1/20})$, the following bound holds for every
deterministic $0\le u\le KR$, with constants independent of $u$:
\begin{equation}
 \sup_k\sup_{\|b\|\le u}
 \|-\nabla^2 L_{n,k}(b)-j_{v,k}I_d\|_{\rm op}
 =O_{P,u}\!\left(\ell^{C_0}\{q^{3/2}(n^{-1/12}+u)+q n^{-1/15}\}\right).
 \label{v2:curv-rate}
\end{equation}
Here the subscript $u$ in $O_{P,u}$ has its uniform-model meaning.
If $q^{r+1/2}=o(n^{9/20})$, where $r$ is the Fourier order in
Lemma~\ref{hdg:fourier}, the implemented algorithm attains a distance
$O_{P,u}(n^{-1})$ from the unique cap maximizer, for every preliminary
and final tuple. The distance of each returned direction from the truth
is $O_{P,u}(\sqrt{q/n})$. In particular,
the same dimension range makes the curvature contribution
$\sqrt q\,\sup\|-\nabla^2 L_{n,k}-j_{v,k}I_d\|_{\rm op}=o_{P,u}(1)$
on any fixed multiple of the $\sqrt{q/n}$ ball.
\end{proposition}

\begin{proof}
\emph{The index and a common integration measure.}
Let $G=P_n(Z-\bar Z)(Z-\bar Z)^T$ and put
\[
 v_b=(1,b),\qquad l_b=v_b/(v_b^TGv_b)^{1/2},\qquad
 Y_i^*=aY_i^w+c .
\]
For every scalar path $b+sw$ the criterion depends on the predictors
only through $T,b^TW,w^TW$, including the sample moments used in
path normalization. The actual working index differences are
$l_b^T(Z_i-Z_j)$, and the actual response has
$a=1/s_{Y^w}$, $c=-\bar Y^w/s_{Y^w}$. In fact a raw direction
becomes its scalar projection divided by its sample standard deviation
after whitening and normalization. Thus a deterministic analysis in
these coordinates includes all training transformations.
The Gaussian sample covariance bound in Lemma~\ref{hdg:fourier} gives
$\|G-I\|_{\rm op}\le C\sqrt{q\ell/n}$ with probability tending to one,
and $|a-1|+|c|\le C\ell/\sqrt n$ with probability tending to one.
Along $b+s w$, $\|w\|=1$, the zeroth, first and second index-vector
jets are $l,l',l''$. The explicit quotient gives
$\|l-e_1\|\le C(u+\sqrt{q\ell/n})$ and
$\|l'\|+\|l''\|\le C$ at $s=0$. For concentration we allow these
three vectors to vary freely within those deterministic bounds. This
avoids a mesh on the $q\times q$ covariance matrix.

Put $B_n=C_A\sqrt{q\ell}$. Outside an event of probability $O(n^{-A})$
all observations satisfy $\|Z_i\|\le B_n$,
$|T_i|+|Y_i^w|+|\varepsilon_{v,i}|\le C_A\sqrt\ell$.
This follows by Gaussian and mixture tails over $n(q+2)$ scalar
coordinates. Use the same bounds with a larger constant to truncate
inner empirical-process summands. Tail expectations of any fixed-degree
polynomial in $\|Z\|$, $|Y^w|$, multiplied by fixed powers of
$n,g^{-1},h^{-1},\lambda^{-1},\rho^{-1}$, can be made $O(n^{-D})$
for any prescribed fixed $D$ by increasing the constant (use
Cauchy--Schwarz and $q\le n$). Write
$p_l(x)=\phi_{\sqrt{\|l\|^2+g^2}}(l^Tx)$,
$p_i=\phi_{\sqrt{1+g^2}}(T_i)$. On the evaluation windows,
$|l^Tx-T_x|\le C(u+\sqrt{q\ell/n})B_n$.
The Gaussian log density formula gives
\[
 \sup_i\sup_l |\log p_l(Z_i)-\log p_i|
 \le C(u+\sqrt{q\ell/n})\{B_n\sqrt\ell+\ell\}=o(1).
\]
The quadratic increment is absorbed because the projected increment is
$o(1)$. Hence all rows and directions have the same baseline
density factors $p_i$ up to bounded multipliers.

\emph{Regression jets.}
Write $\widehat M$ for the ridged leave-one-out fit to $Y^*$ and
$M_{g,\lambda}$ for the population ratio at the deterministic parameters
$(l,a,c)$. For $j=0,1,2$ put
$\sigma_j=(ng^{2j+1})^{-1/2}$.
For both numerator and denominator the directional derivative deviations
of order $j$ satisfy, on the preceding observation window,
\begin{equation}
 C\ell^{C_1} q^{j/2}
 \left\{\sqrt{q p_l(x)}\,\sigma_j+\frac{q}{ng^{j+1}}\right\}
 \label{v2:bern-reg}
\end{equation}
as a simultaneous upper bound, with probability tending to one.
We give the net calculation. Every differentiated summand is a sum of
$K_g^{(t)}(l^T(x-Z'))$, $0\le t\le j$, times at most $j$
linear projections of $x-Z'$ along bounded vectors, optionally times
$Y^{*\prime}$. Its truncated envelope is
$C\ell^C q^{j/2}g^{-j-1}$ and its second moment is at most
$C\ell^Cq^j g^{-2j-1}p_l(x)$, by
$|K_g^{(t)}|^2\le C_tg^{-2t-1}K_g$.
For fixed mesh points these are averages of independent rows.
Mesh $(l,l',l'',x,a,c)$ at Euclidean spacing $n^{-M}$ in each vector
or scalar. Its log cardinality is at most $C_M q\ell$ by the Euclidean
ball covering bound. Bernstein's inequality with threshold
$C_{A,M}q\ell$ and a union bound give \eqref{v2:bern-reg}.
Interpolation is valid on the truncated windows: one more differentiation
of the scalar kernel costs at most $C g^{-5}(1+B_n)^4(1+\sqrt\ell)$,
after enlarging the constant (or replacing $g^{-5}$ by $1+g^{-5}$).
Every population counterpart obeys the same polynomial bound by
integration and the preceding tail estimate. Thus a fixed sufficiently
large $M$ makes interpolation smaller than $n^{-3}$.
The powers bounding these Lipschitz constants are independent of $q$.
The scalar Gaussian density changes by a bounded factor between
neighboring mesh points. Omitting any row costs the envelope divided by
$n-1$, already included in the linear term.

At order zero the relative denominator error after division by
$p_l(x)+\lambda$ is at most
$C\ell^C\{\sqrt{q/(ng\lambda)}+q/(ng\lambda)\}=o(1)$.
This compares two protected denominators; neither is bounded below by
a fixed constant. The displacement moment bound \eqref{op-moments}
applies to the NW weights with floor $(n-1)\lambda$.
It bounds each empirical denominator derivative divided by its ridged
denominator by $C\ell^C q^{j/2}g^{-j}$, $j\le2$.
The population ratio has jets of order $j$ bounded by
$C\ell^C q^{j/2}$. Indeed the conditional mean of the true $T'$ given
$l^TZ'+gZ''=l^Tx$ is $l_1 l^Tx/(\|l\|^2+g^2)$ and its conditional
variance is $1-l_1^2/(\|l\|^2+g^2)$, with $Z''$ an independent
standard normal. The heat derivative formula in Section~\ref{gauss-section}
bounds derivatives of this conditional mean regression and of its
ordinary kernel bias. Derivatives of the ridge factor
$p_l/(p_l+\lambda)$ use logarithmic derivatives of $p_l$; each order
costs at most $C\ell^C\sqrt q$ on the window. The same calculation
bounds the difference between the ridged population ratio and the ideal
conditional mean $M_I=E(Y^*\mid l^TZ=l^Tx)$, through order $j$, by
$C\ell^C q^{j/2}\{g^2+\lambda/(p_l(x)+\lambda)\}$.
For the smoothing part set $g^2$ to zero in the two conditional
parameters and use one further heat derivative, allowed by $C^6$.

Subtract the two numerator identities
$(\widehat D+\lambda)\widehat M=\widehat N$ and
$(D+\lambda)M_{g,\lambda}=N$, and differentiate their difference.
At order $j$ each error of lower order $t$ is multiplied by at most
$C\ell^C q^{(j-t)/2}g^{-(j-t)}$. Induction in $j$ therefore yields
\eqref{v2:bern-reg} divided by $p_i+\lambda$ for the stochastic
profile error at row $i$. The scalar Gaussian integrals
\eqref{gauss-G4} and \eqref{gauss-G11} give
\[
 E\frac{p_i}{(p_i+\lambda)^2}\le C\ell^C,\quad
 E\frac1{(p_i+\lambda)^2}\le C\ell^C/\lambda,\quad
 E\left(\frac{\lambda}{p_i+\lambda}\right)^2\le C\ell^C\lambda .
\]
Markov applied once to each shared empirical average controls every
parameter on the net and in its interpolation. The integrated linear
Bernstein term is bounded by $C\sqrt{q/(ng\lambda)}$ times the
square-root term before logarithms.
For $r=Y^*-\widehat M$, $e_b=Y^*-M_I$ we have proved the directional
jet bounds
\begin{equation}
 U_j:=\sup_{b,w}\|\partial_w^j(r-e_b)\|_{2,n}
 =O_{P,u}\!\left[
 \ell^C q^{j/2}\{g^2+\sqrt\lambda+\sqrt q\,\sigma_j\}\right],
 \qquad j=0,1,2 .
 \label{v2:ured}
\end{equation}
Throughout the suprema, $\|b\|\le u$ and $\|w\|=1$.
On the same event the ideal first-jet empirical $L^2,L^4$ norms are
at most $C\ell^C\sqrt q$ and the second-jet $L^2$ norm is at most
$C\ell^C q$. The proof of \eqref{op-hessian-transfer} tests a fixed Rayleigh
direction. Hence $U_1,U_2$ suffice in place of
the empirical norms of vector and matrix row errors. For the second-jet discrepancy, keep the empirical Cauchy--Schwarz bound
$|DF_r[v]|=|P_n a(r)v|\le\|a(r)\|_{2,n}\|v\|_{2,n}$.
This gives
\begin{align}
 &\sup_{b}\|\nabla^2 F(r)-\nabla^2 F(e_b)\|_{\rm op}\notag\\
 &\quad\le C\ell^C\left\{\frac{qU_0}{h^3}
 +\frac{U_1(\sqrt q+U_1)+qU_0}{h^2}
 +U_2\sup_b\|a(r)\|_{2,n}\right\}.
 \label{v2:fitH}
\end{align}
Here $F$ is \eqref{op-definition}, composed in each Hessian with the
indicated residual map. We next control the last scalar factor together with the ideal Hessian.

\emph{Density rows at ideal residuals.}
Let $f$ be the standardized error density,
$f_h=K_h*f$, $t=n^{-1/5}$, and mark row $i$ as good if
$f_h(\varepsilon_{v,i})\ge t$. Write
$e_b=a(\varepsilon_v+\Delta_b(Z))$ with
$\Delta_b=m_v(T)-E\{m_v(T)\mid v_b^TZ\}$. The scalar conditional
mean depends only on the line through $l_b$, hence no $G$ occurs in
this ideal process. The conditional Gaussian formula gives
$|\Delta_b(x)|\le C\min\{1,u(1+\|x\|)\}$ and
$|\partial^j\Delta_b(x)|\le C(1+\|x\|)^j$ for $1\le j\le3$,
as multilinear operator bounds. Integrating the inner error first gives
\[
 D_b(z)=E K_h\{e_b(z)-e_b(Z',\varepsilon_v')\}
 =a^{-1}E_{Z'}f_{h/a}\big(\varepsilon_v+
           \Delta_b(x)-\Delta_b(Z')\big).
\]
On the predictor windows the shift is $O(uB_n)$, and
$u B_n\sqrt\ell=o(1)$. Gaussian component log-density and scale
derivatives show $D_b(z)\asymp f_h(\varepsilon_v)$ there, uniformly
in $a$. The integral outside the enlarged predictor window is
$O(n^{-D})$ for every prescribed $D$ by increasing its constant.
For $|\varepsilon_v|\le C_A\sqrt\ell$ the comparison density is at
least $n^{-D_A}$ for fixed $D_A$, so this truncation also works for
relative derivative bounds. Differentiating the integrated display
shows, on these windows,
$|\partial_w^jD_b(z)|/D_b(z)\le C\ell^C q^{j/2}$, $j=1,2$,
using the mixture derivative ratios \eqref{gm:mixture-envelope}.

For the empirical average $\widehat D_b(z)$, differentiate its
unintegrated kernel. At order $j\le2$ the truncated envelope is
$C\ell^C q^{j/2}h^{-j-1}$ and the variance is at most
$C\ell^C q^j h^{-2j-1}f_h(\varepsilon_v)$ by the squared-kernel
inequality and the comparison just proved. Mesh $(b,w,x,\varepsilon_v,a)$
as above, with log cardinality $C_Mq\ell$.
The derivative envelopes through order three, the population heat
identities and $h^{-1}=O(n^{1/15})$ give polynomial Lipschitz constants
of fixed degree in $n$; taking $M$ large makes interpolation $O(n^{-3})$.
Deletion again costs the envelope divided by $n-1$. For good rows the
relative errors in the value and directional jets are consequently
bounded by $C\ell^C q^{j/2}V_j$, where
\[
 V_j=\sqrt{q/(nh^{2j+1}t)}+q/(nh^{j+1}t).
\]
This is a bound at deterministic evaluation points before evaluating the
rows of the same sample. Since $\ell^C V_j=o(1)$ for $j\le2$,
expanding $D''/(D+\rho)-[D'/(D+\rho)]^2$ bounds the good-row
Hessian error by $C\ell^C qV_2$.
For a bad row the deterministic outgoing displacement bounds applied to
its log row give $C\ell^C qh^{-2}$ for the empirical Hessian:
each term contains at most two first residual jets or a second jet.
The population log row has bound $C\ell^Cq$. The common bad-row
fraction is $O_{P,u}(t\sqrt\ell)$ by \eqref{gm:tail-integrals} and
Markov. Thus replacing the inner empirical density rows costs, uniformly
as a quadratic form, $O_{P,u}(\ell^C q[V_2+t/h^2])$.

\emph{Integrated size of the complete density score.}
We claim
\begin{equation}
 \sup_b\|a(e_b)\|_{2,n}+\sup_b\|a(r)\|_{2,n}
   =O_{P,u}(\ell^{C}).
 \label{vtwo:densityscoreH}
\end{equation}
To prove it, write $p$ for the density of $e_b$ at deterministic $(b,a)$,
$d_h=K_h*p$ and $D_j=d_h(e_{b,j})$. This $p$ is a mixture of normals
with bounded locations and variances in a fixed positive compact
interval, also when the location mixing distribution is continuous.
Thus $p\le C d_h$ and $|p'|/p+|d_h'|/d_h\le C(1+|y|)$,
by the Gaussian component bounds. Let $\widehat D_j$ be the
leave-one-out density at $e_{b,j}$. Fix a smooth function $0\le\chi\le1$
with $\chi(x)=0$ for $x\le1$ and $\chi(x)=1$ for $x\ge2$,
and put $\chi_j=\chi(D_j/t)$. In what follows the set of rows with
$D_j<2t$ is contained in the single mark
$\{f_h(\varepsilon_{v,j})<Ct\}$, by the common comparison above.
Its empirical fraction is $O_{P,u}(t\sqrt\ell)$.

For the outgoing coefficient at rows with $D_j\ge t$, use the same
Bernstein net for the scalar-argument derivative $K_h'$ (no direction
jets). Its squared envelope integrated over a source is
$Ch^{-3}D_j$ and its maximum is $Ch^{-2}$. Hence, relative to $D_j$,
the error of the derivative is at most
$C\ell^C\{\sqrt{q/(nh^3t)}+q/(nh^2t)\}=o(1)$.
The order-zero relative denominator bound is $C\ell^CV_0=o(1)$,
and $|d_h'|/d_h\le C(1+|e_{b,j}|)$.
The remaining outgoing rows have maximum $C\ell^C/h$ by
\eqref{op-moments}, so their empirical $L^2$ contribution is at most
$O_{P,u}(\ell^C\sqrt{t}/h)$.

For the incoming coefficient, form the nonnegative matrix
\[
 A_{ji}=\frac{\ind\{i\ne j\}|K_h'(e_{b,i}-e_{b,j})|}
                   {(n-1)(\widehat D_j+\rho)}.
\]
Its absolute row and column sums are each bounded by $C\ell^C/h$
by \eqref{op-moments}, so the same bound holds for its
$\ell^2$ operator norm. Deleting the complementary source weights
$1-\chi_j$ costs at most $C\ell^C\|1-\chi\|_{2,n}/h
=O_{P,u}(\ell^C\sqrt{t}/h)$. Replacing $\widehat D_j+\rho$ by
$D_j+\rho$ in the retained incoming terms costs at most
$C\ell^C V_0/h$ in empirical norm. Indeed the absolute error vector
is bounded coordinatewise by $A^T$ applied to the vector
$\chi_j|D_j-\widehat D_j|/(D_j+\rho)$, whose maximum is
$C\ell^C V_0$. Both replacements hold jointly for all locations even
though every empirical denominator uses the same sample.

For the resulting sum take the empirical process of independent
sources with deterministic $(b,a,y)$ and summand
\[
 K_h'(y-e_b(Z',\varepsilon_v'))\,
 \frac{\chi(d_h(e_b(Z',\varepsilon_v'))/t)}
      {d_h(e_b(Z',\varepsilon_v'))+\rho}.
\]
Its maximum is $Ch^{-2}/t$ and its second moment is at most
$Ch^{-3}/t^2$, since $p$ is bounded and $\int (K_h')^2=Ch^{-3}$.
The same $O(q\ell)$ log entropy now uses $(b,a,y)$ only. The cutoff
has bounded derivative at scale $1/t$, and all derivatives for
interpolation on the truncated windows are polynomial in $n$:
differentiate $d_h$ under its expectation, or use its integrated
Gaussian formula. Bernstein, interpolation and leave-one-out deletion
give centered error uniformly at the evaluation rows at most
\[
 C\ell^C\{\sqrt{q/(nh^3t^2)}+q/(nh^2t)\}=o(1).
\]
For its mean, set $w(y)=p(y)\chi(d_h(y)/t)/(d_h(y)+\rho)$.
The component ratio bounds give $|w'(y)|\le C(1+|y|)$: in the
cutoff derivative $1\le d_h/t\le2$, and $p/(d_h+\rho)\le C$.
Integration by parts turns $K_h'*w$ into $K_h*w'$, at most
$C(1+|y|)$ in absolute value. The boundary term vanishes since
$w$ is bounded. The evaluation window has $|e_{b,j}|\le C\sqrt\ell$.
On the stated grid $\sqrt t/h=O(n^{-1/30})$,
$V_0/h=o(1)$ and $nh^3t^2\asymp n^{2/5}$.
This proves the ideal part of \eqref{vtwo:densityscoreH}.
The deterministic score transfer \eqref{op-score-transfer} adds at most
$C\ell^CU_0/h^2=o_{P,u}(1)$ for fitted residuals, proving the claim.

Returning to \eqref{v2:fitH}, $U_1=o_{P,u}(\sqrt q)$.
The three stochastic powers, before their $q^{3/2}$ factor, are now
$\sigma_0/h^3\asymp n^{-13/60}$,
$\sigma_1/h^2\asymp n^{-7/60}$ and $\sigma_2\asymp n^{-1/12}$.
The bias costs at most $Cq\ell^C(g^2+\sqrt\lambda)/h^3$, so the
entire fit-transfer error is $O_{P,u}(\ell^C q^{3/2}n^{-1/12})$.

\emph{The outer average and the scalar target.}
Let $H_{b,a}(z)=\nabla_b^2\log(D_b(z)+\rho)$.
The anchor has an explicit weighted Gaussian Gram form. Set
$F=f_h$, $s=F'/(F+\rho)$, $v_2=F''/(F+\rho)$ and
$A_m=E[m_v'(T)^2]$. At $b=0,a=1$ the first two profile-discrepancy
jets are $-m_v'W$ and
$-m_v''WW^T+(2Tm_v'-m_v'')I_d$, respectively. To see this, in the
conditional Gaussian formula the longitudinal mean is
$(T+b^TW)/(1+\|b\|^2)$ and conditional variance is
$1-1/(1+\|b\|^2)$. Both jets have mean zero by Gaussian integration
by parts. Differentiating the integrated display for $D_b$ gives
\[
 H_{0,1}(z)=\big\{-s m_v''+(v_2-s^2)(m_v')^2\big\}WW^T+
             \big\{s(2Tm_v'-m_v'')+v_2 A_m\big\}I_d ,
\]
with $s,v_2$ evaluated at $\varepsilon_v$ and $m_v',m_v''$ at $T$.
The two scalar weights have uniformly bounded second moments by
\eqref{gm:mixture-envelope} and are $O_{P,u}(\ell^C)$ in sample
maximum norm. Conditionally on the scalar rows, Lemma~\ref{vtwo:weightedgram}
applied to the signed first weight bounds the centered Gram deviation
by $O_{P,u}(\sqrt{q/n}+q\ell^C/n)=O_{P,u}(\sqrt{q/n})$.
The two scalar averages fluctuate by $O_{P,u}(n^{-1/2})$, proving
$\|(P_n-P)H_{0,1}\|_{\rm op}=O_{P,u}(\sqrt{q/n})$.
A bound on the derivatives used to extend this anchor is
\[
 C(1+|\varepsilon_v|)^C e^{C|\varepsilon_v|}
 \{(1+\|x\|)^3+q^{3/2}\}
\]
for the operator norm of $\partial_bH_{b,a}$ and of $\partial_aH_{b,a}$.
To verify it compare each Gaussian component at the two locations in
the integrated display, which differ by at most $4\|m_v\|_\infty$,
to its value at $\varepsilon_v$ at the same scale $h/a$. Their ratios
are between $e^{-C(1+|\varepsilon_v|)}$ and
$e^{C(1+|\varepsilon_v|)}$. Each derivative ratio at these locations
is bounded by a polynomial in $|\varepsilon_v|$. The products in a
relative derivative of total order $j$ contain at most $j$ powers of
$1+\|x\|+\|Z'\|$, by the conditional profile bounds, and
$E\|Z'\|^k\le C_kq^{k/2}$. Products from differentiating the logarithm
have the same total order. The $a$ derivative differentiates the
prefactor and the Gaussian variances, contributing only scalar
polynomials. Independence of $Z$ and $\varepsilon_v$ bounds the expectation of the
displayed envelope by $Cq^{3/2}$ and its empirical average by
$O_{P,u}(q^{3/2})$. Hence the outer empirical
Hessian differs from $PH_{0,1}$ by
$O_{P,u}(\sqrt{q/n}+q^{3/2}(u+\ell/\sqrt n))$ uniformly on the ball.

At the anchor \eqref{eq:oracle-info} gives
$-PH_{0,1}=E[m_v'(T)^2]J_{h,\rho} I_d$. Its difference from $j_{v,k}I_d$
in operator norm is at most
$C\ell^C\{(g^2+\sqrt\lambda)/h^3+h^4+\rho\}$
by \eqref{inf:jbound} and the mixture information comparison.
All bounds are scalar here. Finally
$(nh^5t)^{-1/2}=O(n^{-7/30})$,
$(nh^3t)^{-1}=O(n^{-3/5})$ and $t/h^2=O(n^{-1/15})$.
Combining with \eqref{v2:fitH} proves \eqref{v2:curv-rate}.
The stochastic events for the row replacements and outer envelope can
all be chosen on the largest ball; the last anchor comparison then
depends on each smaller radius through $u$ only.

\emph{The finite calculation.}
The scalar information $i(v)$ is uniformly bounded above and away from
zero by \eqref{hd:info-coefficient}. The practical-grid comparison above
and \eqref{inf:jbound} give the same bounds for $j_{v,k}$ eventually.
The truth score is $O_{P,u}(\sqrt{q/n})$ by the conditional Gaussian
coefficient calculation at \eqref{dir:exactpre}: its squared conditional
norm has mean $d/n$ times the empirical coefficient second moment, which
is $O_{P,u}(1)$ by the scalar score comparison. This uses all training
transformations in that identity. The normalized raw map in
\eqref{hd:raw-map} and its inverse have first and second derivatives of
uniformly bounded norm on a fixed small ball; the extra raw Hessian
term is therefore bounded by a constant times the score norm, which is
$O_{P,u}(\sqrt{q/n}+R)$ throughout the cap by \eqref{v2:curv-rate}.
The full raw curvature hypotheses of Lemma~\ref{hdg:ascent} now follow
from \eqref{v2:curv-rate} applied on fixed multiples of the cap and
$\ell^{C_0}\{q^{3/2}(n^{-1/12}+R)+qn^{-1/15}\}=o(1)$. The four-matrix Fourier error in
Lemma~\ref{hdg:fourier} is $O_{P,u}(q^{r+1/2}/\sqrt n)=o_{P,u}(R)$.
Thus the truth and its $O(\sqrt{q/n})$ neighborhood are interior to the
computed cap. Lemma~\ref{hdg:ascent} then applies to the actual
projection, Armijo halvings, $\lceil\log^3(n+e)\rceil$ budget and
$n^{-1}$ stop. Its contraction proof gives the stated numerical error
for both fits, including the random tuple choices. In the final
assertion of the proposition, substitution of $u=K\sqrt{q/n}$ in
\eqref{v2:curv-rate} gives an upper bound of order
$\ell^{C_0}\{q^2n^{-1/12}+q^{3/2}n^{-1/15}+q^{5/2}/\sqrt n\}=o(1)$.
\end{proof}

\subsection{The case-weight covariance on the implemented shrinking grid}
\label{v2:covariance}

Work conditionally on a correct retained set in the coordinates of
\eqref{hd:scalar-family}, with $n=n_e$, $d=q-1$, and $L=1+\log n$.
Write $f_k=\kappa_k-C_k T$ and $b_k=E f_k^2$. Take all the deterministic
tuples that any stage may use, with $g\asymp n^{-1/6}$,
$h\asymp n^{-1/15}$, $\lambda\asymp n^{-3/4}$ and $\rho=1/n$
(or with fixed positive floor multiples). Use the case derivatives in
\eqref{inf:ijdefinition}, holding the tuple, location, and its tangent
frame fixed in differentiation. The powers of $L$ below are absolute
finite powers. All probability bounds are uniform over individual
correct retained models under \eqref{hd:inherit-spectrum}.

\begin{proposition}[Case-weight covariance near the fits]\label{vtwo:cov}
Let $q\le n^{1/30-\delta}$ for some $0<\delta<1/30$, and fix $M<\infty$.
Use the smooth rotated orthonormal tangent frame at a unit direction
$\gamma$ with $\|\gamma-e_1\|\le M\sqrt{q/n}$. Write $E_{k,\gamma}$
for the $n$ by $d$ matrix of case influences and $E_k^0$ for the matrix
with rows $f_k(T_i,Y_i^w)W_i^T$ in the truth frame. For some $C<\infty$,
simultaneously over this ball and all the tuples,
\begin{align}
 n^{-1/2}\|E_{k,\gamma}-E_k^0\|_{\rm op}
   &=O_P(e_B),\notag\\
 \|\widehat B_{k,\gamma}-b_kI_d\|_{\rm op}
   &=O_P(e_B+e_B^2+\sqrt{q/n}),\label{v2:cov-rate}\\
 e_B&= L^C\{n^{-1/10}+(1+q)n^{-1/30}\}.\notag
\end{align}
Thus if a preliminary or final fit is $O_P(\sqrt{q/n})$ from truth,
the bounds apply at that fit. In raw tangent coordinates replace $b_k I_d$
by $D_{0,S}^{-T}(b_kI_d)D_{0,S}^{-1}$. If the matched curvature has
operator error $O_P(e_J)=o_P(1)$, then the full case-weight sandwich has
operator error $O_P(e_J+e_B+e_B^2+\sqrt{q/n})$ about $D_{0,S}V_kD_{0,S}^T$.
For trace divided by $d$, the respective score-covariance and sandwich
errors are
\begin{align}
 &O_P\{e_B+e_B^2+n^{-1/2}+(nd)^{-1/2}\},\notag\\
 &O_P\{e_J+e_B+e_B^2+n^{-1/2}+(nd)^{-1/2}\}.
 \label{vtwo:cov-trace}
\end{align}
Proposition~\ref{vtwo:hessian} supplies $e_J$ on its dimension range.
\end{proposition}

\begin{proof}
Let $G$ be the $n$ by $q$ matrix of standard Gaussian predictors. For a
unit $\gamma$, let $x=G\gamma$, $Z_\gamma=G U_\gamma$ where
$U_\gamma$ is obtained from $U_0$ by the rotation in the plane of
$\gamma,e_1$, and let $t=(x-P_nx)/s_x$, $y=(Y^w-P_nY^w)/s_Y$.
Write $c$ for the full scalar index coefficient at unit weights, including
both the incoming and outgoing roles as in \eqref{up:weight-matrices}.
The exact weighted coefficient is
\[
 d_w=s_x(w)^{-1}\{c_w-t_w P_w(t_wc_w)\}.
\]
The derivative of the directional index along a tangent is
$s_x(w)^{-1}(I-\boldsymbol1p^T-t_w(p\odot t_w)^T)Z_\gamma$.
Since $P_wc_w=0$ by translation invariance, the score is $Z_\gamma^T
\diag(p)d_w$. With $D_{ij}=\partial_{w_j}d_{w,i}|_{\boldsymbol1}$,
$d_\gamma=d_{\boldsymbol1}$ and
$C_n=I-\boldsymbol1\boldsymbol1^T/n$, differentiation gives
\begin{equation}
 E_{k,\gamma}= C_n\diag(d_\gamma)Z_\gamma+D^TZ_\gamma .
 \label{v2:ij-matrix}
\end{equation}
Here $D\boldsymbol1=0$. The two terms in \eqref{v2:ij-matrix} are,
respectively, the variation of the normalized outer measure and of
all scalar coefficients.

\emph{The scalar Jacobian.}
Write $F_A=n^{-1/2}\|A\|_F$ and use $H,Q,P,S$ from
\eqref{up:density-weight}--\eqref{up:weight-matrices}.
The matrices $H,P$ are the protected LOO regression and density weights,
and $Q,S$ use the respective first Gaussian derivatives in the numerator.
The following inequalities hold on scalar arrays of size at most $L^{C_0}$
with scales in a fixed positive compact interval. Constants may depend
on $C_0$. With $F_H,F_P$ evaluated at the actual standardized inputs,
\begin{align}
 \|D\|_{\rm op}&\le L^C/(g h^2),\notag\\
 F_D&\le L^C\left\{\frac{F_P}{gh}
             +\frac{F_H}{gh^2}+\frac1{\sqrt n gh^2}\right\}.
 \label{v2:ij-ledger}
\end{align}
At fixed raw index $x$, consider two raw response arrays differing by
$e$ whose straight-line interpolation has the indicated scale and
maximum bounds. Then
\begin{equation}
 \|D(x,Y+e)-D(x,Y)\|_{\rm op}
     \le L^C\|e\|_\infty/(g h^3).
 \label{v2:ij-y}
\end{equation}

For a normalized
kernel matrix $H$ or $P$ at bandwidth $s$, multiplication of its entries
by any fixed polynomial of the standardized displacement has absolute
row and column sums bounded by $L^C$, by Lemma~\ref{op-moment-lemma}.
The same matrices have scaled Frobenius norm at most $L^C F_H$ or
$L^C F_P$, respectively, plus $O(n^{-4})$. Split the displacement at
$A\sqrt L$ and use the protected polynomial denominator and Gaussian
tail on the complement, choosing $A$ large enough. A floor or ridge
completion has entries at most $1/(n-1)$ in its added matrix, hence
scaled Frobenius norm $O(n^{-1/2})$. The score map $r\mapsto a(r)$
has sup norm bounded by $L^C/h$, its first derivative has operator and
maximum absolute row-sum norms bounded by $L^C/h^2$, and its second
derivative contracted with $v$ has operator norm at most
$L^C\|v\|_\infty/h^3$. To check the bounds also in absolute row and
column sums, the derivative of a normalized entry $P_{ij}$ in direction
$v$ is $P_{ij}(l_{ij}-\sum_jP_{ij}l_{ij})$,
$l_{ij}=-(r_i-r_j)(v_i-v_j)/h^2$, with zero floor velocity.
Differentiating $-(r_i-r_j)P_{ij}/h^2$ again gives the preceding
bounds by the polynomial displacement estimates. Equivalently
$D_ra$ is a sum of diagonals, row kernels, transposes, and products
of these, each with two derivatives in total. One further derivative
in direction $v$ adds at most $L^C\|v\|_\infty/h$.

The bounds at \emph{fixed} standardized arrays are listed
below. In each column a factor $L^C$ is suppressed. Put
$u_H=F_H+n^{-1/2}$, $u_P=F_P+n^{-1/2}$. The entries follow in the
displayed order from \eqref{up:density-weight}--\eqref{up:weight-matrices}.
\begin{center}
\begin{tabular}{lcc}
matrix & operator bound & scaled Frobenius bound\\ \hline
$D_m$ & $1$ & $u_H$\\
$D_A$ & $1/g$ & $u_H/g$\\
$\partial_w a|_r$ & $1/h$ & $u_P/h$\\
$D_a$ & $1/h^2$ & $u_P/h+u_H/h^2$\\
$D_c$ & $1/(gh^2)$ & $u_P/(gh)+u_H/(gh^2)$
\end{tabular}
\end{center}
For the last row, $B_{ij}=Q_{ij}(y_j-m_i)$ and
$A_i=\sum_j B_{ij}$ have bounds $L^C/g$ in operator norm and
sup norm, respectively. Bound the Frobenius norm of each product by
one Frobenius factor and operator bounds on the others, using the
$B$ factor itself when the term contains $I-M_g$. The derivative in a
standardized response direction $v$ sends $m$ to $Hv$,
$r$ to $(I-H)v$, each bounded by $2\|v\|_\infty$ in sup norm.
It sends $D_m$ to $H\diag(v)-\diag(Hv)M_g$, and $A,B,D_A$ to their
linear response variations, with operator or sup bounds
$L^C\|v\|_\infty/g$ for the latter three. Differentiating $P,S$,
their row sums, $M_D$, and $D_ra$ costs at most one further
$L^C\|v\|_\infty/h$ in their respective operator bounds. The
product rule in the last row therefore gives
$\|\partial_y D_c[v]\|_{\rm op}\le L^C\|v\|_\infty/(gh^3)$.

For the training moments, a weight changes the standardized index by
\[
 \partial_{w_j} t_i
 =-t_j/n-t_i(t_j^2-1)/(2n),
\]
and the same identity holds with $t$ replaced by $y$ for the standardized
response. The index shift cancels from $c$. For the index dilation
$\dot t=t$, the kernel log-velocity is
$-(t_i-t_j)^2/g^2$. Hence the absolute row and column sums of
$\dot H,\dot Q$ are bounded by $L^C,L^C/g$ without an extra $g^{-1}$.
It follows that $\|\dot m\|_\infty\le L^C$ and
$\|\dot c\|_\infty\le L^C/(gh^2)$ by the product rule in
$c=(B^T-\diag(A))a$. The two response directions $\boldsymbol1,y$
give the same bound for $\dot c$. These three variations contribute
rank-one matrices $u v^T/n$ to the full coefficient derivative,
with $\|u\|_\infty\le L^C/(gh^2)$ and $\|v\|_\infty\le L^C$.
A response perturbation of size $\|e\|_\infty$ has a standardized
velocity of size at most $L^C\|e\|_\infty$ by differentiating the
sample mean and scale. In each of these rank terms, one more response
derivative of $\dot c$ has sup norm at most $L^C\|e\|_\infty/(gh^3)$, because
the factors $B,A$ and their index-dilation derivatives are linear
in $y$, the first two residual derivatives are bounded by
$L^C\|e\|_\infty$ when one is a response perturbation, and the
differentiated density factor uses $D_r^2a$. For the last sup norm bound
on $D_r^2a[v_1,v_2]$, use the same normalized-entry formula twice
and the absolute row and column sums, which give
$L^C h^{-3}\|v_1\|_\infty\|v_2\|_\infty$.
Finally $d_w$ adds the scale derivative, the index derivative multiplied
by $P_w(t_wc_w)$, and the derivative of this last scalar. The contribution
from differentiating $c_w$ is $(I-tt^T/n)D_wc/s_x$; the other terms are
rank-one matrices with the same bounds, since the outer probability
derivative of $P_w(t_wc_w)$ has entries
$\{t_jc_j-P_n(tc)\}/n$. All rank contributions have operator bound
$L^C/(gh^2)$ and scaled Frobenius bound $L^C/(\sqrt n gh^2)$.
Their one-response derivative has operator bound
$L^C\|e\|_\infty/(gh^3)$. This proves
\eqref{v2:ij-ledger} and \eqref{v2:ij-y}, including the LOO
mass and training-moment derivatives.

\emph{A shared scalar envelope.}
Take the anchor $\gamma=e_1$, with the original response. Before
training standardization, with $q_g=K_g*\phi$, the conditional row
Bernstein bound for the unprotected kernel denominator shows it differs
from $q_g(T_i)$ by $o(q_g(T_i)+\lambda)$ simultaneously in $i$.
Indeed its conditional variance is bounded by $Cq_g(T_i)/(ng)$
and the bounded-summand term is $C/(ng)$, while
$ng\lambda=n^{1/12}$ up to constants. With this denominator
comparison, integrating the squared entries of $H$ bounds
$F_H^2$ in probability by $L^C/(ng)$:
$K_g^2\le C K_g/g$ and the remaining integral is
$\int \phi q_g/(q_g+\lambda)^2\le L^C$ by Gaussian tail splitting.
For density rows at $R_i=Y_i^w-M(T_i)$ use $t_D=n^{-2/3}$ and
$d_h=K_h*p_R$. Rows with $d_h(R_i)\ge t_D$ have relative denominator
error $o(1)$ simultaneously, because $nht_D\asymp n^{4/15}$.
Here $p_R\le C d_h$ and $P\{d_h(R_i)<t_D\}\le L^Ct_D$ by
\eqref{dir:mixture}. The integrated good-row squared weights are
$O_P(L^C/(nh))$ by the same calculation, while every other row has
squared mass at most one. Thus $F_{P(R)}=O_P(L^C[(nh)^{-1/2}
+\sqrt{t_D}])$. The zeroth scalar jet bound at truth gives
$\|r-R\|_{2,n}=O_P(L^C s_0)$ with
$s_0=(ng)^{-1/2}+(ng\sqrt\lambda)^{-1}$.
In the displayed derivative of $P_{ij}$ above, use $P_{ij}^2\le P_{ij}$
and the two endpoint second displacement moments. This gives
$F_{\partial_rP[v]}\le L^C\|v\|_{2,n}/h$ on any segment, hence
$F_{P(r)}\le F_{P(R)}+O_P(L^Cs_0/h)$.

We next obtain the same envelopes at \emph{every} point of the local
ball for both the true and the auxiliary responses. Set
$Y_i^\gamma=m_v((G\gamma)_i)+\eps_{v,i}$ and
$U=M\sqrt{q/n}$, $a=U(\sqrt q+\sqrt L)$. The Gaussian row maximum
is $O_P(\sqrt q+\sqrt L)$ (apply the chi-square exponential bound
and a union bound over rows), and $\|G\|_{\rm op}/\sqrt n=O_P(1)$ by
the same bound on a $1/4$-net. The maximum index and coupled response
changes from the raw anchor are $O_P(a)$. The anchor means and scales
are $O_P(n^{-1/2})$ from their population values. The same comparisons
after sample standardization cost at most $L^C$ since all index and
response maxima in this ball are $O_P(\sqrt L+1)$ and their scales stay
bounded above and below. This also holds along the response interpolation
between $Y^w$ and $Y^\gamma$. If two index arrays on these events differ
by $\Delta$ in sup norm, the normalized-entry derivative with bandwidth
$g$ bounds the maximum change of their fitted regression means by
$L^C\Delta/g$, plus the maximum response change. Hence, comparing
local standardized arrays to the unstandardized anchor costs
$O_P(L^C(a+n^{-1/2}))$ in the index and
$O_P(L^C(a+n^{-1/2})/g)$ in the fitted residuals. The logarithms of
individual off-diagonal kernels then change by at most
\[
 O_P\{L^C(a+n^{-1/2})/g^2\},\qquad
 O_P\{L^C(a+n^{-1/2})/(gh^2)\},
\]
respectively, by subtracting the two Gaussian exponents. Both are
$o_P(1)$ under the dimension restriction. This compares every positive
entry, its protected denominator, and thus $F_H,F_P$ on a single event.
By \eqref{v2:ij-ledger}, uniformly for the auxiliary fits on the ball,
\begin{equation}
 F_D=O_P\!\left(L^C\left[
 \frac{(nh)^{-1/2}+\sqrt{t_D}+s_0/h}{gh}
 +\frac{(ng)^{-1/2}}{gh^2}+\frac1{\sqrt n gh^2}\right]\right)
 =O_P(L^C n^{-1/10}).
 \label{v2:common-F}
\end{equation}

\emph{Contraction with the tangent design.}
For deterministic $\gamma$ compute $D^\gamma$ using $Y^\gamma$.
Conditional on $(G\gamma,\eps_v)$, $Z_\gamma$ has independent
standard normal entries. For a deterministic unit contrast $v$ and
$z\ge1$, this gives
\[
 P\{\|(D^\gamma)^TZ_\gamma v\|>
      \|D^\gamma\|_F+\sqrt{2z}\|D^\gamma\|_{\rm op}
       \mid G\gamma,\eps_v\}\le e^{-z}.
\]
For example diagonalize $D^\gamma(D^\gamma)^T$ and apply the
weighted chi-square bound obtained from
$-\log(1-x)-x\le x^2/[2(1-x)]$, $0\le x<1$, to the Gaussian
moment generating function. The resulting squared threshold is
$\tr A+2\sqrt{z\tr A^2}+2z\|A\|_{\rm op}$ for
$A=D^\gamma(D^\gamma)^T$, which is no larger than the square of the
displayed bound. Take a $1/2$-net in contrast and a deterministic
$n^{-J}$-net of the location ball. Their joint cardinality is
$\exp(O(JqL))$; take $z=C_JqL$. First union the preceding conditional
tail events after averaging, with their own random thresholds. Then
apply the single envelope \eqref{v2:common-F} and
\eqref{v2:ij-ledger}. This order avoids conditioning on a location
selected using tangent coordinates.

For the interpolation from this location net, the map
$\gamma\mapsto n^{-1/2}(D^\gamma)^TZ_\gamma$ has Lipschitz
constant bounded by a fixed power of $n$ on the maximum and scale event.
Indeed the fixed-order products and sums in the Jacobian formulas have
finite depth independent of $q$, all raw kernel derivatives through
order three are bounded by a fixed power of $g^{-1}+h^{-1}$, and the
unaveraged protected denominators are at least $(n-1)\lambda$ and
$(n-1)\rho$. Each extra summed index costs at most a factor $n$,
conversion from maximum entries to operator norm costs at most $n$,
and derivatives of $G\gamma,G U_\gamma,Y^\gamma$ in a unit direction
are bounded by $O_P(\sqrt q+\sqrt L)$. The rotation can be written
$I+K+K^2/(1+\gamma^T e_1)$, $K=\gamma e_1^T-e_1\gamma^T$.
The same estimates hold on a slightly larger ball, so integration in
a local affine chart proves the assertion. Choose the fixed $J$ larger
than the resulting exponent plus two. Finally \eqref{v2:ij-y} and
$\|Z_\gamma\|_{\rm op}/\sqrt n=O_P(1)$ transfer back to the
observed response. Uniformly on the local ball,
\begin{equation}
 n^{-1/2}\|D^TZ_\gamma\|_{\rm op}
  =O_P\!\left(L^C\left[n^{-1/10}+
       \frac{\sqrt{q/n}}{gh^2}+\frac{a}{gh^3}\right]\right).
 \label{v2:shared-ij}
\end{equation}

\emph{Outer-measure term and the Gram matrix.}
At truth the scalar estimates \eqref{v2:scalar-score} and \eqref{dir:DS} give
$\|d_0-f_k\|_{2,n}=O_P(L^C n^{-7/60})$;
this is a scalar statement on $(T,Y^w)$ alone. Empirical projection of
$\kappa_k$ costs $O_P(L^C/\sqrt n)$ by its polynomial envelope.
Also $\|d_0-f_k\|_\infty=O_P(L^C/(gh))$ by the row and column bounds.
Applying the preceding conditional Gaussian inequality at this one
location to the diagonal matrix of these errors costs
$O_P(L^C[n^{-7/60}+\sqrt{q/n}/(gh)])$ for the scaled operator
norm. The oracle centering has cost
$\|P_n W f_k\|=O_P(\sqrt{q/n})$ by the scalar second moment.
For the observed response and any two local indices within
$\Delta$ in sup norm, the maximum change of $Q$ in absolute row
and column sums is $L^C\Delta/g^2$ by the normalized-entry derivative.
The corresponding maximum change of $m$ is $L^C\Delta/g$; hence the
changes of $A,B$ in sup and absolute row/column sums are bounded by
$L^C\Delta/g^2$. The change in $a(r)$ has sup norm
$L^C\Delta/(gh^2)$. By $c=(B^T-\diag(A))a$ and the formula for $d$,
\begin{equation}\label{vtwo:cov-coef}
 \sup_\gamma \|d_\gamma-d_0\|_\infty
       =O_P\{L^C a/(g^2h^2)\}.
\end{equation}
This coefficient bound holds for any radius $U$ with $a=U(\sqrt q+\sqrt L)=o(1)$, since it uses only the scalar maximum and scale bounds. The change of $s_x$ is also $O_P(a)$. The rotated design differs
from its truth value by $O_P(\sqrt n\, U)$ in operator norm, and
$\|d_0\|_\infty=O_P(L^C/(gh))$. These bounds applied to
\eqref{v2:ij-matrix}, together with \eqref{v2:shared-ij}, prove the
first assertion of \eqref{v2:cov-rate}. In particular
$a/(g^2h^2)\le C_M L^{1/2}(q+1)n^{-1/30}$.

To pass to the oracle Gram matrix use Lemma~\ref{vtwo:weightedgram}
with $a_i=f_k(T_i,Y_i^w)^2$ and
$\mathcal A=\sigma(T_{1:n},Y^w_{1:n},D_s)$. The polynomial envelope
\eqref{dir:kappaibp} and the sub-Gaussian scalar maxima give
$\max_i f_{k,i}^2=O_P(L^2)$, and $E f_k^4$ is uniformly bounded.
Hence its operator error from $b_k I_d$ is
$O_P(\sqrt{q/n}+L^2q/n)=O_P(\sqrt{q/n})$ on the stated range.
The same Gram matrix has normalized trace error
$O_P(n^{-1/2}+(nd)^{-1/2})$ for each bounded deterministic symmetric
multiplier $A$, by \eqref{vtwo:weightedgram-tr}. The first term is
the scalar weight-mean fluctuation. The second is the conditional
Gaussian fluctuation of the trace. In particular
$n^{-1/2}\|E_k^0\|_{\rm op}=O_P(1)$. Replacing $E_k^0$ by
$E_{k,\gamma}$ in its empirical Gram costs $O_P(e_B+e_B^2)$ in
operator norm by submultiplicativity. It costs the same in normalized
weighted trace, since $|\tr(AH)|/d\le\|A\|_{\rm op}\|H\|_{\rm op}$
for symmetric $H$. The raw coordinate Jacobian is
deterministic for a fixed evaluation location and differs from its truth
value by $O(U)$ in operator norm under \eqref{hd:inherit-spectrum}.
The chart is held fixed in each weight derivative, so a multiplication
by this Jacobian gives the case score in the new frame. The raw Hessian
uses its usual gradient connection in the matched curvature bound.
Uniform positive curvature and the inverse identity give an $O_P(e_J)$
cost for replacing the two estimated curvature inverses by their population
versions. The remaining trace uses the bounded deterministic multiplier
obtained by the population raw Jacobian and inverse curvature, to which
\eqref{vtwo:weightedgram-tr} applies. This proves the sandwich and fitted
trace bounds. The finite simultaneous bounds apply to the tuples used
for trace comparison at the preliminary estimate, to the selected
final tuple, and to either LOO fallback selector without a selection
probability argument. For $O_P(\sqrt{q/n})$ evaluation points first
take $M$ large and then apply these uniform bounds.
\end{proof}

\subsection{Consequences for the fitted contrast and the two angular domains}
\label{v2:practical-consequences}
The following argument states separately what is needed beyond the truth
score. Write $u_n=\sqrt{q/n}$, $R_n=\min\{1.5,2n^{-1/20}\}$ and
$D_S=\max_k D_{S,k}$. Suppose $q=o(n)$, $j_k\ge c_*>0$ uniformly, and
the finite Fourier/ascent conditions of Lemmas~\ref{hdg:fourier} and
\ref{hdg:ascent} hold. In particular $q^{r+1/2}=o(n^{9/20})$ places
the four-matrix start inside the cap. Suppose, in population-standardized coordinates, the empirical negative
Hessian has spectrum in a fixed positive compact interval throughout
each fixed multiple of the truth-centered cap, with probability tending
to one. On each fixed multiple of $u_n$ assume its operator error from
$j_kI_d$ is $O_{P,u}(e_J)$, $e_J=o(1)$, uniformly on the grid.
Assume $D_S=o(1)$ and that the final and preliminary
raw case-weight covariance errors are $o_{P,u}(1)$ in matching frames.
These are the score, curvature and fitted case-weight conclusions to
which the dimension and smoothing restrictions are applied.

\begin{corollary}[Linear term and fixed contrasts]\label{v2:contrast}
Let $D$ be \eqref{hd:raw-map}, $V_{S,k}=b_kDD^T/j_k^2$ and
$H_k=D W^T\zeta_k/(j_k\sqrt n)$. At tolerance $1/n$ the raw chart
error of each fit satisfies
\[
 \|\sqrt n\,\widehat\theta_k-H_k\|
 =O_{P,u}\{\sqrt d(D_S+e_J)+\sqrt{d/n}+n^{-1/2}\}.
\]
In particular its line error is $O_{P,u}(u_n)$. For the implemented grid, Propositions~\ref{vtwo:hessian}
and \ref{vtwo:cov} verify these inputs with
$e_J=\log^{C_0}(n+e)\{q^{3/2}(n^{-1/12}+u_n)+q n^{-1/15}\}$ throughout
$q=O(n^a)$, $a<1/30$, subject to the Fourier condition above. Thus
the Euclidean error is $o_P(1)$ throughout the stated range. For any unit raw
contrast $a$ fixed given $D_s$, $(a^T\widehat V_{S,k}a)^{-1/2}
a^T\sqrt n\widehat\theta_k\Rightarrow N(0,1)$ at a deterministic
or consistently selected tuple. On a shrinking finite grid this
conclusion applies to any training selector. On positive grids it applies
at a uniquely separated trace optimum. With tied positive optima the
selected-index expansion and the excess trace bound of
Corollary~\ref{hdg:tuning} apply. In every case $V_{S,k}/n$ is the
covariance of the deterministic-tuple linear approximation on the
unscaled parameter scale.
\end{corollary}
\begin{proof}
In standardized coordinates define $b_k^\circ$ as the unique interior
maximizer of the criterion in a fixed truth-centered Euclidean ball of
radius $K R_n$. Set it to zero if this maximizer fails to exist or is
not unique. This proof variable is equivariant under every orthogonal
rotation of the $W$ columns, including its fallback. Curvature and
\eqref{v2:gaussian-score} place the maximizer at $O_{P,u}(u_n)$, in
the actual Fourier cap, with probability tending to one. Integrating
the Hessian to this maximizer gives
$e_k=\sqrt n b_k^\circ-W^T\zeta_k/(j_k\sqrt n)
=O_{P,u}(\sqrt d(D_S+e_J))$ in norm. A deterministic multiple of $u_n$
may first be chosen with arbitrarily high probability. Given the scalar
rows, both $b_k^\circ$ and $e_k$ have rotationally invariant laws,
by \eqref{hdg:chart}. For any such vector $Z$, any fixed unit $w$, and
any $t>0$, $\E[(w^TZ)^2\boldsymbol1_{\|Z\|\le t}\mid\mathcal A]
=d^{-1}\E[\|Z\|^2\boldsymbol1_{\|Z\|\le t}\mid\mathcal A]$.
Truncating the norm at a sufficiently large multiple of its stochastic
order gives $w^Te_k=O_{P,u}(D_S+e_J)$ and
$w^Tb_k^\circ=O_{P,u}(n^{-1/2})$.

The exact raw chart map is $b\mapsto Db/(1+v_0^Tb)$, where
$v_0^T=\sqrt v\,\beta_S^T\Sigma_S^{-1/2}U_S$ has bounded norm.
Its denominator differs from one by $O_{P,u}(n^{-1/2})$ at $b_k^\circ$.
The numerical iterate is within $O_{P,u}(n^{-1})$ of the exact maximum
in either chart by Lemma~\ref{hdg:ascent}. This proves the vector bound
and the $o_{P,u}(1)$ root-$n$ contrast remainder. Conditionally on the
scalar rows $H_k$ is Gaussian with covariance $(P_n\zeta_k^2/b_k)V_{S,k}$,
and the scalar factor tends to one. Operator consistency supplies
studentization. The common scalar $L^2$ limit of $\zeta_k/j_k$ on a
shrinking grid, proved in Corollary~\ref{hdg:uniform-gain}, makes the
finite list of leading contrasts differ from their common efficient
contrast by $o_{P,u}(1)$. The excess normalized population trace is at
most twice the largest preliminary operator error by Corollary~\ref{hdg:tuning}.
\end{proof}

\begin{corollary}[Angular localization on the shrinking bandwidth path]
\label{v2:radius}
Suppose $d\to\infty$ and use the shrinking smoothing path
\eqref{inf:nonempty} under the preceding estimation and fitted covariance
conditions. Write $a_n=\log(n+e)/\sqrt n$ for the fixed-dimension raw angular
radius, keeping the mapped Fourier cap and global diagnostics. If
$q=o(\log^2 n)$, the conditional mathematical coverage on a successful
screen tends to $1-\alpha$ and its projective diameter is
$O_{P,u}(\sqrt{d/n})$. The pivot at
the true candidate before localization still has the law in
\eqref{hdg:F} whenever its projected coefficient is nonzero.
Without the restriction on $q$, the dimension-dependent radius in \eqref{hdg:radius}
has conditional mathematical coverage $1-\alpha+o(1)$ and diameter
$O_{P,u}(\sqrt{d/n})$. These radius claims concern exact arithmetic.
\end{corollary}
\begin{proof}
The eigenvalues of $V_{S,k}$ lie in a fixed positive compact interval
by \eqref{hdg:geometry} and $b_k\ge j_k^2/i(v)$. Conditional normality
and $\tr(V_{S,k}^2)/(\tr V_{S,k})^2=O(1/d)$ give
$\|H_k\|^2/\tr V_{S,k}\to_P1$. The preceding norm expansion is
$o_{P,u}(\sqrt d)$, and angle is $\arctan\|\widehat\theta_k\|$ in the
raw chart. Therefore on the simultaneous finite-grid event, including
a trace-selected index,
\[
 \frac{\sqrt n\,\operatorname{angle}([\widehat\beta],[\beta_S])}
      {\sqrt{\tr V_{S,\widehat k}}}\ \longrightarrow_P\ 1,
 \qquad
 \frac{\tr\widehat V_{S,\widehat k}}{\tr V_{S,\widehat k}}
 \ \longrightarrow_P\ 1 .
\]
Comparison with $\log(n+e)$ gives truth inclusion in the fixed-dimension
angular ball when $q=o(\log^2 n)$. At the larger trace-based radius the factor two suffices and
the diameter is at most twice the radius. The actual ascent result places
the truth inside the mapped cap, and the curvature and covariance results
imply passage of the global diagnostics with probability tending to one.

For the sharper diameter in the fixed-dimension ball with $q=o(\log^2 n)$,
the entire domain is in a truth neighborhood of radius $C a_n$ with
probability tending to one. Write $t_b$ for its standardized candidate
index. The sample spectral bound and
$\max_i\|X_{i,S}\|=O_{P,u}(\sqrt q+\sqrt{\log n})$ give
$\sup_b\|t_b-t_0\|_\infty=O_{P,u}((\sqrt q+\sqrt{\log n})a_n)$.
Equation \eqref{vtwo:cov-coef}, applied with radius $U=C a_n$ in the
population coordinates, then shows that the empirical norm of the
projected coefficient divided by the index scale is $O_{P,u}(1)$
throughout the domain. Indeed its variation is bounded in maximum norm
by $O_{P,u}(\ell_n^{C_1}a_n(\sqrt q+\sqrt{\log n})/(g^2h^2))=o_{P,u}(1)$,
since $(g^2h^2)^{-1}=O(n^{7/15})$ and $q=o(\log^2 n)$.
The empirical norm at truth is bounded by \eqref{v2:scalar-score};
the raw and population-standardized index scales have uniformly bounded
ratio by the retained spectrum.

For any accepted candidate with nonzero coefficient,
$U=z^T(Z^TZ)^{-1}z$ satisfies $U\le dF_{d,n-d-2;1-\alpha}/
(n-d-2+dF_{d,n-d-2;1-\alpha})=O(d/n)$. Also
$\|Z^TZ\|_{\rm op}/n=O_{P,u}(1)$ uniformly and all candidate index
scales are bounded below. Hence the exact raw tangent score
$Z^T\Pi c/(ns_b)$ has norm $O_{P,u}(\sqrt{d/n})$ for every accepted
candidate, including $\Pi c=0$. The local chart derivative is uniformly
bounded in both directions. Integrating the negative definite Hessian
from the truth and using the simultaneous truth score bound forces every
such candidate to lie within $O_{P,u}(\sqrt{d/n})$ of the truth,
regardless of its tuple or acceptance component.

Finally the candidate-selected tuple uses only $(D_s,X_S\beta_S,Y_e)$,
and the scalar-coefficient bound after Lemma~\ref{v2:truth-score}
makes the zero branch asymptotically negligible. Integrating the exact
conditional test first and then charging the domain exclusion gives the
claimed coverage. For a numerical set the false-exclusion term in
\eqref{hdg:coverage-bound} remains. Integration over the independent
screen adds at most $\delta_{\rm scr}$ to a failure probability by
\eqref{hd:screen-delta}. The zero insertion map preserves
projective distance, and a set with a missed active coordinate cannot
cover the true line.
\end{proof}

\section{Fixed retained dimension: the local efficient ellipsoid}
\label{hd:inference-proofs}
Fix $q\ge2$, put $n=n_e$, $d=q-1$ and use the fixed-dimensional shrinking grids of Section~\ref{hd:inheritance-proofs}. Under independent screening, the proof of Proposition~\ref{hdg:pivot} gives the same exact candidate pivot at $b_0=\beta_{0,S}$, for $n>q+1$. Use the projected quantities $x_b,t_b,\Pi_b,Z_b$ and $z_b,A_b$ of Section~\ref{hdg:inference}, and write $c_k$ for the complete coefficient, $s_b$ for the sample standard deviation of $x_b=X_S b$, and $\overline P_e$ for the estimation-sample average. In this section the final angular radius is $a_e=\log(n+e)/\sqrt n$. Intersect this ball about the fitted line with the mapped initialization cap and the global diagnostic event. The resulting domain and the false-exclusion events $\mathcal X_0,\mathcal N_0$ give the finite-sample bound
\begin{equation}\label{hd:coverage-bound}
 \Pp\{[\beta_0]\in\mathcal C_{\rm screen}^{\rm comp}\}\ge (1-\alpha)\Pp(E_{\rm scr})-\Pp(E_{\rm scr}\cap\mathcal X_0)-\Pp(E_{\rm scr}\cap\mathcal X_0^c\cap\mathcal N_0)
\end{equation}
by the same conditioning argument as \eqref{hdg:coverage-bound}. For the mathematical set the last probability is zero.
\subsection{Uniform local shape and the meaning of efficiency}

The following statement uses the uniform retained-model conditions and inheritance result of Section~\ref{hd:inheritance-proofs}, including uniform positive information and Fourier gaps. Suppose every preliminary, point, and candidate-inference tuple belongs to a fixed finite shrinking grid satisfying the covariance and expansion rates with sample size $n_e$; for example, use \eqref{inf:nonempty} with $n=n_e$. Use the original initialization cap and optimization tolerance with that same sample size. In raw retained coordinates, put $T=b_0^TX_S$, and let $f_\varepsilon$ denote the density of the original, unstandardized error. Define
\begin{equation}
 \mathcal I_S=I_\varepsilon\E\{m_0'(T)^2\}\Omega_S,
 \qquad I_\varepsilon=\int (f_\varepsilon'/f_\varepsilon)^2f_\varepsilon.
 \label{hd:information}
\end{equation}
This follows from Section~\ref{sec:core-efficiency}, since the Gaussian residual tangent $X_S^TF_{b_0}-\E(X_S^TF_{b_0}\mid T)$ is independent of $T$. It is in a raw orthonormal tangent; response standardization does not alter this information after returning to raw units.

\begin{theorem}[Retained-model efficient shape]\label{hd:shape}
Under these uniform conditions and exact arithmetic, conditional on $D_s$ on $E_{\rm scr}$, uniformly over its possible support-containing retained sets, $\Pp(\mathcal X_0\mid D_s)=o(1)$. In a raw local chart about $b_0$, let $\widehat\theta_S$ denote the point estimate and let $\Theta_{S,e}$ be the chart image of the local accepted set before zero embedding. For every $\epsilon>0$, with conditional probability $1-o(1)$,
\begin{align}
 \{v:v^T\mathcal I_Sv\le\chi^2_{q-1,1-\alpha}-\epsilon\}
 &\subseteq\sqrt{n_e}(\Theta_{S,e}-\widehat\theta_S)\notag\\
 &\subseteq
 \{v:v^T\mathcal I_Sv\le\chi^2_{q-1,1-\alpha}+\epsilon\}.
 \label{hd:ellipsoid-inclusions}
\end{align}
Every accepted point is at projective distance $O_P(n_e^{-1/2})$ from the truth, uniformly conditionally on successful screening. If $\delta_{\rm scr}\to0$, the zero-embedded set has projective diameter $O_P(n_e^{-1/2})$ and its mathematical coverage tends to $1-\alpha$. The same coverage conclusion holds for a computed subset if the numerical false-exclusion probability in \eqref{hd:coverage-bound} tends to zero.
\end{theorem}
\begin{proof}
The uniform inheritance bounds give, simultaneously over all tuples,
\begin{equation}
 \sqrt{n_e}\widehat\theta_S
 =\mathcal I_S^{-1}n_e^{-1/2}\sum_{i\in D_e}S_{{\rm eff},S}(O_i)+o_P(1),
 \qquad
 \sup_S\|\mathcal I_S^{-1}\|<\infty,
 \label{hd:common-influence}
\end{equation}
with uniformly bounded score moments. All probability remainders here are uniform retained-model bounds conditional on $D_s$. Hence the truth lies in the final angular cap with conditional probability $1-o(1)$. Given uniform root-$n_e$ Fourier consistency and a broader optimization cap, any bounded root-$n_e$ neighborhood of the truth also lies in the initialization cap with the same probability. By uniform positive curvature, covariance consistency and the original finite-iteration contraction, all inference diagnostics succeed with probability $1-o(1)$. This proves the conclusion for this domain.

The two steps establishing the shape are also uniform. Write
\[
 K_{e,k}(b)=\frac{\overline P_e\{(\Pi_bc_k)^2\}}{s_b^2}
                       \frac{Z_b^TZ_b}{n_e-2}.
\]
The proof of Lemma~\ref{cs:covariance} uses the same moment, coefficient, and preprocessing envelopes verified for the retained models. It gives $K_{e,k}(b)=\mathcal I_S+o_P(1)$, uniformly over all grid points and the $O(a_e)$ domain. Let $U_b=z_b^T(Z_b^TZ_b)^{-1}z_b$. The deterministic identities
\[
 n_e\mathcal S_b^TK_{e,k}(b)^{-1}\mathcal S_b=(n_e-2)U_b,
 \qquad dF_{{\rm CS},S}(b)=(n_e-d-2)U_b/(1-U_b)
\]
show that acceptance forces $U_b=O(n_e^{-1})$. Uniform positive curvature and Taylor's integral identity then imply $\|\theta\|\le C\{\|\nabla L_k(0)\|+\|\nabla L_k(\theta)\|\} =O_P(n_e^{-1/2})$ for every accepted point. Uniform positivity of $K_{e,k}$ also excludes $\Pi_bc_k=0$ throughout the domain with probability $1-o(1)$, so the zero-coefficient convention does not affect localization.

On every bounded $v$-ball, \eqref{hd:common-influence} and uniform Hessian convergence give, for every tuple $k$,
\[
 \sqrt{n_e}\nabla L_k(\widehat\theta_S+v/\sqrt{n_e})
          =-\mathcal I_Sv+o_P(1).
\]
This holds for the candidate-dependent inference selector without continuity of the argmin, because the grid is finite and all tuples share the same efficient influence limit. The quadratic identities above give $dF_{{\rm CS},S}=v^T\mathcal I_Sv+o_P(1)$, with $dF_{d,n_e-d-2;1-\alpha}$ converging to $\chi^2_{d,1-\alpha}$. Localization, uniform eigenvalue bounds on the information matrix and the cap inclusions give \eqref{hd:ellipsoid-inclusions}, as in the proof of Theorem~\ref{cs:efficient}. A fixed-dimensional uniform central limit theorem applied to \eqref{hd:common-influence}, after standardizing by the information matrix, gives conditional coverage $1-\alpha+o(1)$, even though $\mathcal I_S$ varies with $S$ and $n$. Integration over $D_s$ increases these error probabilities by at most $\delta_{\rm scr}$. Finally, $\iota_S$ preserves Euclidean norms, inner products, and projective distances. Hence the diameter bound remains valid after the embedding. The diameter or coverage of a resulting subset cannot exceed those of the original mathematical set. Together with \eqref{hd:coverage-bound}, the vanishing probability of improper exclusion gives the asserted coverage.
\end{proof}

The efficiency statement refers to the retained submodel and the $n_e$ estimation observations; at fixed positive smoothing, the qualifications of Theorem~\ref{up:fixed-shape} apply.

\section{Explicit power-grid bounds and screened inference}
\label{sb:relocated}
The bounds of Section~\ref{vtwo:section} give the sufficient power-grid range $a<1/30$, and the sharper bounds of Theorem~\ref{sb:dimension} give the ranges in main Theorem~\ref*{main-thm:hd-estimation}. For the first argument write $O_P^*(a_n)$ for $O_P((1+\log n)^C a_n)$ with fixed finite $C$, and put $u_n=\sqrt{q/n}$. The bound of Section~\ref{vtwo:section} on the remainder in (\ref*{main-eq:hd-expansion}) is
\[
 O_P^*\!\left(\sqrt q\{n^{-7/60}+\mathcal H_n(u_n)\}+n^{-1/2}\right).
\]
The Euclidean remainder in (\ref*{main-eq:hd-expansion}) is $o_P(1)$ throughout this range. Conditional on the indices and responses, the score at the truth is a weighted sum of independent Gaussian tangent vectors. On a truth-centered ball of radius $u\le K R_n$, with $R_n=\min(1.5,2n^{-1/20})$, the Hessian operator error in standardized coordinates is $O_P^*(\mathcal H_n(u))$, where
\[
 \mathcal H_n(u)=q^{3/2}(n^{-1/12}+u)+q n^{-1/15}.
\]
For the full case-weight score matrix and the matrix of population influence rows, the operator norm of their difference divided by $\sqrt n$ is $O_P^*(n^{-1/10}+(q+1)n^{-1/30})$ within distance $K u_n$ of the truth for each fixed $K$. Integrated tail bounds keep the density floor logarithmic, and Gaussian weighting controls the operator norm and the trace separately. The ridge cost comes from integrating the reciprocal of the projected density plus $\lambda$. The $1/q$ Fourier scale supplies a gap of order at least $q^{-r}$, and the finite ascent reaches the cap maximizer to order $1/n$ (Section~\ref{vtwo:section}).

For example, take iid $N(0,1/4)$ predictors, two equal nonzero coefficients, $m_0(t)=\sin(2t)$, $P=3n$ and $n_s=n$. For an active coordinate the conditional mean is a nonconstant sine, and inactive coordinates are independent of $Y$. This gives a fixed screening gap. Since $\E[T\sin(2\sqrt v T)]=2\sqrt v\exp(-2v)>0$, the Fourier order is $r=0$. Every $q=\max(2,\lfloor n^a\rfloor)$ with $0<a<1/30$ gives an absolute Euclidean expansion on the power grid. For bounded grids of fixed positive tuples with uniformly positive curvature, the alternative sufficient conditions in Theorem~\ref{hdg:sample} apply.

The final radius uses concentration of the squared length of the leading Gaussian vector about $\tr V_{S,k}$ as $d$ grows. The test uses only the candidate's projected index and response to select its tuple, and the fitted radius enters the domain after the test has been calibrated (Theorem~\ref{hdg:coverage}). At fixed retained dimension the angular radius $\log(n+e)/\sqrt n$ and the shrinking-grid efficient ellipsoid of Section~\ref*{main-sec:cs-inference} apply.

With $\iota_S$ the zero embedding, for every $\epsilon>0$ some $M<\infty$ satisfies
\begin{equation}\label{eq:hd-ambient-error}
 \Pp\{\dP(\iota_S\widehat\beta_S,\beta_0)>Mu_n\}\le\delta_{\rm scr}+\epsilon+o(1).
\end{equation}

The leading SI-KDR covariance and the conditional-mean bound both use the $n_e$ observations reserved for estimation, so the criterion $\Lambda(v,k)<G$ applies within each correct retained model. As all four scales shrink, $\Lambda(v,k)\to1$ uniformly for $c\le v\le C$. With non-Gaussian errors, every sufficiently small positive tuple therefore retains a strict gain. At a given $n$ the scalar law, $\Lambda$, and the population ordering of a common grid are identical for all $S$ containing the support. The tangent geometry and number of estimated coordinates enter the trace through $\Omega_S^{-1}$ (Proposition~\ref{hdg:loss} and Corollary~\ref{hdg:uniform-gain}).

A deterministic unit raw contrast with tangent variance bounded away from zero is asymptotically standard normal after studentization at a deterministic tuple, at a uniquely separated population optimum, or after selection from the implemented shrinking grid under the conditions of Theorem~\ref*{main-thm:hd-estimation}. At tied fixed positive optima the simultaneous expansion applies at the selected index.

\section{Independent screening, oracle coupling, and three screens}
\label{sb:interface}
Write $n=n_e$ and let $\mathcal F_s=\sigma(D_s,U_s)$, where $U_s$
is an independent random seed. Both the estimation sample and any fresh
estimation seed are independent of $\mathcal F_s$. Sort retained columns
by their original labels. A screen returns an $\mathcal F_s$-measurable set
$S$, possibly of random size. An inadmissible size can be flagged as a
failure, or mapped to a predetermined admissible output before analysis.
Let $\mathfrak S_n$ be a deterministic class of correct models with
$S_0\subseteq J$, $2\le |J|<n-1$. Put $E=\{S\in\mathfrak S_n\}$
and assume $\Pp(E^c)\le\delta_n$. If correct retention and admissibility
are controlled separately, use the sum of their failure probabilities.

\begin{theorem}[Transfer to a random retained experiment]\label{sb:transfer}
Assume the fixed link and error class and common spectral bounds of
Section~\ref{hd:uniform-assumptions}. In every $J\in\mathfrak S_n$ run the
same specified estimation procedure on $D_e$ using $n$ and $|J|$.
Require the hypotheses of the transferred retained result to hold uniformly
over these individual experiments. In particular the Fourier signal and
start, curvature, grids, training standardization, finite optimization
and numerical tolerances must be those for which the result is proved.
For the power or logarithmic schedule use the conclusion-specific ranges of Theorem~\ref{sb:dimension} with $\sup_J |J|=O(n^a)$ and $\sup_J |J|^{r+1/2}=o(n^{9/20})$; the four-matrix start, cap and finite ascent are part of both procedures. At fixed dimension the alternative conditions
of Theorem~\ref{hd:stage-two} apply. If $p_{n,J}$ is the bad-event
probability in a specified deterministic retained experiment, then
\begin{equation}
 \Pp(E\cap\mathrm{bad})=\E[\ind_E p_{n,S}],\qquad
 \Pp(E^c\cup\mathrm{bad})\le\delta_n+
                 \sup_{J\in\mathfrak S_n}p_{n,J}.
 \label{sb:kernel}
\end{equation}
Consequently the line rate is $O_P(\sqrt{|S|/n})$ when
$\delta_n\to0$. On $E$, conditional on $\mathcal F_s$, the full joint
grid expansions, fitted case-weight covariance bounds, and trace excess
per tangent coordinate hold at their retained rates. Candidate
calibration uses the realized degrees of freedom. For a computed
confidence set,
\begin{equation}
 \Pp\{[\beta_0]\in\mathcal C^{\rm comp}\}\ge
 (1-\alpha)\Pp(E)-\Pp(E\cap\mathcal X_0)
                  -\Pp(E\cap\mathcal X_0^c\cap\mathcal N_0),
 \label{sb:screen-cover}
\end{equation}
where $\mathcal X_0$ includes global failure or exclusion of the truth by the
search domain, and $\mathcal N_0$ is numerical false exclusion within the domain.
\end{theorem}
\begin{proof}
Given $\mathcal F_s$, $S=J$ is fixed. Independence preserves the law of
the estimation sample and seed. On $E$ the regression index in $J$
equals the full index observation by observation, the raw covariance is
$\Sigma_{JJ}$, and its spectrum lies in the stated common interval.
Thus the conditional bad-event probability is exactly $p_{n,J}$.
Integration proves \eqref{sb:kernel}, including events whose tolerance
or normalization depends on $|J|$. This requires no simultaneous
probability bound across subsets. Uniformity over variable sizes
also follows from uniformity along every admissible deterministic
sequence: otherwise a subsequence of sizes and sets violating the
bound would give a contradicting deterministic sequence.

In particular the root-$n$ linear term for tuple $k$ is
$n^{-1/2}\sum_i D_JW_i\zeta_{k,i}/j(v,k)$ with $D_J$ in
\eqref{hd:raw-map}, $v=\beta_0^T\Sigma_n\beta_0$ and
$\zeta_k=\kappa_k-T\E(T\kappa_k)$. The incoming regression refit,
density-center contribution and training scale correction thus remain
in the influence. Equation~\eqref{hdg:gain} supplies its covariance
coefficient. Write $t_{J,k}=\tr V_{J,k}/(|J|-1)$ and $\hat t_{J,k}$
for the corresponding algebraic trace at the preliminary fit. On the
event that all final tuples are eligible, minimization gives
$t_{J,\hat k}-\min_k t_{J,k}\le 2\max_k|\hat t_{J,k}-t_{J,k}|$.
On the two shrinking schedules, Theorem~\ref{sb:dimension} controls this
maximum and gives eligibility with probability tending to one. Its
Frobenius bound suffices on the logarithmic schedule for $a<2/3$,
including the range beyond operator consistency. For other grids the
conditions of Corollary~\ref{hdg:tuning}, when imposed, suffice via its
operator bound. Transfer of the final positive-definite covariance and
studentization uses the respective operator range and, for studentization,
the raw contrast variance floor. The grid expansion
at a selected tuple allows ties. A selected contrast CLT follows at a
consistently chosen tuple or under a common limiting influence, as in
Corollary~\ref{v2:contrast}. A contrast measurable with respect to
$\mathcal F_s$ is fixed in this argument. At the true line,
Proposition~\ref{hdg:pivot} gives the $F_{|S|-1,n-|S|-1}$ pivot
conditionally on the screening information, true index and responses
when its projected coefficient is nonzero. The zero coefficient accepts.
Subtracting domain and numerical exclusions proves
\eqref{sb:screen-cover}. Diameter and asymptotic localization follow under
the fixed- or growing-dimension radius conditions of
Sections~\ref{hd:inference-proofs} and~\ref{hdg:inference}.
\end{proof}

\begin{corollary}[Coupling to the known-support fit]\label{sb:oracle}
Suppose $s=|S_0|\ge2$, the experiment on $S_0$ is admissible, and
$\Pp(S\ne S_0)\le\delta_n^{=}$.
Construct the known-support SI-KDR on the same $D_e$, using the same
preprocessing, all candidate grids, data-dependent rules, initial
direction, seed, tie resolution and failure conventions. Any additional
inputs to these rules must coincide when the support coincides.
The entire zero-embedded output agrees with that oracle with probability
at least $1-\delta_n^{=}$. At a deterministic tuple of positive
curvature its raw tangent covariance coefficient is
\[
 V_{S_0,k}=\frac{\Lambda(v,k)}{G}\,V_{{\rm mean},S_0},\qquad
 V_{{\rm mean},S_0}=\frac{\Var(\varepsilon)}{A(v)}\Omega_{S_0}^{-1}.
\]
Both leading estimator covariances use denominator $n_e$. On a shrinking
schedule satisfying the retained expansion and common-influence
conditions, the oracle efficient limit for the independent-error model
therefore transfers when $\delta_n^{=}\to0$. This is joint at fixed $s$
and applies to each eligible, nondegenerate unit contrast at growing $s$.
\end{corollary}
\begin{proof}
Let $H(J,D_e,U_e)$ be the full implementation, including its failure
return. Both calls have equal arguments on $\{S=S_0\}$. For every
deterministic scaling $a_n$ and every $t>0$ the probability that
$a_n$ times the distance between the two embedded line outputs exceeds
$t$ is at most $\delta_n^{=}$ (use a common coded output on failure).
The coupling applies to any measurable diagnostic and bounds the
difference of the two output laws in total variation by $\delta_n^{=}$.
For $|S|=s$ everywhere, correct retention implies equality, so the
sure-screening bound suffices in this branch. With $s=1$ an identified
support already specifies the line, and the tangent dimension is zero.

In a raw orthonormal tangent the independent-error information is
$A(v)I_\varepsilon\Omega_{S_0}$, by Proposition~\ref{hdg:loss}.
Equation~\eqref{hdg:loss-square} gives the squared $L^2$ distance
between the finite-tuple and efficient scalar parameter influences,
$(\Lambda-1)/i(v)$. On the admitted shrinking path this vanishes.
Bounded singular values of $D_{S_0}$ transfer this convergence to every
fixed unit contrast, jointly in finitely many contrasts by linearity.
Combine the retained remainder and studentization bounds with the
coupling.

For a nonnegative loss bounded by $L_{\max}$ the difference between
screened and oracle expected losses has absolute value at most
$L_{\max}\delta_n^{=}$. More generally the absolute difference is at most
$\E[\ind_{\{S\ne S_0\}}(L_{\rm scr}+L_{\rm or})]$.
For example, a leading risk of order $(s-1)/n_e$ transfers as a
relative expansion under bounded loss if
$\delta_n^{=}=o((s-1)/n_e)$ and the oracle itself has a mean-risk
expansion. For an overscreen, the exact risk identity is
$\E[\ind_E R_n(S)]+\E[\ind_{E^c}L_{\rm scr}]$, where $R_n(J)$
is the unconditional estimation risk for fixed $J$. Holding the
law on $S_0$ fixed, $P$ enters the exact-support coupling only through its
failure probability.
\end{proof}

\subsection{Distance screening, including a correlated even-link example}
\label{sb:dc}
Use the ordinary biased squared distance correlation in
\eqref{hd:screen-stat}, the stated tie rule and top $q$. By
Theorem~\ref{hd:screen-rank} this screen has $\delta_n=\delta_{\rm scr}$
of \eqref{hd:screen-delta} if the weakest active score exceeds the
$(q-s+1)$st largest inactive score by $4A n_s^{-\nu_1}$.
Taking $\nu_2=(1-2\nu_1)/3$ allows
$\log P=o(n_s^{(1-2\nu_1)/3})$. At $q=s$ the comparator is the
strongest inactive. The ordinary row and grand means, diagonal terms and
zero-denominator convention are essential for the cited normalization
\citep{li2012distance}.

There is also a quantitative ordering for correlated Gaussian features.
Let $T$ be the standardized true index and $u(r)$ the population squared
distance correlation between $Y$ and a standard normal feature having
correlation $r$ with $T$, with the response generated conditionally
through $T$. Then $u(-r)=u(r)$ and
\begin{equation}
 u(r_1)\le |r_1/r_2|u(r_2),\qquad 0\le |r_1|<|r_2|\le1.
 \label{sb:dc-attenuate}
\end{equation}
Indeed, after sign changes couple $Z_1=aZ_2+\sqrt{1-a^2}G$ with
$a=|r_1/r_2|$ and $G$ independent of $(Z_2,T,Y)$.
For $0<a<1$, put
$D_j(t,z)=\phi_{Z_j,Y}(t,z)-\phi_{Z_j}(t)\phi_Y(z)$.
Then $D_1(t,z)=e^{-(1-a^2)t^2/2}D_2(at,z)$.
In the scalar distance-covariance integral with measure
$dt\,dz/(\pi^2t^2z^2)$, substitute $u=at$. This gives a factor $a$
and an additional weight $\exp[-(1-a^2)u^2/a^2]\le1$.
The two distance-variance denominators agree. For $a=0$ independence
gives the assertion. Positivity for $r\ne0$ follows from
Lemma~\ref{hd:visibility} for every permitted nonconstant link.
Thus the population ranking is exactly by
$|r_j|=|(\Sigma_n\beta_0)_j|/(\sqrt{\Sigma_{jj}}\sqrt v)$.

For example, let $\Sigma_{ij}=\rho^{|i-j|}$, $\rho=1/2$, and
$\beta_0=(1,1,0,\ldots,0)^T/\sqrt2$. Both active correlations equal
$r_*=\sqrt{(1+\rho)/2}$, and $r_j=\rho^{j-2}r_*$ when $j\ge3$.
The exact-recovery gap is at least $(1-\rho)u(r_*)>0$ for fixed
$m_0(t)=\cos(t)$ as well as for a nonconstant odd link. With independent predictors
every inactive score is zero. Conversely, with the same AR covariance
and $\beta_0\propto(-\rho,1,0,\ldots,0)^T$, the first active
coordinate has $r_1=0$, and overscreening uses the rank gap of that
model.

\subsection{SSIRvRP and its inverse-mean requirement}\label{sb:ssir}
Algorithm~1 of \citet{zhang2026sparse} draws $A$ groups of $B$ uniform
coordinate $k$-subsets, picks the largest leading eigenvalue sum in each
group, averages the eigengap-weighted squared loadings, and keeps the top
$\ell$. In their isotropic, ordinary-eigenproblem Theorem~1 the response
has $H$ fixed slices, $X$ given each slice is Gaussian, the nonzero
inverse-mean eigenvalues obey $\lambda\le\lambda_d\le\lambda_1\le
\kappa\lambda$, and the active row norms have ratio at most $\mu$.
For $k\ge\max(d+1,s)$, $\ell\ge s$ and
$16K\sqrt{k\log P/n_s}\le\lambda_d/(s\mu^2)$,
their bound for the sine-angle loss is
$2K\sqrt{d\ell\log(P)/(n_s\lambda_d^2)}$. Its exceptional probability is at most
\[
 cP^{-3}+P\exp\{-A\lambda_d^2/(50P^2\mu^8\lambda_1^2)\}.
\]
For $d=1$, omission of $j$ by an output-supported unit vector $u$ implies
$\|(I-uu^T)\beta_0\|\ge|\beta_{0j}|$. Hence, if the displayed
angle bound is strictly below $\min_{j\in S_0}|\beta_{0j}|$,
the exceptional probability also bounds screening failure.
Appendix~A of \citet{sbZhang2025supp} proves the stronger
active--inactive weight separation directly, together with containment
for $\ell\ge s$ on its separation event. Their general-covariance
Theorem~2 imposes spectral bounds, sparsity and comparable rows in the
Cholesky-transformed direction, restricted concentration of the Cholesky
and whitened inverse-mean matrices, bounds on ordered Cholesky pivots,
and the sample-size inequalities in that theorem. Its loss is instead
$\|\hat b\hat b^T-bb^T\|_F$ for the covariance-normalized direction
$b=\beta_0/\sqrt v$ at $d=1$. If $j$ is omitted the $j$th row alone
contributes norm $|b_j|\|b\|=|\beta_{0j}|/v$. Thus its loss bound
$C\sqrt{\ell\log(P)/n_s}$ strictly below
$\min_{j\in S_0}|\beta_{0j}|/v$ suffices for inclusion on its good
event.
The result below uses Algorithm~1 with two slices and a prespecified
$\ell$. Slice-wise Gaussian laws are incompatible with a Gaussian
marginal design and a nonzero inverse-mean signal: dividing the
moment-generating function of any projection by that of its marginal
normal and letting $t\to\pm\infty$ forces identical slice laws. We
therefore verify the screen directly for the empirical generalized
eigenproblem in a nonempty SI-KDR class.

\begin{proposition}[Two fixed slices and empirical-covariance SSIRvRP]
\label{sb:ssir-two}
Let $X\sim N(0,I_P)$, $T=X^T\beta_0$, $Z=\ind_{\{Y>c_0\}}$ at a
fixed cut, $p=\Pp(Z=1)\in[p_*,1-p_*]$, and
$g=\Cov(X,Z)/\sqrt{p(1-p)}=\eta\beta_0$ with
$\eta=\E(TZ)/\sqrt{p(1-p)}\ne0$. Put
$b=\min_{j\in S_0}|g_j|$, $b_+=\|g\|_\infty$.
Use Algorithm~1 with $H=2$, $d=1$, empirical centered covariance,
$\max(2,s)\le k<n_s$, $k\le P$, $s\le\ell\le P$, $\ell<n_s$, and deterministic
tie breaking. Set $t=C_{p_*}\sqrt{\log(2P^2/\epsilon)/n_s}$ for
$0<\epsilon<1/2$. If $kt\le c b^2$ for a sufficiently small absolute
constant, then
\begin{equation}
 \Pp(S_0\not\subseteq S)\le \epsilon+
  2P\exp\{-c'A b^4/(P^2 b_+^4)\}.
 \label{sb:ssir-delta}
\end{equation}
Here $B\ge1$ is arbitrary. When $B=1$ the second term can be replaced
by $2P\exp\{-Ak/(8P)\}$. The screening size is $q=\ell$, with exact
recovery for $\ell=s$. The inverse-mean signal here is the nonzero
eigenvalue $\eta^2$ of $gg^T$. A nonempty example has fixed equal
nonzero coefficients, $m_0(t)=\tanh(t)$, independent centered mixture
error, and cut zero. The conditional success probability
$\Pp(Z=1\mid T=t)$ is strictly increasing, so its positive covariance
with $T$ gives $\eta>0$. At fixed $s,k$, $\log P=o(n_s)$ and, for
example, $A/(P^2\log P)\to\infty$ permit consistent support recovery
(take $\epsilon=P^{-1}$ along $P\to\infty$).
\end{proposition}
\begin{proof}
The empirical inverse-mean matrix of two classes is exactly
$\hat g\hat g^T$, with
$\hat g=\{\overline{XZ}-\bar X\bar Z\}/\sqrt{\bar Z(1-\bar Z)}$.
Declare a failure if a class is empty. With probability at least $1-\epsilon$,
\[
 \|\hat g-g\|_\infty\le t,\qquad
 \max_{i,j}|\hat\Sigma_{ij}-I_{ij}|\le t .
\]
Indeed $X_jZ$ has sub-Gaussian tails since $|X_jZ|\le|X_j|$,
$Z$ is bounded, and centered Gaussian coordinate products have uniformly
bounded sub-exponential norms. Apply the sub-Gaussian and Bernstein
bounds to all means and products, including $\bar X$ and $\bar Z$.
On their joint event the empirical probability stays in
$[p_*/2,1-p_*/2]$; the ratio and centering are Lipschitz on the relevant
bounded ranges. Increasing $C_{p_*}$ gives the two displays, since the
hypothesis implies $\log(2P^2/\epsilon)/n_s$ is sufficiently small.
For Gaussian design each covariance restricted to fewer than $n_s$
coordinates is nonsingular almost surely. This also permits the final
generalized eigenvector calculation on the selected $\ell$ coordinates.

For a projection $J$ write $Q=\hat\Sigma_{JJ}$,
$h=\hat g_J$ and $z=Q^{-1}h$. The leading generalized eigenvalue of
$(hh^T,Q)$ is $h^TQ^{-1}h$, the second is zero, and a $Q$-unit leading
vector is $z/(h^TQ^{-1}h)^{1/2}$. The group's coordinate weight is
exactly $z_j^2$ on $J$, zero elsewhere, also when $h=0$.
The row-sum and operator errors of $Q-I$ are at most $kt$, so its inverse
has both norms at most 2. Since $\|g\|_2=|\eta|\le1$, uniformly in $J$,
\[
 \|z-g_J\|_\infty\le2(t+kt b_+)\le b/4,\qquad
 |h^TQ^{-1}h-\|g_J\|^2|\le Ckt<b^2/3 .
\]
The latter follows by inserting $I$ for $Q^{-1}$ and using
$\|h-g_J\|_2\le\sqrt k\,t$. Constants can be chosen strictly smaller
than the stated separation margins. Replacing one inactive member of
$J$ by an absent active member increases the empirical eigenvalue,
since the population increment is at least $b^2$.

If $s=P$, containment is certain. Otherwise condition on these data
and let $\pi_j$ be the probability that coordinate $j$
belongs to a group's winning projection. For active $j$
and inactive $l$, map a tuple whose winner contains $l$ only to the
tuple obtained by replacing $l$ by $j$ in that winner. Its eigenvalue
strictly increases, so the swapped projection wins the new tuple and the mapping
is injective by reversing the swap. The uniform tuple probabilities
give $\pi_j\ge\pi_l$. As $\sum_i\pi_i=k$, $\pi_j\ge
(k-s+1)/(P-s+1)\ge1/P$. The group's expected active weight exceeds
each inactive expectation by at least
$(9\pi_j-\pi_l)b^2/16\ge b^2/(2P)$. Individual weights lie in
$[0,(5b_+/4)^2]$. Conditional Hoeffding over the $A$ independent groups,
with threshold $b^2/(8P)$ for each of the $P$ coordinates, proves
\eqref{sb:ssir-delta}. If $B=1$ all inclusion counts have binomial
marginals with mean $Ak/P$. Multiplicative Chernoff and a union bound
place them all between half and twice that mean except with probability
$2P\exp[-Ak/(8P)]$. The minimum active weight is then at least
$9kb^2/(32P)$ and the maximum inactive weight at most $kb^2/(8P)$.
The stated separation follows in both cases.
\end{proof}

\subsection{Least-squares Lasso for a correlated Gaussian design}
\label{sb:lasso}
On the screening rows alone take the support of a solution of
\[
 \min_b\ (2n_s)^{-1}\sum_{i\in D_s}(Y_i-X_i^Tb)^2+\lambda_L\|b\|_1 .
\]
This is the uncentered loss with mean-zero design in equation (2) of
\citet{sbNeykov2016}. Use a fixed measurable tie rule if necessary.
The ordinary nonlinear conditional mean is allowed. Their unit index
normalization is $\theta=\beta_0/\sqrt v$, so $T=X^T\theta$ has variance
one. Define $c_0=\E(YT)$ and $\xi^2=\E(Y-c_0T)^2$. Gaussian regression
gives $\E(XY)=\Sigma\theta c_0$. Thus the population least-squares
coefficient is $c_0\theta$, whose support is $S_0$ when $c_0\ne0$.
An even link under the centered Gaussian design has $c_0=0$.

We give explicit conditions for Theorem~2.3.4 of \citet{sbNeykov2016}.
Put $C_0=\Sigma_{S_0,S_0}$, $S_0^c=J$, let $D$ be the maximum diagonal
of $\Sigma_{J\mid S_0}$, $a=\lambda_{\min}(C_0)$,
$K_-=\|C_0^{-1/2}\|_{\infty,\infty}$,
$K_+=\|C_0^{1/2}\|_{\infty,\infty}$ (maximum absolute row sums),
and $m=n_s/\{s\log(P-s)\}$ for $P-s>1$.
Require a common positive spectral interval for $C_0$,
$\|\Sigma_{J,S_0}C_0^{-1}\|_{\infty,\infty}\le1-\kappa$ for
$\kappa>0$, bounded fourth response moment, and the source's uniform
moment bounds with $|c_0|$ bounded away from zero. In particular its
second moments and variance parameters for $Y^2,YT,(Y-c_0T)^2$
remain in fixed compact positive intervals. Sufficient tuning and
beta-min inequalities are
\begin{align}
 m&\ge\frac{4D}{\kappa^2}
       \left(\frac4a+\frac{\xi^2+1}{\lambda_L^2s}\right),
       \qquad m\ge a_0>0,\label{sb:lasso-size}\\
 \min_{j\in S_0}|\theta_j|
 &\ge C_1 K_-^2\lambda_L+
 \frac{1}{\sqrt m}\left\{
 C_2K_-K_+\sqrt{\frac{s}{n_s\log(P-s)}}\,\|\theta\|_\infty
 +\frac{C_3K_-}{\sqrt s}\right\}.
 \label{sb:lasso-min}
\end{align}
The positive $C_i$ are the constants of their inequality (7).
Their signed-support conclusion, relative to $c_0\theta$, gives
\begin{equation}
 \Pp(\widehat S\ne S_0)\le
 C\left[e^{-c\min\{s,\log(P-s)\}}+\{\log P\}^{-1}+n_s^{-1}\right].
 \label{sb:lasso-failure}
\end{equation}
On exact recovery the output size is $s$, regardless of
the size on the exceptional event. One can report a flagged failure
when the Lasso size falls outside the fixed admissible range.

For a concrete joint path, let $\Sigma_{ij}=\rho^{|i-j|}$ with
$\rho=1/4$, $S_0=\{1,\ldots,s\}$, and $\beta_{0j}=s^{-1/2}$ on
$S_0$. Take $s=\max\{2,\lfloor\log n_s\rfloor\}$,
$P=\lceil\exp(n_s^{1/4})\rceil$, $n_e\asymp n_s$,
$m_0(t)=\tanh(t)$, and an independent fixed centered nondegenerate
Gaussian-mixture error. For each $j>s$,
$\Sigma_{j,S_0}=\rho^{j-s}e_s^TC_0$, so the irrepresentability
row norm is at most $\rho$. The row-sum norm of $\Sigma-I$ is at most
$2\rho/(1-\rho)=2/3$. This bounds the spectra above and below and,
by the binomial series for $(I+(C_0-I))^{\pm1/2}$ in the row-sum
norm, bounds $K_\pm$ uniformly. Also $D\le1$. The standardized active
magnitudes are all $(sv)^{-1/2}$ with $v$ in a compact positive
interval. The first-order signal $\E[T\tanh(\sqrt v T)]$ has a
positive minimum over that interval. The stated moment bounds hold by
continuity and Gaussian tails, with nondegeneracy from the independent
error. The deterministic penalty $\lambda_L=n_s^{-1/4}$ satisfies
\eqref{sb:lasso-size} since $m\asymp n_s^{3/4}/s$ and
$(\lambda_L^2s)^{-1}=n_s^{1/2}/s$. The right side of
\eqref{sb:lasso-min} is $o(s^{-1/2})$ and
\eqref{sb:lasso-failure} tends to zero. Thus $q=s$ on exact
recovery and $q$ is logarithmic. The nonzero first moment gives
Fourier order $r=0$, so both the dimension and Fourier-basin
conditions hold on the correct branch. This example
has $P\gg n_s+n_e$ and requires no knowledge of $s$ to set the
screening penalty.

\section{Two independent fits on the oracle support and their screened coupling}
\label{sb:crossfit}

Let $D_1,D_2$ be disjoint independent iid blocks of sizes $n_1,n_2$,
$N=n_1+n_2$, $w_r=n_r/N$, with $\min_r w_r$ bounded away from zero
for asymptotics. The selector for fit $r$ uses only $D_{3-r}$ and
its independent seed. Denote its output by $S_r$, and set
$E_r=\{S_r=S_0\}$, $\Pp(E_r^c)\le\delta_r$, $s=|S_0|\ge2$.
In this exact-size branch, screen success means exact recovery.
Use separate independent fit seeds for the two blocks. Order the columns
of each model canonically. Zero-embed each fitted raw unit direction
$\widehat b_r$ in the ambient space and take the top eigenline of
$\sum_r w_r\widehat b_r\widehat b_r^T$. Use the same predeclared
diagnostics and failure convention for a screened fit and its oracle.
Declare a failed point fit or a nonunique pooled leading line a
point-estimation failure in the comparisons below. The spectral rule
requires no sign matching in the implementation.

\begin{proposition}[Weighted pooling in a common raw chart]\label{sb:pool}
Take $F_0$ to be a deterministic raw orthonormal tangent frame on $S_0$
at the oriented unit vector $b_0=\beta_{0,S_0}$. For the oracle fit on
$D_r$, write $t_r=F_0^T\widehat b_r^o/(b_0^T\widehat b_r^o)$ and
suppose that the applicable one-block expansion gives
\[
 t_r=n_r^{-1}\sum_{i\in D_r}\psi_r(O_i)+e_r .
\]
Here $\psi_r$ is the centered parameter influence in this frame and
$V_r=\E(\psi_r\psi_r^T)$. On the regular event $h=\max_r\|t_r\|\le1/4$
the oracle pooled chart satisfies
\begin{equation}
 t_{\rm pool}^o=N^{-1}\sum_{r=1}^2\sum_{i\in D_r}\psi_r(O_i)
       +\sum_{r=1}^2 w_r e_r +v,\qquad \|v\|\le4 h^3 .
 \label{sb:pool-expansion}
\end{equation}
The screened output equals the embedded oracle output on
$E_1\cap E_2$, pathwise. The comparison costs at most
$\delta_1+\delta_2$ in probability, at every tolerance and scaling.
\end{proposition}
\begin{proof}
Construct each oracle on the original probability space with exactly the
corresponding fit rows, fit seed, column order, sample standardization,
tuple rules, optimizer and diagnostics; the selector transmits only its
set. Thus equality holds also for reported failure states. Work with
the independent oracle blocks under the original law. In the orthogonal
basis $(b_0,F_0)$ the weighted projector has blocks
\[
 a=\sum_r\frac{w_r}{1+\|t_r\|^2},\quad
 c=\sum_r\frac{w_rt_r}{1+\|t_r\|^2},\quad
 B=\sum_r\frac{w_rt_rt_r^T}{1+\|t_r\|^2}.
\]
Its leading eigenvalue $\ell$ is in $[a,1]$ by the Rayleigh quotient;
the second eigenvalue is at most $\|B\|\le h^2$ by restriction to
$b_0^\perp$. Hence the top vector has nonzero first coordinate and its
chart is $(\ell I-B)^{-1}c$. Since $1-a\le h^2$,
$\|c-\sum_rw_rt_r\|\le h^3$ and $\|c\|\le h$, the residual is bounded
by $h^3+2h^3/(1-2h^2)\le4h^3$. Zero embedding preserves the orthogonal
decomposition and adds only zero eigenvalues.
\end{proof}

\paragraph*{Common and different influences.}
In the retained notation of \eqref{hdg:estimate} the influence of a
deterministic tuple is $D_{0,S_0}j_{v,k}^{-1}W f_k(T,Y^w)$, transformed
to the single frame $F_0$. The raw coordinate map cancels fold-specific
sample whitening in the standardized index as in
\eqref{hd:map-cancellation}. The law of $(T,W,Y^w)$, the matrix
$D_{0,S_0}$, and the population standardization are common to all
blocks; the tuple itself can differ. At fixed $s$ use the one-fold
remainder $\sqrt{n_r}e_r=o_P(1)$ and $h=O_P(N^{-1/2})$.
For deterministic block influences the covariance of the displayed linear
term is $(w_1V_1+w_2V_2)/N$. Under the one-fold moment and spectral
bounds the triangular-array CLT holds for every fixed contrast, using
the corresponding weighted variance. Simultaneously consistent raw
one-fold covariance estimates can be combined as
$(w_1\widehat V_1+w_2\widehat V_2)/N$ after aligning their frames
with the common tangent (or using ambient matrices and projecting).
For a unit contrast $l$, let $H_N=N^{-1/2}\sum_{r,i}l^T\psi_r(O_i)$; independence gives $\Var H_N=\sum_r w_r l^TV_r l$. A uniform $2+\epsilon$ moment bound and a variance bounded below give the scalar triangular-array Lindeberg condition by truncation. Consistent raw estimates therefore studentize this contrast.

Suppose the oracle using all $N$ rows also has its one-block expansion,
and all three influences can be replaced in $L^2$ by the same centered
$\psi$ (the target may depend deterministically on $N$). For example,
use the same fixed tuple, a consistently selected common separated
positive tuple, or shrinking grids with the common efficient influence
of \eqref{hd:common-influence}. With fixed dimension a centered
$L^2$ discrepancy of $o(1)$ changes its normalized sum by $o_P(1)$
by the second moment identity; take a maximum over the bounded grid
when selection is random. A selector stabilizing with probability
tending to one uses its deterministic branch with that probability.
Subtract their expansions on the same rows. For each deterministic replacement the squared $L^2$ norm of a centered, normalized oracle sum is the per-observation squared $L^2$ error. The two-fold sum and the full-block sum can thus each be replaced on this original probability space. The full oracle and the pool then differ by $o_P(N^{-1/2})$ in their
charts, also after screening if $\delta_1+\delta_2=o(1)$.
The common leading covariance is $V/N$, with
$V=(\Lambda_k/G)V_{\rm mean,S_0}$ at a common fixed tuple, and
$V=\mathcal I_{S_0}^{-1}$ in the shrinking efficient case.
Different positive tuples give the weighted covariance above unless
their influences agree at the required scale. With equal
fold sizes $w_r=1/2$; with sizes $200,400$ the weights are $1/3,2/3$.
For a common influence, equal weights on these unequal folds instead
give covariance $V(1/n_1+1/n_2)/4$.

For increasing $s$, impose the one-fold hypotheses at both sizes and,
for comparison to the full fit, also at $N$. A direction rate
$h=O_P(\sqrt{s/N})$, $s/N=o(1)$ transfers directly. For a deterministic
unit raw tangent contrast $l$, the required remainder is
$\sqrt{n_r}\,l^Te_r=o_P(1)$ and $s^{3/2}/N=o(1)$ suffices for the
nonlinear term. Use the stronger contrast conditions of the applicable
single-block theorem. Its scalar $L^2$ influence replacement then
suffices for common-contrast equivalence and studentization.
Whole-vector root-$N$ equivalence holds under norm remainders $o_P(1)$
on that scale and $\E\|\psi_r-\psi\|^2=o(1)$ for every relevant
grid, including the full fit. For the limiting risk ratio $n_e/N$
relative to a single fit of size $n_e$, take the common
well-conditioned influence, quadratic-mean errors $o(\sqrt{s/N})$ in
the local linearization, uniform integrability of the scaled squared
direction errors on regular fits, and failure probabilities $o(s/N)$
for bounded loss. Then $2-2|\widehat b^Tb_0|=\|t\|^2+O(\|t\|^4)$
locally, and truncation plus uniform integrability gives leading mean
losses $\tr V/N$ and $\tr V/n_e$.

\begin{proposition}[Finite-sample intersection]\label{sb:intersection}
Suppose for each allowed selected model $n_r>|S_r|+1$ and the
Gaussian candidate pivot conditions hold on $E_r$. Compute each
candidate test on its own fit block at level $a=\alpha/2$, using its
own candidate index and response for the tuple and deterministic tie
breaking. Embed the two projective sets in the same $P$ coordinates
and intersect. For fit $r$ let $L_r$ be the union of the following
disjoint losses at the true line on $E_r$: the global diagnostic fails;
diagnostics pass but the line lies outside the mapped optimizer or final
angular cap; both pass, the ideal pivot accepts, and the numerical
implementation excludes the line. Write $p_r=\Pp(E_r^c)$.
The reported intersection has coverage at least
\begin{equation}
 1-2a-(1-a)(p_1+p_2)-\Pp(L_1)-\Pp(L_2).
 \label{sb:intersection-bound}
\end{equation}
For the full mathematical inversion take $L_r=\varnothing$; for
mathematical localized inversion omit numerical exclusion. In particular,
the full inversion has coverage at least $1-\alpha-\delta_1-\delta_2$.
\end{proposition}
\begin{proof}
Condition on the opposite screen and its seed, then the fit block's
true index and response. The complete candidate coefficients and
candidate-only tuning are measurable here. On $E_r$ the Gaussian
projection pivot in Proposition~\ref{hdg:pivot} has denominator
degrees of freedom $n_r-|S_r|-1$, numerator degrees of freedom
$|S_r|-1$, and accepts a zero projected coefficient. Thus the joint
probability of $E_r$ and full-test rejection is at most $a(1-p_r)$.
A union bound for loss in either block gives \eqref{sb:intersection-bound}.
Alternatively, both support-oracle full tests are independent on the
unconditional probability space, for the separate fit seeds above.
Coupling their joint acceptance gives the additional lower bound
$(1-a)^2-p_1-p_2-\sum_r\Pp(L_r)$. If the two recovery errors,
zero-coefficient probabilities, and localization and numerical losses
vanish, the mathematical intersection coverage tends to $(1-a)^2$
(and so does an implementation with vanishing true-line membership
disagreement).
\end{proof}

For subsets of the mathematical localized sets, projective diameter
is at most the smaller fold diameter, with diameter zero by convention
for the empty set. On the coupling event, the conditions of
Theorem~\ref{hd:shape} give $O_P(N^{-1/2})$ at fixed $s$; the
trace-radius conditions of Theorem~\ref{hdg:coverage} give
$O_P(\sqrt{s/N})$ when $s\to\infty$. The probability bounds transfer
unconditionally for $\delta_1+\delta_2=o(1)$.

\section{Coordinate tests from restricted SI-KDR scores}\label{sbvar:section}

Let $D_s$ include any independent screening seed, put $n=n_e$ and condition
on $D_s$. Write $S$ for the retained indices, $q=|S|$, and suppose the
true support is in $S$. For $j\in S$ let $Z=X_{S\setminus j}$,
$x=X_j$, $B=[\boldsymbol 1,Z]$, and $\Pi_j=I-P_B$, where $P_B$ is the
orthogonal projection onto the column space of $B$. All these arrays
have $n$ rows. The null is $H_{0j}:\beta_j=0$. We consider $q\ge2$.
A normalized nonzero direction on a correct singleton support has no
zero coordinate to test.

\paragraph*{A score that can be computed under the null.}
Run the retained SI-KDR algorithm on $(Z,Y)$ from the beginning. Center,
scale and whiten within this reduced data set. Compute the four-matrix
Fourier start, pilot, tuning and final capped ascent there, using the
filtering, tie and stopping rules of Sections~\ref{sec:ascent}
and~\ref{hdg:exp-section}. A prespecified single tuple is also allowed.
For one remaining predictor the sole projective direction is known,
and one uses a prespecified tuple or an index-only tuning rule; our
implementation uses the fixed tuple $(0.25,0.15,0.001,0.005)$ for this
singleton reduced fit. Let $\hat b$
be the resulting reduced raw unit direction. Set $t=(Z\hat b-\overline{Z\hat b})/
s_{Z\hat b}$ and $y=(Y-\bar Y)/s_Y$, with divisor $n$. The following
formulas hold at this $t$ and the selected tuple $(g,h,\lambda,\rho)$.
With $N=n-1$, $\phi_g$ the normalized Gaussian kernel and primes denoting
argument derivatives, put
\begin{align*}
 D_i&=N\lambda+\sum_{l\ne i}\phi_g(t_i-t_l),&
 H_{il}&=\boldsymbol1_{\{i\ne l\}}\phi_g(t_i-t_l)/D_i,&
 Q_{il}&=\boldsymbol1_{\{i\ne l\}}\phi'_g(t_i-t_l)/D_i,\\
 \widehat M_i&=\sum_l H_{il}y_l,&
 r_i&=y_i-\widehat M_i,&
 d_i&=N^{-1}\sum_{l\ne i}\phi_h(r_i-r_l).
\end{align*}
Differentiating $n^{-1}\sum_i\log(d_i+\rho)$ with respect to $r_i$
in the evaluation row and in every other row's center gives $a_i/n$.
The quotient rule for a perturbation $u$ of the index gives
$\dot M_i=\sum_l Q_{il}(y_l-\widehat M_i)(u_i-u_l)$. Thus the
complete scalar-index derivative is $n^{-1}\sum_i c_i u_i$, where
\begin{align}
 a_i&=N^{-1}\sum_{l\ne i}\phi'_h(r_i-r_l)
       \{(d_i+\rho)^{-1}+(d_l+\rho)^{-1}\},\qquad
 A_i=\sum_l Q_{il}(y_l-\widehat M_i),\notag\\
 c_i&=-a_i A_i+\sum_{l\ne i}a_l Q_{li}(y_i-\widehat M_l).
 \label{sbvar:c}
\end{align}
In particular $\sum_i c_i=0$. Even after the nuisance direction has
been fitted, the second term in \eqref{sbvar:c} contributes to a
perturbation of the tested coordinate. It arises from changes in
other rows' fitted links. The raw perturbation $Z\hat b+\theta x$
has standardized-index derivative $(I-P_{[\boldsymbol1,t]})x/s_{Z\hat b}$.
The raw directional derivative is
$x^T(I-P_{[\boldsymbol1,t]})c/(n s_{Z\hat b})$. For the coordinate
test we instead perturb by the regression residual $\Pi_jx$. Since
$\boldsymbol1,t\in\operatorname{col}(B)$, the resulting derivative is
$x^T\Pi_jc/(n s_{Z\hat b})$. The positive scale cancels on studentizing. No whitening
matrix, initialization, bandwidth, stopping diagnostic or restricted
direction may be computed using $x$. Independent fitting randomization
can be included in the conditioning field below.

\begin{proposition}[Conditional coordinate calibration]\label{sbvar:pivot}
Let $U$ contain the independent seeds used by the restricted fit, and
write $\mathcal H_j=\sigma(D_s,Z,Y,U)$.
If $n>q+1$, $B$ has full column rank, the design is nonsingular Gaussian
and $H_{0j}$ holds, then, given $\mathcal H_j$, on the branch $u=\Pi_jc\ne0$,
\begin{equation}
 z=\frac{(\Pi_jx)^T u}{\|\Pi_jx\|\|u\|},\qquad
 T_j=\frac{\sqrt{m-1}\,z}{\sqrt{1-z^2}}\ \sim\ t_{m-1},
 \qquad m=n-q .
 \label{sbvar:t}
\end{equation}
The two-sided $p$-value is $2\Pr(t_{m-1}\ge |T_j|)$. Set $p=1$ for
a failed restricted fit, unusable rank or scale, or $u=0$. The resulting
$p$-value is conditionally super-uniform.
\end{proposition}
\begin{proof}
Under the null, $Y\perp x\mid Z$. Gaussian regression writes
$x=\nu+\sigma_j\xi$ with $\nu=B\gamma$, $\xi\sim N_n(0,I_n)$
independent of $\mathcal H_j$, and
$\sigma_j^2=\Sigma_{jj}-\Sigma_{j,-j}\Sigma_{-j,-j}^{-1}\Sigma_{-j,j}>0$
in the retained covariance. Hence $\Pi_jx=\sigma_j\Pi_j\xi$. Using $u/\|u\|$
as the first member of an orthonormal basis of $\operatorname{ran}(\Pi_j)$
gives $m$ independent standard normals. Their first coordinate divided
by the root mean square of the other $m-1$ yields \eqref{sbvar:t}.
The zero-norm and $|z|=1$ events have probability zero on this branch.
Any additional gate that only changes a rejection into acceptance
preserves the upper bound.
\end{proof}

\paragraph*{Fitted restricted coefficient.}
For power assume the centered Gaussian design, bounded-spectrum retained
law, bounded nonconstant $C^6$ link with bounded derivatives, and
independent centered fixed nondegenerate finite Gaussian-mixture error
of Section~\ref{hd:uniform-assumptions}. Fix a null raw direction
$b_0$ with $(b_0)_j=0$, and choose its sign locally. Write
$T=X^Tb_0/\sqrt v$, $v=b_0^T\Sigma b_0$, and $Y^w=(Y-\E_0Y)/\sigma_v$,
$\sigma_v^2=\Var_0Y$. All expectations in the proof are under the
null. For a tuple $k$, let
\begin{align*}
 f_k&=\kappa_k(T,Y^w)-T\E(T\kappa_k),& b_k&=\E f_k^2,&J_k&=j(v,k),\\
 A(v)&=\E m_0'(\sqrt v T)^2,&\ell&=-m_0'(\sqrt v T)\psi_\varepsilon(\varepsilon),&
 \psi_\varepsilon&=f'_\varepsilon/f_\varepsilon.
\end{align*}
Here $\kappa_k$ is the complete population coefficient in
\eqref{inf:kappa}, $J_k$ the curvature in \eqref{inf:curvature}.
Translation of every index leaves the criterion fixed, so the two
terms of $\kappa_k$ have opposite expectations by exchange of their
indices. Projection removes its $T$ moment. Response and index
standardization give $\zeta_*=\sqrt v\,\ell$, and
\eqref{inf:crossidentity} then gives
\begin{equation}
 \E f_k=0,\quad \E(T f_k)=0,\quad
 \E(f_k\ell)=J_k/\sqrt v,\quad
 \E\ell^2=A(v)I_\varepsilon,\quad
 \Lambda(v,k)=vA(v)I_\varepsilon b_k/J_k^2 .
 \label{sbvar:identity}
\end{equation}
Use a bounded finite grid with $J_k$ uniformly positive, including
pilot tuples. We give two sufficient regimes for actual null fits.
The first has fixed $q$, fixed positive tuples and the reduced Fourier
gap and cap conditions of Lemmas~\ref{hdg:fourier}--\ref{hdg:ascent}
when there is a nuisance direction to estimate. The second uses
\eqref{inf:nonempty}, $q=O(n^a)$ for a fixed $a<1/30$, and the
reduced-model Fourier gap/basin condition \eqref{hdg:basin} with
dimension $q-1$. Corollary~\ref{sbvar:wide-power} gives sharper ranges
for both shrinking schedules.

\begin{lemma}[Removing the fitted nuisance]\label{sbvar:transfer}
Orient each restricted fit toward the null line representative $b_0$
for the analysis. For each possible tuple, let $\hat\gamma$ be its
fitted population-standardized reduced index direction, let $G_-$
be the $n$ by $q-1$ matrix of population-standardized reduced predictors,
and let $s_{\hat\gamma}$ be the sample standard deviation of $G_-\hat\gamma$.
Simultaneously on the grid,
\begin{equation}
 \left\|s_{\hat\gamma}^{-1} \Pi_j c_k(\hat\gamma)-f_k\right\|_{n,2}=o_P(1).
 \label{sbvar:transfer-eq}
\end{equation}
On the right, $f_k$ denotes its vector of observed scalar values.
\end{lemma}
\begin{proof}
Rotate $G_-=[T,W_-]$ at the restricted truth. The columns of $W_-$
are independent $N_n(0,I_n)$ given $(T,Y)$, and
$\operatorname{col}(B)=\operatorname{col}[\boldsymbol 1,T,W_-]$.
Set $d_\gamma=s_\gamma^{-1}(I-P_{[\boldsymbol1,G_-\gamma]})c_k(\gamma)$.
This is the coefficient for the normalized index chart and
$\Pi_jd_\gamma=s_\gamma^{-1}\Pi_jc_k(\gamma)$ exactly.
At $\gamma_0$, Lemma~\ref{v2:truth-score} gives
$\|d_{\gamma_0}-f_k\|_{n,2}=O_P(D_{S,k})=o_P(1)$ on the shrinking
grid. The bound uses both roles of the sample density and the incoming
NW term. At fixed positive scales the same truth approximation follows
from \eqref{dir:crisk}--\eqref{dir:scalarLip} and the empirical
index projection, with error $O_P(n^{-1/2}\log^C n)$.
The preliminary and final restricted estimators are
$O_P(\sqrt{q/n})$ from the truth by Lemmas~\ref{hdg:fourier}
and~\ref{hdg:ascent}, Theorem~\ref{hdg:sample} for the fixed regime,
and Corollary~\ref{v2:contrast} for the implemented regime. In a
one-dimensional reduced model this angular error is zero. The equivalence
of population-standardized and raw angles on the retained spectral
interval follows from \eqref{hd:raw-map}. In particular these rates hold
for the actual training-standardized fit.

If $U=M\sqrt{q/n}$, the maximum fitted-index perturbation on the
standardized ball is $O_P(U(\sqrt q+\sqrt{\log n}))$, including its
sample scale. The scalar coefficient bound \eqref{vtwo:cov-coef} gives
\[
 \sup_{\|\gamma-\gamma_0\|\le U}\|d_\gamma-d_{\gamma_0}\|_{n,2}
 =O_P\!\left(\frac{\log^C n\,U(\sqrt q+\sqrt{\log n})}{g^2h^2}\right)
 =O_P\{\log^{C'}n\,(1+q)n^{-1/30}\}=o_P(1)
\]
on the implemented path. On a fixed positive grid, the bounded inverse
protected denominators and the derivative envelopes for $c,d$ in
the proof of Theorem~\ref{hdg:sample} give
$O_P(\log^C n\,n^{-1/2})$ on the fixed-dimensional root-$n$ ball.
Take $M$ arbitrarily large before substitution.

It remains to project $f_k$. Its population intercept and $T$ slope
are zero in \eqref{sbvar:identity}, so projection onto
$[\boldsymbol1,T]$ has empirical norm $O_P(n^{-1/2})$ by its uniform
fourth moments. Conditional on the scalar rows, the remaining $q-2$
projected Gaussian columns span a uniform $(q-2)$-plane in the
$(n-2)$-dimensional orthocomplement. The conditional expected squared
fraction of any fixed vector there captured by this plane is
$(q-2)/(n-2)$, by rotation invariance and the trace of the projector.
Thus $\|P_B f_k\|_{n,2}=O_P(\sqrt{q/n})$. Contraction of $\Pi_j$ and a
finite union prove \eqref{sbvar:transfer-eq}. The opposite orientation
negates $c$ because the Gaussian regression kernel is even and its
derivative odd. Hence the reported two-sided test is unchanged.
\end{proof}

\begin{theorem}[Local power in raw coordinates]\label{sbvar:power}
Consider a retained set containing $\operatorname{supp}(b_0)\cup\{j\}$.
Hold the link, error law and design fixed and let the raw parameter be
\[
 b_n(\omega)=\sqrt{1-\omega^2/n}\,b_0+\omega e_j/\sqrt n
\]
for fixed $\omega\in\mathbb R$. Take the orientation of
Lemma~\ref{sbvar:transfer} for a signed statistic. At a deterministic
or consistently selected tuple its local normal shift, and that of
the conditional-mean oracle using the observable null-map coefficient
$m_0'(X^Tb_0)\{Y-m_0(X^Tb_0)\}/\Var(\varepsilon)$
in place of $c$, are
\begin{align}
 \mu_K&=\omega\,\sigma_j J_k/\sqrt{v b_k},&
 \mu_M&=\omega\,\sigma_j\sqrt{A(v)/\Var(\varepsilon)} .
 \label{sbvar:shifts}
\end{align}
Both limiting normal laws have unit variance. For triangular arrays the display gives
the corresponding sequence of shifts (with a normal approximation
along each convergent subsequence). For $\omega\ne0$,
\begin{equation}
 |\mu_K|/|\mu_M|=\sqrt{G/\Lambda(v,k)},\qquad
 \mu_K^2/\mu_M^2=G/\Lambda(v,k),\qquad G=\Var(\varepsilon)I_\varepsilon .
 \label{sbvar:ratio}
\end{equation}
The level-$\alpha$ two-sided local power is
$1-\Phi(z_{1-\alpha/2}-\mu)+\Phi(-z_{1-\alpha/2}-\mu)+o(1)$.
Any restricted selector on the implemented shrinking grid has the
same limit with $\Lambda\to1$. At a fixed positive grid a restricted
selector concentrating on tuples with the same limiting $J_k/\sqrt{b_k}$
also has their common shift. This includes tied positive-curvature
population trace optima when $q\ge3$. If $q=2$, the reduced tangent
has rank zero, so its trace cannot identify a tuple. Use a fixed or
consistent tuple, or verify a common shift for the index-only rule.
Any additional numerical acceptance rule is assumed to activate with
probability tending to zero.
\end{theorem}
\begin{proof}
Under the null write $x=\nu+\sigma_j\xi$ as above. Along
$b(\theta)=\sqrt{1-\theta^2}\,b_0+\theta e_j$ the derivative of the
conditional log density is $x_i\ell_i$. Its conditional second
derivative at zero is
\[
 x_i^2 m_0'(X_i^Tb_0)^2\psi_\varepsilon'(\varepsilon_i)
 -\{x_i^2m_0''(X_i^Tb_0)-(X_i^Tb_0)m_0'(X_i^Tb_0)\}
      \psi_\varepsilon(\varepsilon_i).
\]
Its mean is $-I_{\rm path}$, where $I_{\rm path}=\E(x_i^2\ell_i^2)$,
by $\E\psi_\varepsilon=0$ and $\E\psi'_\varepsilon=-I_\varepsilon$.
For a fixed nondegenerate finite Gaussian mixture, derivatives of
$\log f_\varepsilon$ through order three have polynomial envelopes:
each $f^{(l)}/f$ is a mixture-posterior average of component
polynomials. The link derivatives are bounded, and
$(x_i,X_i^Tb_0,\varepsilon_i)$ has uniform finite moments of every
fixed order. Thus Taylor expansion, the row law of large numbers
and the ordinary triangular-array central limit theorem give
\[
 \log(dP_{\omega,n}/dP_{0,n})=\omega\Delta_n-\omega^2 I_{\rm path}/2+o_P(1),
 \quad \Delta_n=n^{-1/2}\sum_i x_i\ell_i .
\]
This likelihood calculation is for the estimation sample given the
screen.

Put $w=\Pi_j c/\|\Pi_j c\|$, oriented as above. With probability tending
to one it exists, by \eqref{sbvar:transfer-eq} and
$b_k\ge J_k^2/\{vA(v)I_\varepsilon\}>0$. Under the null
\[
 T_j=\xi^Tw\left\{\frac{\|\Pi_j\xi\|^2-(\xi^Tw)^2}{n-q-1}\right\}^{-1/2}
     =\xi^Tw+o_P(1).
\]
The denominator is a conditional normalized chi square. Given
$\mathcal H_j$, $(\xi^Tw,\sigma_j n^{-1/2}\sum_i\xi_i\ell_i)$ is
jointly Gaussian. Its variances are $1$ and
$\sigma_j^2 P_n\ell^2$, and its cross covariance is
\[
 \frac{\sigma_j w^T\ell}{\sqrt n}
 =\sigma_j\,\frac{P_n f_k\ell}{(P_n f_k^2)^{1/2}}+o_P(1)
 =\sigma_j J_k/\sqrt{v b_k}+o_P(1).
\]
Here $\ell$ is $\mathcal H_j$-measurable at the null. The first equality
uses empirical Cauchy--Schwarz and \eqref{sbvar:transfer-eq}, even for
a coefficient fitted to all estimation responses. The empirical moment
limits are simultaneous on the finite grid. The remaining term
$n^{-1/2}\sum_i\nu_i\ell_i$ in $\Delta_n$ is
$\mathcal H_j$-measurable and centered, with variance $\E(\nu_i^2\ell_i^2)$
and an ordinary row CLT (a vanishing variance gives a degenerate limit).
The conditional Gaussian characteristic function converges to a
deterministic one whenever the displayed cross covariance has a common
limit. Multiplication by the characteristic function factor of this
remaining term and bounded convergence then give joint normality of
$(T_j,\Delta_n)$ with the stated cross covariance and variances
$1,I_{\rm path}$. Exponential tilting by the local likelihood, or
Le Cam's third lemma, gives the first shift in \eqref{sbvar:shifts}.
This argument also handles a restricted random selector at tied
optima when $q\ge3$. Their $J_k/\sqrt{b_k}$ agree because the
reduced trace is $b_k/J_k^2$ times the same strictly positive geometry
factor, and $\xi$ is independent of the selector
under the null. On a shrinking grid \eqref{hdg:loss-square} gives the
common limit. Contiguity transfers all null $o_P(1)$ events.

For the mean oracle put $f_M=m_0'(\sqrt v T)\varepsilon/\Var(\varepsilon)$.
It has zero mean and zero $T$ slope. As in the last paragraph of
Lemma~\ref{sbvar:transfer}, $\|\Pi_j f_M-f_M\|_{n,2}=o_P(1)$. Independence
and $\E(\varepsilon\psi_\varepsilon)=-1$ give
$\E(f_M\ell)=\E f_M^2=A(v)/\Var(\varepsilon)$. The same conditional
Gaussian and likelihood calculation gives $\mu_M$. Division proves
\eqref{sbvar:ratio}. The unknown conditional scale
cancels only in the comparison. Curvature $J_k>0$ fixes the signed
alignment, while the two-sided power is invariant to reversing the
representative of the line.
\end{proof}

\paragraph*{Two computable comparators.}
Use the same restricted fitted direction for a feasible mean test.
Recompute $H,Q,\widehat M,A$ at $g_M=n^{-1/6}$,
$\lambda_M=n^{-3/4}$, using only $(Z,Y)$. Its coefficient is
$c^M_i=A_i(y_i-\widehat M_i)$ in \eqref{sbvar:t}.
In general, exact calibration follows by measurability. The mean
oracle shift follows if the projected coefficient estimates a positive
multiple of the oracle coefficient in empirical norm, namely
$\|\Pi_j(c^M-a_M f_M)\|_{n,2}=o_P(1)$ for $a_M$ bounded away from zero
and infinity. For the displayed choices this holds in the two regimes
above with $a_M=\sqrt v\,\Var(\varepsilon)/\sigma_v^2$. Indeed the
truth-index leave-one-out query derivative and fit approximate $m_v'$
and $m_v$ in empirical norm up to logs with errors
$g_M^2+\sqrt{\lambda_M}+(ng_M^3)^{-1/2}$ and
$g_M^2+\sqrt{\lambda_M}+(ng_M)^{-1/2}$,
respectively, by \eqref{eq:gaussian-primitives} and the normalized
Gaussian profile formula \eqref{inf:gaussianprofile}. Training response
centering/scaling and truth-index standardization introduce a scalar
perturbation $O_P(n^{-1/2}\log^C n)$ before smoothing. For two
standardized indices at maximum distance $\Delta$, the row-derivative
bounds used in \eqref{vtwo:cov-coef} give changes of
$\log^C n\,\Delta/g_M^2$ for $A$, and $\log^C n\,\Delta/g_M$ for
$\widehat M$. At the actual restricted fit,
$\Delta=O_P(\log^C n(1+q)/\sqrt n)$, so both bounds tend to zero
at a polynomial rate. Response affine perturbations alone cost at
most $\log^C n/(\sqrt n\,g_M)$ for $A$ and $\log^C n/\sqrt n$ for
$\widehat M$, by the linearity of the row kernel in the response.
To pass to products, use $|y_i-\widehat M_i|\le2\max_i|y_i|
=O_P(\sqrt{\log n})$ and bounded $m_v'$. Thus $c^M$ approaches
$m_v'(T)\varepsilon/\sigma_v=a_M f_M$ in empirical norm, and the
projection contracts.

For an inverse-mean comparator use the marginal coordinate statistic
of \citet[][equations (10)--(11)]{sbCook2004}. Take a prespecified
number of response-rank slices with deterministic ties, let $D$ contain
their indicators and let $P_D$ project onto their span. If $r=\Pi_jx$,
the single-coordinate statistic is $n\,r^TP_Dr/(r^Tr)$. To check
the reduction, the full centered-predictor whitening sends the
coordinate-contrast direction to the normalized whitened contrast
$\widehat\Sigma^{-1/2}e_j/
\|\widehat\Sigma^{-1/2}e_j\|$. The vector of its row scores is
$\sqrt n\,r/\|r\|$, by residual regression (both vectors belong to
the centered predictor span, are orthogonal to the remaining centered
columns and have positive inner product with $x$).
Summing their squared slice means weighted by slice counts gives the
displayed statistic. A reference law
can be obtained conditionally on $(D_s,Z,Y)$ by substituting independent
$\Pi_j\xi^{(b)}$, $\xi^{(b)}\sim N_n(0,I)$, for $r$. With $B_{\rm MC}$
draws, the upper rank $(1+\#\{T^{(b)}\ge T^{\rm obs}\})/(B_{\rm MC}+1)$
is super-uniform by exchangeability. Calibration averages over the
independent simulation draws, and ties enter conservatively.
Residual-subspace rotation \citep{perry2010rotation} and model-X
conditional tests with nuisance-only fitting \citep{sbLiu2022} supply
the calibration principle; the SI-KDR-specific ingredients are the complete
restricted score in \eqref{sbvar:c} and its information comparison in
\eqref{sbvar:ratio}.

\paragraph*{Multiplicity and screen entry.}
Condition first on $D_s$ with correct retention. For each truly null
retained coordinate integrate its own $\mathcal H_j$-conditional
$p$-value inequality. This gives marginal super-uniformity given
$D_s$ for every true null without jointly conditioning on
$\{\mathcal H_j:j\in S\}$. Choose the tested family before accessing
the estimation split, e.g. the $M=q$ retained indices or $M=P$ indices
with unselected $p$-values set to one. The two conventions give
different cutoffs. For ordered $p$-values use Holm's step-down cutoffs
$\alpha/(M-l+1)$ and the Benjamini--Yekutieli step-up cutoffs
$t_l=\alpha l/(M H_M)$, $H_M=\sum_{l=1}^M1/l$
\citep{sbHolm1979,sbBY2001}. For Holm, the first true null rejected
must have $p\le\alpha/M_0$ if $M_0$ true nulls are in the family.
Their union bound gives FWER at most $\alpha$. For the step-up rule
write $N_R$ for the number rejected. Each true null contributes at most
$\sum_{l=1}^M\Pr(t_{l-1}<p\le t_l\mid D_s)/l$ to
$\E[V/(N_R\vee1)\mid D_s]$, where $t_0=0$. Summation by parts and
super-uniformity bound this by
$t_M/M+\sum_{l<M}(1/l-1/(l+1))t_l=\alpha/M$.
Thus FDR is at most $\alpha M_0/M$. Both proofs allow arbitrary
dependence. On a screen miss the error indicator and the false
discovery proportion are at most one. If the screen miss probability
is at most $\delta_n$, each unconditional error criterion is therefore
at most $\alpha+(1-\alpha)\delta_n$.

A rejected null asserts a nonzero raw coordinate on correct retention.
When the procedure sets $p_j=1$ outside $S$, end-to-end rejection
is bounded by the entry probability $\Pr(j\in S)$. More precisely,
its contribution from correct retained models is the expectation over
$D_s$ of $\boldsymbol1_{\{j\in S,\ \operatorname{supp}(b_n)\subseteq S\}}$
times the retained-model power; the remaining rejection probability
is at most the screen miss probability. A target forced in by a
rule specified for all samples has entry probability one; with the
other active coordinates retained, its power is the estimation-split
power.

\section{Retained dimension on two finite grids}\label{sb:rates}
Let $n=n_e$, $d=q-1$, $L=1+\log(n+e)$, $u_n=(q/n)^{1/2}$ and
$R=\min(1.5,2n^{-1/20})$. Condition on a correct, independently retained
model with the spectral and scalar-family conditions of
Section~\ref{hd:inheritance-proofs}. Constants below are uniform in that
model, including $v$ in its compact spectral interval. The union of
preliminary, final and candidate grids has bounded cardinality, with
positive multipliers in a fixed compact subinterval of $(0,\infty)$.
The grid determines four scalar quantities in the same twice-refitted
leave-one-out objective. In this section the upright labels P and L denote schedules; $L$ in formulas denotes the logarithmic factor above. The power schedule P uses
$(n^{-1/6},n^{-1/15},n^{-3/4},n^{-1})$ for $(g,h,\lambda,\rho)$.
The logarithmic schedule L uses $(L^{-1},L^{-1},L^{-1},L^{-1})$.
For both schedules use the four empirical Fourier matrices, the
cap $R$, first step $1/\log(n+e)$, projected Armijo ascent, budget
$\lceil\log^3(n+e)\rceil$ and $1/n$ score tolerance, at both fits.
In particular the floor mass for L is $(n-1)\rho$ at unit weights
and the sum of the other case weights times $\rho$ under perturbation.

\begin{theorem}[Retained dimension and the four tuning parameters]\label{sb:dimension}
Suppose $2\le q=O(n^a)$ and
$q^{r+1/2}=o(n^{9/20})$ for the Fourier order of Lemma~\ref{hdg:fourier}.
Every contrast below is a deterministic unit raw vector, with its tangent
variance bounded away from zero uniformly in $n$.
For P, line estimation, fitted case-weight sandwich operator consistency,
normalized trace consistency and fixed-contrast studentization hold for
$a<1/18$. For L, line estimation and normalized trace consistency for the algebraic sandwich hold
for $a<2/3$. This includes the preliminary raw trace selector on L. The final covariance is suppressed if its positive-definiteness
diagnostic fails; its availability with probability tending to one is
established on the operator range below. Sandwich operator consistency and studentization of a fixed
raw contrast hold on L for $a<1/2$. The indicated line error is
$O_P(u_n)$ for preliminary and final fits, and the excess population
trace per tangent coordinate of the implemented selector is $o_P(1)$.
Studentization uses the complete influence in Corollary~\ref{v2:contrast}.
It applies to deterministic tuples and training-selected fits on these
shrinking finite grids, for deterministic raw contrasts whose tangent variance is bounded below
by a fixed positive constant (including contrasts fixed given $D_s$).
The complete finite-tuple root-$n$ raw chart expansion has an absolute
$o_P(1)$ Euclidean remainder on P if $a<1/24$, and on L if $a<1/2$.
Throughout the respective estimation ranges its remainder is $o_P(\sqrt d)$.
\end{theorem}

Population convergence on L follows by dominated convergence in
\eqref{dir:kappaibp}--\eqref{dir:poplimit}: the bounded-location
Gaussian-mixture envelope is uniform on the compact retained spectral
family, and these population limits impose no relative decay rate on the four
parameters.

For the cosine link ($r=1$), both schedules additionally require $q=o(n^{3/10})$. For example $q=\max(2,\lfloor c_0 n^{1/6}\rfloor)$, for every fixed
$c_0>0$, satisfies the L conditions for $r=0,1,2$, including the cosine
($r=1$). Under L this path also gives the absolute vector expansion at
the finite tuple. A broader L
example for estimation and the preliminary trace is
$q=\max(2,\lfloor n^{3/5}\rfloor)$ with $m_0(t)=\tanh(t)$, for which
$r=0$. Its Fourier ratio is $O(n^{-3/20})$, its local curvature cost
$L^C n^{-1/5}$, and its normalized Frobenius cost
$O_P(L^C[n^{-1/5}+n^{-3/10}])$. At the truth the
per-contrast empirical score error itself has no dimension multiplier.

\paragraph{Independent-row comparison.}
\citet[Theorems 2.1--2.2]{he2000increasing} study convex sums of
independent one-observation losses. Their bounds on centered score
increments are \(A(n,m)=o(n/\log n)\) for squared error
\(O_P(m/n)\), and \(A(n,m)=o(n/(m\log n))\) for an absolute
\(o_P(n^{-1/2})\) linearization remainder under their other conditions.
Their \(p^2\log p=o(n)\) contrast example in Corollary~2.1 also uses
its smooth-score and design conditions. In SI-KDR, perturbing one case
changes its role in the fits of other observations. The estimates below
control these derivatives directly.

\paragraph{Rates.}
In the following display a common factor $L^C$ is suppressed, and
$e_J(u)$ denotes the Hessian operator error on $\|b\|\le u$. Write $e_E$
for the $1/\sqrt n$-scaled operator error of the full matrix of case
scores at distance at most $M u_n$, for fixed $M$. Each rate is relative
to its finite-tuple population object.
\[
\begin{array}{l|c|c|c}
 &\text{P, Section~\ref{vtwo:section}}&\text{P, this section}&\text{L}\\ \hline
 \text{scalar truth score}& n^{-7/60}& n^{-7/60}&n^{-1/2}\\
 e_J(u)&q^{3/2}(n^{-1/12}+u)+q n^{-1/15}
 &q^{3/2}n^{-1/12}+q n^{-1/15}+u & \sqrt{q/n}+u\\
 e_E & n^{-1/10}+(1+q)n^{-1/30}
 & n^{-1/10}+\sqrt q\,n^{-1/30}+q n^{-2/15}&q/\sqrt n
\end{array}
\]
The Hessian rates hold on every fixed multiple of $R$; input
standardization costs at most the displayed orders. In the respective operator ranges the Gram error is
$O_P(e_E+e_E^2+\sqrt{q/n}+L^Cq/n)$, with the suppressed logarithms
restored in $e_E$. In normalized trace, the oracle Gram contributes
$O_P(n^{-1/2}+(nd)^{-1/2})$, also after any deterministic bounded raw
multiplier. The fitted sandwich adds $O_P(e_J(Mu_n)+u_n)$ in either
comparison. Thus trace consistency does not intrinsically require
operator convergence of the score covariance. Indeed for any two
$n\times d$ arrays $E,E^0$ with $\|E^0\|_F/\sqrt{nd}=O_P(1)$,
the error of their normalized Gram traces against any bounded multiplier
is $O_P(t+t^2)$ if $\|E-E^0\|_F/\sqrt{nd}=O_P(t)$. This follows by
Frobenius Cauchy--Schwarz after expanding the product. On the operator ranges the proved
operator bound implies this sufficient Frobenius bound. The
rates for all three tasks use curvature control for the inverse Hessian or
the local fit. The L case-score operator row applies for $a<1/2$. Above
that range a separate Frobenius argument below gives the normalized
trace.

\begin{proof}
\emph{Projected moments.} Let $Z$ have rows $(T_i,W_i^T)$, independent
$N_q(0,I_q)$, in population coordinates. For $p\ge2$ fixed and
$q=o(n^{2/p})$,
\begin{equation}\label{sb:moment}
 \sup_{\|v\|\le1}\|Zv\|_{p,n}=O_P(1).
\end{equation}
For fixed unit $v$, the map $z\mapsto n^{-1/p}\|z\|_p$ on
$\mathbb R^n$ is $n^{-1/p}$-Lipschitz in Euclidean distance and its
mean is at most $(E|N(0,1)|^p)^{1/p}$. Gaussian concentration gives a
tail $\exp(-c n^{2/p})$ for a fixed positive excess. A $1/2$-net of
size $5^q$ and the triangle inequality for the operator from $\ell^2$
to empirical $L^p$ give \eqref{sb:moment}. Centering the rows changes
the bound by at most a factor two. All the bounds hold simultaneously
for directions chosen from the data once the indicated supremum is taken.

\emph{P: integrated curvature.}
Use the regression jets $U_j$ and common density marks in the proof of
Proposition~\ref{vtwo:hessian}. We check where its range was used.
On the full cap, $(R+\sqrt{qL/n})\sqrt{qL}\sqrt L=o(1)$, so the
protected common-row density comparison holds. Also $q/(ng\lambda)
\asymp q n^{-1/12}=o(L^{-C})$ for every fixed $C$.
The covariance of the design and the response scale give exactly the
same free index jets $(l,l',l'')$ as in \eqref{v2:bern-reg}. Hence
the mesh (log size $O(qL)$), relative denominator comparison,
integrated linear Bernstein term and ridge bias give, simultaneously,
\[
 U_j=O_P\{L^Cq^{j/2}(g^2+\sqrt\lambda+\sqrt q\,(ng^{2j+1})^{-1/2})\},
 \qquad j=0,1,2 .
\]
This is the power-grid bound for the fitted regression jets, including
the $q$ price on the second jet. For density rows take $t=n^{-1/5}$.
The relative $j$th-derivative fluctuation after its $q^{j/2}$ factor
is $V_j=\sqrt{q/(nh^{2j+1}t)}+q/(nh^{j+1}t)$.
The good-row and bad-row errors are $L^C q(V_2+t/h^2)$, with
$V_2=O(\sqrt q\,n^{-7/30}+qn^{-3/5})$ and $t/h^2=n^{-1/15}$.
For the complete density score the largest separate requirements in
the cutoff argument are $\sqrt t/h=n^{-1/30}$,
$V_0/h=o(L^{-C})$ and $q/(nh^3t^2)\asymp q n^{-2/5}=o(L^{-C})$;
also $U_0/h^2=o_P(1)$. Thus \eqref{vtwo:densityscoreH} and
\eqref{v2:fitH} apply. Substitution in the latter gives the three
stochastic terms $L^C q^{3/2}$ times $n^{-13/60},n^{-7/60},n^{-1/12}$
and the smaller bias $L^C q(g^2+\sqrt\lambda)/h^3$.
Only the extension of the \emph{integrated ideal} row Hessian from zero
needs a different argument.

For that extension, the conditional profile at $x$ is the Gaussian
expectation of $m_v$ with mean
$(T+b^TW)/(1+\|b\|^2)$ and variance
$\|b\|^2/(1+\|b\|^2)$. Put
$V_x=1+|T|+|W^T\bar b|+|W^Tw_1|+|W^Tw_2|+|W^Tw_3|$,
where $\bar b=b/\|b\|$ (take zero at $b=0$) and the $w_i$ are
three fixed tangent unit velocities. Every mixed directional derivative
of order $j\le3$ of the discrepancy $\Delta_b(x)$ has magnitude
at most $C V_x^j$. Indeed all derivatives of the mean are linear in
the displayed projections with bounded scalar coefficients, the variance
has bounded derivatives, and its $k$th heat derivative uses at most
$2k$ bounded link derivatives. These facts also hold at zero variance
by the Gaussian semigroup from the right. With $a$ the response scale
in the integrated density formula of Proposition~\ref{vtwo:hessian},
each differentiated Gaussian component, divided by its value, costs a
polynomial in $1+|\varepsilon|$ and at most one profile jet of the
corresponding order per derivative. The shifts in its argument are
bounded by $4\|m_v\|_\infty$ uniformly in $x,x'$.
Ratios of shifted and unshifted mixture densities at the same scale
are bounded above and below by $\exp(\pm C(1+|\varepsilon|))$.
The same comparison bounds the ratios after integrating the inner row.
Hence every mixed trilinear entry of the ideal log-density row and
its order-two entry with one additional scale derivative is bounded by
$C K(\varepsilon)(1+V_x)^3$, where
$K(e)=(1+|e|)^C\exp(C(1+|e|))$. In this step integrate the at most
third powers of the corresponding $V_{x'}$ for the inner row. Every
specified set of at most five unit projections has uniformly bounded
moments under the Gaussian law, including when the directions coincide.
Derivative products of the logarithm have total directional order at
most three, and the protected denominator can only reduce each ratio.
This is a pointwise inequality for each set of directions, uniform in its
constants. Cauchy--Schwarz, $P_n K(\varepsilon)^2=O_P(1)$ and
\eqref{sb:moment} for $p=6$ now bound the supremum of its outer empirical
\emph{average} by $O_P(1)$. The same holds for the population average.
This replaces the contribution $q^{3/2}(u+L/\sqrt n)$ of
Proposition~\ref{vtwo:hessian} by $O_P(u+L/\sqrt n)$.
At the anchor the signed conditional Gaussian Gram of the ideal integrated
row costs $O_P(\sqrt{q/n}+L^Cq/n)$. Its population difference from
$j_{v,k}I_d$ costs $L^C[(g^2+\sqrt\lambda)/h^3+h^4+\rho]$, by the
last scalar comparison in Proposition~\ref{vtwo:hessian}. Both are
absorbed by the P Hessian rate in the display.

\emph{P: the fitted case scores.}
In \eqref{v2:ij-matrix}, write $d(t)$ for the complete scalar normalized
coefficient as a function of the raw index array, holding the raw
response fixed. On the local ball the standardized arrays have maxima
$O_P(L^C)$ and scales bounded above and below, including along
straight segments to the anchor. For $p=4$ we claim
\begin{equation}\label{sb:lpcoef}
 \|d(x')-d(x)\|_{p,n}\le
 C L^C(g^2h^2)^{-1}\|x'-x\|_{p,n}.
\end{equation}
Indeed, a nonnegative matrix with both row and column sums at most
$K$ has empirical $L^p$ operator norm at most $K$ by Jensen and
summation. This applies, in either orientation, to the protected
$H,P$ times any fixed polynomial in the absolute standardized pair
displacement by Lemma~\ref{op-moment-lemma}; a derivative $Q$ costs
one $g^{-1}$. If $v$ is the standardized index velocity, differentiating
a normalized $H_{ij}$ multiplies it by the pair log velocity minus its
row mean, with zero velocity at the added ridge mass. After absolute
values the envelope consists of row kernels applied to $|v_i|+|v_j|$,
and their row means, times $L^C/g$. Acting on a bounded input $z$
therefore costs $L^C\|z\|_\infty\|v\|_{p,n}/g$, also for transposes:
for a row-mean factor use the transposed absolute kernel on the vector
of row means. For $Q$ the same bound has $g^{-2}$, by differentiating
the Gaussian first derivative itself. Thus the fitted $m$-velocity
has norm $\le L^C\|v\|_{p,n}/g$, and the velocities of $A$ and the
action of $B$ or $B^T$ on a bounded input have cost at most
$L^C\|v\|_{p,n}/g^2$ times that input bound. Here
$B_{ij}=Q_{ij}(y_j-m_i)$ and the extra $m$ derivative is controlled
by the same Schur bound. The derivative of $a(r)$ costs at most
$L^C h^{-2}\|\dot r\|_{p,n}$ since its absolute Jacobian has both
sum bounds, and $\|a\|_\infty\le L^C/h$. The product rule in
$c=(B^T-\diag(A))a$ gives the claimed power $g^{-2}h^{-2}$.
Finally centering and scaling send a raw velocity to
$(v-\bar v-tP_n(tv))/s_x$, bounded in $L^p$ on these events.
The outer $s_x^{-1}$ and projection $c-tP_n(tc)$ have the same
property, including their derivatives since $t,c$ have the stated
maximum bounds ($\|c\|_\infty\le L^C/(gh)$).
Integrating gives \eqref{sb:lpcoef}.

For $\|\gamma-e_1\|\le M u_n$, \eqref{sb:moment} gives
$\|Z(\gamma-e_1)\|_{4,n}=O_P(u_n)$ uniformly. Hence
$n^{-1/2}\|\diag(d_\gamma-d_0)Z_\gamma\|_{\rm op}
\le\|d_\gamma-d_0\|_{4,n}\sup_w\|Z_\gamma w\|_{4,n}
=O_P(L^C u_n/(g^2h^2))=O_P(L^C\sqrt q\,n^{-1/30})$,
where the supremum is over unit vectors. The unmodified rotation
error is $O_P(L^C u_n/(gh))$.
For the other matrix term retain the auxiliary-response proof of
\eqref{v2:shared-ij}. In that proof the maximum local increment is
$A_n=Mu_n(\sqrt q+\sqrt L)=O(L^Cq/\sqrt n)$.
The common Frobenius envelopes remain valid since
$L^C(A_n+n^{-1/2})/g^2$ and
$L^C(A_n+n^{-1/2})/(g h^2)$ tend to zero. Specifically the
$D^TZ_\gamma$ bound is $L^C[n^{-1/10}
+\sqrt{q/n}/(gh^2)+A_n/(gh^3)]$.
Conditional Gaussian concentration is used separately at each
\emph{deterministic} mesh location for the response
$m_v(Z\gamma)+\varepsilon_v$; the union is taken before replacing its
response by the observed one via \eqref{v2:ij-y}. The final term is
$O(L^C qn^{-2/15})$. At the actual truth only,
$L^C[n^{-7/60}+\sqrt{q/n}/(gh)]$ suffices for the diagonal coefficient
error by the scalar comparison. Combining these bounds gives
the P rate for $e_E$. Projection of the outer measure,
derivatives of the response moments, both LOO masses, density centers
and incoming regression refit are all present in \eqref{v2:ij-matrix}
and \eqref{up:weight-matrices}.

\emph{L: curvature at and away from truth.}
For L every reciprocal protected bandwidth and mass is $O(L)$.
At the anchor, the coordinate-entry empirical comparison in the
proof of Theorem~\ref{hdg:sample} gives
$\max_{ij}|(\widehat J(0)-jI)_{ij}|=O_P(L^C/\sqrt n)$ and
$\max_i|d_i(0)-\zeta_{k,i}|=O_P(L^C/\sqrt n)$.
The anchor comparison uses the empirical node replacement described
below and includes differentiation of sample moments and LOO masses.
We now sharpen the conversion of entries to operator norm. Extend the
criterion to its raw index input $x\in\mathbb R^n$ with the response
fixed and its training standardization retained. At $x=T$ put
$A=-n\partial_x^2 L_{n,k}(T)$. In the linear chart
$x=T+Wb$, its negative Hessian is exactly $W^TAW/n$.
Conditionally on the scalar rows, $A$ is fixed. We claim
$\|A\|_{\rm op}=O_P(L^C)$. Indeed, along raw input velocities of
empirical $L^2$ norm one, each derivative of $(x-P_nx)/s_x$ of first
order has bounded $L^2$ norm. Each second mixed derivative has bounded
$L^1$ norm by differentiating the scalar variance, for which first and
second mixed derivatives are bounded by Cauchy--Schwarz.
Differentiate the NW ratio, residual difference, density row and outer
log in this order. At orders one and two use $L^2,L^1$ on rows and on
ordered distinct pairs; row averaging contracts these norms, products
of first derivatives obey Cauchy--Schwarz, scalar kernel derivatives
and inverse denominator powers cost $L^C$, and $\max_i |y_i|=O_P(L^C)$.
This bounds the second derivative of the full scalar objective by
$O_P(L^C)$ uniformly in the two input velocities, proving the claim.

Diagonalize $A$ conditionally; left multiplication of $W$ by the
diagonalizing rotation leaves its rows iid Gaussian. Applying
Lemma~\ref{vtwo:weightedgram} with its eigenvalues gives
$\|W^TAW/n-(\tr A/n)I\|_{\rm op}
=O_P(L^C[\sqrt{q/n}+q/n])$. The first diagonal entry costs
$O_P(L^C/\sqrt n)$ from $\tr A/n$ by the one-column version of the
same bound. The anchor entry comparison above places this entry within
$O_P(L^C/\sqrt n)$ of $j_{v,k}$. Thus the anchor Hessian operator
error is $O_P(L^C\sqrt{q/n})$. The entry proof at the anchor uses only
$(T,Y^w)$ and at most two specified Gaussian columns, so it holds for
$a\ge1/2$ without the global hypothesis $H_n(R)=o(1)$. At each of its finitely many integration nodes split
$Q_n F_n-P F_P=Q_n(F_n-F_P)+(Q_n-P)F_P$. Remove the finitely many
fixed free rows from the second term. Conditional Hoeffding on the
remaining independent truncated rows, followed by a union over
$n^{O(1)}q^2$ row and coordinate choices, costs $O_P(L^C/\sqrt n)$.
The Gaussian tail outside $C\sqrt L$ costs any prescribed inverse
power of $n$ on increasing $C$. Products and protected quotients
propagate these errors through the finite-depth derivatives, and the
LOO deletion costs $O(L^C/n)$. This is the anchor-only argument of
Theorem~\ref{hdg:sample}, with no neighborhood in its hypothesis.

We obtain curvature on the \emph{actual} cap directly. For all unit
velocities the first three derivatives of the standardized index
$t_b$ in \eqref{hdg:chart} have $L^3(P_n)$ norm $O_P(1)$, also jointly
for mixed velocities. To see this write
$t_b=(Z-\bar Z)(1,b)^T/\{(1,b)G(1,b)^T\}^{1/2}$ with $G$ the
centered sample covariance in these coordinates. The inverse square
root here is scalar. On the sample spectral event its denominator is
bounded away from zero on a fixed small neighborhood, and all first
three derivatives of its scalar multiplier are bounded. Thus the row
jets are sums of bounded multiples of at most four centered unit
projections and \eqref{sb:moment} for $p=3$ applies.

Use the uniform probability measure on ordered distinct
pairs $(i,j)$. Both marginals are uniform on the rows. Each row
average over $j\ne i$ contracts $L^p$ by conditional Jensen. For
order $l=1,2,3$ give an $l$th mixed derivative the norm $L^{3/l}$
on rows or pairs as appropriate. In a product, positive orders adding
to $l$ satisfy H\"older's inequality in exactly those exponents.
A lower-order norm also bounds every lower required integrability
exponent. All undifferentiated kernels through order three, their
derivatives as scalar functions, the denominator inverses and the
maximum absolute response are $O_P(L^C)$ on this grid.
The standardized-index undifferentiated value need not be bounded in
maximum norm to apply a Gaussian kernel derivative, which is bounded
over the whole real line. Pair differences of any jet have at most
twice its row norm. Differentiate first the numerator and denominator
of the NW ratio. The chain and product rules and the row contraction
just proved bound every positive-order derivative in its assigned norm
by $O_P(L^C)$. The reciprocal derivative is a sum of products of
lower derivatives divided by powers of a denominator $\ge\lambda$,
so the identical conclusion holds for the fitted mean, residual and
its pair differences. Repeat for the density kernel, sum over
$j\ne i$, then take its logarithm with denominator $\ge\rho$.
Averaging its third derivative now bounds the full trilinear
operator norm by $O_P(L^C)$ uniformly on the cap. The index, regression,
residual, density and outer log have all been differentiated at the
same dataset. Integration from zero gives
$e_J(u)=O_P(L^C[\sqrt{q/n}+u])$ whenever $a<2/3$, the condition
used by \eqref{sb:moment} with $p=3$. This calculation also bounds curvature
on any fixed multiple of $R$, although $R\sqrt q$ need not be small.

\emph{L: the matrix of case scores.}
In the local ball write $A_n=Mu_n(\sqrt q+\sqrt L)$ as above.
There and on the segments from the anchor all index and response
maxima are $O_P(L^C)$ and scales are bounded above and below, since
$A_n=o(1)$ for $q=O(n^a)$, $a<1/2$. At these scalar arrays every
off-diagonal entry of $H,Q,P,S,M_g,M_D$ has magnitude $\le L^C/n$,
using the global kernel bounds and the protected sums in their
denominators. In the raw index velocity $v$ each entry derivative is
bounded by $L^C\|v\|_\infty/n$: the standardized and residual
velocities have maximum $\le L^C\|v\|_\infty$, and differentiation
of an unnormalized denominator costs at most $n L^C\|v\|_\infty$.
The following closed algebra makes the dimension cost explicit. A
dense matrix of type $\mathcal M$ has maximum entries $L^C/n$ and
their velocity bound as above. Transpose, sums, products of two such
matrices, or multiplication by a diagonal of scalar entries and
velocities bounded by $L^C,L^C\|v\|_\infty$ preserve
$\mathcal M$, with changed powers. $D_ra$ is a diagonal of the
indicated type plus a matrix in $\mathcal M$. Indeed write
$a_i=\sum_j(S_{ij}-S_{ji})$ with
$S_{ij}=K'_h(r_i-r_j)/D_i$. In a residual velocity $u$,
differentiating a numerator contributes a weighted $(u_i-u_j)$
with entries $L^C/n$. The denominator variation divided by $D_i$
is another row sum of this form, multiplied by $S_{ij}$.
Row summation and transposition give only diagonals, dense matrices
or their products. One further index derivative keeps these types
by the same entry and row-sum bounds, now using $K_h'''$.
It follows term by term in \eqref{up:weight-matrices} that
$D_m,D_A,D_a,D_c$ are in $\mathcal M$: in particular the term with
$I-M_g$ has a dense left factor. The input training-moment derivatives
are rank-one matrices $u v^T/n$ of this type, with the index dilation
and response translation/dilation coefficients and their one extra
index derivatives bounded by $L^C$ and $L^C\|v_{\rm raw}\|_\infty$.
To check this directly apply one or two scalar-array derivatives to
the normalized kernel entries. Their derivatives of order $j=1,2$
have maximum $L^C/n$ times the product of the input maximum velocities,
because the unnormalized denominators are at least $(n-1)\lambda$
or $(n-1)\rho$, and each differentiated denominator divided by $n$
is bounded by the same product times $L^C$. Row sums preserve
the $L^C$ bound. This also handles differentiating the directions of
dilation themselves. Finally $d=s_x^{-1}\{c-tP_n(tc)\}$ adds precisely
such rank terms and bounded products to the full weight Jacobian $D$.
Hence $D$ is in $\mathcal M$. In particular $\|D_0\|_F\le L^C$,
$\|D_0\|_{\rm op}\le L^C$, and
$\|D_\gamma-D_0\|_{\rm op}\le L^C A_n$ uniformly. The same
entry computation for the nondifferentiated scalar coefficient gives
$\|d_\gamma-d_0\|_\infty\le L^C A_n$.

At the anchor $D_0,d_0$ use only scalar data. The conditional
Gaussian contraction with the contrast net used to prove
\eqref{v2:shared-ij}, now needed only once at that anchor, gives
$n^{-1/2}\|D_0^TW\|_{\rm op}=O_P(L^C(1+\sqrt q)/\sqrt n)$.
The change in the tangent design costs $O_P(\sqrt n\,u_n)$
in operator norm, and its own operator norm is $O_P(\sqrt n)$.
The changes in $d,D$ thus add at most $O_P(L^C[A_n+u_n])$
after scaling. The scalar anchor estimation error contributes
$O_P(L^C/\sqrt n)$ by the maximum bound. Centering the leading
array costs $O_P(u_n)$ since the $\zeta_{k,i}$ are scalar.
This proves $e_E=O_P(L^Cq/\sqrt n)$ on the local ball.

\emph{L: normalized trace when the operator comparison is unnecessary.}
For $q\le n^{1/4}$ the preceding local operator argument suffices.
Suppose now $q>n^{1/4}$ and $q=O(n^a)$ for $a<2/3$.
The same Gaussian net proof as for \eqref{sb:moment}, using an excess
$C(1+\sqrt q/n^{1/p})$ for each fixed vector, gives
$\sup_{\|v\|=1}\|Zv\|_{p,n}=O_P(1+\sqrt q/n^{1/p})$ for fixed $p$.
Hence, on every $M u_n$ ball the raw and standardized index
arrays and their straight-line interpolations with the anchor have
bounded empirical fourth norms. In fact the raw index change costs
$O_P(u_n[1+\sqrt q/n^{1/4}])$, tending to zero since $q/n^{3/4}\to0$;
the scales on the segments are bounded above and below by the sample
spectral event. Their maximum is not required. The response is fixed.

On these scalar arrays $\|d_\gamma-d_0\|_{2,n}\le
O_P(L^C)\|x_\gamma-T\|_{2,n}=O_P(L^C u_n)$ uniformly.
For this bound, all normalized kernel entries and their scalar argument
derivatives before normalization cost $L^C/n$ after division by the
protected sum. In a velocity $v$, applying a differentiated row or
transposed row to a bounded input costs $L^C\|v\|_{2,n}$ by the
two marginal sums, as in \eqref{sb:lpcoef}. This bounds the changes
in $m,a,c$; their unperturbed maxima cost $L^C$, even if an index
has a large value, since Gaussian kernel derivatives are globally
bounded at their respective scales. The derivative of raw
standardization as a map $L^2\to L^2$ is bounded, and in
$d=s_x^{-1}(c-tP_n(tc))$ every undifferentiated $t$ costs its
empirical second norm, equal to one. The Gaussian sample maximum
$\max_i\|Z_i\|/\sqrt d=O_P(1)$ here; multiplying a row-scalar error
by $Z_{\gamma,i}$ and dividing the resulting Frobenius norm by
$\sqrt{nd}$ therefore costs only its empirical second norm.
Rotating $Z_\gamma$ back to $W$ costs $O_P(L^C u_n)$ by its
operator estimate and $\max_i|d_{0,i}|=O_P(L^C)$.

For the other term of \eqref{v2:ij-matrix}, at fixed scalar arrays
the matrix identities \eqref{up:weight-matrices} give
$\|D_c\|_F\le L^C$. For these fixed-array weight derivatives this
follows from the dense-entry argument above, independently of
$\max_i|t_i|$: every Gaussian kernel derivative is globally bounded,
and all residual, response and coefficient maxima cost $L^C$.
Training standardization gives additional rank-one terms $uv^T/n$.
For the case-weight derivative of $c$, an index translation vanishes;
the index-dilation coefficient is bounded by $L^C$ in maximum norm
even for large index entries. Indeed each differentiated regression
kernel for this dilation contains
$(t_i-t_j)K_g^{(j)}(t_i-t_j)$, uniformly bounded by $L^C$.
This bounds the resulting fitted-mean, density-score and incoming
refit derivatives by the two marginal sums. The response translation
and dilation likewise give coefficient maximum $O_P(L^C)$. The other
factors in these ranks are $t_j^2-1$, $y_j$, $y_j^2-1$, divided by $n$,
and have bounded empirical second norms. In the derivative of $d$ the
term involving $D_wc$ is multiplied by $(I-tt^T/n)/s_x$. The remaining
ranks from $s_x^{-1}$, $t$ and the direct outer weight in $P_w(t_wc_w)$
have factors of empirical second norm $O_P(L^C)$, by
$\|t\|_{4,n}=O_P(1)$ and $\|c\|_\infty=O_P(L^C)$.
Thus the full $D$ satisfies $\|D\|_F=O_P(L^C)$ uniformly here.
It follows that
$\|D^T Z_\gamma\|_F/\sqrt{nd}\le
\|D\|_F\|Z_\gamma\|_{\rm op}/\sqrt{nd}=O_P(L^C/\sqrt d)$.
Combining the scalar anchor comparison and oracle centering with
these two bounds yields
\[
 \sup_{\|\gamma-e_1\|\le M u_n}
 \frac{\|E_{k,\gamma}-E_k^0\|_F}{\sqrt{nd}}
 =O_P\{L^C(u_n+n^{-1/2}+d^{-1/2})\}=o_P(1).
\]
The Frobenius trace inequality of the Rates paragraph applies also to the
bounded random matched inverse curvature multiplier. Replacing this
multiplier by its population version costs $o_P(1)$ since the oracle
score Gram has bounded normalized trace. The weighted normalized oracle
trace estimate of Lemma~\ref{vtwo:weightedgram} then yields normalized
sandwich trace consistency throughout this larger L range. Positive
curvature makes every tuple eligible in the trace selector with
probability tending to one. In the larger range the algebraic sandwich
is defined whenever the curvature is invertible, and preliminary trace
selection uses finite positive-semidefinite $B$; this proves
consistency of the algebraic final trace and of the implemented
preliminary selector.

\emph{From curvature to the computed iterates, selected contrasts and trace.}
On either grid $\inf j>0$ eventually by Corollary~\ref{hdg:uniform-gain}.
The exact truth score is $W^T d_0/n$ by Lemma~\ref{v2:truth-score}.
Its norm is $O_P(u_n)$ from conditional Gaussianity and
$\|d_0\|_{2,n}=O_P(1)$. Also, if $D_s^*$ denotes the scalar truth
score error in the Rates paragraph (distinct from the first-stage data),
its root-$n$ norm error is $O_P(\sqrt d D_s^*)$.
The Hessian error on each fixed multiple of the true cap tends to zero:
for P the restrictive stochastic fit term is
$q^{3/2}n^{-1/12}$, and for L it is $L^C\sqrt{q/n}$ at the anchor and $L^C R$ across the cap.
The score throughout this cap is $O_P(u_n+R)$ by integration.
Thus the gradient times the second derivative of the raw
chart change also vanishes. The raw map and its inverse have uniformly
bounded first and second derivatives and nonsingular tangent
Jacobians by the spectral condition. Lemmas~\ref{hdg:fourier}
and \ref{hdg:ascent} now apply without using consistency of the fitted
sandwich. The Fourier error $O_P(q^{r+1/2}/\sqrt n)=o_P(R)$ puts
the truth in the computed cap, and strict curvature puts the unique
maximizer at $O_P(u_n)$ from truth. The projection is nonexpansive
and the first Armijo step is accepted eventually, giving the
$O_P(n^{-1})$ distance of each final iterate from its maximizer.
This applies simultaneously to all tuples in both fits. Evaluate
the appropriate operator or Frobenius case error and $e_J$ on $M u_n$ only after this localization, taking $M$
large for arbitrary fixed probability. Lemma~\ref{vtwo:weightedgram}
at the truth and Gram factorization then yield the sandwich
and weighted normalized trace bounds of the Rates paragraph, including the $O_P(u_n)$
coordinate and gradient connection at the fitted frame.

For each \emph{deterministic} tuple use the proof variable $b_k^\circ$
equal to the unique interior maximizer in a deterministic round
truth-centered ball of radius $K R$, and zero if there is no such
unique maximum. In standardized coordinates it rotates with $W$,
because \eqref{hdg:chart} depends on $b,W$ only through their
projections and sample moments. With probability tending to one it
is the computed cap's exact maximum: choose $K$ fixed large enough
to enclose the latter cap on the spectral and pilot event, then use
the $O_P(u_n)$ location and interiority. The error
$e=\sqrt n b_k^\circ-W^T\zeta_k/(j_k\sqrt n)$ therefore has norm
$O_P(\sqrt d[D_s^*+e_J(Mu_n)])$ by integrated curvature.
It rotates with $W$ conditionally on $(T,Y^w,D_s)$, without any
restriction to a success or diagnostics event. For any fixed unit
$w$, symmetry gives $E[(w^Te)^2\boldsymbol1_{\|e\|\le t}] =
d^{-1}E[\|e\|^2\boldsymbol1_{\|e\|\le t}]$, conditionally on
that scalar information. Norm truncation and Markov therefore give
$w^Te=O_P(D_s^*+e_J(Mu_n))$, and likewise
$w^Tb_k^\circ=O_P(n^{-1/2})$. The raw chart equals
$Db/(1+v_0^Tb)$ with bounded deterministic $D,v_0$;
its denominator differs from one by $O_P(n^{-1/2})$. Passing from
the equivariant proof variable to the potentially nonequivariant
finite iterate costs $O_P(n^{-1/2})$ on the root-$n$ scale. An
ambient coordinate contrast of the sign-aligned unit direction also
has a normalization remainder $O_P(q/\sqrt n)=o_P(1)$ on the stated
studentization ranges. Its tangent variance has the stated uniform lower bound.

All these bounds are first simultaneous for a finite list of
deterministic tuples. The complete leading contrasts are conditionally
Gaussian, with scalar variance multiplier $P_n\zeta_k^2/E\zeta_k^2
\to_P1$. The common $L^2$ limit of $\zeta_k/j_k$ on the shrinking
grids (Corollary~\ref{hdg:uniform-gain}) makes each fixed contrast
differ from its efficient leading contrast by $o_P(1)$. Taking the
finite maximum permits a training selector, including one based on
nonspherical raw traces or data-dependent tie breaking. The empirical trace minimization
inequality bounds its excess normalized trace by twice the maximum
normalized trace estimation error. On the larger L range the curvature
is positive and the fitted case traces finite with probability tending
to one. Positivity of the full fitted case covariance and the resulting
inference diagnostics use the narrower operator range. For a positive fixed grid the same
reasoning yields a selected Gaussian limit at a consistently chosen
tuple and a selected-index expansion at a positive tie.

For an absolute Euclidean root-$n$ expansion multiply the local
Hessian and scalar errors by $\sqrt q$. On P the new binding condition
is $L^Cq^2n^{-1/12}=o(1)$; also $L^C q^{3/2}n^{-1/15}=o(1)$
and $L^C\sqrt q n^{-7/60}=o(1)$ for $a<1/24$.
On L the condition is $L^C q/\sqrt n=o(1)$, using the sharpened
conditional Hessian bound $L^C(\sqrt{q/n}+u_n)$.
The raw fractional denominator costs only $O_P(\sqrt{q/n})$ in the
vector expansion at the equivariant maximizer. For each direction the rate $O_P(u_n)$ uses only the cap bounds and
truth score, before any covariance calculation.

Finally, the exact direction pivot before restricting its domain
requires its stated design ranks ($n>d+2$) and candidate-only
measurability. The truth coefficient is nonzero with probability
tending to one by its scalar norm bound and $E\zeta_k^2\ge j_k^2/i(v)$.
Within the respective sandwich operator ranges and at growing $d$,
conditional Gaussian concentration for the norm of
the leading vector with a bounded-spectrum raw covariance, and the
$o_P(\sqrt d)$ remainder above, give the dimension-dependent
trace radius and $O_P(\sqrt{d/n})$ diameter as in \eqref{hdg:radius}.
For bounded $d$, the fixed-dimensional raw radius
$\log(n+e)/\sqrt n$ places the truth in the ball with probability tending
to one by the direction rate. For a computed confidence region the local
domain and numerical false-exclusion terms of \eqref{hdg:coverage-bound}
apply.
\end{proof}

\subsection{Coordinate power on the two shrinking grids}
\label{sbvar:wide-grid}
\begin{corollary}\label{sbvar:wide-power}
Use the Gaussian model, coordinate procedure and raw local path of
Theorem~\ref{sbvar:power}, replacing its sufficient dimensional and grid
restrictions as follows. For the deleted-column model assume the
spectral and Fourier-basin conditions of
Theorem~\ref{sb:dimension}. On grid P the shifts in \eqref{sbvar:shifts}
hold for $q=O(n^a)$, $a<1/18$; on grid L they hold for $a<2/3$.
On both grids the SI-KDR shift has the common shrinking-grid limit, including
the restricted trace selector. The feasible mean coefficient of
Section~\ref{sbvar:section} has its oracle shift throughout the indicated
P range and on L for $a<1/3$. When $q=2$, the reduced line is known and its
trace cannot rank tuples. The shrinking-grid statements then require a
prespecified shrinking singleton tuple on the stated schedule; the fixed
tuple of Section~\ref{sbvar:section} has its finite-tuple shift. The test
uses a finite restricted point and its score without a positive-definite
case-covariance gate, and its finite-sample calibration is
Proposition~\ref{sbvar:pivot}.
\end{corollary}
\begin{proof}
Work under the null with the notation of Lemma~\ref{sbvar:transfer}.
For one reduced predictor there is no nuisance angle. Otherwise the
actual restricted point lies in the $U=M\sqrt{q/n}$ ball with probability
arbitrarily close to one for sufficiently large fixed $M$, by
Theorem~\ref{sb:dimension}. The truth coefficient $d_0$ has empirical
$L^2$ error $O_P(\log^C n\,n^{-7/60})$ on P and
$O_P(\log^C n/\sqrt n)$ on L about its own finite-tuple $f_k$.
On P apply \eqref{sb:lpcoef} and the uniform fourth projected moment
\eqref{sb:moment} to the actual raw index increment. They give, on
this ball,
\[
 \|d_\gamma-d_0\|_{2,n}\le O_P(\log^C n)U/(g^2h^2)
 =O_P(\log^C n\sqrt q\,n^{-1/30})=o_P(1).
\]
On L the $L^2$ scalar-coefficient bound in the normalized-trace proof
of Theorem~\ref{sb:dimension} gives $O_P(\log^C n\,U)$; that bound
also applies when $q\le n^{1/4}$. It uses bounded protected row and
transposed-row derivatives, the bounded $L^2$ operator of index
standardization, and $\|t\|_{2,n}=1$ in $d=s_x^{-1}\{c-tP_n(tc)\}$.
In both regimes, projecting $f_k$ onto the reduced
nuisance columns costs $O_P(\sqrt{q/n})$, as proved in
Lemma~\ref{sbvar:transfer}. Thus \eqref{sbvar:transfer-eq} holds
simultaneously, $n-q-1\to\infty$, and the conditional Gaussian and LAN
argument in Theorem~\ref{sbvar:power} applies. Its mean oracle requires
only the same Gaussian projection estimate. Both schedules shrink, so
Corollary~\ref{hdg:uniform-gain} gives the common limiting ratio.

For the feasible mean, keep $g_M=n^{-1/6}$ and $\lambda_M=n^{-3/4}$.
The truth-index regression value and query derivative have the scalar
rates of \eqref{eq:gaussian-primitives}; $ng_M\lambda_M=n^{1/12}$.
To transfer them to the fitted index, differentiate the row sum
$A_i=\sum_l Q_{il}(y_l-\widehat M_i)$ and the fitted mean along a
standardized index velocity $v_1$. The two endpoint absolute sums for
the Gaussian weights in Lemma~\ref{op-moment-lemma} bound the empirical
$L^2$ action in both orientations. A derivative of $H$ costs one inverse
$g_M$, and a derivative of $Q$ costs two; the remaining factor
$y_l-\widehat M_i$ costs only a logarithmic maximum. Differentiating
$\widehat M_i$ inside the sum costs the first inverse $g_M$ in addition
to the undifferentiated $Q$. Hence
\[
 \|\dot A\|_{2,n}\le O_P(\log^C n)\|v_1\|_{2,n}/g_M^2,
 \qquad \|\dot{\widehat M}\|_{2,n}\le
 O_P(\log^C n)\|v_1\|_{2,n}/g_M.
\]
The sample-standardization derivative is bounded on $L^2$ along the
local raw-index segment, whose empirical $L^2$ increment is $O_P(U)$.
The resulting derivative error $O_P(\log^C n\sqrt q\,n^{-1/6})$
vanishes for $a<1/3$. The anchor scale and affine response changes cost
$O_P(\log^C n/(\sqrt n g_M^2))$ overall by the same derivative bound
and response linearity. Multiplying by the bounded-derivative population
link and logarithmic response maximum as in Section~\ref{sbvar:section}
therefore yields $\|\Pi_j(c^M-a_M f_M)\|_{2,n}=o_P(1)$ with its stated
positive $a_M$.
\end{proof}

\FloatBarrier
\section{Moderate- and high-dimensional comparison}
\label{hd:protocol}\label{sec:hd-protocol}

\paragraph*{Design.}
For support $S_0=\{1,\ldots,s\}$ put $\beta_{0j}=s^{-1/2}$ on $S_0$ and zero elsewhere, with $s=P$ in D07 and D08. With $R=I_P$ or $R_{jk}=r^{|j-k|}$, generate
\[
 X\sim N_P(0,\Sigma),\qquad
 \Sigma=\frac{v}{\beta_0^TR\beta_0}R,\qquad
 Y=m_0(\beta_0^TX)+\varepsilon,
\]
so that the index has variance $v$. The links are $\sin(2t)$ and $\cos(2t)$ and the errors are those of Section~\ref{sec:sim-details}. Direct settings use $n=600$, the rescaled AR(.5) design and $v=1.5$. Screened settings split $600$ observations into $n_s=200$ and $n_e=400$, retain $q=10$ or $20$ coordinates by the distance-correlation statistic of Section~\ref{sec:hd-screen-proofs}, and use $v=0.25$, since in the independent design with equal nonzero coefficients the regression of $Y$ on an active $X_j$ is damped by the factor $\exp\{-2v(s-1)/s\}$.

\paragraph*{Methods.}
SI-KDR, mean MAVE, csMAVE and Ma--Zhu are fitted to the same data, in the screened settings to the retained coordinates of $D_e$ with all constants based on $n_e$. Ma--Zhu is implemented as in Section~\ref{sec:sim-details}. SI-KDR starts from the mean-MAVE direction, and the initialization cap used for optimization and for localizing the score-inversion set is centered at this direction; the other steps of Section~\ref{sec:algorithm} are unchanged. The expansion of Section~\ref*{main-sec:rates} requires only a $\sqrt n$-consistent start. In the screened settings the reweighted SSIRvRP-BIC of \citet{zhang2026sparse} is fitted to all $600$ observations by its Algorithm~2 with $d=1$, $k=20$, $l'=50$, $(A_1,B_1)=(900,300)$, $(A_2,B_2)=(600,200)$ and ten equal-size response slices; BIC chooses the support size from $\{2,3,5,10,20,30\}$. An oracle SI-KDR fit to the true support of $D_e$ is reported for reference.

\paragraph*{Evaluation.}
Each setting has $200$ point-estimation samples and $300$ independent inference samples. Risk is the mean line loss, with loss two for a numerical failure. After screening we report the screening failure rate and the coverage of the score-inversion set among the samples whose screen retained the support. The area of this set, relative to the cap of radius $\log(n_e+e)/\sqrt{n_e}$, is estimated on the first $50$ inference samples by importance sampling with $8192$ draws, as in Section~\ref{sec:sim-details} with $d=q-1$.

\paragraph*{Results.}
SI-KDR has the lowest risk among the methods other than the oracle in all fourteen settings (Tables~\ref{tab:hd-point-direct} and~\ref{tab:hd-point-screen}). In the direct settings its risk is $35$--$89\%$ below that of mean MAVE, $28$--$90\%$ below csMAVE and $27$--$70\%$ below Ma--Zhu. After screening, its risk is $22$--$95\%$ lower than that of mean MAVE, csMAVE, Ma--Zhu and SSIRvRP-BIC, including the settings with $P=1000$, where the number of predictors exceeds the sample size. Whenever the screen retains the support, the score-inversion set covers the truth in $94$--$96\%$ of the samples (Table~\ref{tab:hd-coverage}).

\begingroup
\footnotesize
\setlength{\tabcolsep}{3pt}
\begin{longtable}{lrrrlrrrr}
\caption{Direct settings: mean line risk ($\times10^3$) with Monte Carlo standard error, 200 samples. The lowest risk is in bold.}\label{tab:hd-point-direct}\\
\toprule
Setting & $P$ & $q$ & $s$ & Link/error & SI-KDR & mean MAVE & csMAVE & Ma--Zhu \\
\midrule\endfirsthead
\toprule
Setting & $P$ & $q$ & $s$ & Link/error & SI-KDR & mean MAVE & csMAVE & Ma--Zhu \\
\midrule\endhead
\bottomrule\endlastfoot
D01 & 10 & 10 & 3 & sin/A & \textbf{2.58 (0.11)} & 24.3 (1.1) & 24.6 (7.6) & 3.83 (0.14) \\
D02 & 10 & 10 & 3 & sin/M & \textbf{3.81 (0.41)} & 22.8 (1.0) & 39.5 (9.1) & 11.1 (0.4) \\
D03 & 10 & 10 & 3 & cos/A & \textbf{2.53 (0.10)} & 20.9 (0.8) & 15.4 (0.6) & 3.80 (0.13) \\
D04 & 10 & 10 & 3 & cos/M & \textbf{3.18 (0.27)} & 21.1 (0.8) & 28.0 (1.1) & 10.7 (0.4) \\
D05 & 20 & 20 & 3 & cos/A & \textbf{63.6 (22.1)} & 102 (20) & 88.4 (20.8) & 99.5 (28.2) \\
D06 & 20 & 20 & 3 & cos/M & \textbf{66.6 (21.4)} & 103 (21) & 254 (36) & 91.8 (23.6) \\
D07 & 10 & 10 & 10 & sin/A & \textbf{3.58 (0.14)} & 32.5 (1.1) & 32.8 (9.3) & 5.90 (0.19) \\
D08 & 20 & 20 & 20 & cos/M & \textbf{32.6 (10.3)} & 93.3 (12.1) & 289 (38) & 66.7 (17.0) \\
\end{longtable}
\endgroup

\begingroup
\scriptsize
\setlength{\tabcolsep}{1.6pt}
\begin{longtable}{lllrrrrrrr}
\caption{Screened settings: retention (\%) and mean line risk ($\times10^3$) with Monte Carlo standard error, 200 samples. Design gives the link, the error and $R$. The lowest risk among the methods other than the oracle is in bold.}\label{tab:hd-point-screen}\\
\toprule
 & & & & \multicolumn{4}{c}{Screened} & & \\
\cmidrule(lr){5-8}
Setting & $P/q/s$ & Design & Ret. & SI-KDR & mean MAVE & csMAVE & Ma--Zhu & SSIRvRP & Oracle \\
\midrule\endfirsthead
\toprule
Setting & $P/q/s$ & Design & Ret. & SI-KDR & mean MAVE & csMAVE & Ma--Zhu & SSIRvRP & Oracle \\
\midrule\endhead
\bottomrule\endlastfoot
H01 & 200/10/3 & sin/A, I & 98.0 & \textbf{14.0 (4.0)} & 51.7 (4.3) & 24.6 (4.2) & 60.2 (18.7) & 67.6 (2.9) & 1.15 (0.08) \\
H02 & 200/10/3 & sin/M, I & 92.5 & \textbf{36.8 (8.0)} & 74.2 (7.6) & 92.5 (11.3) & 200 (34) & 180 (5) & 1.11 (0.11) \\
H03 & 1000/10/3 & sin/A, I & 85.0 & \textbf{73.5 (12.5)} & 112 (13) & 94.7 (15.2) & 218 (37) & 142 (4) & 1.27 (0.11) \\
H04 & 1000/10/3 & sin/M, I & 78.0 & \textbf{103 (13)} & 137 (13) & 154 (14) & 346 (44) & 325 (5) & 0.92 (0.07) \\
H05 & 1000/10/3 & sin/A, AR(.2) & 99.5 & \textbf{9.73 (1.86)} & 61.5 (3.0) & 25.5 (2.3) & 94.2 (25.2) & 173 (5) & 1.96 (0.14) \\
H06 & 1000/20/3 & sin/A, I & 94.5 & \textbf{43.7 (11.8)} & 123 (11) & 69.7 (8.5) & 806 (57) & 138 (4) & 1.35 (0.10) \\
\end{longtable}
\endgroup

\begingroup
\footnotesize
\setlength{\tabcolsep}{5pt}
\begin{longtable}{lrrr}
\caption{Score-inversion sets after screening, 300 samples: percentage of screening failures, coverage (\%) among samples whose screen retained the support, and mean relative area on the first 50 samples.}\label{tab:hd-coverage}\\
\toprule
Setting & Screening failure & Coverage given retention & Area \\
\midrule\endfirsthead
\toprule
Setting & Screening failure & Coverage given retention & Area \\
\midrule\endhead
\bottomrule\endlastfoot
H01 & 1.7 & 95.9 & 0.0324 \\
H02 & 6.7 & 95.7 & 0.0820 \\
H03 & 12.7 & 95.4 & 0.1316 \\
H04 & 23.0 & 96.1 & 0.2181 \\
H05 & 2.3 & 94.9 & 0.0510 \\
H06 & 5.0 & 94.4 & 0.0547 \\
\end{longtable}
\endgroup

\FloatBarrier
\section{Further experiments after screening}\label{sb:newexperiments}
\subsection{Ambient dimension}
\label{sec:x23-confirm}
Each data set has $N=600$ or $1200$ observations, split into $n_s=2N/3$ for screening and $n_e=N/3$ for estimation. The first two coordinates have equal positive coefficients, their index has variance $0.25$, and the errors follow the standardized asymmetric mixture of Section~\ref{sec:sim-details}. At each $N$ the three values of $P$ use successive prefixes of the same Gaussian arrays. The biased distance-correlation statistic of \citet{li2012distance}, computed by ordinary double centering with the sorting algorithm of \citet{x23Huo2016,x23Huo2017}, retains $q=s=2$ coordinates. SI-KDR, mean MAVE and csMAVE are fitted to the same retained estimation data. SI-KDR uses the four-matrix Fourier start, the power bandwidth grid and full case-weight trace selection; mean MAVE and csMAVE use the computational core of the \textsf{MAVE} package, version 1.3.12. A failed direction incurs loss two. A further cell with $N=600$ and $P=1000$ uses the sine link and the AR(.5) design. Its two active coordinates have index correlation $r_*=\sqrt{(1+.5)/2}$, followed by $r_*(.5)^{j-2}$ for $j\ge3$, so \eqref{sb:dc-attenuate} gives a strict population rank gap.

All screens retained both active coordinates in all 600 samples of every cell. For each $N$, the three values of $P$ selected identical coordinates in every sample, so the fitted directions and losses coincide across $P$ (Table~\ref{x23:risk}).

\begin{table}[htbp]\centering\small
\caption{Mean line losses ($\times10^3$) and within-sample SI-KDR-minus-comparator differences on the same scale, 600 samples per row, with Monte Carlo standard errors in parentheses. The sine link is used throughout.}\label{x23:risk}
\setlength{\tabcolsep}{4pt}
\begin{tabular}{rrlrrrrr}\toprule
$N$ & $P$ & $R$ & SI-KDR & mean MAVE & csMAVE & vs.\ mean MAVE & vs.\ csMAVE\\\midrule
600 & 1000 & I & 1.94 (0.16) & 9.46 (0.55) & 3.78 (0.24) & $-7.52$ (0.54) & $-1.83$ (0.27) \\
600 & 5000 & I & 1.94 (0.16) & 9.46 (0.55) & 3.78 (0.24) & $-7.52$ (0.54) & $-1.83$ (0.27) \\
600 & 10000 & I & 1.94 (0.16) & 9.46 (0.55) & 3.78 (0.24) & $-7.52$ (0.54) & $-1.83$ (0.27) \\
600 & 1000 & AR(.5) & 5.45 (0.46) & 26.95 (1.54) & 11.29 (0.73) & $-21.51$ (1.50) & $-5.84$ (0.77) \\
1200 & 1000 & I & 0.66 (0.04) & 4.51 (0.24) & 1.64 (0.09) & $-3.85$ (0.23) & $-0.98$ (0.08) \\
1200 & 5000 & I & 0.66 (0.04) & 4.51 (0.24) & 1.64 (0.09) & $-3.85$ (0.23) & $-0.98$ (0.08) \\
1200 & 10000 & I & 0.66 (0.04) & 4.51 (0.24) & 1.64 (0.09) & $-3.85$ (0.23) & $-0.98$ (0.08) \\
\bottomrule\end{tabular}
\end{table}

Table~\ref{x23:time} gives mean wall times in seconds per single-worker fit and per complete screen, measured on AMD EPYC 9V74 CPUs with one BLAS thread per worker. Screening includes input conversion, response preprocessing, all $P$ features and the final ordering. Fit times exclude compilation and screening.

\begin{table}[htbp]\centering\small
\caption{Mean wall time in seconds.}\label{x23:time}
\setlength{\tabcolsep}{4pt}
\begin{tabular}{rrlrrrr}\toprule
$N$ & $P$ & $R$ & Screen & SI-KDR & mean MAVE & csMAVE\\\midrule
600 & 1000 & I & 0.095 & 0.062 & 0.035 & 0.409 \\
600 & 5000 & I & 0.475 & 0.062 & 0.035 & 0.409 \\
600 & 10000 & I & 0.949 & 0.062 & 0.035 & 0.409 \\
600 & 1000 & AR(.5) & 0.097 & 0.063 & 0.035 & 0.410 \\
1200 & 1000 & I & 0.207 & 0.223 & 0.066 & 0.732 \\
1200 & 5000 & I & 1.028 & 0.223 & 0.066 & 0.732 \\
1200 & 10000 & I & 2.046 & 0.223 & 0.066 & 0.732 \\
\bottomrule\end{tabular}
\end{table}

\subsection{The cosine link after screening}\label{sb:cosine-new}
\begingroup
\renewcommand{\theparagraph}{\thesubsection.\arabic{paragraph}}
\paragraph{Population information in the cosine design.}
With $v=1/4$ the standardized index is $T=2\beta_0^TX$. The raw
error has variance one and $m_v(t)=(\cos t-e^{-1/2})/\sigma_Y$,
where $\sigma_Y^2=1+(1-e^{-1})^2/2$. Hence
$A(v)=2(1-e^{-2})$, $G=12.7764$ and $i(v)=vA(v)I_\varepsilon=5.5237$.
Writing $\alpha=(1+g^2)^{-1}$ and $q_g=\phi_{\sqrt{1+g^2}}$,
the ridge profile is
\[
 M(t)=\frac{q_g(t)}{q_g(t)+\lambda}
       \frac{e^{-\alpha g^2/2}\cos(\alpha t)-e^{-1/2}}{\sigma_Y}.
\]
Conditional Gaussian integration of $R=Y^w-M(T)$ gives the two
density contributions to $a$ in \eqref{inf:fulla}. With
$H(t)=\phi(t)\E[a(R)\mid T=t]/(q_g(t)+\lambda)$ the incoming
regression term is $-y(K'_g*H)(t)+(K'_g*(HM))(t)$. We used the
projected score variance $b=\E\{(\kappa-T\E[T\kappa])^2\}$ and
$j$ in \eqref{inf:curvature}. In centered raw response units
$(h,\rho)$ becomes $(\sigma_Y h,\rho/\sigma_Y)$ and the same
$j,b$ result. Since $\Omega_S=I/4$, the raw trace per tangent
coordinate is $b/j^2=\Lambda/i(v)$.

At $n_e=600$ the smallest $j$ among the 20 final power-grid tuples
is $2.582$, and $\Lambda$ ranges from $1.022$ to $1.882$, far below
$G$. The raw population trace is minimized at the smallest
$(g,h)=(0.1437,0.0646)$, with $(j,b)=(5.420,5.432)$; the second-best
trace lies $0.15\%$ above the minimum. Thus every fixed final tuple
has a leading covariance below the conditional-mean bound in the
correctly retained model. The population four-matrix Fourier start
identifies the true line, with eigenvalues $0.008419$ and zero in our
normalization.

\paragraph{Screening and estimation.}
With $P=1000$, $q=s=2$, independent $X_j\sim N(0,0.25)$, the cosine
link and the standardized asymmetric mixture error, the population
squared distance correlation is $0.00908838$ for each active variable
and zero for every inactive one. We computed the active value by
conditioning on the Gaussian index and integrating folded-normal
moments, and checked it by integrating the squared difference of the
joint and product marginal distribution functions. Each of 500
independent data sets has $n_s=2400$ screening and $n_e=600$
estimation observations. Screened SI-KDR, mean MAVE and csMAVE use the
same selected pair and estimation rows. The support-only SI-KDR fit uses
those rows, the same Fourier start and the same trace selector; it
coincides with the screened SI-KDR fit whenever the selected pair is the
true support.

The screen retained both coordinates in 498 of the 500 data sets and
one coordinate in the other two. On the retained data sets SI-KDR had
lower loss than both competitors and coincided with the support-only
fit (Table~\ref{x3:cos-risk}); the two screening misses account for
the entire difference between screened and support-only SI-KDR.

\begin{table}[htbp]\centering\small
\caption{Cosine link after screening $P=1000$ predictors. Mean embedded
line losses or within-sample differences ($\times10^3$), with Monte
Carlo standard errors in parentheses. Numerical failures count as
loss two; none occurred.}\label{x3:cos-risk}
\begin{tabular}{lrr}\toprule
& All 500 & Retained, 498 \\ \midrule
Screened SI-KDR & 3.417 (1.907) & 0.740 (0.059) \\
Mean MAVE & 6.972 (2.034) & 4.156 (0.285) \\
csMAVE & 4.405 (2.010) & 1.594 (0.113) \\
Support-only SI-KDR & 0.738 (0.059) & 0.740 (0.059) \\
\midrule
SI-KDR minus mean MAVE & $-3.555$ (0.306) & $-3.416$ (0.278) \\
SI-KDR minus csMAVE & $-0.988$ (0.157) & $-0.854$ (0.112) \\
\bottomrule\end{tabular}
\end{table}

\endgroup
\subsection{Two folds: pooled point estimation}
\label{sb:x4-results}
Each of 500 independent data sets has $P=1000$, $N=600$ and
$S_0=\{1,2\}$. The raw predictor law is $N(0,0.25I_P)$, the true
active coefficients are $1/\sqrt2$, and the response is the sine link
$\sin(2X^T\beta_0)$ plus the standardized asymmetric mixture error of
Section~\ref{sec:sim-details}. The two blocks contain 200 and 400 rows,
and each biased distance-correlation screen keeps two coordinates. The
single split screens on the 200-row block and fits the 400 rows; the
reverse fit exchanges the roles of the blocks. Both oracles fit only
the two true coordinates. All fits use the four-matrix Fourier start,
the full double-refitted criterion and capped finite ascent, with
preliminary tuple $(.6,.15,.001,.005)$ and common final tuple
$(.45,.15,.001,.005)$ in the order $(g,h,\lambda,\rho)$, so that the
two fits share a finite-smoothing influence
(Proposition~\ref{sb:pool}). The pool gives its 200-row fit weight
$1/3$ and its 400-row fit weight $2/3$.

\begin{table}[ht]
\centering
\caption{Single split, weighted pool and support oracles. Mean squared
projective loss ($\times10^3$) with Monte Carlo standard error in
parentheses. The last column uses the same jointly recovered pairs for
every method. A numerical point failure receives loss two.}
\label{sb:x4-risk}
\begin{tabular}{lcc}
\toprule Method and fitting size & All 500 & Joint recovery, 476\\
\midrule
Reverse screened, 200 & 1.64 (0.12) & 1.59 (0.12) \\
Single split, 400 & 29.08 (5.67) & 0.66 (0.05) \\
Weighted pool, 600 & 15.42 (2.98) & 0.48 (0.03) \\
Support oracle, 400 & 0.67 (0.05) & 0.66 (0.05) \\
Support oracle, 600 & 0.43 (0.03) & 0.42 (0.03) \\
\bottomrule
\end{tabular}
\end{table}

The 400-row screen supplying the 200-row fit and the 200-row screen
supplying the 400-row fit recovered $S_0$ in 500 and 476 data sets,
respectively. The paired pool-minus-single risk difference
($\times10^3$) was $-13.66$ (2.69) overall and $-0.18$ (0.04) on the
476 jointly recovered data sets, where the ratio of mean risks was
0.722 (0.042), close to the ratio $400/600$ for a common influence.

\subsection{Coordinate tests at and near a zero coordinate}\label{sbvar:x5}
The study uses 512 independent split samples at each of four local
alternatives, with $n_s=200$, $n_e=400$, $P=1000$ and $q=4$. Within each
replication all alternatives and all methods use the same predictors,
error-component labels and Gaussian innovations. The design is
$X\sim N(0,I_P/4)$, $m_0(u)=\sin(2u)$, and the error is the standardized
asymmetric mixture of Section~\ref{sec:sim-details}. The unit-norm path
in Theorem~\ref{sbvar:power} starts at $(e_1+e_2)/\sqrt2$ and tests the
third coordinate. The ordinary biased distance screen keeps its top four
coordinates. The forced screen, specified for all samples, keeps
coordinate three plus the top three remaining coordinates. Every tested
raw column is removed before whitening, initialization and tuple
selection. The feasible mean statistic uses the restricted SI-KDR direction
and the shrinking mean fit of Section~\ref{sbvar:section}, and the mean
oracle uses the true null link and direction. The SIR comparator uses
Cook's statistic with five balanced response slices and 999 conditional
draws. A missed target or a numerical test failure counts as a
nonrejection. Both multiple-testing procedures correct within the four
retained hypotheses.

\begin{table}[!htbp]\centering\small
\caption{Two-sided rejection rates at level 0.05 for the forced target,
among the $m$ samples whose screen retained the active coordinates, with
Monte Carlo standard error $\{\hat p(1-\hat p)/(m-1)\}^{1/2}$ below
0.023.}\label{sbvar:xtarget}
\begin{tabular}{c r rrrr}\toprule
$\omega$ & $m$ & SI-KDR & Mean & Oracle & SIR\\\midrule
0 & 505 & 0.053 & 0.036 & 0.034 & 0.040\\
0.6 & 504 & 0.198 & 0.056 & 0.052 & 0.042\\
1.2 & 504 & 0.665 & 0.127 & 0.117 & 0.069\\
1.8 & 500 & 0.904 & 0.246 & 0.258 & 0.102\\
\bottomrule\end{tabular}\end{table}

\begin{table}[!htbp]\centering\scriptsize
\caption{Error rates (percent) over all 512 independent samples per row, with between-sample Monte Carlo standard errors in parentheses. Single-null error averages the within-sample proportions. FWER and FDR use Holm and BY, respectively.}\label{sbvar:xfamily}
\setlength{\tabcolsep}{3pt}\begin{tabular}{c l rrr rrr rrr}\toprule
&&\multicolumn{3}{c}{Single-null error}&\multicolumn{3}{c}{Holm FWER}&\multicolumn{3}{c}{BY FDR}\\
\cmidrule(lr){3-5}\cmidrule(lr){6-8}\cmidrule(lr){9-11}
$\omega$ & Screen & SI-KDR & Mean & SIR & SI-KDR & Mean & SIR & SI-KDR & Mean & SIR\\\midrule
0 & Top four & 4.6 (0.6) & 5.6 (0.7) & 6.2 (0.8) & 3.5 (0.8) & 4.5 (0.9) & 7.0 (1.1) & 0.8 (0.2) & 1.2 (0.3) & 1.7 (0.3)\\
0 & Forced & 5.0 (0.7) & 4.9 (0.7) & 5.8 (0.7) & 4.1 (0.9) & 4.1 (0.9) & 6.2 (1.1) & 1.0 (0.3) & 1.1 (0.3) & 1.9 (0.4)\\
0.6 & Top four & 4.9 (0.7) & 5.6 (0.7) & 6.5 (0.8) & 3.1 (0.8) & 4.3 (0.9) & 7.4 (1.2) & 0.8 (0.2) & 1.0 (0.3) & 2.0 (0.4)\\
0.6 & Forced & 4.4 (0.9) & 6.4 (1.1) & 6.4 (1.1) & 1.6 (0.5) & 3.1 (0.8) & 4.3 (0.9) & 0.3 (0.1) & 0.8 (0.2) & 1.1 (0.3)\\
1.2 & Top four & 4.8 (0.7) & 5.4 (0.7) & 6.8 (0.8) & 4.3 (0.9) & 4.5 (0.9) & 7.2 (1.1) & 1.0 (0.2) & 1.2 (0.3) & 1.8 (0.3)\\
1.2 & Forced & 4.1 (0.9) & 6.4 (1.1) & 8.0 (1.2) & 2.9 (0.7) & 2.7 (0.7) & 4.1 (0.9) & 0.3 (0.1) & 0.8 (0.2) & 1.1 (0.3)\\
1.8 & Top four & 4.9 (0.7) & 5.7 (0.7) & 6.4 (0.8) & 5.3 (1.0) & 4.9 (1.0) & 7.8 (1.2) & 1.3 (0.3) & 1.4 (0.3) & 1.9 (0.4)\\
1.8 & Forced & 3.6 (0.8) & 7.0 (1.1) & 7.8 (1.2) & 3.1 (0.8) & 3.5 (0.8) & 4.5 (0.9) & 0.3 (0.1) & 0.9 (0.2) & 1.0 (0.2)\\
\bottomrule\end{tabular}
\end{table}

\begin{table}[!htbp]\centering\small
\caption{Forced-target mean signed statistics with Monte Carlo standard errors. Paired columns use samples that are correctly retained and finite both under the null and at the alternative. The ratios compare mean SI-KDR and mean-oracle signed increments, with paired delta-method standard errors.}\label{sbvar:xshift}
\setlength{\tabcolsep}{5pt}\begin{tabular}{c rrr rrr}\toprule
& \multicolumn{3}{c}{Mean signed $t$}&\multicolumn{3}{c}{Paired with the null}\\
\cmidrule(lr){2-4}\cmidrule(lr){5-7}
$\omega$&SI-KDR&Mean&Oracle& $m$ & Shift ratio & Squared ratio\\\midrule
0 & -0.06 (0.04) & -0.03 (0.04) & -0.06 (0.04) & -- & -- & --\\
0.6 & 1.13 (0.05) & 0.40 (0.04) & 0.39 (0.04) & 504 & 2.65 (0.02) & 7.02 (0.09)\\
1.2 & 2.31 (0.05) & 0.83 (0.04) & 0.84 (0.04) & 504 & 2.64 (0.02) & 6.95 (0.08)\\
1.8 & 3.46 (0.05) & 1.25 (0.04) & 1.28 (0.04) & 500 & 2.62 (0.01) & 6.86 (0.08)\\
\bottomrule\end{tabular}
\end{table}

At every nonzero amplitude the mean signed SI-KDR statistic is about 2.6 times that of the mean oracle (Table~\ref{sbvar:xshift}), and its rejection rate is 3.5 to 5.7 times that of the oracle mean test (Table~\ref{sbvar:xtarget}).

\subsection{Growing retained dimension on two schedules}\label{sb:x6-results}
Each of 300 independent data sets has $P=600$, $n_s=180$ and
$n_e=300$. The first two coordinates have equal nonzero coefficients.
The design is independent Gaussian with coordinate variance $1/4$, the
link is $\sin(2t)$, and the errors follow the standardized asymmetric
mixture of Section~\ref{sec:sim-details}. On each screening split the
biased distance-correlation ranking gives nested models with
$q=5,10,20,40$, all fitted to the same estimation rows. Both schedules
use the four empirical Fourier matrices, full trace selection and the
same finite ascent. The constants of the L schedule
$c/\{1+\log(n_e+e)\}$ equal those of the P grid at the reference size
600. At $n_e=300$ the smallest final $(g,h)$ are $(.1613,.0677)$ on P
and $(.1585,.0713)$ on L, and the respective $(\lambda,\rho)$ are
$(.000191,.003333)$ and $(.000125,.001838)$.

\begin{center}\small
\begin{tabular}{lrrrr}\toprule
Grid & $q$ & Retained & $10^3$ risk (SE) & $n_e$ risk/$(q-1)$ (SE)\\\midrule
P & 5 & 298 & 8.465 (3.274) & 0.635 (0.246) \\
P & 10 & 300 & 8.445 (0.259) & 0.282 (0.009) \\
P & 20 & 300 & 17.499 (0.408) & 0.276 (0.006) \\
P & 40 & 300 & 40.919 (1.320) & 0.315 (0.010) \\
\midrule
L & 5 & 298 & 8.415 (3.251) & 0.631 (0.244) \\
L & 10 & 300 & 8.424 (0.264) & 0.281 (0.009) \\
L & 20 & 300 & 17.414 (0.415) & 0.275 (0.007) \\
L & 40 & 300 & 41.656 (1.603) & 0.320 (0.012) \\
\bottomrule\end{tabular}\end{center}
Retention counts are out of 300, and every fit returned a valid
direction. Risk is the unconditional raw embedded line loss, with
standard errors from independent replicates. On the 298 correct
screens at $q=5$ the scaled loss is 0.291 (0.013) on P and 0.289
(0.013) on L. The paired differences in overall risk (L minus P, in
units $10^{-3}$) are $-0.049$ (0.045), $-0.021$ (0.060), $-0.085$
(0.146) and 0.737 (1.019) for $q=5,10,20,40$, respectively.

On L we also studentized the fixed raw contrast $(e_1-e_2)/\sqrt2$,
whose true value is zero. The numerator is $\sqrt{n_e}$ times the
contrast of the sign-aligned fitted unit direction, and the denominator
is the square root of the final raw case-weight covariance projected
onto this contrast. Eligible fits have a correctly retained model,
positive fitted raw curvature and a positive-definite case covariance.
\begin{center}\small
\begin{tabular}{rrrrrr}\toprule
$q$ & Eligible & Mean $z$ & SD $z$ & Accepted/eligible & Accepted/300\\\midrule
5 & 298 & -0.029 & 0.975 & 0.950 (0.013) & 0.943 (0.013) \\
10 & 300 & -0.023 & 0.969 & 0.953 (0.012) & 0.953 (0.012) \\
20 & 300 & 0.002 & 0.941 & 0.960 (0.011) & 0.960 (0.011) \\
40 & 300 & 0.059 & 1.098 & 0.953 (0.012) & 0.953 (0.012) \\
\bottomrule\end{tabular}\end{center}
Acceptance is at the nominal 95\% level, and ineligible outcomes count
as rejections in the last column.

\FloatBarrier
  \bibliography{references}